\documentclass[a4paper,fleqn]{cas-sc}
\usepackage{amsmath}
\usepackage{amssymb}
\usepackage{graphicx}
\usepackage{mathrsfs}
\usepackage{amsfonts}
\usepackage{amsthm}
\usepackage{color}
\usepackage{colordvi}
\usepackage{titletoc}
\usepackage{physics}
\usepackage{comment}

\titlecontents{section}
              [1.1cm]
              { \normalsize}%
              {\contentslabel{2em}}%
              {}%
              {\titlerule*[0.3pc]{$\cdot$}\normalsize\contentspage\hspace*{0cm}}%
\titlecontents{subsection}
              [2cm]
              {\normalsize}%
              {\contentslabel{2em}}%
              {}%
              {\titlerule*[0.3pc]{$\cdot$}\normalsize\contentspage\hspace*{0cm}}%
\titlecontents{subsubsection}
              [2.9cm]
              {\normalsize}%
              {\contentslabel{2.5em}}%
              {}%
              {\titlerule*[0.3pc]{$\cdot$}\normalsize\contentspage\hspace*{0cm}}%

\usepackage{lipsum}
\usepackage{gensymb}
\usepackage[numbers,sort&compress]{natbib}
\usepackage[numbers]{natbib} 
\def\tsc#1{\csdef{#1}{\textsc{\lowercase{#1}}\xspace}}
\tsc{WGM}
\tsc{QE}
\tsc{EP}
\tsc{PMS}
\tsc{BEC}
\tsc{DE}

\renewcommand\({\begin{equation}}	
\renewcommand\){\end{equation}}

\renewcommand\[{\begin{eqnarray}}	
\renewcommand\]{\end{eqnarray}}
\newcommand{\al}[1]{\begin{aligned}#1\end{aligned}}

\UseRawInputEncoding
\begin{document}
\let\WriteBookmarks\relax
\def\floatpagepagefraction{1}
\def\textpagefraction{.001}
\shorttitle{Tao Yu, Ji Zou, Bowen Zeng,  J. W. Rao, and Ke Xia}
\shortauthors{Tao Yu et~al.}

\title [mode = title]{Non-Hermitian Topological Magnonics}                      

\author[a]{Tao Yu}[orcid=https://orcid.org/0000-0001-7020-2204]
\cormark[1]
\ead{taoyuphy@hust.edu.cn}
\address[a]{School of Physics, Huazhong University of Science and Technology, Wuhan 430074, China}

\author[b]{Ji Zou}
[orcid=https://orcid.org/0000-0003-0828-1153]
\address[b]{Department of Physics, University of Basel, Klingelbergstrasse 82, 4056 Basel, Switzerland}

\author[a,c]{Bowen Zeng}
[orcid=https://orcid.org/0000-0003-1084-3813]
\cormark[1]
\ead{zengbowen@csust.edu.cn}
\address[c]{School of Physics and Electronic Science, Changsha University of Science and Technology, Changsha 410114, China}

\author[d]{J. W.~Rao}[orcid=https://orcid.org/0000-0002-7199-1302]

\address[d]{School of Physical Science and Technology, ShanghaiTech University, Shanghai 201210, China}

\author[e]{Ke Xia}[orcid=https://orcid.org/0000-0002-0246-3944]

\address[e]{School of Physics, Southeast University, Jiangsu 211189, China}

\cortext[cor1]{Corresponding author at: School of Physics, Huazhong University of Science and Technology, Wuhan 430074, China.}

\begin{abstract}
Dissipation in mechanics, optics, acoustics, and electronic circuits is nowadays recognized to be not always detrimental but can be exploited to achieve non-Hermitian topological phases or properties with functionalities for potential device applications, ranging from sensors with unprecedented sensitivity, energy funneling, wave isolators, non-reciprocal signal amplification, to dissipation induced phase transition. As elementary excitations of ordered magnetic moments that exist in various magnetic materials, magnons are the information carriers in magnonic devices with low-energy consumption for reprogrammable logic, non-reciprocal communication, and non-volatile memory functionalities. Non-Hermitian topological magnonics deals with the engineering of dissipation and/or gain for non-Hermitian topological phases or properties in magnets that are not achievable in the conventional Hermitian scenario, with associated functionalities cross-fertilized with their electronic, acoustic, optic, and mechanic counterparts, such as giant enhancement of magnonic frequency combs, magnon amplification, (quantum) sensing of the magnetic field with unprecedented sensitivity, magnon accumulation, and perfect absorption of microwaves. In this review article, we address the unified approach in constructing magnonic non-Hermitian Hamiltonian, introduce the basic non-Hermitian topological physics, and provide a comprehensive overview of the recent theoretical and experimental progress towards achieving distinct non-Hermitian topological phases or properties in magnonic devices, including exceptional points,  exceptional nodal phases, non-Hermitian magnonic SSH model, and non-Hermitian skin effect. We emphasize the non-Hermitian Hamiltonian approach based on the Lindbladian or self-energy of the magnonic subsystem but address the physics beyond it as well, such as the crucial quantum jump effect in the quantum regime and non-Markovian dynamics. We provide a perspective for future opportunities and challenges before concluding this article.
	
\end{abstract}


\maketitle

\tableofcontents

\section{Introduction}

\subsection{Non-Hermitian topological phenomena}

In a closed quantum system the dynamics are governed by the unitary time evolution under a Hermitian Hamiltonian. However, the interaction between a quantum subsystem and its environment is usually inevitable. The ``bath'' can either extract energy and information from the subsystem or supply them to it, thereby breaking the Hermiticity of the subsystem~\cite{bender2005introduction,moiseyev2011non,heiss2012physics,konotop2016nonlinear,el2018non,miri2019exceptional,ozdemir2019parity,ashida2020non,bergholtz2021exceptional,yuan2022quantum,rameshti2022cavity,ding2022non,hurst2022perspective,lin2023topological,chang2018colloquium,okuma2023non}. On one hand, in nature the systems of interest often exhibit a loss or leakage of energy or information to the bath, which results in their non-Hermiticity. 
One textbook example might be the radiation-induced damping of electric and magnetic dipoles in an open electromagnetic environment or the ``radiation damping'' in classical electrodynamics, which in the quantum language contributes to a complex frequency with an imaginary component referred to as the ``loss''. In magnetism, the damping of magnetization fluctuation is governed by the magnon-electron or magnon-phonon interaction but in the waveguide, the radiation damping in magnetic insulators via the microwaves can dominate as well due to the Purcell effect~\cite{lodahl2015interfacing}. On the other hand, by external experimental interventions, one can counteract the loss, which also leads to the non-Hermiticity of the systems of interest.
Efforts are made by engineering the devices to achieve the amplification channels to the subsystem, namely the ``gain'' that acts as the inverse process of the loss \cite{el2007theory,kottos2010broken,lin2011unidirectional,regensburger2012parity,brody2012mixed,lee2015macroscopic,liu2016metrology,shi2016accessing,zhang2017observation,hodaei2017enhanced,sakhdari2019experimental,xiao2019enhanced,tang2020exceptional,wang2021coherent,stegmaier2021topological}, e.g., the amplifier in LRC circuits and the escapement system of the simple pendulum. Several quantum subsystems or objects may interact with the same bath, which may mediate an effective coupling with distinguished features, such as the coherent coupling, the dissipative coupling \cite{liu2016metrology,asenjo2017exponential,chang2018colloquium,harder2018level,grigoryan2019cavity,zhang2019theory,zhang2020subradiant,yu2020chiral,yu2020magnon_accumulation,harder2021coherent,rao2021interferometric,rameshti2022cavity,zou2022prb,zeng2023radiation}, the non-reciprocal coupling \cite{wang2018observation,brandenbourger2019non,lee2019hybrid,fruchart2021non,yu2020nonreciprocal,zhang2020unidirectional,yamamoto2020non,gou2020tunable,wang2021nonreciprocal}, as well as the chiral coupling without equal backaction between two objects~\cite{lodahl2017chiral,yu2019chiral,yu2019chiral_2,yu2020magnon,yu2020chiral,yu2020magnon_accumulation,yu2022giant,yu2023chirality,zeng2023radiation}. Thereby engineering the dissipation and/or gain promises opportunities to realize novel functionalities beyond those in the Hermitian scenario.

An effective non-Hermitian Hamiltonian is a convenient and widely exploited theoretical instrument for describing the dynamics of quantum subsystem   \cite{breuer2002theory,breuer2016colloquium,chang2018colloquium,yuan2022quantum,rameshti2022cavity}, but several approximations are often presumed when integrating out the degree of freedom of the bath, such as the Born-Markov approximation and disregarding the probabilistic quantum jump effect \cite{daley2014quantum,Plenio1998rmp,PhysRevA.36.5543,PhysRevLett.70.548}. In this review article, we shall address the conditions of these approximations in the context of bosonic dynamics. With the non-Hermitian Hamiltonian, the eigenvalues are generally complex with the imaginary components, which denote the reciprocal lifetime of states. Similar to the Hermitian scenario, the symmetries associated with the non-Hermitian Hamiltonian are crucial in determining the frequency spectra and the eigenmodes. For example, when there exists the parity $\mathcal{P}$ and time-reversal $\mathcal{T}$ symmetries, namely the $\mathcal{PT}$-symmetry,  the eigenvalues of a non-Hermitian Hamiltonian become real as long as the $\mathcal{PT}$-symmetry is respected by the wavefunction, implying an infinite long lifetime although the quantum subsystem is open, which can be achieved in the experiments by several strategies such as balancing the gain and loss \cite{guo2009observation,ruter2010observation,makris2010pt,hodaei2014parity,bendix2009exponentially,joglekar2010robust,zhu2014p,jing2014pt,fleury2015invisible,liu2016metrology,sakhdari2019experimental,mao2023non}. This ever motivated the generalization of the Hermiticity restriction to quantum mechanics two decades ago \cite{bender2007making,bender1998real,bender1999pt,bender2002complex}. Recent years witness tremendous advancements in the non-Hermitian topological phases or properties in the optics \cite{feng2014single,hodaei2014parity,chen2017exceptional,weidemann2020topological}, phononics \cite{christensen2016parity,liu2022experimental,del2022non}, mechanics \cite{kane2014topological,susstrunk2015observation,wang2018observation,trainiti2019time}, electronic circuits \cite{albert2015topological,ningyuan2015time,chang2018colloquium,hofmann2019chiral,liu2016metrology}, and hybrid systems \cite{yuan2022quantum,rameshti2022cavity,rao2021interferometric} such as optomechanics \cite{aspelmeyer2014cavity,xu2016topological,liu2016metrology,el2018non}, optomagnonics \cite{xu2020floquet,yang2020Unconventional}, and light-atom interaction in a cavity \cite{ashida2020non,li2022non,rameshti2022cavity}, in which the effective non-Hermitian Hamiltonian is demonstrated to successfully characterize many exotic non-Hermitian topological states or properties achieved via engineering the dissipation and/or gain.

Exotic properties exist for the non-Hermitian Hamiltonian. Among them, the coalescence of eigenvalues and eigenvectors of a non-Hermitian Hamiltonian is referred to as the exceptional points (EPs) in the parameter space \cite{heiss2012physics,wiersig2014enhancing,zhen2015spawning,liu2016metrology,xu2016topological,chen2017exceptional,hodaei2017enhanced,zhou2018observation,miri2019exceptional,xiao2019enhanced,sakhdari2019experimental,PhysRevLett.123.237202,ozdemir2019parity,wang2021coherent,grigoryan2022pseudo}. The lowest order EPs is of two-fold degeneracy. For the higher ranking $N\times N$ non-Hermitian Hamiltonian matrix with $N>2$, the  $N$-fold degeneracies of the eigenvalues and the corresponding coalesce of $N$ eigenvectors into a  single one lead to higher-order EPs \cite{hodaei2017enhanced,wang2019arbitrary,zhang2019higher,zhong2020hierarchical,wang2021enhanced,yu2020higher,mandal2021symmetry}. The sensitivity of the system is significantly improved via the EPs \cite{wiersig2014enhancing,liu2016metrology,hodaei2017enhanced,lai2019observation,cao2019exceptional,wang2021enhanced}. Tuning parameters across the EPs brings intriguing physical phenomena and potential applications such as unidirectional invisibility \cite{lin2011unidirectional,feng2013experimental,fruchart2021non,castaldi2013p}, a stable entangled state \cite{yuan2020steady,zou2022prb,lee2022exceptional}, single-mode lasing \cite{peng2014loss,brandstetter2014reversing,wong2016lasing,liu2017integrated}, coherent perfect absorption \cite{sweeney2019perfectly,wang2021coherent,rao2021interferometric} and enhancement of spontaneous emission \cite{lin2016enhanced,ferrier2022unveiling}. 
In the two and even higher dimensional parameter space, the EPs may become lines or surfaces, e.g., links or knots~\cite{heiss2012physics,kozii2017non,zhou2018observation,carlstrom2019knotted}.

For the non-Hermitian band structure with wave vector acting as the parameter, the exceptional points, lines, or surfaces are energy degeneracies in the reciprocal space that define the non-Hermitian nodal phases \cite{yang2019non,bergholtz2021exceptional,carlstrom2019knotted,liu2022experimental}. 
Typically, in the two-dimensional reciprocal space parametrized by wave vectors ${\bf k}=(k_x,k_y)$,  the second-order EPs are contained in a square root $E({\bf k})=\sqrt{{ C}({\bf k})}$, where ${ C}({\bf k})$ is a complex number. The isolated EPs governed by ${\rm Re}[C({\bf k})]={\rm Im}[C({\bf k})]=0$ appear in pairs (refer to Sec.~\ref{Nodal_phase}). $E({\bf k})=\{\sqrt{|{ C}({\bf k})|}e^{i\arg\left[C({\bf k})\right]/2},\sqrt{|{ C}({\bf k})|}e^{i(\arg\left[C({\bf k})\right]+2\pi)/2}\}$ are multi-valued and is thereby not analytic, where the polar angle $-\pi< \arg\left[C({\bf k})\right]\le \pi$. When $\arg\left[C({\bf k})\right]\rightarrow \arg\left[C({\bf k})\right]+2\pi$ the two branches swap between a two-sheeted Riemann surface. Such EPs in two dimensions form the branch points and the directional curve with ${\rm Re}C({\bf k})=0$ connecting a pair of EPs corresponds to a branch cut. It is called the non-Hermitian Fermi arc \cite{kozii2017non,zhou2018observation,bergholtz2021exceptional} in the non-Hermitian nodal phase, resembling the surface Fermi arc that connects two surface projected Weyl points in the three-dimensional Weyl semimetal \cite{yan2017topological,armitage2018weyl}, but the bulk Fermi arc no longer corresponds to a surface state. Moreover, in the two-dimensional reciprocal space by choosing a path around one EP, the winding number is a half-integer, which implies the nontrivial topological robustness of these EPs to the perturbations \cite{shen2018topological,yang2019non,carlstrom2019knotted}.

Even in the absence of exceptional degeneracies such as EPs in the wave vector space, there also exist nontrivial non-Hermitian topological states or properties in the band theory such as topological edge states \cite{yao2018edge,kawabata2018anomalous,ghatak2020observation,parto2018edge,comaron2020non,flebus2020non,gunnink2022nonlinear}, similar to their Hermitian counterpart. Intrinsic symmetries are crucial to identify the gapped and gapless non-Hermitian topological phases
\cite{shen2018topological,zhou2019periodic,2019Symmetry}.  Alternatively, there exist nontrivial categories in the non-Hermitian system that no longer corresponds to their Hermitian counterpart, i.e., the non-Hermitian ``skin effect'' with a macroscopic number of bulk eigenstates piling up at one boundary \cite{xiong2018,shen2018topological,kunst2018biorthogonal,yao2018edge,yokomizo2019non,lee2019hybrid,okuma2020,yang2020non,zhang2020correspondence,weidemann2020topological,li2020topological,kawabata2020higher,zhang2022review,liang2022dynamic,deng2022non,yu2022giant,zhang2022universal,zeng2023radiation,chen2023topological}. 
These states are very sensitive to the boundary at which the energy leaks out, a merit of open systems. In such an open system, the energy spectra with the open boundary condition are no longer approximated by those with the periodic boundary condition, very different from the Hermitian band theory. Here we focus on one dimension. Compared to the extended Bloch state taking the plane-wave form with wave vector $k\in[-\pi,\pi]$ under periodic boundary conditions, these edge-localized states under open boundary conditions can be described by a similar Bloch state but with complex wave vector $\kappa$, corresponding to amplification  (Im$\kappa<0$)  or attenuation (Im$\kappa>0$) during propagation in the positive direction. Such a  distribution of $\kappa$ on the complex plane is referred to as the generalized Brillouin zone \cite{yao2018edge,yao2018non,yokomizo2019non,yang2020non}. The associated topologically nontrivial state cannot be described by the conventional topological invariant defined by the Bloch wavefunction under the periodic boundary condition, i.e., the conventional bulk-boundary correspondence fails in the non-Hermitian scenario~\cite{lee2016anomalous,xiong2018,kunst2018biorthogonal}. But these states are topologically characterized  by the winding number of the \textit{energy spectra} under the periodic boundary condition \cite{lee2016anomalous,gong2018topological,shen2018topological,yao2018edge}. Short-range asymmetric or chiral coupling becomes popular in the study of the non-Hermitian skin effect
\cite{hatano1996localization,yao2018edge,yao2018non,bergholtz2021exceptional,kunst2018biorthogonal,lee2019hybrid,ashida2020non,xiong2018,zhu2020photonic,zhang2021acoustic,borgnia2020non,brody2013biorthogonal}.  This effect has been successfully observed in systems with relatively short-range asymmetric hopping \cite{xiao2020non}, such as light funnel in photonic lattice \cite{weidemann2020topological}, non-local response in electric circuit \cite{helbig2020generalized}, and enhanced sensitivity in classical and quantum metamaterials \cite{ghatak2020observation,budich2020non}.

We review and elucidate the unified properties of various non-Hermitian topological properties or phases in Sec.~\ref{Exceptional_topology}, i.e., unconventional topological characterizations or properties that are distinguished from those in the Hermitian scenario~\cite{bergholtz2021exceptional}. This review article focuses on the non-Hermitian topological phenomena of magnons, i.e., elementary excitations in ordered magnets. Magnons are information carriers with low-energy consumption that hold potential applications with such as reprogrammable logic \cite{lenk2011building,grundler2016nanomagnonics}, non-reciprocal transmission  \cite{jamali2013spin,yu2023chirality}, and non-volatile
memory \cite{chumak2015magnon,zhang2015magnon,demidov2017magnetization,brataas2020spin,baumgaertl2023reversal} functionalities. In comparison with other information carriers, particularly electrons used in CMOS technology, magnons hold the potential to realize similar functionalities but with much lower energy consumption in information processing and quantum technologies \cite{yuan2022quantum,rameshti2022cavity,yu2023chirality}.   Basic magnonic structures are micro- and nano-waveguides \cite{duerr2012enhanced,yu2019chiral,wang2020steering} and
heterostructures combined with magnetic and nonmagnetic materials \cite{myers1999current,yang2018antiferromagnetism,Liu2019Observation,ruckriegel2020long}. Such structures exploiting magnons 
operate at a frequency range that lies between gigahertz (GHz) and terahertz (THz) \cite{stancil2009spin,baltz2018antiferromagnetic,kim2022ferrimagnetic}. This is compatible with conventional CMOS technology that is
restricted by the GHz-frequency range \cite{nguyen2004cmos,laskin2009nanoscale}. On the other hand, magnons hold unique chirality \cite{yu2020magnon_accumulation,hurst2022perspective,yu2023chirality}, can
propagate with little damping in dielectric ferro-, antiferro-, and ferrimagnetic materials over distances of
micrometers \cite{liu2018long,cornelissen2015long,liu2018long,ruckriegel2020long}, and can interact strongly with magnons, electrons \cite{bender2012electronic,auerbach2012interacting,flebus2020non}, phonons \cite{weiler2011elastically,li2018magnon,casals2020generation,yu2020nonreciprocal,berk2019strongly,zhang2020unidirectional,PhysRevX.12.011060,cai2023acoustic,zeng2023radiation}, photons \cite{soykal2010strong,zhang2014strongly,tabuchi2014hybridizing,zhang2016cavity,bai2017cavity,PhysRevLett.123.237202,yu2020magnon_accumulation,yu2020nonreciprocal,yuan2020steady,rameshti2022cavity}, Cooper-pair supercurrents \cite{tabuchi2015coherent,johnsen2021magnon,li2022coherent,barman20212021,yu2022efficient,borst2023observation,zhou2023gating}, and even spin qubits \cite{trifunovic2013long,neuman2020nanomagnonic,fukami2021opportunities}. These bring various control dimensions and efficient energy transduction ways to design magnon modes and control the magnetic damping or gain in magnonic devices. For example, its intrinsic damping can
be easily influenced, e.g., by parametric pumping~\cite{bracher2017parallel,demokritov2006bose} and/or by the spin transfer torque~\cite{slavin2005current,apalkov2005slonczewski,galda2016parity,galda2018parity}.  The progress in the study of non-Hermitian topological physics in terms of magnons \cite{harder2021coherent,rameshti2022cavity,hurst2022perspective} is providing ways to engineer the dissipation or gain for useful functionalities in future spintronic and magnonic devices, which will fertilize the other research fields as well.

\subsection{Non-Hermitian topological magnonics}

Similar to their electronic, acoustic, optic, and mechanic counterparts, the nontrivial role of dissipation should be emphasized in magnonics that may hold functionalities fertilized with and even beyond the other systems.  For example, via interaction with other quasiparticles, hybrid magnonics holds several remarkable
advantages over other systems, such as high frequency and dissipation tunability \cite{PhysRevLett.123.237202,yang2020Unconventional,nair2021enhanced,yuan2022quantum}, rich nonlinearity \cite{tabuchi2015coherent,lachance2017resolving,lachance2019hybrid}, enhanced coupling strength \cite{tabuchi2014hybridizing,rameshti2022cavity,rao2021interferometric}, intrinsic chirality \cite{yu2020magnon_accumulation,hurst2022perspective,yu2023chirality}, and non-reciprocity \cite{osada2016cavity,wang2019nonreciprocity,zhang2020unidirectional}.

\begin{figure}[!htp]
	\centering
	\includegraphics[width=13.8cm]{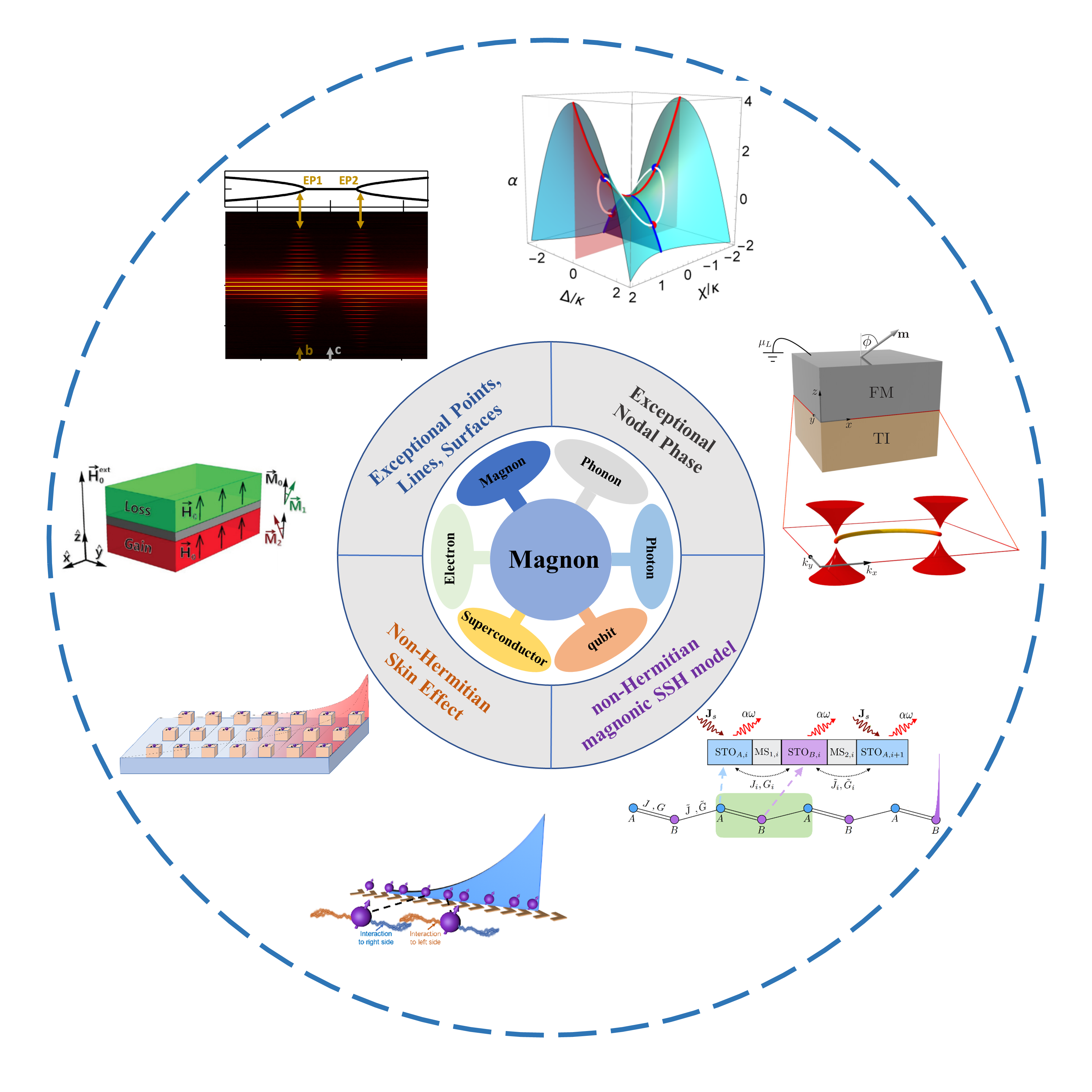}
	\caption{Engineering the dissipation of magnonic systems on demand for non-Hermitian topological phenomena. The interaction between the magnonic subsystem  with the other degrees of freedom such as electrons, photons, phonons, and the other magnons contributes to the magnon self-energy or Lindbladian that is generally complex, resulting in a non-Hermitian Hamiltonian. Via engineering the magnon self-energy, one may obtain the non-Hermitian Hamiltonian on demand towards the non-Hermitian topological properties or states such as the EPs, exceptional nodal phases, non-Hermitian magnonic SSH model, and non-Hermitian skin effect in magnonic devices for different functionalities.}
	\label{fig:introduction}
\end{figure}

Topological magnon states have been well proposed in the magnetic systems
\cite{mook2014edge,fransson2016magnon,kim2016realization,mcclarty2017topological,boyko2018evolution,li2021topological,park2021hinge,wang2021topological,mcclarty2022topological,zhuo2023topological}. The exceptional topological properties or phases of non-Hermitian magnonic system, however, belong to a different category \cite{flebus2020non,yu2022giant,deng2022non,hurst2022perspective}. 
Such non-Hermitian topological properties or states in magnonic devices can be driven by the coupling between the magnons and the other degrees of freedom such as the electrons \cite{bergholtz2019non,flebus2020non,deng2022non,deng2023exceptional}, photons \cite{zhang2017observation,wen2019non,yao2019microscopic,yuan2020steady,rameshti2022cavity,wang2022pt}, phonons \cite{huai2019enhanced,lu2021exceptional,hurst2022perspective}, and the other magnons  \cite{yang2018antiferromagnetism,mcclarty2019non,Liu2019Observation,yu2020higher,PT_bilayer,yu2022giant}, which contributes to the magnon self-energy or Lindbladian that is generally not Hermitian. Here we are allowed to emphasize the new progress in handling the unavoidable dissipation of magnons towards useful functionality in magnonics and magnetism since inspired by the non-Hermitian topology, contemporary new breakthroughs have been achieved in these fields. The \textit{Perspective}~\cite{hurst2022perspective} focused on the non-Hermitian Hamiltonian of magnons with EPs and non-Hermitian skin effect. In this review article, we emphasize a unified methodology in constructing the non-Hermitian Hamiltonian of magnons (Sec.~\ref{general_approach}) and the general topological characterization and properties (Sec.~\ref{Exceptional_topology}) that well describes the non-Hermitian topological phenomena such as the EPs~\cite{lee2015macroscopic,galda2016parity,PhysRevB.95.214411,zhang2017observation,galda2018parity,yang2018antiferromagnetism,wang2018magnon,xiao2019enhanced,PhysRevLett.123.237202,cao2019exceptional,huai2019enhanced,Liu2019Observation,zhang2019higher,yuan2020steady,yang2020Unconventional,zhao2020observation,tserkovnyak2020,yu2020higher,wang2020steering,wang2021coherent,lu2021exceptional,wang2022pt,nair2021enhanced,rao2021interferometric,PT_bilayer,deng2023exceptional,wang2023MFC}, non-Hermitian nodal phases \cite{mcclarty2019non,bergholtz2019non,li2022multitude,yang2021exceptional}, non-Hermitian magnonic SSH model \cite{flebus2020non,gunnink2022nonlinear}, and non-Hermitian skin effect \cite{deng2022non,yu2022giant,yu2020magnon_accumulation,zeng2023radiation,cai2023corner} in magnonic systems, as summarized in Fig.~\ref{fig:introduction} for an overview.

Magnonic systems are highly tunable with many control dimensions, e.g., magnetic fields, driving power, and damping, which thereby provide a powerful platform for engineering EPs and exceptional surfaces.
Realization of the EPs in the magnonic devices has been pursued in the magnetism community for years since the exotic properties produced by such magnetic excitation are distinguished with promising applications in coherent/quantum information processing, such as scattering enhancement of magnons~\cite{zhang2017observation,Liu2019Observation}, magnon lasing or amplification \cite{wang2018magnon,wang2020steering,wang2022pt}, and  (quantum) sensing with unprecedented sensitivity~\cite{cao2019exceptional,wang2021enhanced}.  So far, tremendous efforts have been made to achieve the magnonic EPs with progress reviewed in Sec.~\ref{Magnonic_EPs}. One representative approach to realize the EPs is to facilitate the magnetic heterostructures \citep{yu2020higher, Liu2019Observation,PT_bilayer}, a pure magnetic setup, by delicately tuning the magnon-magnon coupling and the gain and loss in different magnetic layers. 
The experimental realization of such theoretical proposals in pure magnetic systems remains wanting.  Another route that already achieves the EPs experimentally is to utilize the hybridized system such as the magnon-photon coupling in the cavity magnonics \cite{soykal2010strong,huebl2013high,tabuchi2015coherent,bai2017cavity,goryachev2014high,cao2015exchange,boventer2018complex,grigoryan2018synchronized,yu2019prediction,rao2019analogue,zhang2019higher,lachance2019hybrid,bhoi2020roadmap,rao2021interferometric,li2022coherent,rameshti2022cavity,wang2023MFC}. 
In such hybridized cavity-magnon systems, the strong magnon-photon coupling can be easily and precisely controlled by adjusting the field spatial overlap between the cavity photon and magnon modes. Compared to the pure magnetic setup, complicated device design and fabrication can be largely avoided, and hence the realization of EPs in the cavity magnonic systems is feasible. Along this path, both theoretical and experimental studies are flourishing. Many unique functionalities of the EPs, including the topological mode switching \cite{doppler2016dynamically,PhysRevB.95.214411}, giant enhancement of magnonic frequency combs \cite{rao2023unveiling,wang2023MFC}, the polariton coherent perfect absorption \cite{zhang2017observation}, and the exceptional surface \cite{PhysRevLett.123.237202}, have been successfully observed in the experiments. In recent experiments~\cite{rao2023unveiling, wang2023MFC}, the researchers found that the magnonic frequency combs can be strongly enhanced from several to tens of tones when a magnetic sphere in a waveguide is driven to a specific nonlinear regime, which was attributed to the emergence of EPs in such a magnonic system. These findings demonstrate useful functionalities of the EPs in information processing and promote the further exploration of the EPs in magnonic devices.  
In the reciprocal space, a non-Hermitian
perturbation on the magnon Dirac and Weyl points drives a pair of EPs connected by the topologically protected bulk Fermi
arcs, which are predicted
in magnetic junctions \cite{bergholtz2019non} and a spin-1/2 ferromagnet of
the honeycomb lattice~\cite{mcclarty2019non}, as reviewed in Sec.~\ref{exceptional_nodal_phases_magnons}.

The generalization of the SSH model \cite{su1979solitons,su1980soliton,fradkin1983phase,berry1984quantal,heeger1988solitons,zak1989berry,atala2013direct,meier2016observation}
to the non-Hermitian magnetic system in terms of an array of spin-torque oscillators promises the topological magnonic lasing edge modes, which can be excited by spin current injection \cite{flebus2020non,gunnink2022nonlinear}. 
On the other hand, the non-Hermitian skin effect \cite{hatano1996localization,lee2016anomalous,kunst2018biorthogonal,xiong2018,yao2018edge,yokomizo2019non,okuma2020,zhang2020,yang2020non,budich2020non,borgnia2020non,bergholtz2021exceptional,zhang2022universal} stems from the high sensitivity of the bulk modes to the boundary, which leads to the piling up 
of a macroscopic number of magnonic states at one boundary. Although the non-reciprocal hoping in the Hatano-Nelson model needs a special design, chirality is a common ingredient in magnetic orders.
Chiral coupling, also known as asymmetric or nonreciprocal coupling, is very common in the interaction between magnons and other quasiparticles~\cite{yu2019chiral,yu2020nonreciprocal,zhang2020unidirectional,yamamoto2020non,yu2020chiral,yu2020magnon_accumulation,yu2022giant,yu2023chirality}. Facilitated the chirality,
interesting non-Hermitian skin effects are predicted in an array of magnetic wires coupled with the magnetic films via the dipolar interaction, where the combination of chirality and dissipation of traveling
waves drive all the modes to one edge \cite{yu2022giant,zeng2023radiation}. A strong accumulation
of magnon modes at one boundary significantly enhances the sensitivity in the detection of small magnetic-field signals \cite{yu2022giant}. 
Further, recent works show that both the edge and corner skin effects can appear in higher dimensions, which also raises theoretical
challenges and urgent issues in the topological characterization of different skin modes~\cite{kawabata2020higher,zhang2022universal,wang2022amoeba,hu2023nonhermitian}.  Recently, Deng \textit{et al.} predicted in two-dimensional van der Waals ferromagnetic monolayer honeycomb lattice
\cite{deng2022non} that the edge skin effect can be driven by Dzyaloshinskii-Moriya interaction and nonlocal magnetic dissipation. Both the corner and edge skin effects are recently predicted to be observable in the two-dimensional magnetic array on a magnetic substrate by changing the direction of the in-plane magnetization~\cite{cai2023corner}. In such non-reciprocal two-dimensional non-Hermitian systems,
the two winding numbers defined along two normal directions can precisely
distinguish different edge and corner skin effects, i.e., a
precise prediction of the edge or corner on which the
modes localize~\cite{cai2023corner}, which is a straightforward generalization
of the one-dimensional winding number \cite{zhang2020correspondence,okuma2020,ashida2020non}.   These theoretical proposals that await future experimental observations are reviewed in Sec.~\ref{Non_Hermitian_skin_effect_magnon}.

We conclude and discuss the future opportunities and challenges existing in the non-Hermitian topological magnonics in Sec.~\ref{summary_outlook}.

\section{General approaches for magnon non-Hermitian dynamics}

\label{general_approach}

\subsection{Magnon: quanta of spin wave}
\label{Magnon}

 In solids, the spin and orbital motion of electrons contribute to the magnetic moments that are spontaneously ordered due to the Coulomb exchange interaction $E_{\rm{ex}}$. The specific form of orders is dominated by the interplay and competition of the much weaker interactions such as the Dzyaloshinskii-Moriya (DM) exchange interaction $E_{\rm{DM}}$ \cite{dzyaloshinsky1958thermodynamic,moriya1960anisotropic,yang2023first,kuepferling2023measuring}, the magnetic dipolar interaction $E_{\rm{dip}}$, and the crystal anisotropies $E_{\rm{ani}}$ \cite{johnson1996magnetic,dieny2017perpendicular,hu2018engineering}. 
 The ground magnetic states are governed by the competition of various interactions that lead to rich states such as ferromagnetic, anti-ferromagnetic, and textured magnetization configurations \cite{yoshimura2016soliton,vousden2016skyrmions,kim2017fast,tserkovnyak2018energy,jones2020energy,psaroudaki2021skyrmion,zou2022domain,xia2022qubits,xia2023universal}.

 Including the Zeeman interaction to the applied field ${{\bf H}}_{\rm{ext}}({\bf r})$, the total free energy of a ferromagnet  
$F_{\rm{FM}}=E_{\rm{ex}}+E_{\rm{Z}}+E_{\rm{dip}}+E_{\rm{ani}}+E_{\rm{DM}}$. The classical magnetization dynamics can be described by the  Landau-Lifshitz-Gilbert (LLG) phenomenology \cite{landau1992theory,gilbert2004phenomenological}. 
In the continuum limit of ferromagnet \cite{kittel1949physical}, $E_{\rm{ex}}=(\alpha_{\rm ex}/2)\int d{\bf {r}}\left(\nabla{\bf{M}}({\bf r})\right)^2$,
where $\alpha_{\rm ex}$ is the exchange stiffness constant, and ${\bf M}({\bf r})$ is the magnetic moment density or magnetization. An external magnetic field ${{\bf H}}_{\rm{ext}}({\bf r})$ biases the magnetization via the Zeeman interaction 
$E_{{\rm Z}}=-\mu_0\int d{\bf r} {\bf M}({\bf r})\cdot{\bf H}_{\rm{ext}}({\bf r})$,
where  $\mu_0$ is the vacuum permeability. Much weaker electromagnetic interaction contributes to the dipolar interaction between the magnetization
\begin{equation}
E_{\rm dip} = -\frac{\mu_{0}}{{8\pi}}\int d\mathbf{r}\left({\mathbf{M}(\mathbf{r})}\cdot\nabla\int d\mathbf{r}^{\prime}\frac{\nabla^{\prime}\cdot\mathbf{M} (\mathbf{r}^{\prime})}{|\mathbf{r} - \mathbf{r}^{\prime}|}\right),
  \label{free_energy}
\end{equation}
which together with the relativistic spin-orbit interaction also affects the magnetization anisotropy, e.g., the uniaxial anisotropy $E_{\rm{ani}}=-({K}/{M_s})\int d\boldsymbol{\rm{r}} (\boldsymbol{\rm{M}}\cdot\hat{\bf n})^2$,
where $M_s$ is the saturated magnetization, 
favors the magnetization along $\hat{\bf n}$ by the (temperature dependent) constant $K>0$.
On the other hand, the spin-orbit interaction leads to asymmetric DM exchange coupling in non-centrosymmetric lattice structure due to the broken inversion symmetry \cite{dzyaloshinsky1958thermodynamic,moriya1960anisotropic,yang2023first,kuepferling2023measuring}, which is conveniently addressed in terms of the local spins ${\bf S}_i$ at different sites $\{i,j\}$: $E_{\rm{DM}}=\sum\limits_{ij}\boldsymbol{\rm{D}}_{ij}\cdot(\boldsymbol{\rm{S}}_i\times\boldsymbol{\rm{S}}_j)$,
where $\boldsymbol{\rm{D}}_{ij}$ is the so-called DM vector. 
The torque provided by the magnetic interaction ${\bf H}_{\rm eff}({\bf r})=-({1}/{\mu_{0}})\delta F[{\bf M}]/\delta {\bf M}({\bf r})$ drives the magnetization precession in the LLG equation  
\cite{landau1992theory,gilbert2004phenomenological} 
\begin{equation}
\partial {\bf M}({\bf r})/{\partial t}=-\mu_0\gamma{\bf M}({\bf r})\times {\bf H}_{\rm eff}+\alpha({\bf M}/M_s)\times{\partial {\bf M}}/{\partial t},
\label{eq:LLG}
\end{equation} 
where $\gamma$ is the gyromagnetic ratio 
and $\alpha$ is
the phenomenological Gilbert damping coefficient.

The magnetic excitations around the static-ordered magnetic moments are spin waves with frequencies ranging from gigahertz to terahertz scales. 
The quantization of spin waves into their quanta, i.e., magnons, involves magnon bosonic operators and their eigenmodes or ``wavefunction'' with \textit{proper} orthonormalization procedure.  
Using the Holstein-Primakoff transformation \cite{holstein1940field} with the bosonic operator $\hat{a}({\bf r})$ that obeys the commutation relation
$\left[ a(\mathbf{r}), a^{\dagger}(\mathbf{r}') \right] = \delta\left(\mathbf{r}-\mathbf{r}'\right)$, the spin operators are quantized in the linear regime as 
\begin{subequations}
\begin{align}
&\hat{S}_+(\mathbf{r})=\hat{S}_x({\bf r})+i\hat{S}_y({\bf r}) \approx \sqrt{2S} \hat{a}(\mathbf{r}),\\
&\hat{S}_-(\mathbf{r})=\hat{S}_x({\bf r})-i\hat{S}_y({\bf r}) \approx \sqrt{2S} \hat{a}^{\dagger}({\mathbf{r}}),\\ 
&\hat{S}_z(\mathbf{r}) = S - \hat{a}^{\dagger}({\mathbf{r}}) \hat{a}({\mathbf{r}}),
\end{align}
\end{subequations}
where $S=M_s/(\gamma\hbar)$ with the saturated magnetization $M_s$. The magnon operator $\hat{\alpha}_p$ in mode ``$p$'' is defined in terms of the ``wavefunction'' $A_p$ and $B_p$: 
\begin{align}
\hat{\alpha}_p \equiv \int d{\bf r}\left[A_p^*({\bf r})\hat{a}({\bf r})-B_p({\bf r})\hat{a}^{\dagger}({\bf r})\right].
\end{align}
Their commutators
\begin{subequations}
\begin{align}
\left[\hat{\alpha}_{p'},\hat{\alpha}_p^{\dagger}\right] &= \int d{\bf r} \left[A_p({\bf r})A_{p'}^*({\bf r})-B_p^*({\bf r})B_{p'}({\bf r}) \right]=\delta_{pp'}, 
\label{orthogonal_1}\\
\left[\hat{\alpha}_{p'},\hat{\alpha}_p\right] &=-\int d{\bf r} \left[B_p({\bf r})A_{p'}^*({\bf r})-A_p^*({\bf r})B_{p'}({\bf r})\right]=0.
\label{orthogonal_2}
\end{align}
\label{condition_1}
\end{subequations}
On the other hand, the completeness relation ensures that $\hat{\alpha}_p$ form a complete set: 
\begin{subequations}
\begin{align}
& \sum_p \left[A_p({\bf r})A_p^*({\bf r'})-B_p({\bf r})B_p^*({\bf r'})\right] {=} \delta\left({\bf r}-{\bf r'}\right), \\
& \sum_p \left[A_p({\bf r})B_p({\bf r'})-B_p({\bf r})A_p({\bf r'})\right] {=} 0,
\end{align}
\label{condition_2}
\end{subequations}
such that 
$\sum_p \left[ A_p({\bf r}) \hat{\alpha}_p + B_p({\bf r}) \hat{\alpha}_p^{\dagger} \right] = \hat{a}(\mathbf{r})$ and $\sum_p \left[ A_p^*({\bf r}) \hat{\alpha}_p^{\dagger} + B_p^*({\bf r}) \hat{\alpha}_p \right] = \hat{a}^{\dagger}(\mathbf{r})$.

Magnetization obeys its equation of motion. Without loss of generality, here we consider its coupling to the applied magnetic field $H_0$ along the $-\hat{\bf z}$-direction and its dipolar stray field in the  Hamiltonian
\begin{align}
\hat{H} = -\mu_0 \gamma \hbar \int d{\bf r}\hat{S}_z({\bf r})H_0 + \frac{\mu_0 \gamma \hbar}{2} \int d{\bf r}\hat{\bf S}(\bf r)\cdot\nabla\phi,
\label{Hamiltonian}
\end{align}
where the magnetic scalar potential 
\begin{align}
\nonumber
\phi&= \int d{\bf r'}\frac{\partial_{\alpha}'\hat{S}_{\alpha}}{|{\bf r}-{\bf r'}|}= \frac{1}{2}\int d{\bf r'}\frac{\partial_-'\hat{S}_++\partial_+'\hat{S}_-}{|{\bf r}-{\bf r'}|}\\
&=\frac{\sqrt{2S}}{2}\sum_p\int\frac{\partial_-'A_p({\bf r'})+\partial_+'B_p^*({\bf r'})}{|{\bf r}-{\bf r'}|}d{\bf r'}\hat{\alpha}_p+\frac{\sqrt{2S}}{2}\sum_p\int\frac{\partial_+'A^*_p({\bf r'})+\partial_-'B_p({\bf r'})}{|{\bf r}-{\bf r'}|}d{\bf r'}\hat{\alpha}_p^{\dagger}. \label{magnetic_potential}
\end{align}
The linearized equation of motion reads
\begin{equation}
{\partial \hat{S}_-}/{dt} = -i\omega_0 \hat{S}_- - i\eta_s \partial_-\phi,
\end{equation}
where $\omega_0 = \mu_0\gamma H_0$ and $\eta_s = \mu_0 \gamma S$. As magnons, $\hat{\alpha}_p \sim e^{-i\omega_pt}$, so we require
\begin{equation}
\sum_q \left[ \omega_q A_q({\bf r}) \hat{\alpha}_q - \omega_q B_q({\bf r}) \hat{\alpha}_q^{\dagger} \right] - \omega_0 \sum_q \left[ A_q({\bf r}) \hat{\alpha}_q + B_q({\bf r}) \hat{\alpha}_q^{\dagger} \right]  = \frac{\eta_s}{\sqrt{2S}} \partial_-\phi(\mathbf{r}),
\label{phiLLG}
\end{equation}
leading to 
\begin{align}
\left[\partial_-\phi(\mathbf{r}), \hat{\alpha}_p^{\dagger} \right] = \sqrt{2S}\ \frac{\omega_p - \omega_0}{\eta_s} A_p(\mathbf{r}),~~~
\left[\partial_-\phi(\mathbf{r}), \hat{\alpha}_p \right] = \sqrt{2S}\ \frac{\omega_p + \omega_0}{\eta_s} B_p(\mathbf{r}).
\label{commutation_2}
\end{align}
On the other hand,  we can calculate the commutations in Eq.~(\ref{commutation_2}) in terms of Eq.~(\ref{magnetic_potential}), leading to the relations 
\begin{subequations}
\begin{align}
&\int\frac{\partial_-'A_p({\bf r'})+\partial_+'B_p^*({\bf r'})}{2|{\bf r}-{\bf r'}|}d{\bf r'} = \frac{\omega_p - \omega_0}{\eta_s} A_p(\mathbf{r}), \\
&\int\frac{\partial_+'A^*_p({\bf r'})+\partial_-'B_p({\bf r'})}{2|{\bf r}-{\bf r'}|}d{\bf r'} = -\frac{\omega_p + \omega_0}{\eta_s} B_p(\mathbf{r}).
\end{align}
\label{condition_3}
\end{subequations}

Let us prove that the Hamiltonian (\ref{Hamiltonian}) is diagonalized under conditions (\ref{condition_1}), (\ref{condition_2}), and (\ref{condition_3}). With Eq.~(\ref{phiLLG}),
\begin{align}
\nonumber
\hat{H} &= E_0 + \sum_p \hbar\omega_p \int \frac{d{\bf r}}{2} \left[ \hat{a}^{\dagger}(\mathbf{r}) \left( A_p({\bf r}) \hat{\alpha}_p - B_p({\bf r}) \hat{\alpha}_p^{\dagger} \right) + \hat{a}(\mathbf{r}) \left( A_p^*({\bf r}) \hat{\alpha}_p^{\dagger} - B_p^*({\bf r}) \hat{\alpha}_p \right) \right]\\
&=E_0 + \sum_p \frac{\hbar\omega_p}{2} \left[ \hat{\alpha}_p^{\dagger} \hat{\alpha}_p + \hat{\alpha}_p \hat{\alpha}_p^{\dagger} \right], 
\end{align}
which is $\sum_p \hbar\omega_p \hat{\alpha}_p^{\dagger}\hat{\alpha}_p$ up to constant energy. All the magnon Hamiltonian in magnets can be obtained in principle by this procedure.

The magnons in the magnets or hybrid magnetic nanostructures interact with many quasiparticles, such as the phonons, microwave or optical photons, electrons, as well as other magnons. Such interaction can be generally divided into two categories, i.e., the bilinear and nonlinear couplings. To properly deal with the interaction between the magnons and the other degrees of freedom, a convenient way is to ``integrate out'' the other degrees of freedom such that one can effectively describe the magnon subsystem in terms of an effective non-Hermitian Hamiltonian, but some information may be disregarded in this procedure as well. Below we discuss the universal master-equation (Sec.~\ref{master_equation_approach}) and Green-function (Sec.~\ref{Green_function}) approach that can arrive at the effective non-Hermitian Hamiltonian with a perturbation treatment of the other degrees of freedom.

\subsection{Master-equation approach for magnon non-Hermitian dynamics}

\label{master_equation_approach}

\subsubsection{Lindblad master equation and its application to magnonic systems}
\label{bilinear_coupling}

The Lindblad master equation approach provides a powerful and universal description for the non-Hermitian dynamics of the magnon subsystem when their interaction with the other quasiparticles or ``environment'' can be dealt with perturbatively.
This approach is applicable to the quantum regime and can be linked to the non-Hermitian Hamiltonian description and Green-function approach, provided the \textit{quantum jump} effect is negligible \cite{Plenio1998rmp,rivas2012open,de2017dynamics,chang2018colloquium} or only the \textit{mean-field} dynamics are of interest \cite{Heinz,Lidar2019,zou2023dissipative}. We will delve deeper into these topics in Sec.~\ref{sec:qjump}.

We present a comprehensive derivation of the Lindblad master equation from a microscopic model. This allows us to obtain microscopic expressions for all parameters that describe the environment-induced coherent and dissipative dynamics of the magnonic system. We then apply this master equation approach to the magnonic system interacting with a bosonic bath (which can naturally be the phonon bath). We emphasize the Born and Markov approximations behind this master equation approach and discuss the breakdown of these approximations, highlighting specific physical scenarios in which this approach does not apply to the magnetic system.

\begin{figure}[!htp]
	\centering
	\includegraphics[width=13.6cm]{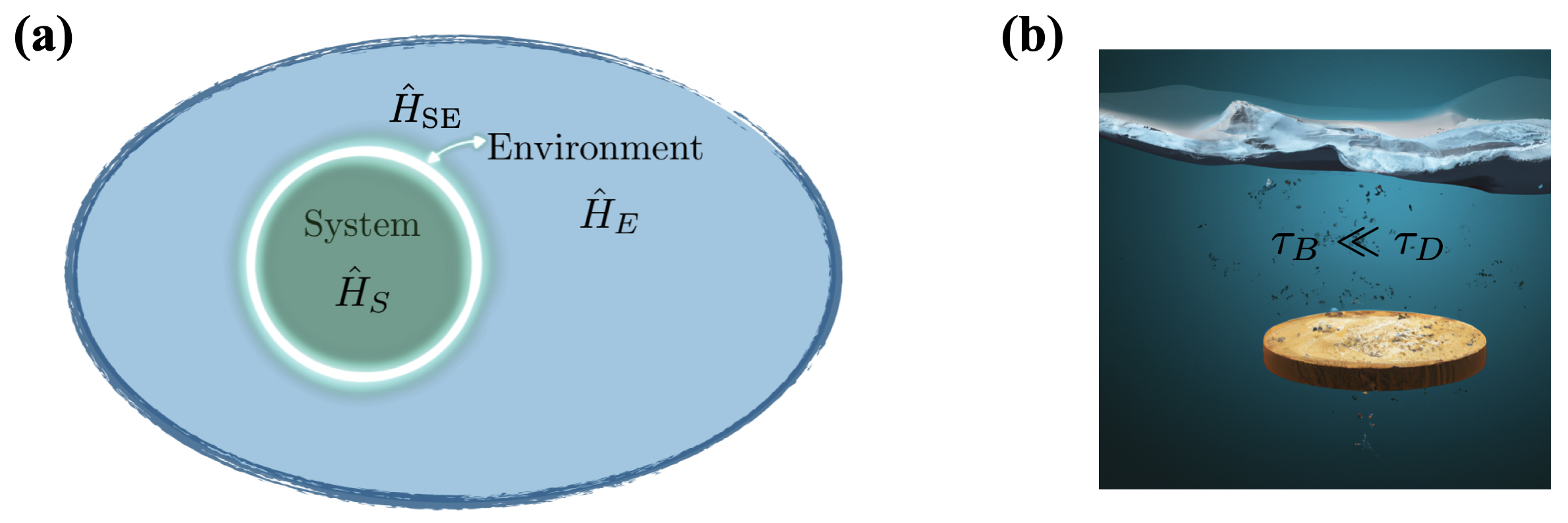}
	\caption{(a) A system of interest described by $\hat{H}_S$ interacts with an environment $\hat{H}_E$ through interaction $\hat{H}_{\text{SE}}$. (b) Markovian approximation. When a hot coin is put into a lake at temperature $T$, there are two important timescales. One is $\tau_B$, during which the excited water molecules near the hot coin return back to temperature $T$ by many collisions with other molecules in the lake; the second is $\tau_D$, during which the coin reaches the same temperature $T$ as the lake. It is generally true that $\tau_B\ll \tau_D$. In the Markovian approximation, we neglect all dynamics happening at the timescale $\tau_B$, which effectively takes this bath correlation time to zero, $\tau_B\rightarrow 0.$ }
	\label{fig:331}
\end{figure}

\textbf{Lindblad master equation.} We consider a closed quantum system,  governed by a time-independent Hamiltonian $\hat{H}$, which consists of a subsystem ``$S$'' and an environment ``$E$''. The subsystem $S$ is open since it interacts with the environment, as shown in Fig~\ref{fig:331}(a). We assume the  Hamiltonian of the whole system to be
\( \hat{H}=\hat{H}_S+\hat{H}_E+\hat{H}_{\text{SE}}, \)
where $\hat{H}_S$ and $\hat{H}_E$ are Hamiltonians of the system and the environment, respectively, and  $\hat{H}_{\text{SE}}$ is the interaction between them. We start with a uncorrelated state $\hat{\rho}_{\text{SE}}(0)=\hat{\rho}_S(0)\otimes \hat{\rho}_E(0)$. Typically, the environment is taken to be in thermal equilibrium at some temperature $T$, $\hat{\rho}_E(0)=\hat{\rho}_E^{\text{eq}}=e^{-\beta \hat{H}_E}/\mathcal{Z}_E$,  where $\mathcal{Z}_E=\tr e^{-\beta \hat{H}_E}$ with $\beta=1/(k_BT)$. The effect of the environment is to cause irreversible decay of the system.

As in standard time-dependent perturbation theory, it is convenient to work in the interaction picture, where 
\( \dv{\hat{\rho}^I_{\text{SE}}}{t}=- \frac{i}{\hbar} [\hat{H}^I_{\text{SE}}(t), \hat{\rho}^I_{\text{SE}}(t)],\;\;\; \text{with}\;\;\; \hat{H}^I_{\text{SE}}=\hat{U}^\dagger_0(t) \hat{H}_{\text{SE}} \hat{U}_0(t) \;\;\text{and} \;\; \hat{\rho}^I_{\text{SE}}(t)=\hat{U}_0^\dagger(t) \hat{\rho}_{\text{SE}}(t) \hat{U}_0(t). \)
Here $\hat{U}_0(t)=e^{-i\hat{H}_S\,t/\hbar}\otimes e^{-i\hat{H}_B\, t/\hbar}$. 
By tracing out the degrees of freedom of the environment, we obtain the equation of motion for the density matrix $\hat{\rho}^I$ of the subsystem ``S'':
\( \frac{d\hat{\rho}^I(t)}{dt}=- \frac{i}{\hbar} \tr_E [\hat{H}^I_{\text{SE}}(t), \hat{\rho}^I_{\text{SE}}(t)]. \)
Let us integrate for a short time interval, giving us:
\( \al{\hat{\rho}^I(t+\Delta t)&=\hat{\rho}^I(t)+\tr_E\Big\{  - \frac{i}{\hbar} \int_t^{t+\Delta t} dt^\prime \; [\hat{H}^I_{\text{SE}}(t^\prime), \hat{\rho}^I_{\text{SE}}(t^\prime)]   \Big\} \\
 &= \hat{\rho}^I(t) - \frac{i}{\hbar} \int_t^{t+\Delta t} dt^\prime \; \tr_E [\hat{H}^I_{\text{SE}}(t^\prime), \hat{\rho}^I_{\text{SE}}(t)]  \\
   &  + \frac{(-i)^2}{\hbar^2}   \int_t^{t+\Delta t} dt^\prime\int^{t^\prime}_tdt^{\prime\prime} \tr_E\big[ \hat{H}^I_{\text{SE}}(t^\prime), [\hat{H}^I_{\text{SE}}(t^{\prime\prime}), \hat{\rho}^I_{\text{SE}}(t)] \big] +\cdots,  }  \label{chap1_hh}\)
and it is clear how this continues.  The requisite criterion for sufficient convergence is $||\hat{H}^I_{\text{SE}}||\Delta t\ll 1$, which is attainable by ensuring a weak coupling between the system and the environment.
We point out that this equation \eqref{chap1_hh} is exact.  To proceed, we now start to apply Born and Markov approximation. 
 We will only keep the terms we write down above. This is known as the \textit{Born approximation}~\cite{Heinz}. This approximation also implies that the frequency scales associated with the dynamics induced by the system-environment coupling are significantly smaller in magnitude compared to the relevant dynamical frequency scales of the system.
Before introducing the \textit{Markovian approximation}, let us develop some intuitions. Assuming there is a characteristic time scale $\tau_B$ (environment correlation time), 
it is unlikely that information from the environment will return to the system after this time scale. By neglecting (coarse-graining) the dynamics within this correlation time, the dynamics of the system become irreversible. To better understand this concept, let us consider a classic example. The scenario involves placing a hot coin into a lake at temperature $T$ [see Fig.~\ref{fig:331}(b)], where we have two important timescales.
The first one is $\tau_B$, during which the ``excited water molecules'' rethermalize due to many collisions with other molecules in the lake. The other timescale is $\tau_{D}$, the time it takes for the coin to reach the same temperature as the lake.  It is clear that the thermalization of water molecules happens much faster than $\tau_D$. As we are interested in how the coin reaches the thermal equilibrium, when we construct a theory for it, it is natural to coarse grain over a time scale $\Delta t$: $ \tau_B\ll \Delta t\ll \tau_D$.  Let us now apply this argument to the quantum master equation. For a sufficiently large bath that is, in particular, much larger than the system, it is reasonable to assume that while the system undergoes nontrivial evolution, the bath remains unaffected, and hence that the state of the composite system at time $t$ is $\hat{\rho}^I_{\text{SE}}\approx \hat{\rho}^I(t)\otimes \hat{\rho}_E^{\text{eq}} $,
since the environment returns to equilibrium after the coarse-graining timescale $\Delta t$. In the Markovian limit, we have the master equation in the interaction picture: 
\( \dv{\hat{\rho}^I}{t}(t)=- \frac{1}{\hbar^2} \int^\infty_0d\tau \,  \tr_E\big[ \hat{H}^I_{\text{SE}}(t), [\hat{H}^I_{\text{SE}}(t-\tau), \hat{\rho}^I(t)\otimes \hat{\rho}^{\text{eq}}_E] \big]. \label{lindblad} \)
 This  expression is valid for a general interaction
$ \hat{H}_{\text{SE}}=\sum_\alpha \hat{S}_\alpha\otimes \hat{B}_\alpha  $,
with $\langle \hat{B}_\alpha\rangle_{\text{eq}}=0$~\cite{Heinz}. Here, $\hat{S}_\alpha$ and $\hat{B}_\alpha$ are operators acting on Hilbert spaces of the system and the environment, respectively. For $\langle \hat{B}_\alpha\rangle_{\text{eq}}\neq 0$, we can always redefine $ \hat{B} \rightarrow \hat{B}- \langle \hat{B}\rangle_{\text{eq}}$. 
This master equation is well-known \textit{Redfield equation}. 
It is straightforward to work out the master equation for $\hat{\rho}^I$ once the interaction is specified. 
We use the example of bilinear magnon-quasiparticle interaction to illustrate this. We remark that the Redfield equation above does not warrant the positivity of the evolution in general, and it sometimes gives rise to density matrices that are non-positive. We sometimes need to perform  one further approximation, the rotating wave approximation, which can be achieved by simply neglecting rapidly oscillating terms in the master equation~\cite{Heinz}, to obtain a master equation in the Lindblad form:
\( \dv{\hat{\rho}^I}{t} =-\frac{i}{\hbar} [\hat{H}, \hat{\rho}^I] +\sum_{n,m}h_{nm}\left( \hat{L}_n\hat{\rho}^I \hat{L}^\dagger_m - \frac{1}{2} \left\{ \hat{L}_m^\dagger \hat{L}_n, \hat{\rho}^I \right\} \right).  \)
 The first and the second term are related to the environment-induced coherent and dissipative dynamics, respectively.  $\{\hat{L}_m\}$ are arbitrary operators acting on the Hilbert space of the system ``S'' and the matrix $h_{nm}$ is a positive semidefinite matrix due to the complete positivity of the dynamics. 

In practical applications, a quick assessment of the applicability of the Born and Markov approximations can be conducted by comparing various  timescales: $\Omega^{-1}$, the timescale  associated with the dynamics of the system; $\tau$, the timescale of the dynamics induced by the environment on the system; $\tau_B$, the relaxation timescale of the environment  (e.g., the lifetime of quasiparticles within the environment).  The Born approximation requires the environment-induced dynamics (such as magnon damping or effective magnon-magnon coupling) to be much slower compared to intrinsic system dynamics (magnon resonance frequency): $\tau^{-1}\ll \Omega$. The Markov approximation firstly requires that the system-environment coupling should be independent of frequency, to remove any back-action of the environment on the system that is not local in time. This is again justified by the large system frequency compared to the environment-induced damping.  Since the system only couples to the environment around the system frequency $\Omega$ with a bandwidth $\tau^{-1}$, when $\tau^{-1}\ll \Omega$, the variation in the coupling strength over the narrow frequency window is very small and thus we can always approximate it with a constant. To apply the Markov approximation, we also require that the environment returns rapidly to equilibrium in a manner essentially unaffected by its coupling to the system, which is ensured by a short environment relaxation time compared to the environment-induced system dynamics (whose timescale is basically the inverse of the effective system-environment coupling), $\tau_b\ll \tau$. As a result, the environment can always rethermalize quickly, and the dynamics of the system are not affected by its coupling to the environment at earlier times.

\textbf{Application to bilinear magnon-quasiparticle interaction.}
As an example, we consider a combined system consisting of  a magnon mode with frequency $\omega_m$ and  an environment, with the following Hamiltonian: 
\( \hat{H}=\hbar \omega_m \hat{m}^\dagger \hat{m} +\sum_{\vb k} (g_{\vb k} \hat{m}^\dagger \hat{b}_{\vb k} + g_{\vb k}^* \hat{m} \hat{b}^\dagger_{\vb k} ) + \sum_{\vb k} \epsilon_{\vb k} \hat{b}_{\vb k}^\dagger \hat{b}_{\vb k}, \label{Eq:3321}  \)
where $\hat{m}$ and $\hat{m}^\dagger$ are magnon annihilation and creation operator, and $\hat{b}_{\vb k}$ stands for modes of quasiparticles in the environment, satisfying the bosonic algebra $[\hat{b}_{\vb k}, \hat{b}^\dagger_{\vb k^\prime}]=\delta_{\vb k \vb k^\prime}$. In the interaction picture, the second term (interaction part) of Hamiltonian~\eqref{Eq:3321} is
\( \hat{H}^I_{\text{SE}}(t)=\sum_{\vb k} [g_{\vb k} e^{i\omega_m t} \hat{m}^\dagger \hat{b}^I_{\vb k}(t)  + \text{H.c.} ], \;\;\;\text{and}\;\;\; \hat{H}^I_{\text{SE}}(t-\tau)= \sum_{\vb k} [g_{\vb k} e^{i\omega_m (t-\tau)} \hat{m}^\dagger \hat{b}^I_{\vb k}(t-\tau)  + \text{H.c.} ].   \)
Taking the first term of $\hat{H}^I_{\text{SE}}(t)$ and the second term of $\hat{H}^I_{\text{SE}}(t-\tau)$, the integral on the right-hand side of the master equation~\eqref{lindblad} gives us
\( \al{ \int^\infty_0d\tau \,  \tr_E\big[ \hat{H}^I_{\text{SE}}(t), [\hat{H}^I_{\text{SE}}(t-\tau), \hat{\rho}^I(t)\otimes \hat{\rho}^{\text{eq}}_E] \big] &\supset  \sum_{\vb k, \vb k^\prime } g_{\vb k} g^*_{\vb k^\prime} \int^\infty_0 d\tau \; e^{i\omega_m \tau} \tr_E\big[ \hat{m}^\dagger \hat{b}^I_{\vb k}(t), [ \hat{m} \hat{b}^{I\dagger}_{\vb k^\prime}(t-\tau), \hat{\rho}^I]\otimes \hat{\rho}^{\text{eq}}_E ]  \big] \\
&= \sum_{\vb k, \vb k^\prime } g_{\vb k} g^*_{\vb k^\prime} \int^\infty_0 d\tau \; e^{i\omega_m \tau} \Big[ \langle \hat{b}^I_{\vb k}(t) \hat{b}^{I\dagger}_{\vb k^\prime}(t-\tau)\rangle (\hat{m}^\dagger \hat{m} \hat \rho^I -\hat{m} \hat{\rho}^I \hat{m}^\dagger) \\
& \;\;\;\;\;\;\;\;\;\;\;\;\;\;\;\;\;\;\;\;\;\;\;\;\;\;\;\;\;\;\;\;\;\;\;\;\; +  \langle \hat{b}^{I\dagger}_{\vb k^\prime} (t-\tau) \hat{b}^I_{\vb k}(t) \rangle (\hat{ \rho }^I \hat{m} \hat{m}^\dagger -\hat{m}^\dagger \hat{\rho}^I \hat{m} ) \Big].  }\)
One can also take the second term of $\hat{H}^I_{\text{SE}}(t)$  and the first term of $\hat{H}^I_{\text{SE}}(t-\tau)$, which is given by the Hermitian conjugate of the above result. We remark that we only used the fact that the environment is U(1) invariant; thus correlators $\langle \hat{b}^{I\dagger}_{\vb k}(t) \hat{b}^{I\dagger}_{\vb k^\prime}(t^\prime) \rangle, \langle \hat{b}^I_{\vb k}(t) \hat{b}^I_{\vb k^\prime}(t^\prime) \rangle$ vanish. We note that the environment is time-translational invariant and also conserves momentum. Thus, we obtain: 
\( \al{ \int^\infty_0d\tau \,  \tr_E\big[ \hat{H}^I_{\text{SE}}(t), [\hat{H}^I_{\text{SE}}(t-\tau), \hat{\rho}^I(t)\otimes \hat{\rho}^{\text{eq}}_E] \big] &= \sum_{\vb k} |g_{\vb k}|^2 \int^\infty_0 d\tau \; e^{i\omega_m \tau}   \langle \hat{b}^I_{\vb k}(\tau) \hat{b}^{I\dagger}_{\vb k}  \rangle (\hat{m}^\dagger \hat{m} \hat{\rho}^I -\hat{m} \hat{\rho}^I \hat{m}^\dagger ) \\
&+  \sum_{\vb k} |g_{\vb k}|^2 \int^\infty_0 d\tau \; e^{i\omega_m \tau}  \langle \hat{b}^{I\dagger}_{\vb k} \hat{b}^I_{\vb k}(\tau) \rangle (\hat{\rho}^I \hat{m}\hat{m}^\dagger -\hat{m}^\dagger \hat{\rho}^I \hat{m} ) \\
&+  \sum_{\vb k} |g_{\vb k}|^2 \int^\infty_0 d\tau \; e^{-i\omega_m \tau}  \langle \hat{b}^I_{\vb k} \hat{b}^{I\dagger}_{\vb k}(\tau) \rangle (\hat{\rho}^I \hat{m}^\dagger \hat{m} - \hat{m} \hat{\rho}^I \hat{m}^\dagger  ) \\
&+  \sum_{\vb k} |g_{\vb k}|^2 \int^\infty_0 d\tau \; e^{-i\omega_m \tau} \langle \hat{b}^{I\dagger}_{\vb k}(\tau) \hat{b}^I_{\vb k} \rangle (\hat{m} \hat{m}^\dagger \hat{\rho}^I - \hat{m}^\dagger \hat{\rho}^I \hat{m} ).    } \label{3.3.2_89}  \)
The first and third terms can be written into [we also restore the factor $1/\hbar^2$ of Eq.~\eqref{lindblad}]: 
\(  \Gamma_{\downarrow} \left( \frac{1}{2}\hat{m}^\dagger \hat{m} \hat{\rho}^I + \frac{1}{2} \hat{\rho}^I \hat{m}^\dagger \hat{m} - \hat{m}\hat{\rho}^I \hat{m}^\dagger  \right) + \omega_1 i[\hat{m}^\dagger \hat{m}, \hat{\rho}^I], \)
with 
\( \Gamma_{\downarrow}= \frac{1}{\hbar^2} \sum_{\vb k} |g_{\vb k}|^2 \int^\infty_{-\infty} d\tau \, e^{i\omega_m\tau } \langle \hat{b}^I_{\vb k}(\tau) \hat{b}_{\vb k}^{I\dagger} \rangle, \;\;\;  \omega_1= 
  \frac{1}{\hbar^2} \sum_{\vb k} |g_{\vb k}|^2 \Im{ \int^\infty_0 d\tau\, e^{i\omega_m \tau} \langle \hat{b}^I_{\vb k}(\tau) \hat{b}_{\vb k}^{I\dagger} \rangle}  .  \)
Here, $\Gamma_{\downarrow}$ is the magnon decay rate, and the Lamb shift $\omega_1$ renormalizes the magnon frequency. Similarly, the second and fourth terms in Eq.~\eqref{3.3.2_89} read 
\(  \Gamma_{\uparrow} (\frac{1}{2} \hat{m} \hat{m}^\dagger  \hat{\rho}^I + \frac{1}{2} \hat{\rho}^I \hat{m} \hat{m}^\dagger  - \hat{m}^\dagger \hat{\rho}^I \hat{m}  ) + \omega_2 i[\hat{m}^\dagger \hat{m}, \hat{\rho}^I],  \)
with
\( \Gamma_{\uparrow}=  \frac{1}{\hbar^2} \sum_{\vb k} |g_{\vb k}|^2\int^\infty_{-\infty} d\tau \, e^{i\omega_m\tau } \langle \hat{b}_{\vb k}^{I\dagger}  \hat{b}^I_{\vb k}(\tau) \rangle, \;\;\;  \omega_2=  \frac{1}{\hbar^2}\sum_{\vb k} |g_{\vb k}|^2 \Im{ \int^\infty_0 d\tau\, e^{-i\omega_m \tau} \langle \hat{b}^{I\dagger}_{\vb k}(\tau) \hat{b}_{\vb k}^I  \rangle }.     \)
Here, $\Gamma_{\uparrow}$ is the magnon pumping rate due to the environment, and $\omega_2$ is again a Lamb shift of the magnon frequency. Therefore, the Lindblad master equation of the magnon mode is (in the interaction picture): 
\(  \dv{\hat{\rho}^I}{t}=-i[(\omega_1+\omega_2)\hat{m}^\dagger \hat{m}, \hat{\rho}^I] +  \Gamma_{\downarrow} \mathcal{L}_{\hat{m}}[\hat{\rho}^I] + \Gamma_{\uparrow} \mathcal{L}_{\hat{m}^\dagger}[\hat{\rho}^I],  \)
where $\mathcal{L}_{\hat{\mathcal{O}}}[\hat{\rho}^I]=\hat{\mathcal{O}}\hat{\rho}^I\hat{\mathcal{O}}^\dagger-\hat{\{\mathcal{O}}^\dagger \hat{\mathcal{O}}, \hat{\rho}^I\}/2$ is the dissipator.

Let us first look at the dissipative process. The rates $\Gamma_{\uparrow}$ and $\Gamma_{\downarrow}$ can be evaluated explicitly when the spectrum $\varepsilon_{\vb k}$ of the quasiparticle in the environment is given:
\(  \Gamma_{\downarrow}=  \frac{1}{\hbar^2} \sum_{\vb k} |g_{\vb k}|^2 \int^\infty_{-\infty} d\tau \, e^{i\omega_m\tau } \langle \hat{b}^I_{\vb k}(\tau) \hat{b}_{\vb k}^{I\dagger} \rangle = \frac{1}{\hbar^2} \sum_{\vb k}|g_{\vb k}|^2 \int^\infty_{-\infty} d\tau\, e^{i(\omega_m -\varepsilon_{\vb k})\tau} (\langle n(\varepsilon_{\vb k}) \rangle +1) =\gamma_0(\omega_m)(\langle n({\omega_m}) \rangle +1),   \)
where $\langle n({\omega_m}) \rangle$ is the Bose-Einstein distribution and $\gamma_0(\omega_m)\equiv 2\pi \sum_{\vb k}|g_{\vb k}|^2 \delta(\omega_m-\varepsilon_{\vb k}) /\hbar^2$ depends on the spectrum of the environment. Similarly, we have $\Gamma_{\uparrow}=\gamma_0(\omega_m)  \langle n(\omega_m) \rangle.$
Since $\langle n({\omega_m}) \rangle +1=e^{\beta \omega_m} \langle n(\omega_m) \rangle$, we obtain the relation:
\( \Gamma_{\downarrow}=e^{\beta \omega_m} \Gamma_{\uparrow},  \)
which is the detailed balance condition and is independent of the spectrum of the environment. At zero temperature $\beta\rightarrow \infty$, only the decay process survives.

The frequency correction due to the environment can also be evaluated explicitly when the spectrum of the environment $\varepsilon_{\vb k}$ is specified. For example, 
\( \omega_1 = \frac{1}{\hbar^2} \sum_{\vb k}|g_{\vb k}|^2 \Im \int^\infty_0 d\tau \; e^{i(\omega_m - \varepsilon_{\vb k})\tau} \langle \hat{b}_{\vb k} \hat{b}^\dagger_{\vb k} \rangle  = P \frac{1}{\hbar^2} \sum_{\vb k} \frac{|g_{\vb k}|^2 [1+ n(\varepsilon_{\vb k}) ] }{\omega_m-\varepsilon_{\vb k}}.  \)
Here we have used the identity \(  \int^\infty_0 ds\; e^{ixs}=\pi \delta(x) +iP \frac{1}{x},   \)
where $P$ stands for the Cauchy principal value. Similarly, we have: 
\( \omega_2= -P \sum_{\vb k} \frac{|g_{\vb k}|^2n(\varepsilon_{\vb k})}{\omega_m -\varepsilon_{\vb k}}.  \)
We remark that the discussion above can be easily generalized to $N$ magnon modes $\hat{m}_i, \hat{m}_i^\dagger$ ($i=1,2,\cdots, N$). The generic Lindblad master equation in this case takes the form of 
\( \dv{\hat{\rho}^I}{t} = - \frac{i}{\hbar} \Big[\sum_{i,j} J_{ij} \hat{m}^\dagger_i \hat{m}_j + \text{H.c.} ,\hat{\rho}^I  \Big] +\sum_{\mu\nu} h_{\mu\nu} \big[ \hat{\mathcal{O}}_\mu \hat{\rho}^I \hat{\mathcal{O}}_\nu^\dagger - \frac{1}{2} \{ \hat{\mathcal{O}}_\nu^\dagger \hat{\mathcal{O}}_\mu, \hat{\rho}^I  \}   \big],  \)
where $\hat{\mathcal{O}}_\mu=[\hat{m}_1,\cdots, \hat{m}_N, \hat{m}_1^\dagger, \cdots, \hat{m}_N^\dagger]$, and  $J_{ij}$ quantifies the environment-mediated  coherent interactions between different magnon modes (or frequency renormalizations when $i=j$). $h_{\mu\nu}$ is a positive semidefinite matrix such that we always have non-negative decay rates~\cite{Heinz}, where $h_{ii} \hat{m}_i\hat{\rho}^I \hat{m}_i^\dagger$ and $h_{ij} \hat{m}_i\hat{\rho}^I \hat{m}_j^\dagger$ stand for local and collective magnon decay, and  
$h_{ii} \hat{m}_i^\dagger\hat{\rho}^I \hat{m}_i$ and $h_{ij} \hat{m}_i^\dagger\hat{\rho}^I\hat{m}_j$ represent local and collective magnon pumping.

\begin{figure}[!htp]
	\centering
	\includegraphics[width=14.5cm]{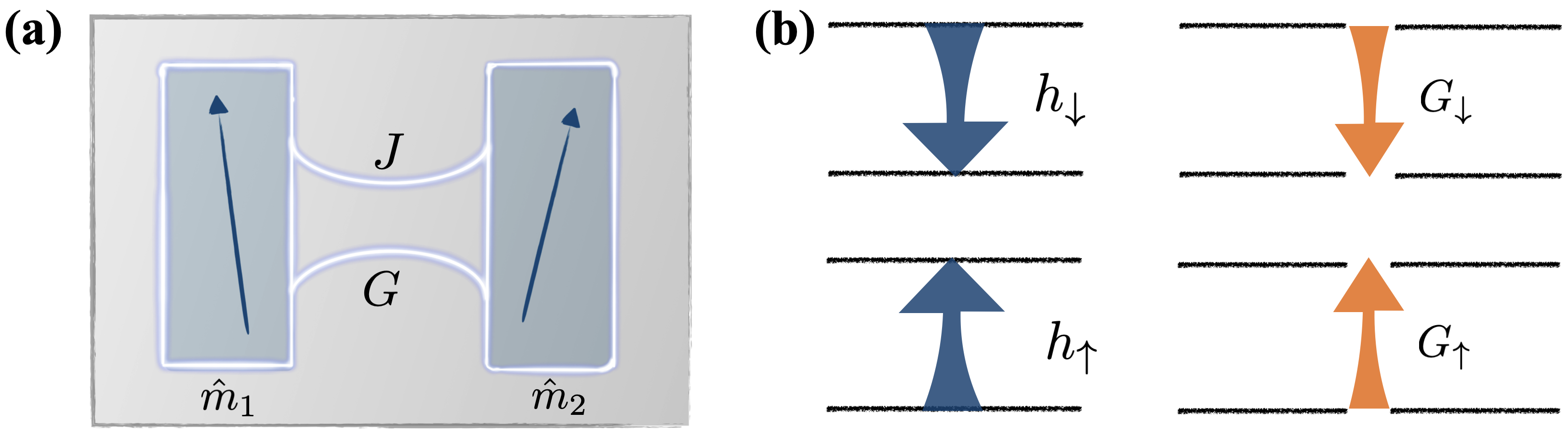}
	\caption{(a) Two macrospins described by magnon operators $\hat m_1$ and $\hat m_2$ with frequency $\omega_m$. When they are coupled to the same environment, such as a common phonon bath, there is bath-induced coherent coupling $J$  as well as dissipative coupling $G$ between the two macrospins. (b) Dissipative processes of a system consisting of two magnon modes that are interacting with the same environment. There are two types of processes. One is the local process, which is described by $h_{\uparrow}$ (local magnon pumping) and $h_{\downarrow}$ (local decay). The other is non-local process, captured by $G_{\uparrow}$ (collective pumping) and $G_{\downarrow}$ (collective decay). }
	\label{fig:332}
\end{figure}

Let us now examine the case of two magnon modes in detail, $N=2$, as shown in Fig~\ref{fig:332}(a). The Lindblad master equation is given by
\( \al{ \dv{\hat{\rho}^I}{t} &=  - \frac{i}{\hbar} [ J \hat{m}_1^\dagger \hat{m}_2 +J^* \hat{m}_1 \hat{m}_2^\dagger, \hat{\rho}^I  ] +  h_{11}\Big[ \hat{m}_1\hat{\rho}^I\hat{m}_1^\dagger - \frac{1}{2}\{ \hat{m}_1^\dagger \hat{m}_1, \hat{\rho}^I   \}   \Big]  + h_{22}\Big[ \hat{m}_2\hat{\rho}^I\hat{m}_2^\dagger - \frac{1}{2}\{ \hat{m}_2^\dagger \hat{m}_2,\hat{\rho}^I   \}   \Big] \\
& +  h_{33}\Big[ \hat{m}_1^\dagger \hat{\rho}^I\hat{m}_1 - \frac{1}{2}\{ \hat{m}_1 \hat{m}_1^\dagger ,\hat{\rho}^I   \}   \Big] +  h_{44}\Big[ \hat{m}_2^\dagger \hat{\rho}^I \hat{m}_2 - \frac{1}{2}\{ \hat{m}_2 \hat{m}_2^\dagger ,\hat{\rho}^I   \}   \Big]  + h_{12} \Big[ \hat{m}_1 \hat{\rho}^I \hat{m}_2^\dagger - \frac{1}{2}\{ \hat{m}_2^\dagger \hat{m}_1, \hat{\rho}^I \} \Big]  \\
&+  h_{21} \Big[ \hat{m}_2 \hat{\rho}^I  \hat{m}_1^\dagger - \frac{1}{2}\{ \hat{m}_1^\dagger \hat{m}_2, \hat{\rho}^I \} \Big] + h_{34}\Big[ \hat{m}_1^\dagger \hat{\rho}^I \hat{m}_2 -\frac{1}{2}\{ \hat{m}_2\hat{m}_1^\dagger, \hat{\rho}^I \}  \Big]  +  h_{43}\Big[ \hat{m}_2^\dagger \hat{\rho}^I \hat{m}_1 -\frac{1}{2}\{ \hat{m}_1\hat{m}_2^\dagger, \hat{\rho}^I \}  \Big] ,   }\)
where we have neglected the renormalization of the magnon frequency due to the environment. Here, $J \hat{m}_1^\dagger \hat{m}_2 +J^* \hat{m}_1 \hat{m}_2^\dagger$ is the environment-induced coherent coupling between two magnon modes~\cite{zou2022prb,zou2023spatially}. $J$ is generally complex and comprises both a real part, describing the symmetric exchange coupling strength, and an imaginary part, which accounts for the induced Dzyaloshinskii-Moriya interaction. The latter is nonzero only when the inversion symmetry of the environment is broken.
Assuming the system is translational invariant, we have $h_{\downarrow}\equiv h_{11} =h_{22}$, which is a real parameter describing the local magnon decay into the environment. $h_{\uparrow}\equiv h_{33} =h_{44}$ is also a real parameter describing the local magnon pumping. As we derived previously, these two processes are not independent; they are related by $h_{\downarrow}=e^{\beta \omega_m} h_{\uparrow}$ (assuming all magnon modes have the same frequency $\omega_m$). Similarly, $G_{\downarrow}\equiv h_{12}=h_{21}^*$ describes the collective magnon decay, while $G_{\uparrow}\equiv h_{34}=h_{43}^*$ represents the collective magnon pumping process from the environment, see Fig.~\ref{fig:332}(b).
One can also show that these two processes are also related to each other through $G_{\downarrow}=e^{\beta \omega_m} G_{\uparrow}$~\cite{zou2022prb}.  We remark that $G_{\uparrow,\downarrow}$ are complex in general, whose complex parts are the dissipative version of the Dzyaloshinskii-Moriya interaction. Assuming the environment is thermodynamically stable (meaning that the dissipation power of the whole system is always non-negative when the environment is subjected to external drives), we obtain the constraint~\cite{zou2022prb}:
\( h_{\uparrow}\geq |G_{\uparrow}|, \;\;\; h_{\downarrow}\geq |G_{\downarrow}|.  \)
It indicates that the local process is always stronger than the nonlocal process, as one may expect. This constraint also ensures the matrix $[h_{\mu\nu}]$
is positive-semidefinite, and thus, the system has a non-negative decay rate. 

We have thus far employed the Lindblad master equation in our treatment of the magnonic system, without immediate concern for the Born and Markov approximations. As we have previously discussed, the Born approximation requires that the frequency associated with bath-induced dynamics, including the effective magnon-magnon interaction strength denoted as $J$ and the magnon damping rate $h_{ij}$, should be smaller than the magnon resonance frequency,  $\omega_m$. In a typical spintronic heterostructure consisting of a magnetic layer, a nonmagnetic spacer, and another magnetic layer, the Gilbert damping parameter typically falls within the range of $10^{-3} \sim 10^{-2}$, placing the damping rate within the MHz regime. Besides, experimentally reported effective magnon-magnon coupling strengths also tend to be in the MHz regime~\cite{PhysRevLett.64.2304,PhysRevLett.66.2152}. Thus, in such cases, the Born approximation is well-justified. For the Markov approximation, we require the phonon lifetime (assuming we have a phonon bath for concreteness) to be shorter than the bath-induced magnon dynamics (in $\mu$s regime if we assume the frequency scale is about MHz). The phonon lifetime usually varies from picoseconds to nanoseconds. In this case, the Markov approximation applies. However, we point out that, for certain high-quality materials, the phonon lifetime may reach sub-microsecond [such as gadolinium gallium garnet (GGG)]~\cite{an2020coherent,PhysRevX.12.011060}. As a result, a clear hybrid mode (magnetoelastic mode) forms, allowing information to oscillate between the magnon and phonon modes, consequently leading to the breakdown of the Markov approximation. One may also enhance the effective magnon-phonon coupling to the GHz regime~\cite{berk2019strongly}. In this scenario, both the Born and Markov approximations would be invalidated.

\subsubsection{From magnon master equation to non-Hermitian magnon Hamiltonian}
\label{sec:qjump}

\begin{figure}[!htp]
	\centering
	\includegraphics[width=13.6cm]{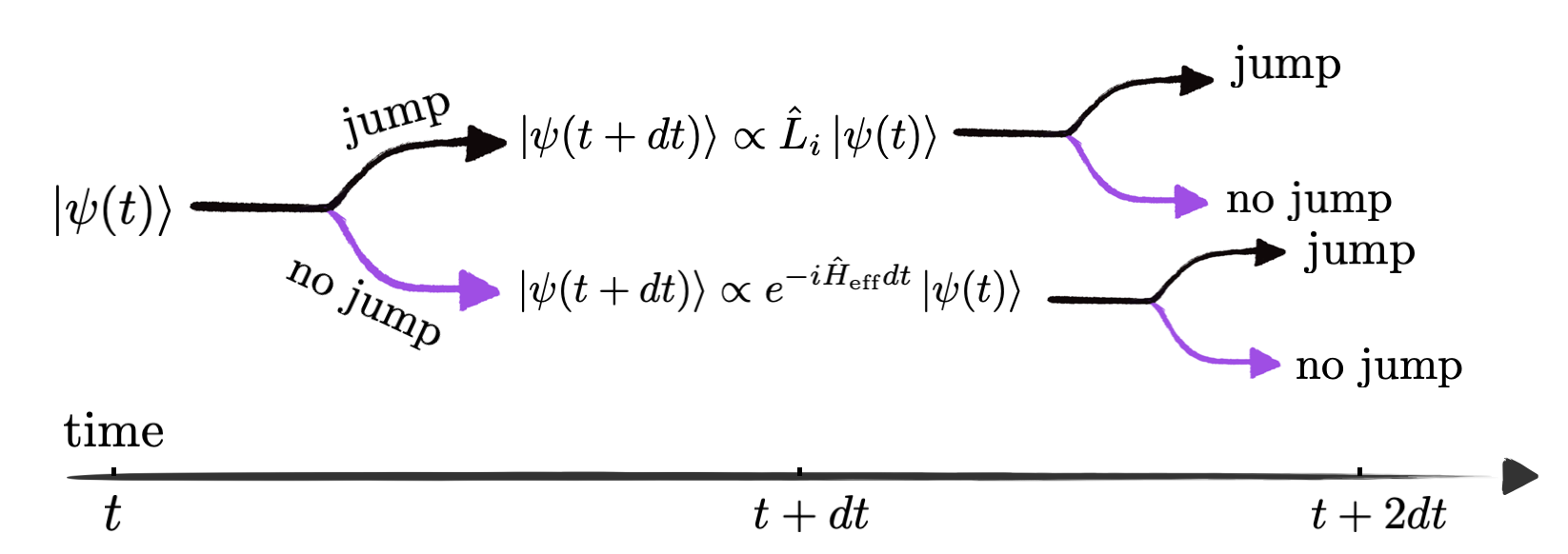}
	\caption{Quantum trajectories. A state $\ket{\psi(t)}$ at time $t$ may evolve to a state $\ket{\psi(t+dt)}\propto L_i \ket{\psi(t)} $ at time $t+dt$ with probability $p_i=dt \bra{\psi(t)}L_i^\dagger L_i \ket{\psi(t)}$ or to the state $\ket{\psi(t+dt)} \propto e^{-iH_{{\rm eff}} dt }\ket{\psi(t)}$ with probability $1-\sum_i p_i$. The evolution from time $t+dt$ to $t+2dt$ follows the same rule.  }
	\label{fig:3332}
\end{figure}

In this section, we show how to obtain a non-Hermitian magnon Hamiltonian from its Lindblad master equation. To this end, we first introduce the quantum trajectory theory, which is an interpretation of the Lindblad master equation from  a quantum  measurement perspective~\cite{daley2014quantum,Plenio1998rmp,PhysRevA.36.5543,PhysRevLett.70.548}.
Let us consider the following Lindblad equation in Schr{\"{o}}dinger picture (we will drop $\hbar$ hereafter in this section for notational simplicity): 
\( \dv{\hat{\rho}}{t}=-i[\hat{H},\hat{\rho}] +\sum_i \big[ \hat{L}_i\hat{\rho} \hat{L}_i^\dagger - \frac{1}{2}\{\hat{L}^\dagger_i \hat{L}_i, \hat{\rho}\} \big], \label{eq:lindblad} \)
which can be regarded as a differential map with operators $\hat{K}_i$: 
\( \hat{\rho}(t+dt)=\hat{K}_0(dt) \hat{\rho}(t) \hat{K}_0(dt) +\sum_{i>0} \hat{K}_i(dt) \hat{\rho}(t) \hat{K}_i^\dagger(dt),   \)where $ \hat{K}_i(dt)=\sqrt{dt} \hat{L}_i$, and $\hat{K}_0(dt)=\hat{I}-i\hat{H}_{\text{eff}}dt$. Here we have written
$ \hat{H}_{\text{eff}}=\hat{H}-i\sum_i\hat{L}^\dagger_i \hat{L}_i/2,$
which is non-Hermitian. The real part is the Hamiltonian, and the imaginary part accounts for the decay. 
Let us suppose that, at time $t$, the state is pure, $\ket{\psi(t)}$. Then, at time $t+dt$, the state is given by 
\( \hat{\rho}(t+dt)=(1-i\hat{H}_{\text{eff}} dt) \ket{\psi(t)}\bra{\psi(t)} (1+i \hat{H}^\dagger_{\text{eff}} dt) +dt \sum_i \hat{L}_i \ket{\psi(t)}\bra{\psi(t)}  \hat{L}_i^\dagger.  \)
From the generalized measurement theory,  we have the following interpretation of the differential map. With  probability $p_i=\bra{\psi(t)}\hat{K}_i^\dagger(dt)\hat{K}_i(dt)\ket{\psi(t)}=dt \bra{\psi(t)}\hat{L}^\dagger_i \hat{L}_i\ket{\psi(t)}$, the state jumps: $\ket{\psi(t)}\rightarrow \hat{L}_i\ket{\psi(t)}$, as shown in Fig.~\ref{fig:3332}.
    $\hat{L}_i$ are thereby known as jump operators. We observe that the probability $p_i$ is directly proportional to the time interval $dt$, implying that the jump process is less likely to occur in short time scales with $\Delta t\ll 1/\bra{\psi(t)}\hat{L}_i^\dagger \hat{L}_i\ket{\psi(t)}$. In this case, the dynamics of the system are mainly dictated by the effective Hamiltonian $\hat{H}_{\text{eff}}$. One can also eliminate quantum jumps by using continuous measurements and postselections~\cite{wiseman2009quantum,hume2007high,negnevitsky2018repeated}. For instance, in the case of magnon modes coupled to the environment that we previously discussed, we can eliminate the decay of magnons by post-selecting the absence of any emitted bosons (such as phonons or photons) in the environment. The probability of no jump is  $p_0=1-\sum_i p_i$.
      In this case, the evolution of the state is governed by the effective non-Hermitian Hamiltonian: $\ket{\psi(t)} \Rightarrow \hat{K}_0(dt)\ket{\psi(t)} =e^{-i\hat{H}_{\text{eff}}dt}\ket{\psi(t)}$ (see Fig.~\ref{fig:3332}),  
where we have used $e^{-i\hat{H}_{\text{eff}}dt}\approx 1-i\hat{H}_{\text{eff}}dt$.

\textbf{Non-Hermitian magnon Hamiltonian by disregarding the quantum jump.}
The quantum jump effect is insignificant in a short-time regime, as we detailed above, or can be eliminated by employing post selections~\cite{wiseman2009quantum,hume2007high,negnevitsky2018repeated}. In these cases, the Lindblad master equation is reduced to the non-Hermitian formalism. To this end, we rewrite the Lindblad master equation~\eqref{eq:lindblad} into the form of  
\(   \dv{\hat{\rho}}{t} = -i [\hat{H}_{\text{eff}}, \hat{\rho}] + \sum_i  \hat{L}_i \hat{\rho} \hat{L}_i^\dagger, \;\;\;\text{with}\;\;\; \hat{H}_{\text{eff}} = \hat{H}-\frac{i}{2} \sum_i \hat{L}^\dagger_i \hat{L}_i. \label{eq:jump}  \)
Here, $\hat{H}_{\text{eff}}$ is non-Hermitian and the commutator should be understood as
$[\hat{H}_{\text{eff}}, \hat{\rho}] \equiv \hat{H}_{\text{eff}} \hat{\rho} -\hat{\rho} \hat{H}_{\text{eff}}^\dagger$. It is clear that in the absence of quantum jumps [the last term in ~\eqref{eq:jump}], the dynamics of the system are governed by $\hat{H}_{\text{eff}}$. It has complex eigenvalues in general, whose real parts are effective values of energies, while the imaginary parts stand for the rates at which the corresponding eigenstates decay. 

It is useful to introduce the evolution operator $\hat{U}(t)=e^{-i\hat{H}_{\text{eff}} t}$. We note that  $\hat{U}^\dagger(t)=e^{-i\hat{H}^\dagger_{\text{eff}} t}$ and the evolution is not unitary in general. The differential equation for $\hat{U}(t)$ is given by
\(  \dv{\hat{U}(t)}{t} = -i \hat{H}_{\text{eff}} \hat{U}, \;\;\; \text{and}\;\;\; \dv{\hat{U}^\dagger(t)}{t} = i \hat{H}^\dagger_{\text{eff}} \hat{U}^\dagger,  \)
from which one can verify that
\(  \dv{\hat{\rho}}{t}\equiv \dv{}{t} \hat{U}(t)\hat{\rho}_0 \hat{U}^\dagger(t)=-i( \hat{H}_{\text{eff}} \hat{\rho} - \hat{\rho} \hat{H}^\dagger_{\text{eff}} )=-i[\hat{H}_{\text{eff}}, \hat{\rho} ].    \)
We remark that this dynamics does not preserve the trace (as we neglect the quantum jump). Its integration is appropriately normalized to give
\( \hat{\rho}(t) = \frac{ \hat{U}(t) \hat{\rho}_0 \hat{U}^\dagger(t) }{ \tr \big[ \hat{U}(t) \hat{\rho}_0 \hat{U}^\dagger(t)  \big]  }.     \)
 
As an example,  let us take the example of $N$ magnon modes coupled with an environment, which we discussed in detail in Sec.~\ref{bilinear_coupling}. In the absence of quantum jumps, the dynamics are governed by the  following non-Hermitian Hamiltonian 
\(  \hat{H}_{\text{eff}}= \big(\omega_m - \frac{i}{2}h\big) \sum_i \hat{m}_i^\dagger \hat{m}_i    + \big( J-\frac{i}{2}G \big) \sum_{i}   \hat{m}_i^\dagger \hat{m}_{i+1} + \big(J^*- \frac{i}{2}G^*\big) \sum_{i}  \hat{m}_{i}\hat{m}_{i+1}^\dagger, \)
where $h\equiv h_\downarrow + h_\uparrow >0$ describes the local dissipative effect  and $G=G_\uparrow+G^*_\downarrow \in \mathbb{C}$ stands for the dissipative coupling mediated by the environment. Here we have only taken the environment-induced nearest coherent and dissipative couplings into account, while the generalization to long-ranged couplings is straightforward. 
An interesting observation is that the local decay factor $h_\downarrow$ and the local pump factor $h_\uparrow$ (as well as $G_\downarrow$ and $G_\uparrow^*$) in the presented Hamiltonian exhibit similar effects when quantum jumps are disregarded.

\textbf{Mean-field non-Hermitian Hamiltonian.}
We can also obtain an effective non-Hermitian Hamiltonian description for the mean-field dynamics from a full Lindblad master equation. 
To illustrate this,  we consider the two-macrospin case (two magnon modes) with the following Lindblad master equation (in the Schr\"{o}dinger picture) for concreteness: 
\( \dv{\hat{\rho}}{t}=-i[\hat{H}, \hat{\rho}] +\mathcal{L}[\hat{\rho}],  \)
where the coherent Hamiltonian is given by 
\(   \hat{H}= \omega_m(\hat{m}_1^\dagger \hat{m}_1 + \hat{m}_2^\dagger \hat{m}_2) +  J \hat{m}_1^\dagger \hat{m}_2 +J^* \hat{m}_1 \hat{m}_2^\dagger,   \)
and the dissipative Lindbladian is
\( \mathcal{L}[\hat{\rho}] = h_{\downarrow} \sum_{i=1,2}\Big[ \hat{m}_i{\rho} \hat{m}_i^\dagger - \frac{1}{2}\{ \hat{m}_i^\dagger \hat{m}_i, \hat{\rho}   \}   \Big]  + G_{\downarrow} \Big[ \hat{m}_1 \hat{\rho} \hat{m}_2^\dagger - \frac{1}{2}\{ \hat{m}_2^\dagger \hat{m}_1, \hat{\rho} \} \Big]+  G_{\downarrow}^* \Big[ \hat{m}_2 \hat{\rho} \hat{m}_1^\dagger - \frac{1}{2}\{ \hat{m}_1^\dagger \hat{m}_2, \hat{\rho} \} \Big]. \label{eq:335lind}  \)
Here, we have assumed zero temperature for simplicity such that only local and nonlocal decay processes survive. Let us now explore the dynamics of the mean values of $\hat{m}_1, \hat{m}_2$, defined by $\langle \hat{m}_i \rangle \equiv \tr(\hat{\rho} \hat{m}_i)\propto \langle \hat{m}_i^x\rangle-i \langle \hat{m}_i^y\rangle $.  To this end, we only need to evaluate the following terms:
\( \dv{\langle \hat{m}_i \rangle}{t} = -i\langle [\hat{m}_i, \hat{H}]\rangle +\tr( \hat{m}_i \mathcal{L}[\hat{\rho}] ).  \)
Here we provide the details for the mode $\hat{m}_1.$ The contribution from the coherent part is
\( -i \langle [\hat{m}_1, \hat{H}] \rangle =-i\omega_m \langle \hat{m}_1 \rangle -i J\langle \hat{m}_2 \rangle.  \)
In $\tr( \hat{m}_1 \mathcal{L}[\hat{\rho}] )$, we have four terms corresponding to the four terms in the Lindbladian~\eqref{eq:335lind}. The first term is 
\(  h_\downarrow \tr\Big[ \hat{m}_1 \Big(  \hat{m}_1 \hat{\rho} \hat{m}_1^\dagger - \frac{1}{2}\{ \hat{m}_1^\dagger \hat{m}_1, \hat{\rho} \}    \Big)   \Big] = h_\downarrow \Big( \langle \hat{m}_1^\dagger \hat{m}_1 \hat{m}_1  \rangle - \frac{1}{2} \langle \hat{m}_1 \hat{m}_1^\dagger \hat{m}_1 \rangle  - \frac{1}{2} \langle \hat{m}_1^\dagger \hat{m}_1 \hat{m}_1 \rangle \Big) = \frac{h_\downarrow}{2} \langle [\hat{m}_1^\dagger, \hat{m}_1] \hat{m}_1\rangle= - \frac{h_\downarrow}{2} \langle \hat{m}_1 \rangle.   \)
It describes the damping of $\langle \hat{m}_1\rangle$, where $h_\downarrow$ is the local Gilbert damping. It is important to note that the effect of the quantum jump is not disregarded in this treatment.
The second term is 
\(  h_\downarrow \tr\Big[ \hat{m}_1 \Big(  \hat{m}_2 \hat{\rho} \hat{m}_2^\dagger - \frac{1}{2}\{ \hat{m}_2^\dagger \hat{m}_2, \hat{\rho} \}    \Big)   \Big] = h_\downarrow \Big(  \langle \hat{m}_2^\dagger \hat{m}_2 \hat{m}_1  \rangle - \frac{1}{2}\langle \hat{m}_2^\dagger \hat{m}_2 \hat{m}_1  \rangle - \frac{1}{2}\langle \hat{m}_2^\dagger \hat{m}_2 \hat{m}_1  \rangle  \Big)=0,  \)
which implies that the local decay process of the second magnon mode does not impact the dynamics of $\langle \hat{m}_1\rangle$, as one may expect.  The third term is given by
\( G_\downarrow \tr\Big[ \hat{m}_1 \Big( \hat{m}_1 \hat{\rho} \hat{m}_2^\dagger - \frac{1}{2} \{ \hat{m}_2^\dagger \hat{m}_1, \hat{\rho} \}   \Big)   \Big] =G_\downarrow \Big( \langle \hat{m}_2^\dagger \hat{m}_1 \hat{m}_1 \rangle  - \frac{1}{2} \langle \hat{m}_2^\dagger \hat{m}_1 \hat{m}_1 \rangle - \frac{1}{2} \langle \hat{m}_2^\dagger \hat{m}_1 \hat{m}_1 \rangle   \Big)=0,  \)
and the fourth term is 
\( G^*_\downarrow \tr \Big[  \hat{m}_1 \Big( \hat{m}_2 \hat{\rho} \hat{m}_1^\dagger - \frac{1}{2} \{ \hat{m}_1^\dagger \hat{m}_2, \hat{\rho} \}   \Big)   \Big]  =  G^*_\downarrow  \Big(  \langle \hat{m}_1^\dagger \hat{m}_1 \hat{m}_2\rangle - \frac{1}{2} \langle \hat{m}_1 \hat{m}_1^\dagger \hat{m}_2 \rangle - \frac{1}{2} \langle \hat{m}_1^\dagger \hat{m}_1 \hat{m}_2  \Big)= \frac{G_\downarrow^*}{2} \langle [\hat{m}_1^\dagger, \hat{m}_1] \hat{m}_2 \rangle=- \frac{G_\downarrow^*}{2} \langle \hat{m}_2 \rangle.  \)
Therefore, the dynamics of the mean-field $\langle \hat{m}_1\rangle$ is governed by
\(  \dv{\langle \hat{m}_1\rangle}{t} = \Big( -i\omega_m - \frac{h_\downarrow}{2}  \Big)\langle \hat{m}_1\rangle +\Big( -iJ - \frac{G^*_\downarrow}{2}  \Big) \langle \hat{m}_2\rangle , \)
and, similarly, the dynamics of $\langle \hat{m}_2\rangle$ is 
\(  \dv{\langle \hat{m}_2\rangle}{t} =  \Big( -i\omega_m - \frac{h_\downarrow}{2}  \Big)\langle \hat{m}_2\rangle    +\Big( -iJ^* - \frac{G_\downarrow}{2}  \Big) \langle \hat{m}_1\rangle.   \)
We point out that they are the linearized Landau-Lifshitz-Gilbert equations~\cite{landau2013statistical}. We can recast the two equations above into a more compact Schor\"{o}dinger-like equation with a non-Hermitian Hamiltonian: 
\( i\dv{\vec \psi}{t} = \mathcal{H}  \vec \psi, \)
with $\Vec{\psi}=(\langle \hat{m}_1\rangle, \langle \hat{m}_2\rangle  )^T$ and  
\(   \mathcal{H} = \mqty[ \omega_m - \frac{ih_\downarrow}{2} &  J- \frac{iG^*_\downarrow}{2} \\ J^*-i \frac{G_\downarrow}{2} & \omega_m - \frac{ih_\downarrow}{2}  ].  \)

Finally, we remark that the spin-pumping process can also be modeled by a Lindbladian. For example, let us pump the first macroscopic magnetic spin
 by subjecting it to spin-transfer~\cite{slonczewski1996current,berger1996emission} or spin Seebeck  torques~\cite{bauer2012spin}, which can be described by 
 \(  \mathcal{L}^\text{pump}[\hat\rho]= \gamma_{\uparrow} \Big[  \hat{m}_1^\dagger \hat\rho \hat{m}_1 - \frac{1}{2} \{ \hat{m}_1 \hat{m}_1^\dagger, \hat\rho \} \Big],   \) 
with the pumping rate being $\gamma_{\uparrow}>0$. This would effectively shift the local damping $h_\downarrow$ to a smaller value, which can be seen from its contribution to the equation of motion of $\langle \hat{m}_1\rangle$: 
\( \tr(\hat{m}_1 \mathcal{L}^{\text{pump}} [\hat\rho]) = \gamma_{\uparrow} \Big[  \langle \hat{m}_1 \hat{m}_1 \hat{m}_1^\dagger \rangle - \frac{1}{2}\langle \hat{m}_1 \hat{m}_1 \hat{m}_1^\dagger \rangle - \frac{1}{2} \langle \hat{m}_1 \hat{m}_1^\dagger \hat{m}_1\rangle  \Big] =\frac{\gamma_{\uparrow}}{2} \langle \hat{m}_1 [\hat{m}_1, \hat{m}_1^\dagger]\rangle=\frac{\gamma_{\uparrow}}{2} \langle \hat{m}_1\rangle.   \)
Then the equation for $\langle \hat{m}_1 \rangle$ is given by
\(  \dv{\langle \hat{m}_1\rangle}{t} = \Big( -i\omega_m - \frac{h_\downarrow-\gamma_{\uparrow}}{2}  \Big)\langle \hat{m}_1\rangle +\Big( -iJ - \frac{G^*_\downarrow}{2}  \Big) \langle \hat{m}_2\rangle , \)
where we see that the local damping $h_\downarrow$ is reduced to $h_\downarrow - \gamma_{\uparrow}$ as expected~\cite{bender2012electronic,bender2014dynamic}. We  highlight that the signs before the parameters $h_\downarrow$ and $\gamma_{\uparrow}$ are directly linked to the jump terms in the respective Lindbladians. The opposite signs of these parameters are dictated by their physical interpretations, with $h_\downarrow$ representing the decay process and $\gamma_{\uparrow}$ corresponding to the pumping process.

\subsection{Green-function approach for magnon non-Hermitian dynamics}
\label{Green_function}

In this review article, we focus on the non-Hermitian topological states of magnons when they interact with other degrees of freedom. Thus we mainly discuss Green's function of bosons, which allows us to describe the response of the system at any point due to the excitation at any other point.
By employing the Green-function approach, we can effectively capture the effects of magnon-bath interactions through the self-energy, which can be viewed as an effective Hamiltonian arising from the interaction of the magnon subsystem with the bath. In Sec.~\ref{sec:3.4.1}, we first review the magnon Green's function, where an effective non-Hermitian Hamiltonian is introduced by considering the magnon self-energy. 
Following this, in Sec.~\ref{sec:3.4.2}, we  study an example with bilinear magnon-quasiparticle interaction, and compare this Green's function approach with the Lindblad master equation approach that we discussed before.

\subsubsection{Green's function of magnons}
\label{sec:3.4.1}
We consider the following Hamiltonian 
\( \hat{H}=\hat{H}_0+\hat{H}_E+\hat{H}_{\text{SE}},  \)
where $\hat{H}_0=\sum_{\vb k} h(\vb k) \hat{m}_{\vb k}^\dagger \hat{m}_{\vb k}$ is the free magnonic Hamiltonian, $\hat{H}_E$ describes the environment, and $\hat{H}_{\text{SE}}$ stands for their interactions. Let us focus on the retarded Green's function of single-magnon defined as~\cite{bruus2004many,altland2010condensed}
\( G^R(\vb r, \vb r^\prime; t, t^\prime)=-i\theta(t-t^\prime) \langle [\hat{m}_{{\bf r}}(t), \hat{m}_{{\bf r}^\prime}^\dagger (t^\prime)] \rangle,  \)
where $\theta(t)$ is the step function,  the magnon operator is in the Heisenberg picture $\hat{m}_{\vb r}(t)=e^{i\hat{H} t} \hat{m}_{\vb r} e^{-i\hat{H} t}$, and $\langle \hat{\mathcal{O}}\rangle\equiv \tr (e^{-\beta \hat{H}} \hat{\mathcal{O}} )/Z$ with $Z=\tr e^{-\beta \hat{H}}$. We will refer to such retarded Green's function simply as ``Green's function''. 
It is worth noting that we define Green's function in the real space and time domain. When the system exhibits time and space translational symmetries, it is convenient to work with $G^R(\vb k, \omega)$ in momentum and frequency space through Fourier transformations. 
It is well-known that the complex poles of Green's function $G^R(\vb k, \omega)$ of magnons determine the magnon spectrum. For example, in the case of free magnon gas $\hat H=\hat H_0$, we can evaluate the Green's function easily with the method of the equation of motion~\cite{bruus2004many,altland2010condensed}: 
\( i\partial_t G^R(\vb k; t, t^\prime)=\delta (t-t^\prime) -i\theta(t-t^\prime) \langle [i\partial_t  \hat{m}_{\vb k}(t), \hat{m}^\dagger_{\vb k}(t^\prime)] \rangle = \delta(t-t^\prime) +h(\vb k) G^R(\vb k, t, t^\prime), \)
where $G^R(\vb k; t, t^\prime)=-i\theta(t-t^\prime) \langle [\hat{m}_{\vb k}(t), \hat{m}_{\vb k}^\dagger (t^\prime)] \rangle,$ and we have used $i\partial_t \hat{m}_{\vb k}(t)=[\hat{m}_{\vb k}(t), \hat{H}_0]$. When we perform the Fourier transformation for the equation above, we obtain
\(  (\omega+i\eta)  G^R(\vb k, \omega)=1+ h(\vb k) G^R(\vb k, \omega),  \)
which  gives us the free  Green's function of magnons: 
\( G^R(\vb k, \omega) = \frac{1}{\omega-h(\vb k)+i\eta}.  \)
Here $G^R(\vb k, \omega) = \int^\infty_{-\infty}dt\, e^{i\omega t} G^R(\vb k, t)$, and   we remark that in order to ensure convergence of the Fourier transform of the Green's function, which may suffer from some ringing in the future, we need to make the replacement $\omega\rightarrow \omega+i\eta$, where $\eta$ is a positive infinitesimal.

In general, in the presence of the environment and interactions, the magnon Green's function takes the form of 
\(  G^R(\vb k, \omega)= \frac{1}{\omega - h(\vb k) -\Sigma^R(\vb k, \omega) },   \)
where $\Sigma^R(\vb k, \omega) $ is known as \textit{self-energy}, whose real part renormalizes the bare magnon spectrum and the imaginary part gives rise to finite magnon lifetimes. 
Evaluating the self-energy exactly is typically a difficult task. Instead, a perturbative expansion can be employed in terms of the magnon-bath interaction Hamiltonian. The self-energy is defined as the sum of all diagrams in the perturbative expansion that have two external magnon lines and any number of internal bath lines. The self-energy diagram can be represented as a loop with internal magnon and bath propagators, which can be computed using standard diagrammatic techniques such as Feynman diagrams~\cite{bruus2004many,altland2010condensed}.

We can now introduce an effective Hamiltonian: 
\( H_{\text{eff}}(\vb k, \omega)= h(\vb k)+ \Sigma^R(\vb k, \omega),  \)
which is non-Hermitian in general. Then the magnon Green's function can be written as 
\( G^R(\vb k, \omega)=  \frac{1}{\omega - H_{\text{eff}}(\vb k, \omega)}.  \)
In contrast to the Lindblad master equation approach, which relies on the Born-Markov approximation, the Green-function approach discussed here can capture non-Markovian effects of the environment by noting the frequency dependence in the self-energy. It can also go beyond the second order in the magnon-environment coupling in general. However, it is important to note that the Green-function approach presented here only captures the spectrum properties of magnons. To also capture the statistical properties, one needs to invoke the Keldysh Green's function, with which one can establish the equivalence between the Lindblad master equation approach and the Keldysh formalism in the limit of weak magnon-environment coupling~\cite{sieberer2016keldysh,kamenev2023field}.

\subsubsection{Bilinear magnon-quasiparticle interaction}
\label{sec:3.4.2}
To illustrate the Green-function method and compare it with the Lindblad master equation approach, we consider the same model that we previously studied using the Lindblad master equation method in Sec.~\ref{bilinear_coupling}, consisting of a single magnon mode with frequency $\omega_m$ that interacts with an environment with Hamiltonian:
\( \hat{H}=\omega_m \hat{m}^\dagger \hat{m} + \sum_{\vb k} ( g_{\vb k} \hat{m}^\dagger \hat{b}_{\vb k} +g^*_{\vb k} \hat{m} \hat{b}^\dagger_{\vb k} ) +\sum_{\vb k} \epsilon_{\vb k} \hat{b}^\dagger_{\vb k} \hat{b}_{\vb k}.   \)
We introduce two Green's functions: 
\(  G^R(t)=-i\theta(t)\langle [\hat{m}(t), \hat{m}^\dagger]\rangle, \;\;\; \text{and} \;\;\; \tilde{G}^R(\vb k, t)=-i\theta(t) \langle [ \hat{b}_{\vb k}(t), \hat{m}^\dagger] \rangle.    \)
Since the Hamiltonian is quadratic we can solve the Green's function of magnons exactly. 
To this end,  let us write down the equations of motion for these two Green's functions: 
\( i\partial_t G^R(t)=\delta(t)+\omega_m G^R(t) +\sum_{\vb k} g_{\vb k} \tilde{G}^R(\vb k, t),\;\;\;\;\;\;  i\partial_t \tilde{G}^R(\vb k, t) = \epsilon_{\vb k} \tilde{G}^R(\vb k, t) +g^*_{\vb k} G^R(t).  \)
We perform Fourier transformations on these two equations, which results in two algebraic equations that can be exactly solved:
\(   G^R(\omega)= \frac{1}{\omega-\omega_m -\Sigma^R(\omega)}, \;\;\; \text{with}\;\;\;\Sigma^R(\omega) = \sum_{\vb k} \frac{|g_{\vb k}|^2}{\omega -\epsilon_{\vb k} +i\eta } \equiv  \Sigma^\prime(\omega) -i\Sigma^{\prime\prime}(\omega),     \)
with 
\(  \Sigma^\prime(\omega) = P \sum_{\vb k} \frac{|g_{\vb k}|^2}{\omega-\epsilon_k}, \;\;\;  \Sigma^{\prime\prime}(\omega) =\pi \sum_{\vb k} |g_{\vb k}|^2 \delta (\omega-\epsilon_k).   \)
So far, we do not make any approximation. Assuming the coupling between the magnon mode and the environment is weak, we can approximate $\Sigma^R(\omega) \approx \Sigma^R(\omega_m)$.
Then we can write the effective Hamiltonian of the magnon mode as: 
\( \hat{{H}}_{\rm eff}=\big[\omega_m +\Sigma^\prime(\omega_m) -i \Sigma^{\prime\prime}(\omega_m)\big] \hat{m}^\dagger \hat{m}, \)
implying that the lamb shift of the frequency is $\Sigma^\prime(\omega_m)$ and the decay rate of this mode is $\Sigma^{\prime\prime}(\omega_m)$.
We point out that this result is consistent with what we have obtained from the Lindblad master equation in the zero temperature case by noting that $\Sigma^\prime(\omega_m)=\omega_1$ and $\Sigma^{\prime\prime}(\omega_m)=\Gamma_{\downarrow}/2$. As discussed in the previous section, the Green-function approach accurately captures the spectral properties of the magnon mode subjected to an environment.

Throughout this review article, we assume that the dissipative (magnonic) subsystem can be well addressed in terms of the non-Hermitian Hamiltonian $\hat{H}_{\mathrm{eff}}$, with the associated dynamics governed by the  Heisenberg equation of motion. Since $\hat{H}_{\mathrm{eff}}$ is non-Hermitian, it has to be diagonalized by introducing the left and right
eigenvectors. The right eigenvectors, say $\{\psi_{\zeta}\}$ with corresponding eigenvalues $\{\gamma_{\zeta}\}$ for a mode with label $\zeta$, satisfy 
\begin{align}
\hat{H}_{\rm eff} \psi_{\zeta}=\gamma_{\zeta}\psi_{\zeta}.
\end{align}
Here $\mathrm{Re} (\gamma_{\zeta})$ and $\mathrm{Im}(
\gamma_{\zeta})$ are the resonance frequency and the reciprocal lifetime, respectively. $\{\phi_{\zeta}\}$ are the eigenvectors of $\hat{H}_{\rm eff}^{\dagger}$ with eigenvalues $\{\gamma_{\zeta}^{\ast}\}$:
\begin{align}
\hat{H}^{\dagger}_{\rm eff} \phi_{\zeta}=\gamma^*_{\zeta}\phi_{\zeta}.
\end{align}
In the absence of degeneracies in $\{\gamma
_{\zeta}\}$ the (normalized) modes are \textquotedblleft
bi-orthonormal\textquotedblright, i.e.,
\begin{align}
    \phi_{\zeta}^{\dagger}\psi
_{\zeta^{\prime}}=\delta_{\zeta\zeta^{\prime}}.
\end{align}
Since the eigenvalues $\gamma_{\xi}$ of the non-Hermitian matrix are generally complex and the eigenvectors have additional freedom, there are many exotic properties that do not exist in the Hermitian realm.

\section{Topological characterization: from Hermitian to non-Hermitian systems}

\label{Exceptional_topology}

This review article focuses on the exceptional topological phases or properties in the non-Hermitian magnonic systems or devices, including the EPs (Sec.~\ref{Magnonic_EPs}), the exceptional nodal phases (Sec.~\ref{exceptional_nodal_phases_magnons}), the non-Hermitian topological edge state (Sec.~\ref{Non_Hermitian_skin_effect_magnon}), and the non-Hermitian skin effect (Sec.~\ref{Non_Hermitian_skin_effect_magnon}). Here we provide a comprehensive introduction to the general topological characterization of these non-Hermitian topological phases or properties, highlighting their unconventional aspects in comparison with those in the Hermitian scenario.

Topology in mathematics studies the intrinsic properties that remain unchanged under continuous deformation. These preserved intrinsic properties correspond to topological invariants, which are typically related to the integrals of some local quantities over a closed parameter space. An illustrative example from the textbook is the number of holes (genus) that remains unchanged as a donut is smoothly reshaped into a handle coffee cup. 
In this case, the genus is the topological invariant, given by an integration of the Gaussian curvature over a closed two-dimensional surface. Therefore, closed surfaces with different numbers of holes, such as a sphere and a donut, are topologically distinct and cannot be deformed into each other smoothly. The study of topology began in the twentieth century, while many other fields of mathematics, including calculus, had already been developed three hundred years earlier. The evolution of physical science has underscored the vital significance of topology in modern condensed matter physics. Its influence spans from Dirac's pioneering investigations into magnetic monopoles~\cite{dirac1931quantised} to the classification of topological defects within ordered phases~\cite{mermin1979topological}, the Berezinskii-Kosterlitz-Thouless transitions~\cite{kosterlitz2018ordering}, the dynamics of spin chains~\cite{haldane1983continuum,haldane1983nonlinear}, the quantum Hall effects~\cite{von1986quantized}, topological insulators~\cite{hasan2010colloquium,qi2011topological} and semimetals~\cite{armitage2018weyl} in Hermitian systems over the past decade, and, notably, the recent surge of interest in the exceptional topology in non-Hermitian systems, i.e., unconventional topological characterizations or properties that are distinguished from the Hermitian counterpart~\cite{bergholtz2021exceptional}.

The Hermitian topology concerns topological properties and phenomena in closed systems governed by Hermitian Hamiltonians~\cite{hasan2010colloquium,qi2011topological,bernevig2013topological}, while the non-Hermitian topology explores the topological features of open systems, characterized by non-unitary evolutions~\cite{kawabata2019symmetry,ashida2020non,bergholtz2021exceptional,ding2022non}. Before elucidating the detailed mechanisms behind various topological phenomena, we start by offering an overview of Hermitian and non-Hermitian topological characterization from two fundamental aspects of the Hamiltonian: \textit{eigenvalues and eigenvectors}. We emphasize the correspondence between topological invariants and topological phenomena, as summarized in Table~\ref{category}.
We list representative models of  Hermitian and non-Hermitian systems in the second column, which we use to elucidate topology-related terminologies and concepts. 
For instance, the Bogoliubov-de-Gennes (BdG) Hamiltonian is addressed in Sec.~\ref{BDGHamiltonian}. We discuss the Hermitian and non-Hermitian SSH model in Sec.~\ref{SSH_model_summary}, explore the Hatano-Nelson model in Sec.~\ref{Nonhermitian_skin_effect}, delve into the two-sheeted Riemann surface involving second-order EPs in Sec.~\ref{exceptionaltopology}, and examine the non-Hermitian Weyl semimetal in Sec.~\ref{Nodal_phase}. In the third column, we outline two types of topological characterization: wavefunction topology, associated with eigenvectors, which is common to both Hermitian and non-Hermitian systems; and spectral topology, associated with eigenvalues, exclusive to non-Hermitian systems. In the fourth and fifth columns, we present the correspondence between topological invariants and physical phenomena.

\begin{table}
\caption{Comparison of Hermitian and non-Hermitian topological characterization.}
\begin{tabular}{|c|c|c|cc|}
\hline
Systems                        & Representative model                    & Characterization                             & \multicolumn{2}{c|}{Correspondence}   \\ \hline
Hermitian                      & \begin{tabular}[c]{@{}c@{}}SSH model,\\ BdG Hamiltonian\\ Weyl semimetal, etc.\end{tabular}                          & \multirow{2}{*}{\begin{tabular}[c]{@{}c@{}}\textbf{Wavefunction} \\ \textbf{topology}\end{tabular}} & \multicolumn{1}{c|}{\multirow{2}{*}{\begin{tabular}[c]{@{}c@{}}Winding numebr,\\ Chern number,\\$Z_2$ invariant, \\ etc.\end{tabular}}} & \multirow{2}{*}{\begin{tabular}[c]{@{}c@{}}Topological \\ edge state, \\ Surface Fermi arcs, \\ etc. \end{tabular}} \\ \cline{1-2}
\multirow{4}{*}{Non-Hermitian} & \begin{tabular}[c]{@{}c@{}}Non-Hermitian SSH \\ model, etc.\end{tabular}     &                                       & \multicolumn{1}{c|}{}                 &                                       \\ \cline{2-5} 
& \begin{tabular}[c]{@{}c@{}}Hatano-Nelson \\ model, etc.\end{tabular}            & \multirow{3}{*}{\textbf{Spectral topology}}    & \multicolumn{1}{c|}{\begin{tabular}[c]{@{}c@{}}Spectral winding \\ number\end{tabular}}                      & \begin{tabular}[c]{@{}c@{}}Non-Hermitian \\ skin effect\end{tabular}               \\ \cline{2-2} \cline{4-5} 
& \multirow{2}{*}{\begin{tabular}[c]{@{}c@{}}Two-sheeted Riemann surface \\  involving  2-order EPs, \\ Non-Hermitian Weyl semimetal, etc. \end{tabular}} & & \multicolumn{1}{c|}{\multirow{2}{*}{Energy vorticity}}                     & \begin{tabular}[c]{@{}c@{}}Swapping of\\ eigenvalues\end{tabular}                  \\ \cline{5-5} 
&                                       &                                       & \multicolumn{1}{c|}{}                 & \begin{tabular}[c]{@{}c@{}}Bulk Fermi arcs\\  and surfaces\end{tabular}            \\ \hline
\end{tabular}\label{category}
\end{table}

A well-known example of a Hermitian system demonstrating the nontrivial topological states is the topological insulator~\cite{hasan2010colloquium,qi2011topological,bernevig2013topological}. While it possesses a band gap in its bulk, it features robust conducting states at its boundaries~\cite{hasan2010colloquium,qi2011topological,bernevig2013topological}. The manifestation of these edge states is dictated by the nonzero bulk topological invariants, known as the bulk-boundary correspondence. Every energy band can be linked to a topological index by examining the topology of its associated wavefunctions. In other words, the nontrivial topology of their bulk wavefunctions in momentum space ensures the presence of conducting edge states. These basic concepts of topological insulators can be understood through one-dimensional Su-Schrieffer-Heeger (SSH) models (Sec.~\ref{SSH_model_summary}) \cite{su1979solitons}. Such ``wavefunction topology'' and bulk-boundary correspondence have counterparts in magnonic systems with the bosonic Bogoliubov-de-Gennes (BDG) Hamiltonian \cite{kondo2020non} and in non-Hermitian systems \cite{yao2018edge}, though some modifications need to be carried out (Sec.~\ref{BDGHamiltonian} and Sec.~\ref{SSH_model_summary}).

Different from real eigenvalues found in the Hermitian systems, the non-Hermitian systems commonly exhibit complex eigenvalues that may give rise to nontrivial ``spectral topology''~\cite{ding2022non}, which is absent in the Hermitian counterparts. 
For example, as we shall detail in Sec.~\ref{Nonhermitian_skin_effect}, one can define a spectral winding number even for a single band because, in this case, the eigenenergies lie on a complex plane rather than on a real axis as in the Hermitian system. This spectral winding number is suggested to be associated with the non-Hermitian skin effect in the Hatano-Nelson model (Sec.~\ref{Nonhermitian_skin_effect})~\cite{zhang2020correspondence,okuma2020,okuma2023non}. In the non-Hermitian systems, there is another spectral topology where the eigenvalues on different Riemann branches can swap with each other when the parameter encircles a single EP once \cite{heiss2012physics,bergholtz2019non}. Note the EP is the band singularities where both the eigenvalues and eigenvectors coalesce. The swapping of eigenvalues arises from the nature of multi-valued spectra, which is captured by topological invariant ``energy vorticity'' in terms of the EP~\cite{shen2018topological,ding2022non} (Sec.~\ref{exceptionaltopology}). A non-Hermitian nodal phase with a pair of EPs in the reciprocal space is called non-Hermitian Weyl semimetal  \cite{kozii2017non,zhou2018observation,bergholtz2019non} since it shares many similarities with Hermitian Weyl semimetal \cite{yan2017topological,armitage2018weyl}  such as nontrivial topological charge (Sec.~\ref{Nodal_phase}). There exist bulk Fermi arcs with vanished real part of energy in the non-Hermitian Weyl semimetal, which connects a pair of EPs \cite{kozii2017non,zhou2018observation,bergholtz2019non}.

\subsection{Wavefunction topology:  Hermitian \textit{vs}. non-Hermitian systems}\label{nonHermitianedgestate}

The Berry phase, also known as the geometric phase, is a phase factor acquired by the wavefunction of a quantum system as it undergoes a cyclic and adiabatic evolution in its parameter space~\cite{berry1984quantal,berry1989quantum,bernevig2013topological}. It is geometric in nature because it depends only on the path taken in the parameter space but not on the specifics of how or when the system traverses that path. In the topological band theory, the Berry phase plays a central role in defining and understanding topological invariants, such as the Chern number~\cite{thouless1982quantized} and the $\mathbb{Z}_2$ invariant~\cite{kane2005z}, of electronic band structures. These invariants characterize different topological phases of matter, which are distinct from the conventional phases described by symmetry breaking. 
 Since such topology is closely related to the wavefunction of a quantum system, it is also known as wavefunction topology. In this part, we compare the topological characterization of Hermitian and non-Hermitian systems when the wavefunction topology applies.
 
\subsubsection{Berry phase and Chern number}
\label{BDGHamiltonian}
 
 \textbf{Fermion system}.---For a comparison with the bosonic system, we first introduce the Berry phase and Chern number in the fermion lattice systems. A free (fermionic) Hermitian Hamiltonian in periodic lattice systems  can be typically written in the momentum space as~\cite{bernevig2013topological,shen2012topological}
\begin{align}\label{eq:Hermitian1}
  \hat{H} = \sum_{\bf k} \hat{C}_{\bf k}^{\dagger} H({\bf k})  \hat{C}_{\bf k},
\end{align}
where $H({\bf k})$ is a Hermitian matrix and $\hat{C}_{\bf k}=\left(\hat{c}_{{\bf k},1},\hat{c}_{{\bf k},2},\cdots,\hat{c}_{{\bf k},N}\right)$ represents the Fermion annihilation operators with $\{1,2,\cdots, N\}$ denoting the degree of freedoms within a unit cell. The Hamiltonian is diagonalized by
\begin{align}
\label{eq:Hermitiandiagon}
  \langle \Phi_n({\bf k})| H({\bf k}) |\Phi_n({\bf k})\rangle = E_n ({\bf k}),
\end{align}
with $|\Phi_n({\bf k})\rangle$ and $E_n ({\bf k})$ being the eigenvector and eigenvalue of the $n$-th band. For simplicity, we assume there is no degeneracy in energy bands. One can straightforwardly generalize the Berry phase when the band has N-fold degeneracy~\cite{bernevig2013topological}. As the lattice momentum ${\bf k}$  adiabatically evolves in the Brillouin zone (BZ), besides the conventional dynamical phase the wavefunction acquires a geometric Berry phase~\cite{berry1984quantal,shapere1989geometric}
\begin{equation}\label{eq:berry phase}
  \gamma_n=\oint_{BZ}   A_n({\bf k}) \cdot d {\bf k},
\end{equation}
 where  $A_n({\bf k}) = i  \langle \Phi_n({\bf k}) | \nabla_{\bf k}   |\Phi_n({\bf k})\rangle $ defines the Berry connection. In the two-dimensional case,  using Stokes theorem the above integral can be rewritten into the surface integral of Berry curvature  $\Omega_n({\bf k})=\hat{z}\cdot[ \nabla_{{\bf k}} \times  A_n({\bf k})]$ over the (first) Brillouin zone~\cite{bernevig2013topological,shen2012topological}
\begin{align}\label{eq:surface integral}
 \gamma_n=\oiint_{\rm BZ} \Omega_n({\bf k})  \cdot  d^2 {\bf k}.
\end{align}
The Chern number for the $n$-th band is given by $\mathcal{C}_n=\gamma_n/(2 \pi)$.

When there are degenerate points in the Brillouin zone, the Chern number can be also defined for topological characterization.
As an example, we address the Weyl physics that appears in three-dimensional space. Generally speaking, the two-band Hermitian Hamiltonian
\begin{align}\label{twoband}
    H({\bf k})=h_0({\bf k})\sigma_0+h_x({\bf k})\sigma_x+h_y({\bf k})\sigma_y+h_z({\bf k})\sigma_z,
\end{align}
where
$\pmb{\sigma}=(\sigma_z,\sigma_y,\sigma_z)$
are the Pauli matrices,
leads to the energy dispersion
\begin{align}
E_{\pm }({\bf k})=h_0({\bf k})\pm\sqrt{|h_x({\bf k})|^2+|h_y({\bf k})|^2+|h_z({\bf k})|^2}.
\end{align}
The energy degeneracy appears at those momenta that satisfy
\begin{align}\label{cnfordiracorweyl}
    h_x({\bf k})=h_y({\bf k})=h_z({\bf k})=0,
\end{align}
where there are three variables ${\bf k}=(k_x,k_y,k_z)$ to be determined. Three equations with three variables determine several \textit{isolated} points in the Brillouin zone as the solutions \cite{burkov2011topological,armitage2018weyl}. The stability of such energy degeneracies is characterized by the Chern number $\mathbb{Z}$, which is the integration of Berry curvature ${\bf \Omega}({\bf k})$ over a closed contour enclosing the Weyl point~\cite{moore2010birth,bernevig2013topological,asboth2016short,qi2011topological}:
\begin{equation}
    \mathbb{Z} = \frac{1}{2 \pi} \oint_{S} {\bf \Omega}({\bf k})\cdot d {\bf S}({\bf k}).
\end{equation}
The properties of Chern number $\mathbb{Z}$ are as follows:
\begin{itemize}
\item a), when $\mathbb{Z}=0$, the energy degeneracies are easily relieved by the perturbation, which is referred to as the Dirac points;
\item b), when $\mathbb{Z}=\pm 1$, the energy degeneracies are stable that is referred to as the Weyl points;
\item c), when $|\mathbb{Z}|>1$, the energy degeneracies are easy to be split to several more stable Weyl points with $|\mathbb{Z}|=1$ by the perturbation.
\end{itemize}
Further, the summation of the Chern number is zero in such Weyl semimetal such that the Weyl points always appear in pairs \cite{nielsen1981absence,nielsen1981absencetwo,chiu2016classification}. Such Weyl points do not need the protection of the symmetries, in contrast to the Dirac point in the two-dimensional space with the stability protected by the time-reversal and inversion symmetries or chiral symmetry \cite{hasan2010colloquium}.

\textbf{Bogoliubov-de-Gennes magnonic Hamiltonian}.---This review article mainly focuses on the magnonic system---a bosonic system. Driven by both the fundamental curiosity and potential for promising device applications, there has been a burgeoning interest in the realization of topological phases with bosonic (quasi)particles, such as photons~\cite{lu2014topological,khanikaev2017two,ozawa2019topological}, phonons~\cite{susstrunk2015observation,susstrunk2016classification}, excitons~\cite{wu2017topological,kwan2021exciton}, as well as magnons~\cite{mook2014edge,mook2021interaction,nakata2017magnonic,mook2021chiral,murakami2017thermal,nakatsuji2022topological,bonbien2021topological,shindou2013topological,li2021topological,wang2021topological,zhuo2023topological,mcclarty2022topological}. Here we address the basic knowledge about topological magnon band theory to understand topological phases in Hermitian magnonic systems, such as   magnon Chern insulators~\cite{zhang2013topological,shindou2013topological,chernyshev2016damped,owerre2016topological}, Dirac (Weyl) magnon semimetals~\cite{su2017magnonic,mook2017magnon,owerre2017magnonic}, and higher-order topological magnons~\cite{sil2020first,hirosawa2020magnonic,mook2021chiral}.  For simplicity, we consider two-dimensional collinear (ferro- or antiferro-) magnetic systems described by the following quadratic Hamiltonian
\(  \hat{H}=\sum_{\vb k} \hat{\Psi}^\dagger_{\vb k} H({\bf k}) \hat{\Psi}_{\vb k}, \label{top_1} \)
 where the Bloch Hamiltonian takes the form of
 \(  H({\bf k})=\mqty[ \vb h_{\vb k}  & \vb*\Delta_{\vb k} \\  \vb*\Delta_{-\vb k}^* & \vb h_{-\vb k}^*  ],   \)
and the vector magnon operator is
\(  \hat{\Psi}_{\vb k}= \left(  \hat{m}^\dagger_{\vb k, 1},\cdots, \hat{m}^\dagger_{\vb k, N},\; \hat{m}_{-\vb k, 1}, \cdots, \hat{m}_{-\vb k, N}     \right).  \)
Here $\hat{m}^\dagger_{\vb k, i}$ and $\hat{m}_{\vb k, i}$ are the magnon creation and annihilation operators, obeying the standard bosonic algebra, with $\vb k=(k_x, k_y)$ being the wave vector (running over the first Brillouin zone) and $i\in \{1,2,\cdots, N\}$ ($N$ is the number of orbitals or degree of freedoms within a unit cell). We remark that the Hamiltonian $H_{\vb k}$ is similar to a BdG Hamiltonian for superconductivity, and  $\vb h_{\vb k} $ and  $\vb*\Delta_{\vb k}$ are $N\times N$ matrices, related to hopping
and pairing, respectively. This Hamiltonian can describe various interactions, such as Heisenberg exchange interaction, Dzyaloshinskii- Moriya interaction, magnetostatic dipolar interaction, etc~\cite{wang2021topological}.
We note that there are generally two methods for deriving the quadratic Hamiltonian \eqref{top_1}. One approach involves linearizing the LLG equation, while the other begins with a quantum spin Hamiltonian, subsequently applying the Holstein-Primakoff transformations and retaining terms up to the second order (Sec.~\ref{Magnon}). One can also study the topological phases due to magnon-magnon interactions which are beyond the quadratic Hamiltonian~\cite{mcclarty2022topological,mcclarty2019non,mook2021interaction}.

Distinct from the fermionic case, to preserve the bosonic algebra a bosonic BdG Hamiltonian can be diagonalized by a paraunitary Bogoliubov transformation $T_{\vb k}$ satisfying~\cite{li2021topological,wang2021topological,zhuo2023topological,mcclarty2022topological}
\(  T_{\vb k} \sigma_z T^\dagger_{\vb k}=\sigma_z.   \)
$\sigma_z$ is a $2N\times 2N$  diagonal matrix, taking $+1$ in the particle sector while $-1$ in the hole sector. In the fermionic scenario, $\sigma_z$ should be replaced by an identity matrix. To obtain $T_{\vb k}$ for a given Hamiltonian, one can simply solve the eigenvalue problem for $\sigma_zH_{\vb k}$ whose eigenvectors provide a paraunitary matrix $T_{\vb k}$ which diagonalizes the Hamiltonian:
\(  T_{\vb k}^\dagger H({\bf k}) T_{\vb k}=\mqty[E_{\vb k} & \\  & E_{-\vb k}].   \)
We note that, since $\sigma_zH_{\vb k}$ is non-Hermitian,  the inner product of two states $|\phi({\bf k})\rangle$, $|\psi({\bf k})\rangle$ should be defined as $\langle \phi({\bf k}) | \sigma_z |\psi({\bf k}) \rangle$. We have a $U(1)$ gauge freedom in specifying the state, corresponding to its phase factor. This allows us to introduce a $U(1)$ gauge field (or Berry connection) associated with the $n$-th band
\begin{align}\label{eq:Berrycurformagnon}
  A_n({\bf k}) = i  \frac{\langle \Phi_n({\bf k}) | \sigma_z \nabla_{\bf k}   |\Phi_n({\bf k})\rangle }{\langle \Phi_n({\bf k}) | \sigma_z |\Phi_n({\bf k})\rangle},
\end{align}
similar to the above electronic case~\cite{hasan2010colloquium,qi2011topological,bernevig2013topological}. 
The corresponding Berry curvature is given by \(  \Omega_n(\vb k)=\hat{z}\cdot( \nabla_{\vb k} \times  \vb A_n ). \)
Here $|\Phi_n({\bf k})\rangle$ is the $n$-th eigenvector of $\sigma_zH_{\vb k}$.
One can also write this curvature in a more compact way by introducing a projection operator~\cite{li2021topological,wang2021topological,zhuo2023topological,mcclarty2022topological}:
\(  P_n(\vb k)= T_{\vb k} \Gamma_n \sigma_z T^\dagger_{\vb k} \sigma_z,  \)
where $\Gamma_n$ is a diagonal matrix with $+1$ for the $n$-th diagonal component and zero otherwise. The Berry curvature then can be recast into the following form:
\(   \Omega_n(\vb k)= i\epsilon_{ij} \Tr\left[  P_n (\partial_{k_i} P_n)(\partial_{k_j} P_n)  \right]=i\Tr\left[ P_n \left( \pdv{P_n}{k_x} \pdv{P_n}{k_y} - \pdv{P_n}{k_y} \pdv{P_n}{k_x} \right)  \right].     \)
Then the Chern number in the magnonic system is given by
\begin{align}
\mathcal{C}_n=\frac{1}{2 \pi}\oiint_{\rm 
 BZ} \Omega_n({\bf k})  \cdot  d^2 {\bf k},
\end{align}
where the integration is over the first Brillouin zone.
Similar to electronic systems, we require a ``spin-orbit''-like interaction in magnetic systems to ensure that the Berry curvature is nonzero. Typically, this role is fulfilled by the Dzyaloshinskii-Moriya interaction or magnetic dipolar interaction (which depends on the directions and relative positions of the magnetic moments)~\cite{li2021topological,wang2021topological,zhuo2023topological,mcclarty2022topological}.

The discussion above about the Berry curvature of magnon bands serves as a good starting point to understand various Hermitian topological phenomena in magnetic systems. One well-known example is the magnon thermal Hall effect~\cite{onose2010observation,mook2014magnon,murakami2017thermal}. Similar to the electronic Hall effect (or anomalous Hall effect in the metallic ferromagnets), one could expect that magnons acquire an anomalous velocity (due to nonzero Berry curvature) perpendicular to the external force that drives the motion of magnons. As a result,  a longitudinal temperature gradient in a two-dimensional magnet would lead to a transverse thermal magnon current with a finite thermal Hall conductivity. This effect has been experimentally observed, for example, in the insulating ferromagnet $\text{Lu}_2\text{V}_2\text{O}_7$ of pyrochlore lattice structures~\cite{onose2010observation}. Other phenomenon based on the topology of magnon bands are also extensively studied, such as spin Nernst effect (magnonic version of the spin Hall effect)~\cite{meyer2017observation,cheng2016spin,zyuzin2016magnon}, (high-order) topological magnon insulators~\cite{zhang2013topological,shindou2013topological,chernyshev2016damped,owerre2016topological},  topological magnon semimetals~\cite{su2017magnonic,mook2017magnon,owerre2017magnonic}, etc.

\subsubsection{Hermitian and non-Hermitian Su-Schrieffer-Heeger models}

\label{SSH_model_summary}

In this part, we address the SSH model \cite{su1979solitons,heeger1988solitons,fradkin1983phase,li2014topological,lieu2018topological,yao2018edge,obana2019topological} to exemplify the calculation of topological invariants in the one-dimensional case and demonstrate the bulk-boundary correspondence. We compare with its non-Hermitian generalization by emphasizing their difference in the topological characterization.

\begin{figure}[!htp]
    \centering
    \includegraphics[width=0.92\textwidth]{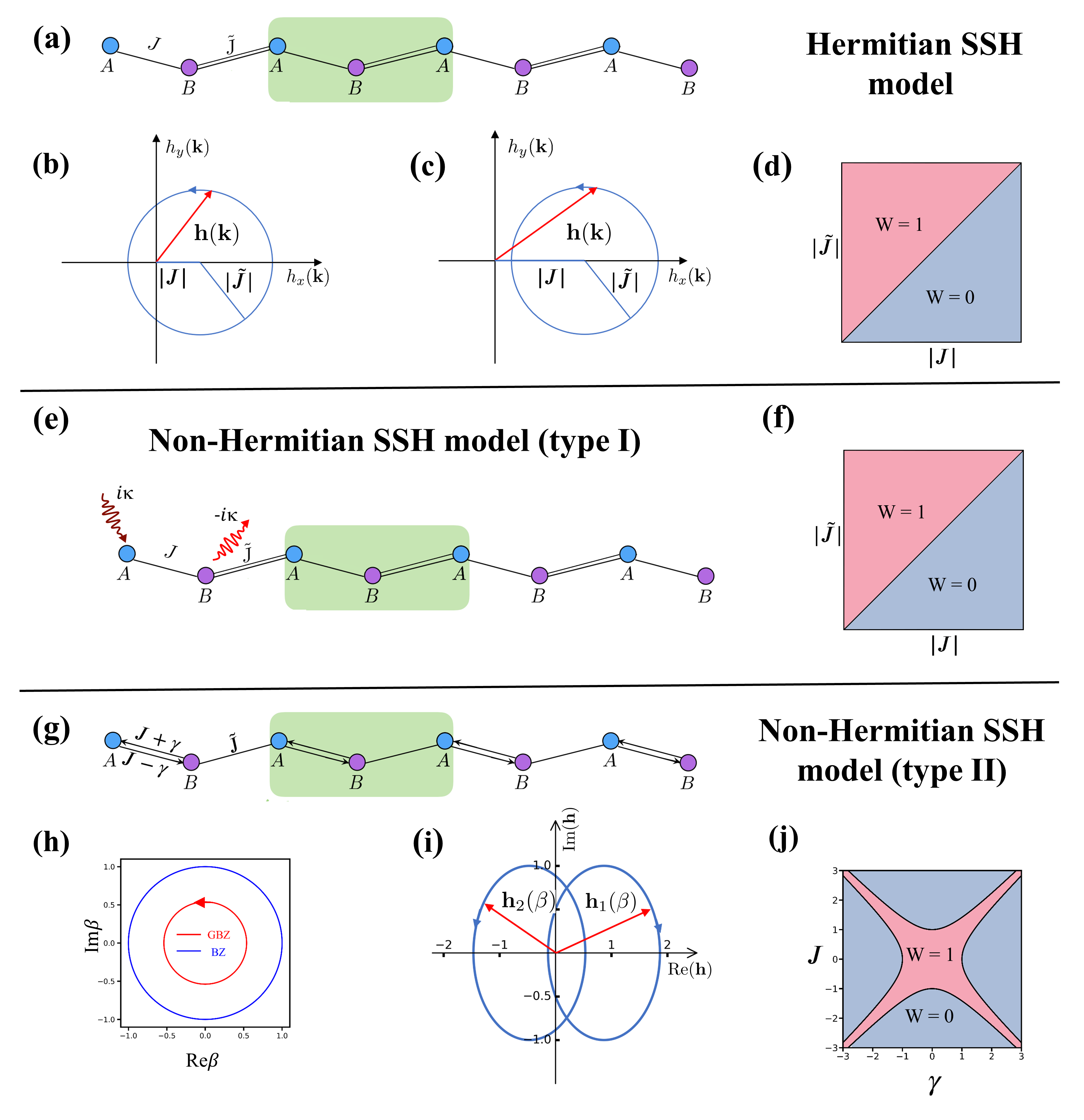}
    \caption{Comparison of the Hermitian and non-Hermitian SSH models. (a) shows the Hermitian SSH model with staggered coupling $J$ and $\tilde{J}$. When  $\left|J\right|>\left|\tilde{J}\right|$ [(b)]  $\left|J\right|<\left|\tilde{J}\right|$ [(c)], the evolution of $h(\bf k)$ when ${\bf k}$ evolves in the Brillouin zone can or cannot encircle the origin point, corresponding to the winding number $W = 1$ and $W=0$ as shown in (d). (e) The non-Hermitian SSH model (type I), where balanced gain $i\kappa$ and loss $-i\kappa$ are introduced in ``A'' and ``B'' sites, respectively. (f) plots the dependence of the topological phase diagram on coupling. (g) The non-Hermitian SSH model (type II), where the intercell coupling is chiral with the hopping strength $J+\gamma$ ($J-\gamma$) from right to left (from left to right).  (h) shows the GBZ on the complex plane with $J=2/3$, $\gamma=0.2$, and $\tilde{J}=1$. As $\beta$ evolves along the GBZ, indicated by the red arrow in (h), ${h}_1(\beta)$ and ${ h}_2(\beta)$ (defined in the text) encircle the origin point with the winding number $W = 1$ as shown in (i). (j) plots the dependence of the winding number on $J$ and $\gamma$ when $\tilde{J}=1$.}
  		\label{fig:SSH1}
	\end{figure}

\textbf{Hermitian Su-Schrieffer-Heeger model}.---The Hermitian SSH model is first considered in the pioneering works \cite{su1979solitons,heeger1988solitons} for the motion of solitons in polyacetylene, where the hopping is staggered since it is different from sublattice ``B'' to two neighboring ``A'' sites [Fig.~\ref{fig:SSH1}(a)]. It turns out to be one of the simplest topological lattice models that are widely used to demonstrate the basic topological concepts and properties in Hermitian and non-Hermitian scenarios \cite{fradkin1983phase,li2014topological,lieu2018topological,yao2018edge,obana2019topological}. From a modern viewpoint, this staggered hopping potential makes an analogy to a dimer lattice with distinct intercell and intracell hoppings; there are two topologically distinct phases depending on whether the intercell hopping is larger than the intracell one. The edge states exist between topologically distinct phases that are characterized by the bulk-boundary correspondence \cite{asboth2016short,hasan2010colloquium,qi2011topological}.

With the intracell and intercell couplings $J$ and $\tilde{J}$, the Bloch Hamiltonian for the Hermitian SSH model 
\begin{align}
H({\bf k})=h_x({\bf k})\sigma_x+h_y({\bf k})\sigma_y,
\label{Hermitian_SSH}
\end{align}
 where $h_x({\bf k}) = J+\tilde{J}\cos{k}$ and $h_y({\bf k}) = \tilde{J}\sin{k}$ \cite{li2014topological,asboth2016short,yao2018edge}.  The eigenequation $H({\bf k})\left|\Phi({\bf k})\right> = E ({\bf k})\left|\Phi({\bf k})\right>$ solves the eigenvalue $E ({\bf k})$ and eigenstates $|\Phi({\bf k})\rangle$:
\begin{subequations}
\label{eigen}
 \begin{align}
   & E_{\pm} ({\bf k})= \pm \left|{\bf h}({\bf k})\right|= \pm \sqrt{h_x^2({\bf k})+h_y^2({\bf k})},\\
   &|\Phi({\bf k})\rangle_{\pm} = \frac{1}{\sqrt{2}}\left( \begin{array}{c}
\pm e^{-i\theta_{\bf k}} \\
 1\end{array}\right),
 \end{align}
 \end{subequations}
where $\tan \theta_{\bf k}=h_y({\bf k})/h_x({\bf k})$. The topological property is well characterized by the Berry phase \cite{berry1984quantal}, which acts as a topological invariant when the wave vector evolves along a closed loop.   For a one-dimensional periodic lattice, the Brillouin zone is periodic by $2\pi$ \cite{zak1989berry} with $H({\bf k})= H({\bf k}+2\pi)$, thus forming a closed loop when ${\bf k}$ evolves in one period.  For example,  the Berry phase over the Brillouin zone, also known as ``Zak phase'' \cite{zak1989berry,atala2013direct}, for state $|\Phi({\bf k})\rangle_{-} $ reads
 \begin{equation}\label{berry}
        \gamma=\frac{i}{2}\oint_{\rm BZ} d{\bf k}\left( -e^{i\theta({\bf k})},1 \right)  \frac{\partial}{\partial {\bf k}} \left(\begin{array}{c}
                -e^{-i\theta_{\bf k}} \\
                1
              \end{array}\right) = \frac{\theta  ({\bf k}=2\pi)- \theta  ({\bf k}=0)}{2},
\end{equation}
which depends on whether the parameter evolution surrounds the energy degeneracy point, \textit{i.e.}, Dirac point with $h_x({\bf k})=h_y({\bf k})=0$ can be $0$ or $\pi$, as plotted in Fig.~\ref{fig:SSH1}(b) and (c) for the evolution of $\theta_{\bf k}$ denoted by the red vector.
When $\left|J\right|<\left|\tilde{J}\right|$ [Fig.~\ref{fig:SSH1}(b)] and $\left|J\right|>\left|\tilde{J}\right|$ [Fig.~\ref{fig:SSH1}(c)], $\theta_{\bf k}$ is changed by $2\pi$ and unchanged, respectively. Only the former case contributes a nonzero Berry phase, which also corresponds to a nonzero winding number $W=\gamma/\pi=1$. The topological nontrivial and trivial phases are thereby bounded by $\left|J\right|=\left|\tilde{J}\right|$, as shown in Fig.~\ref{fig:SSH1}(d). When $\left|J\right|<\left|\tilde{J}\right|$ and $W=\gamma/\pi=1$, an edge state emerges, which can be understood as the intracell coupling that favors the ``dimer'' between adjacent cells, leaving two unpaired boundary states. This implies a bulk-boundary correspondence in the Hermitian topological description \cite{asboth2016short,hasan2010colloquium,qi2011topological}.

The Hermitian SSH model is widely used to study the topological edge states in such as phononic and photonic crystals \cite{xiao2014surface,xiao2015geometric,zangeneh2019topological} and a chain of plasmonic nanoparticles or cold atoms \cite{atala2013direct,poddubny2014topological,meier2016observation}.
Its magnonic version is proposed in a chain of magnetic spheres loaded in a waveguide \cite{pirmoradian2018topological,pirmoradian2023topological}, where distinct intracell and intercell couplings can be tuned by the separation between magnetic spheres and external magnetic inductions, providing an experimentally feasible and tunable platform for engineering the magnonic soliton states.

\textbf{Non-Hermitian Su-Schrieffer-Heeger model}.---The established bulk-boundary correspondence in the Hermitian scenario is based on a prerequisite that the bulk spectrum and associated eigenstates under open boundary condition (OBC) can be approximated by those under periodic boundary condition (PBC). However, in the non-Hermitian cases, it was reported that the spectrum and eigenstates can be strongly altered under the OBC and PBC, which comes as a surprise since the bulk-boundary correspondence breaks down \cite{lee2016anomalous,kunst2018biorthogonal,xiong2018,yao2018edge,zhou2020renormalization}.
By allowing the complex wave vectors under the OBC, the edge state can be again characterized by the bulk topological quantity \cite{yao2018edge,yokomizo2019non,zhang2020correspondence,yang2020non}. Here we address this possibility by extending the Hermitian SSH model to the non-Hermitian scenario.

The non-Hermiticity can be introduced in various ways. Here we focus on two well-studied types. 
In type I, the non-Hermitian terms are introduced via the balanced gain $i\kappa$ and loss $-i\kappa$ in the neighboring ``A'' and ``B'' sites \cite{schnyder2008classification,esaki2011edge,lieu2018topological,flebus2020non,halder2022properties}, while the staggered hoppings $J$ and $\tilde{J}$ remain unchanged, as shown in Fig.~\ref{fig:SSH1}(e). In the momentum space the Bloch Hamiltonian (\ref{Hermitian_SSH}) becomes
\begin{equation}
\label{eq:SSH2hamiltonian}
  H({\bf k})=\left( \begin{array}{cc}
    i\kappa  & {h}_1({\bf k})\\
    {h}_1^{\dagger}({\bf k})  & -i\kappa
    \end{array}\right) = \left( \begin{array}{cc}
    i\kappa   & \left|{h}_1({\bf k})\right| e^{-i\theta({\bf k})}\\
   \left|{h}_1({\bf k})\right| e^{i\theta({\bf k})} & -i\kappa
    \end{array}\right),
\end{equation}
where  ${h}_1({\bf k})= J + \tilde{J}e^{-i {\bf k}}$ with $\theta({\bf k})$ denoting its phase angle.
 The non-Hermitian Hamiltonian \eqref{eq:SSH2hamiltonian} is diagonalized by biorthogonal eigenvectors
\begin{subequations}
\begin{align}
 &  \langle \Psi ({\bf k}) |_{+} = \left( e^{i\theta({\bf k})}\cos{\phi({\bf k})}, \sin{\phi({\bf k})}  \right),~~~\langle \Psi ({\bf k}) |_{-} = \left(- e^{i\theta({\bf k})}\sin{\phi({\bf k})}, \cos{\phi({\bf k})}  \right),\\
 &  | \Phi ({\bf k})   \rangle_{+} =  \left( \begin{array}{c}
   e^{-i\theta({\bf k})}\cos{\phi({\bf k})}\\
  \sin{\phi({\bf k})}
 \end{array}    \right),~~~| \Phi ({\bf k})   \rangle_{-} =  \left( \begin{array}{c}
  - e^{-i\theta({\bf k})}\sin{\phi({\bf k})}\\
  \cos{\phi({\bf k})}
 \end{array}    \right),
\end{align}
\end{subequations}
where $\phi({\bf k}) = \tan^{-1}\left(\sqrt{({E_{+}-i\kappa})/({E_{+}+i\kappa})}\right)$, leading to 
the eigenvalues 
\begin{equation}
    E_{\pm} ({\bf k}) = \pm  \sqrt{\left|{h}_1({\bf k})\right|^2-\kappa^2}.
\end{equation} 
The eigenvalues are purely imaginary when $\kappa>|{h}_1({\bf k})|$.
The winding number can characterize the topological edge modes, but we need to notice that the eigenvectors are biorthogonal. With a biorthogonal basis, the complex Berry phase \cite{liang2013topological} is introduced for calculating the geometrical phase for a non-Hermitian Hamiltonian. Here for the two bands, the complex Berry phases are  
\begin{subequations}
\begin{align}
    \gamma_{+} &= i \oint_{\rm BZ} d{\bf k} \left( e^{i\theta({\bf k})}\cos{\phi({\bf k})}, \sin{\phi({\bf k})}  \right) \frac{\partial}{\partial {\bf k}} \left( \begin{array}{c}
   e^{-i\theta({\bf k})}\cos{\phi({\bf k})}\\
  \sin{\phi({\bf k})}
 \end{array}    \right) = \oint_{\rm BZ} d{\bf k} \frac{\partial \theta({\bf k})}{\partial {\bf k}}  \cos^2{\phi({\bf k})} , \\
\gamma_{-} &= i \oint_{\rm BZ} d{\bf k} \left(- e^{i\theta({\bf k})}\sin{\phi({\bf k})}, \cos{\phi({\bf k})}  \right)\frac{\partial}{\partial {\bf k}} \left( \begin{array}{c}
  - e^{-i\theta({\bf k})}\sin{\phi({\bf k})} \\
  \cos{\phi({\bf k})}
 \end{array}    \right)= \oint_{\rm BZ} d{\bf k} \frac{\partial \theta({\bf k})}{\partial {\bf k}}  \sin^2{\phi({\bf k})}.  
\end{align} 
\end{subequations}
It should be noted that the introduction of on-site gain and dissipation breaks the inversion symmetry of the Hermitian SSH model, which leads to a not-quantized Berry phase in the above equations. However, the global Berry phase, \textit{i.e.}, the sum of Berry phase of two bands,
\begin{equation}
    \gamma = \gamma_{+} + \gamma_{-} = \oint_{\rm BZ} d{\bf k} \frac{\partial \theta({\bf k})}{\partial {\bf k}}  = \theta  ({\bf k}=2\pi)- \theta  ({\bf k}=0)
\end{equation} 
is quantized, which thereby accurately captures the topological transition \cite{liang2013topological}. The corresponding winding number for global Berry phase may be defined as $W = \gamma/\left(2 \pi\right)$. The winding number is 1 and the global Berry phase is $2 \pi$ when $|J| < |\tilde{J}|$ and both of them become zero when $|J| > |\tilde{J}|$, as shown in Fig.~\ref{fig:SSH1}(f). Clearly, the first type of non-Hermitian SSH model exhibits a similar feature of topological transition to that of the Hermitian SSH model.

In the type-II non-Hermitian SSH model, the intercell coupling $J$ in Fig.~\ref{fig:SSH1}(a) is replaced by the non-reciprocal hopping with strength $J+\gamma$  from the right to left, but $J-\gamma$ from the left to right, as shown in Fig.~\ref{fig:SSH1}(g). We allow the complex wave vector under the OBC and perform the mapping from ${\bf k}$ to ${\bf \beta}=e^{i{k}}$ on the complex plane. The distribution of $\beta$ on the complex plane is known as the generalized Brillouin zone (GBZ) \cite{yao2018edge,zhang2020correspondence,yang2020non}, while the distribution of real wave vector is composed of a unit circle, as illustrated in Fig.~\ref{fig:SSH1}(h). 
In terms of $\beta$ and the non-reciprocal hopping between two neighboring sites, the Hamiltonian (\ref{Hermitian_SSH})  is modified in the non-Hermitian SSH model [Fig.~\ref{fig:SSH1}(h)] to be \cite{lee2016anomalous,yao2018edge,zhang2022review}
\begin{equation}
\label{eq:transformation}
  H({\bf \beta})=\left( \begin{array}{cc}
    0   & {h}_1({\bf \beta})\\
    {h}_2({\bf \beta})  & 0
    \end{array}\right) = \left( \begin{array}{cc}
    0   & \left|{h}_1({\bf \beta})\right| e^{i\theta_1({\bf \beta})}\\
    \left|{h}_2({\bf \beta}) \right|  e^{i\theta_2({\bf \beta})} & 0
    \end{array}\right),
\end{equation}
where $h_1({\bf \beta})=J+\gamma+\tilde{J}/{\bf \beta}$ and  $h_2({\bf \beta})=J-\gamma+\tilde{J}{\bf \beta}$ with $\theta_1({\bf \beta})$ and $\theta_2({\bf \beta})$ denoting their respective phase angles.
The off-diagonal terms in (\ref{eq:transformation}) are not conjugated with each other. The eigenvalue equation
\begin{align}
E^2({\bf \beta})=h_1({\bf \beta})h_2({\bf \beta}) ,
\label{zero_energy}
\end{align}
and with two roots $\beta_1$ and $\beta_2$,  $\beta_1\beta_2 = \left(J-\gamma\right)/ \left(J+\gamma\right)$.
The continuum bands form in the sufficiently long chain when $\left|\beta_1\right|=\left|\beta_2\right|$ \cite{yao2018edge,yokomizo2019non}, leading to $\left|\beta_{1,2}\right|=\sqrt{\left|(J-\gamma)\right|/\left|(J+\gamma)\right|}<1$.

Under OBC $\left|{\bf \beta}\right|=\sqrt{\left|(J-\gamma)\right|/\left|(J+\gamma)\right|}<1$ when $J>\gamma>0$ \cite{yao2018edge,yokomizo2019non,yang2020non}, implying the amplification of the eigenstate when approaching the left boundary of the chain, i.e., a manifestation of skin effect. On the other hand, for the type-I non-Hermitian SSH model, $|\beta|=1$ under the OBC. In this case, the GBZ coincides with the conventional Brillouin zone, so there is no non-Hermitian skin effect. More details for the calculation of the GBZ and the underlying mechanism of the non-Hermitian skin effect will be addressed later in the Hatano-Nelson model (Sec.~\ref{Nonhermitian_skin_effect}).

The biorthogonal basis diagonalize the non-Hermitian matrix \eqref{eq:transformation} via $ \langle \Psi ({\bf \beta}) | H({\bf \beta}) | \Phi ({\bf \beta}) \rangle = E ({\bf \beta})$ \cite{banach1987theory}, with the
eigenvalues and eigenvectors
\begin{subequations}
\begin{align}
\label{biorth}
&E_{\pm} ({\bf \beta}) = \pm  h ({\bf \beta}) = \pm \sqrt{\left|{ h}_1({\bf \beta}){ h}_2({\bf \beta})\right|e^{i\left[\theta_1({\bf \beta})+\theta_2({\bf \beta})\right]}},\\
 &  \langle \Psi ({\bf \beta}) |_{\pm} = \frac{1}{\sqrt{2} h ({\bf \beta})} \left( \left|{h}_2({\bf \beta}) \right|  e^{i\theta_2({\bf \beta})},\pm h ({\bf \beta})\right), \\
 &  | \Phi ({\bf \beta})   \rangle_{\pm} =  \frac{1}{\sqrt{2} h ({\bf \beta})} \left( \begin{array}{c}
   \left|{ h}_1({\bf \beta})\right| e^{i\theta_1({\bf \beta})} \\
    \pm h ({\bf \beta})
 \end{array}    \right) .
\end{align}
\end{subequations}
To characterize the topological invariant (in both the Hermitian and non-Hermitian systems), a convenient tool is the so-called ``Q-matrix'' \cite{chiu2016classification}.
For a Hermitian system, with the eigenvalue equation 
$H({\bf k})\left|\Phi({\bf k})\right>_{\alpha} = E_{\alpha} ({\bf k})\left|\Phi({\bf k})\right>_{\alpha}$, the Bloch Hamiltonian $H({\bf k})=\sum_{\alpha}E_{\alpha} ({\bf k})|\Phi({\bf k})\rangle_{\alpha} \langle \Phi({\bf k}) |_{\alpha} $ and 
the corresponding Q-matrix 
$
    Q({\bf k}) = \sum_{\alpha}\lambda_{\alpha} ({\bf k})|\Phi({\bf k})\rangle_{\alpha} \langle \Phi({\bf k})|_{\alpha}$
is constructed by projection weights $\pm 1$ for different bands $\alpha$ with $\lambda_{\alpha}=1$ for unoccupied bands and $-1$ for occupied bands. Analogously, for the non-Hermitian Hamiltonian $H(\beta)=\sum_{\alpha}E_{\alpha} (\beta)|\Phi(\beta)\rangle _{\alpha} \langle \Psi(\beta) |_{\alpha} $, where $\langle \Psi(\beta) |_{\alpha}$ and $|\Phi(\beta)\rangle_{\alpha}$ are the left and right eigenvectors, the Q-matrix is defined as $Q(\beta) = \sum_{\alpha}\lambda_{\alpha} (\beta)|\Phi(\beta)\rangle_{\alpha} \langle \Psi(\beta) |_{\alpha}$. For a Hamiltonian matrix that possess chiral symmetry  $\sigma_z H(\beta) \sigma_z = -H(\beta)$, the Q-matrix holds the same symmetry $\sigma_z  Q(\beta) \sigma_z = - Q(\beta)$. 
Then the Q-matrix can be used to calculate the winding number with parameters $\beta$ that define the GBZ in one-dimensional non-Hermitian systems~\cite{yao2018edge,yang2020non}.  For a non-Hermitian Hamiltonian \eqref{eq:transformation} holding chiral symmetry,
the Q-matrix is given by 
\begin{equation}
  Q(\beta) =  | \Phi ({\bf \beta}) \rangle_{+}  \langle \Psi ({\bf \beta}) | _{+} - | \Phi ({\bf \beta}) \rangle_{-}   \langle \Psi ({\bf \beta}) |_{-}
   = \left(\begin{array}{cc}
    0 & q(\bf \beta)\\
    q^{-1}(\bf \beta) & 0
    \end{array}\right),
\end{equation}
with $q(\beta)=\sqrt{h_1(\beta)/h_2(\beta)}$ governed by two non-Hermitian couplings $h_1({\bf \beta})$ and  $h_2({\bf \beta})$. Its left and right eigenvectors with eigenvalues $\pm 1$ are, respectively,  $1/\sqrt{2}\left(\pm q^{-1}(\beta),1\right)$ and $1/\sqrt{2}\left(\pm q(\beta),1\right)$. Accordingly, the winding number of occupied state with $\lambda=-1$ can be calculated by
\begin{equation}\label{eq:doublewinding}
        W=\frac{i}{2\pi}\oint_{\rm GBZ} \left( -q^{-1}(\beta),1 \right)  \frac{\partial}{\partial \beta} \left(\begin{array}{c}
                -q(\beta) \\
                1
              \end{array}\right)d{\beta} = \frac{i}{2\pi} \oint_{\rm GBZ} 
 q^{-1}(\beta)dq=-\frac{1}{2\pi}\frac{(\delta\theta_1- \delta\theta_2)|_{\rm GBZ}}{2},
\end{equation}
where $\delta\theta_1|_{\rm GBZ}$ and $\delta\theta_2|_{\rm GBZ}$ are the phase changes of $\theta_1({\beta})$ and $\theta_2({\beta})$ when ${\bf \beta} $ evolves on a closed curve in a counterclockwise fashion on the GBZ. As shown in Fig.~\ref{fig:SSH1}(i), when $J=2/3$, $\gamma=0.2$, and $\tilde{J}=1$, the evolution of $h_1({\bf \beta})$ and  $h_2({\bf \beta})$ on closed curves denoted by the blue arrows acquire the phase accumulations $\delta\theta_1|_{\rm GBZ} = -2\pi$ and $\delta\theta_2|_{\rm GBZ} = 2\pi$, respectively, thus contributing a nonzero winding number $W = 1$ by Eq.~\eqref{eq:doublewinding}.

To determine the critical parameter of the chiral coupling for a topological phase transition, we solve the \textit{zero-energy} edge states \cite{su1979solitons,heeger1988solitons,yao2018edge} via $h_1({\bf \beta})h_2({\bf \beta})=0$ [Eq.~(\ref{zero_energy})], leading to two solutions ${\beta_{1,2}= -\left(J-r\right)/\tilde{J}}$ and $-\tilde{J}/\left(J+r\right)$. With $\left|\beta_1\right|=\left|\beta_2\right|$, we find the critical chiral coupling $\left|J^2-r^2\right|=\left|\tilde{J}^2\right|$, which divides the region of $W = 0$ and $W = 1$ in the parameter space, as shown in Fig.~\ref{fig:SSH1}(j)  for the topological phase diagram for this non-Hermitian SSH model when $\tilde{J}=1$.

However, the Berry phase in Eq.~\eqref{berry} or the winding number in Eq.~\eqref{eq:doublewinding} is not well-defined when the evolution path of the couplings contains the energy degeneracy points in the parameter space, such as the Dirac point with $h_1(\beta)=0$ and $h_2(\beta)=0$ \cite{armitage2018weyl,hasan2021weyl,liu2021higher}. These energy degeneracy points in the non-Hermitian case correspond to the EPs \cite{bergholtz2021exceptional} such that the matrix becomes defective and the eigenstates are collapsed. Thereby, the winding number is not well-defined when the EPs are involved in the parameter space, an issue well addressed in Ref.~\cite{yokomizo2019non}.

\subsection{Spectral topology in non-Hermitian systems}
\label{nonHeigen}

\begin{figure}[!htp]
\centering
\includegraphics[width=0.95\textwidth]{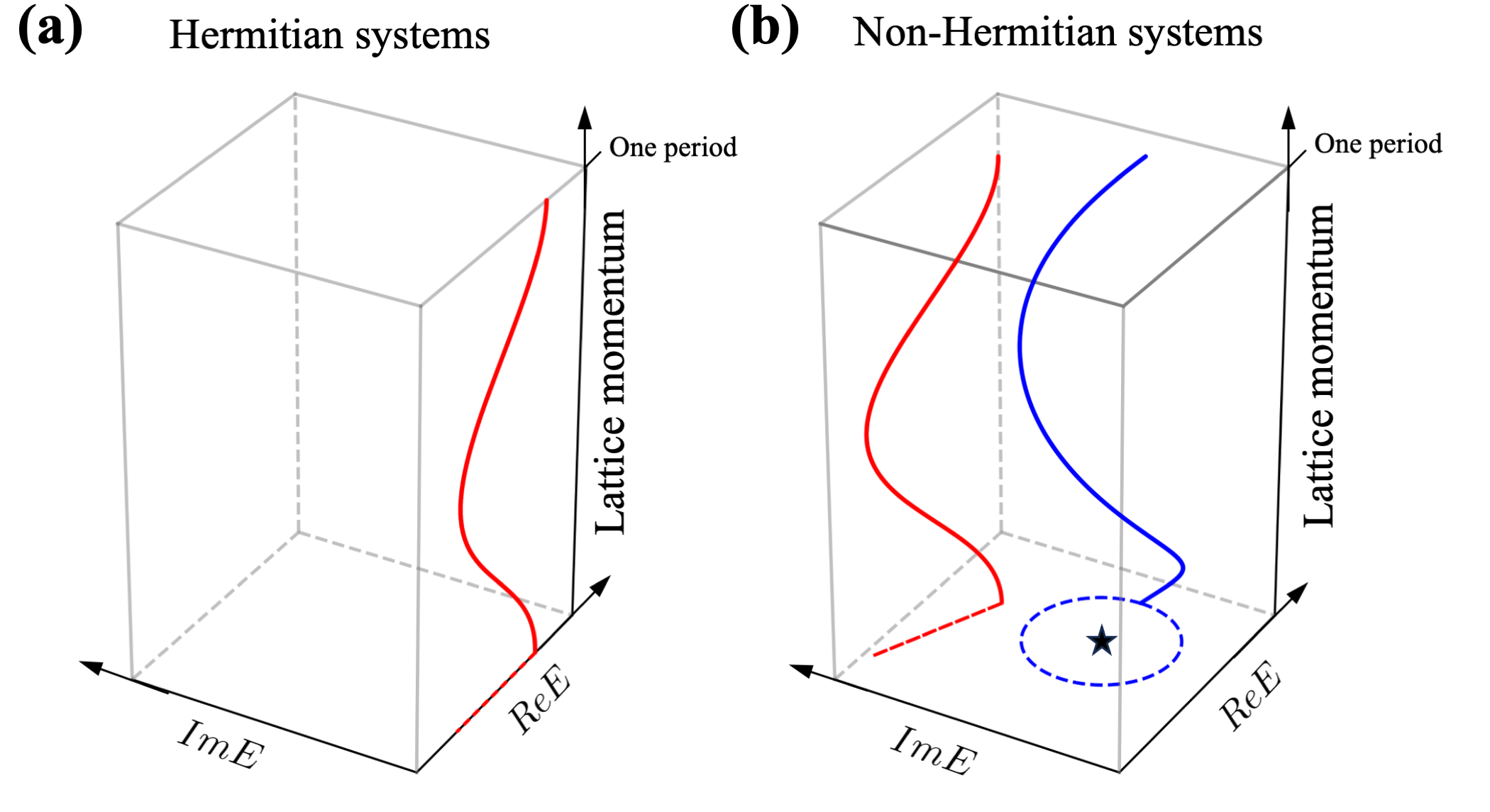}
\caption{Representative energy spectra for a single band in (a) Hermitian and (b) non-Hermitian systems. The reference energy is represented by the star in (b).}\label{fig:nonHeigen}
\end{figure}

The Hermitian topology only focuses on the wavefunction topology. The reason is that the eigenvalue is real in Hermitian systems and the evolution of eigenvalues of a single band on the real axis is trivial without allowing a topological structure, as shown in Fig.~\ref{fig:nonHeigen}(a). Differently, the eigenvalue can be complex that contains an imaginary component in a non-Hermitian system. As shown in Fig.~\ref{fig:nonHeigen}(b), the evolution of eigenvalue on the complex plane can acquire a spectral zone (blue curve) even for a single band or just be a segment (red curve), as the lattice momentum evolves along a period. These two kinds of spectra are topologically different for specified energy denoted by a star in Fig.~\ref{fig:nonHeigen}(b): the spectral zone can give rise to spectral winding number while the segment can not. Such spectrum brings a unique band gap in the non-Hermitian systems, called ``point gap''. Mathematically, it appears when the spectrum cannot include a specified energy under continuous deformation of the Hamiltonian with parameters~\cite{kawabata2019symmetry,ashida2020non,ding2022non}. In the Hermitian scenario, the well-known band gap is the difference between the valence and conduction bands in semiconductors that can be extended to a line on the imaginary axis of the complex plane. Such a band gap has a counterpart in non-Hermitian systems such that all the energy on the real or imaginary axis acts as the band gap, which is called a ``line gap'' \cite{ashida2020non}. Different from such line-gapped systems, the point-gap in the non-Hermitian systems can exist for a single band when the reference energy cannot be included in the spectrum \cite{kawabata2019symmetry,ashida2020non,ding2022non}, as shown in Fig.~\ref{fig:nonHeigen}(b). This allows to exploitation of the spectral topology to characterize the topological states or properties in non-Hermitian systems. In the following, we first discuss the correspondence between the spectral winding number and the non-Hermitian skin effect in the Hatano-Nelson model in Sec.~\ref{Nonhermitian_skin_effect}. Then we address energy vorticity, another topological invariant in terms of spectral topology and corresponding topological phenomena in Sec.~\ref{exceptionaltopology} and Sec.~\ref{Nodal_phase}.

\subsubsection{Hatano-Nelson model}\label{Nonhermitian_skin_effect}

In some non-Hermitian periodic systems, where lots of states are localized at the edge, the conventional bulk-boundary correspondence is broken; that is, the bulk wavefunction topology no longer governs the appearance of the edge localization \cite{xiong2018,yao2018edge,kunst2018biorthogonal,yokomizo2019non,xiao2020non,helbig2020generalized,ghatak2020observation,yang2020non}.
The energy spectra of such systems turn out to be very distinct under the periodic and open boundary conditions~\cite{zhang2020correspondence,okuma2020}, which can be well captured by the spectral winding number in the one-dimension, as explained in this part. A nonzero spectral winding number leads to the pile-up of a macroscopic number of bulk states at the boundary, not simply several edge states. Such an exotic non-Hermitian topological phenomenon is known as the non-Hermitian skin effect  \cite{hatano1996localization,xiong2018,kunst2018biorthogonal,yao2018edge,yokomizo2019non,song2019non,longhi2019probing,li2020topological,zhang2020correspondence,zhang2020,ashida2020non,kawabata2020higher,gou2020tunable,bergholtz2021exceptional,zhang2022universal,zhang2022review,Franca2022Non,lin2023topological,okuma2023non}. For such non-Hermitian systems, the wave vector in the Brillouin zone is allowed to be complex, which define the generalized Brillouin zone (GBZ). 
Despite the fact that the flourishing investigation of the non-Hermitian skin effect has brought many novel physics perspectives, efforts remain needed for a deep understanding and implementation of these effects~\cite{zhang2022review,lin2023topological,zhang2022universal}, in particular in higher dimensional systems, as reviewed in Sec.~\ref{high_dimension_skin_effect}.

\begin{figure}[htp]
\centering
\includegraphics[width=0.8\textwidth]{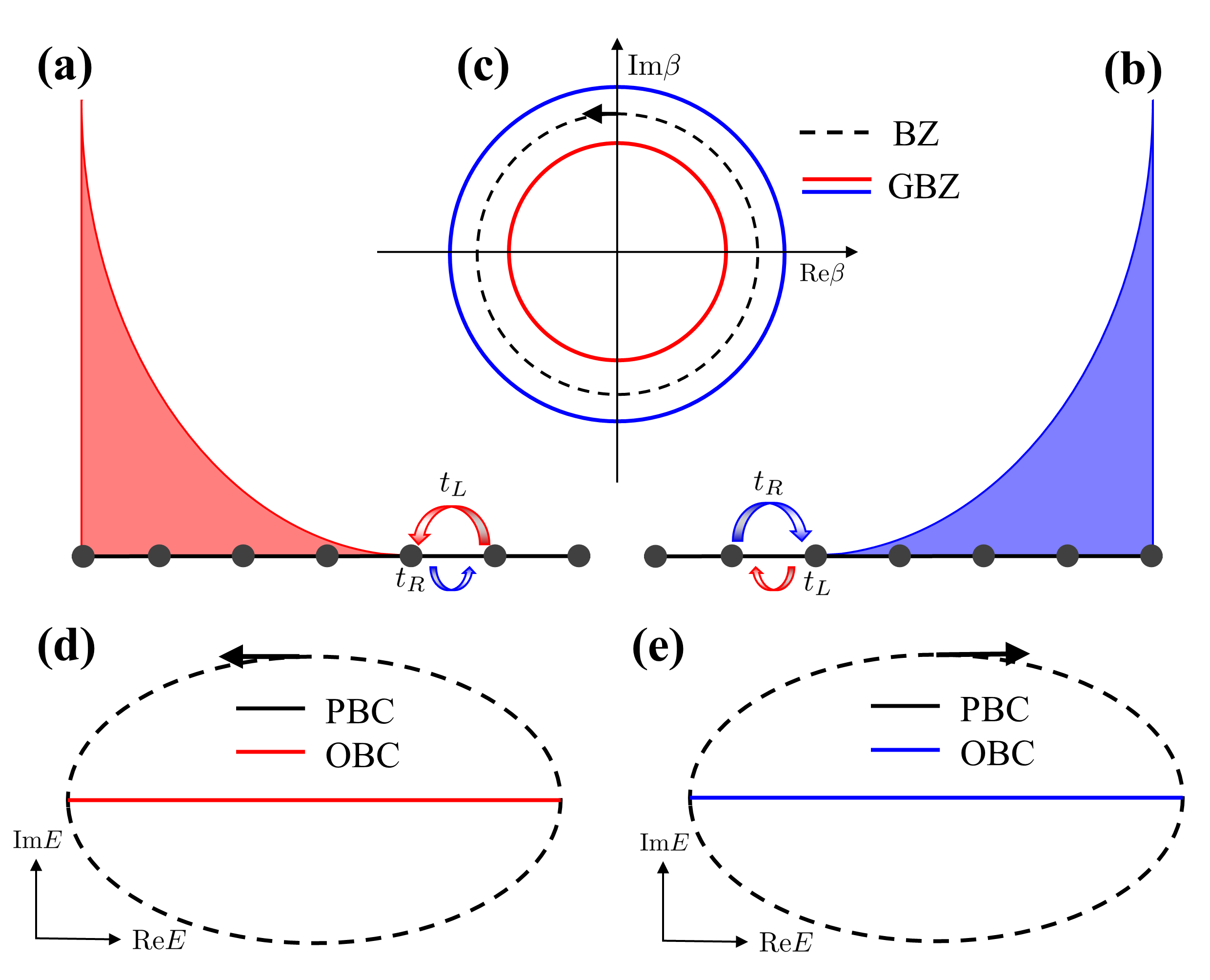}
\caption{Non-Hermitian skin effect and spectral topology in the one-dimensional Hatano-Nelson model with asymmetric couplings $t_L>t_R$ [(a)] and $t_L<t_R$ [(b)]. By the non-Hermitian skin effect, all the eigenstates aggregate at the left boundary when $t_L>t_R$ [(a)], but become skewed at the right boundary when $t_L<t_R$ [(b)].  In (c) with the complex wave vectors $\left|\beta \right| <1 $ when $t_L>t_R$ and $\left|\beta \right| >1 $ when $t_L<t_R$, with a distribution deviating from conventional Brillouin zone $\left|\beta \right| =1 $. (d) and (e) plot the energy spectra of the model under the PBC and OBC, respectively, with opposite asymmetric couplings $t_L>t_R$ [(d)] and $t_L<t_R$ [(e)].  As the wave vector evolves counterclockwise in (c) along the black dashed circle, the winding number of the energy spectra is $1$ when $t_L>t_R$ [(d)] and $-1$ when $t_L<t_R$ [(e)].}
\label{fig:NHlattice}
\end{figure}

In the following, we utilize the Hatano-Nelson model  \cite{hatano1996localization}, which may act as the simplest non-Hermitian discrete lattice model, to illustrate the non-Hermitian skin effect and its relation to the spectral topology. As shown in Fig.~\ref{fig:NHlattice}, the Hatano-Nelson model in one dimension is described by the non-reciprocal hopping between neighboring lattices:
\begin{equation}
\hat{H}=\sum\limits_{i=1}^{N-1}\left(t_L\hat{c}_i^{\dagger}\hat{c}_{i+1}+t_R\hat{c}_i\hat{c}_{i+1}^{\dagger}\right),\;\;\;t_L,t_R\in\mathbb{R},
\label{Hatano_Nelson}
\end{equation}
where $\hat{c}_i$ denotes the annihilation operator of particles at the site $i$, and $t_{L}$ and $t_{R}$ are the hopping amplitudes to the left and to the right, respectively.  $t_{L} \neq t_{R}$ is known as asymmetric coupling, chiral coupling, or non-reciprocal coupling, depending on which research fields it is concerned.

For a periodic lattice, the allowed wave vector $k$ is limited in the range $[-\pi,\pi]$. For a long chain with $N$ sites, the allowed wave vectors under the open boundary condition (OBC) are no longer real numbers. Accordingly, it is convenient to build a mapping from the complex $k$ to $\beta=e^{ik}$ and construct a solution for Eq.~\eqref{Hatano_Nelson}  with the Bloch basis $| \beta\rangle=\left(\beta,\beta^{2},...,\beta^{N}\right)^{T}$.
Note that the eigenvalue equations for the bulk and the boundary are different due to their different environment.
The eigenstate $\left|\Phi_{\beta}\right\rangle = \sum_{n} \alpha_n |{\beta}_n\rangle$ at energy $E$ is a superposition of the Bloch basis with amplitudes $\alpha_n$, which obeys $\hat{H}\left|\Phi_{\beta}\right\rangle=E\left|\Phi_{\beta}\right\rangle$. For the bulk, it leads to the eigenvalue equation
\begin{equation}
    t_R\sum_{n}\alpha_n\beta_n^{-1} +t_L\sum_{n}\alpha_n\beta_{n} = E \sum_{n}\alpha_n,
\end{equation}
so each $\beta_n=e^{ik_n}$ obeys
\begin{equation}
\label{bulk_equation}
t_{R}\beta_n^{-1}+t_{L}\beta_n-E=0.
\end{equation}
There are only two roots $\beta_{1}$ and $\beta_{2}$ for Eq.~\eqref{bulk_equation} with a degenerate energy, which implies $n=2$.  So $\left|\Phi_{\beta}\right\rangle=\alpha_{1}\left|\beta_{1}\right\rangle+\alpha_{2}\left| \beta_{2}\right\rangle$. Therefore, the eigenvalue equation for this problem reads
\begin{equation}
  \left(
  \begin{array}{ccccc}
    -E & t_L &  \cdots & 0  & 0 \\
    t_R & -E &  \cdots & 0 & 0 \\
    \cdots &  \cdots &  \cdots &  \cdots &  \cdots \\
    0 & 0 &  \cdots & -E & t_L \\
    0 & 0 &  \cdots & t_R & -E \\
  \end{array}
\right)\left(
  \alpha_1 \left(
             \begin{array}{c}
               \beta_1 \\
               \beta_1^2 \\
              \cdots \\
               \beta_1^{N-1} \\
               \beta_1^{N} \\
             \end{array}
           \right)
   +\alpha_2 \left(
               \begin{array}{c}
                 \beta_2 \\
                 \beta_2^2 \\
                 \cdots\\
                 \beta_2^{N-1} \\
                 \beta_2^{N} \\
               \end{array}
             \right)
     \right)=0.
\end{equation}
At the left and right boundaries, we find \begin{subequations}
\begin{align}
&t_L\left(\alpha_1\beta_1^2+\alpha_2\beta_2^2\right)=E\left(\alpha_1\beta_1+\alpha_2\beta_2\right), \\
&t_R\left(\alpha_1\beta_1^{N-1}+\alpha_2\beta_2^{N-1}\right)=E\left(\alpha_1\beta_1^{N}+\alpha_2\beta_2^{N}\right),
\end{align}
\end{subequations}
which with Eq.~\eqref{bulk_equation} leads to $t_R\left(\alpha_1 + \alpha_2\right)=0$ and $t_L\left(\alpha_1\beta_1^{N+1}+\alpha_2\beta_2^{N+1}\right)=0$, i.e.,
the matrix equation
\begin{equation}
\label{boundary_condition}
	\left(\begin{array}{cc}
		1 & 1\\
		\beta_{1}^{N+1} & \beta_{2}^{N+1}
\end{array}\right)\left(\begin{array}{c}
		\alpha_{1}\\
		\alpha_{2}
	\end{array}\right)=0.
\end{equation}
The secular equation requires $\beta_{1}^{N+1}=\beta_{2}^{N+1}$. Combined with $\beta_{1}\beta_{2}=t_R/t_L$ from Eq.~\eqref{bulk_equation}, the wave vectors obey  $\beta_{1,2}^{2(N+1)}=\left(t_R / t_L\right)^{N+1}$, leading to
\begin{equation}
	\beta_{1,2}= \sqrt{{t_R}/{t_L}}e^{ik_m},
\end{equation}
where $k_m = {m\pi}/({N+1})$ with  $m = \pm 1, \pm 2, \cdots, \pm N $, noting $\beta_1\ne \beta_2$. When $t_R>t_L$ ($t_R<t_L$), $\left|\beta\right|>1$ ($\left|\beta\right|<1$), so the wavefunction of all states exponentially increases when approaching the right (left) boundary, which is known as the non-Hermitian skin effect as illustrated in Fig.~\ref{fig:NHlattice}(a) and (b), respectively, for $t_R>t_L$ and $t_R<t_L$.

Figure~\ref{fig:NHlattice}(c) plots the distribution of $\beta_{1,2}$ on the complex plane, which is defined as the GBZ \cite{yao2018edge,yang2020non}, while $\beta$ for the conventional Brillouin zone (BZ) locates at a unit circle on the complex plane.  The GBZ may act as a useful tool for understanding the non-Hermitian phenomena such as the non-Hermitian skin effect \cite{yao2018edge,yang2020non}, the breakdown of the bulk-boundary correspondence \cite{yao2018edge,yokomizo2019non}, and the anomalous evolution of wavepackets \cite{longhi2019probing,mao2021boundary,zhang2022universal}.
The energy spectra  $E=2\sqrt{t_Rt_L}\cos{k_m}$ under the OBC is real and is thus simply a line as in Fig.~\ref{fig:NHlattice}(d) and (e). While under the PBC the energy spectra $E=t_R e^{-ik} + t_Le^{ik}$ form a closed loop on the complex plane when $k$ evolves in a period of $[-\pi,\pi]$. The manner of the energy spectra under PBC that encircles the reference energy is captured by the spectral topology, \textit{viz}. the spectral winding number    \cite{gong2018topological,shen2018topological}
\begin{equation}
	{\cal W}(E_r)=\frac{1}{2\pi}\text{\ensuremath{\oint_{{\rm BZ}}\frac{d}{d\beta}\text{arg}\left[E-E_r\right]d\beta}},
	\label{eq:wn}
\end{equation}
which counts the times of the energy encircles the reference energy $E_r$. Figure~\ref{fig:NHlattice}(d) and (e) plot the energy spectra  under the PBC and OBC, respectively, with opposite asymmetric couplings $t_L>t_R$ [Fig.~\ref{fig:NHlattice}(d)] and $t_L<t_R$ [Fig.~\ref{fig:NHlattice}(e)].  As the wave vector evolves counterclockwise in Fig.~\ref{fig:NHlattice}(c) along the black dashed circle, the winding number of the energy spectra is $1$ when $t_L>t_R$ [Fig.~\ref{fig:NHlattice}(d)] and $-1$ when $t_L<t_R$ [Fig.~\ref{fig:NHlattice}(e)]. Such a winding number has an exact correspondence to the non-Hermitian skin effect in that ${\cal W} =1$ and $-1$ correspond to the emergence of the non-Hermitian skin effect at the left and right boundaries, respectively.

\subsubsection{Two-sheeted Riemann surface involving second-order EPs}
\label{exceptionaltopology}

One common characteristic of the above-mentioned wavefunction topology and spectral topology is that the eigenvalue returns to its initial value after the parameter evolves along a closed path in the parameter space. However, this is not always true in the non-Hermitian scenario since the eigenvalue may take the form of multi-valued functions with different branches distributed in several Riemann surfaces.  For example, on the two-sheeted Riemann surface involving the second-order EPs, the eigenvalue of different branches can swap with each other\cite{heiss2012physics,shen2018topological,ding2022non}, as addressed in the following.

\begin{figure}[!htp]
\centering
\includegraphics[width=0.96\textwidth]{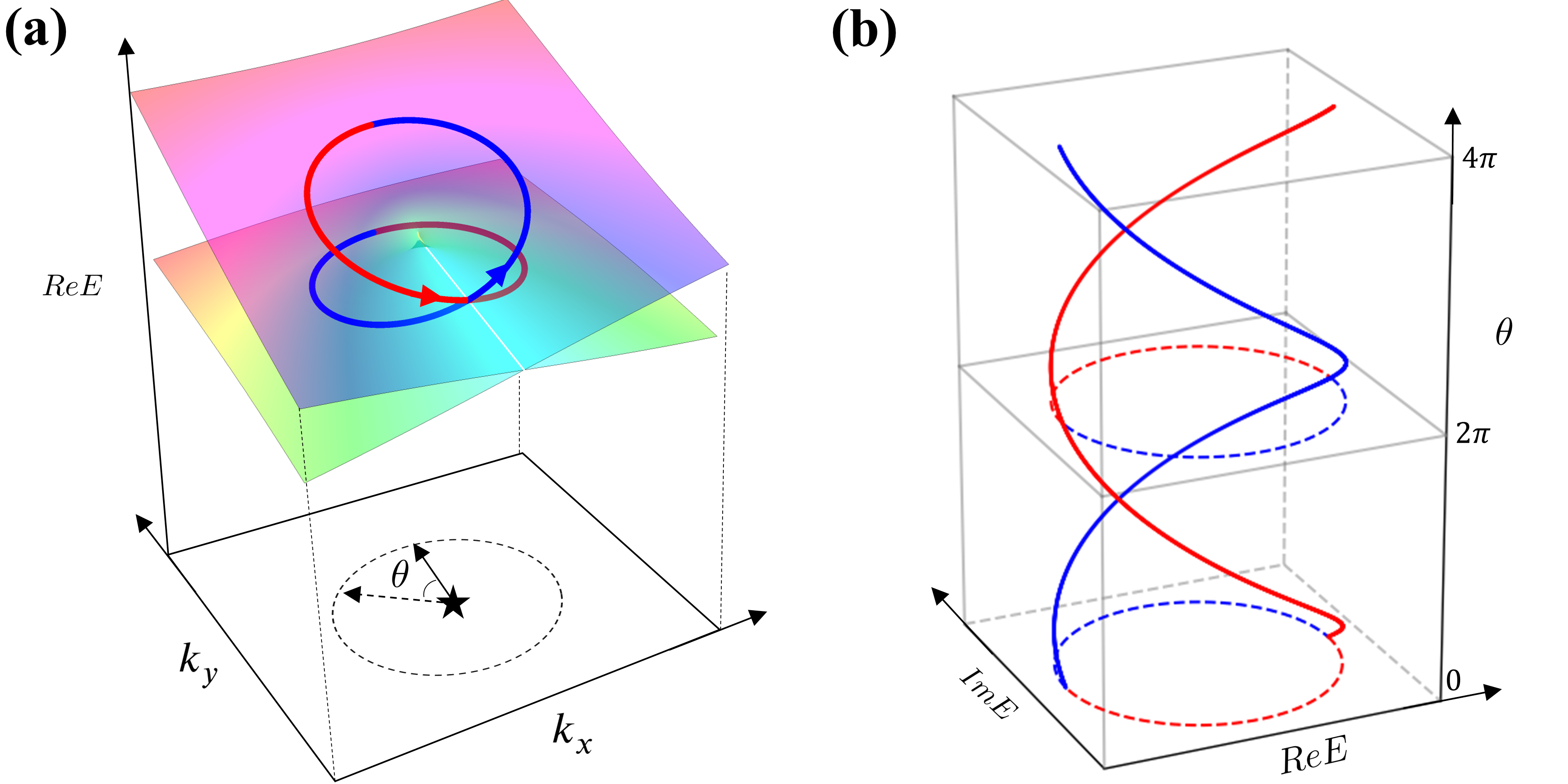}
    \caption{In (a), when the parameter $k_+$ evolves along a closed loop $\Gamma$ encircling the EPs denoted by the black star once, the real part of the eigenvalue switches from one Riemann surface to another, where the two evolution paths for the two eigenvalues are denoted by the red or blue curves. (b) addresses the swapping of eigenvalue on the complex plane when $k_+$ evolves on a circle.}
    \label{epswitch}
\end{figure}

To elucidate the topological properties of the EPs, we focus on a simple (chiral) Hamiltonian
\begin{equation}\label{eq:chiralHami}
      {\cal H}= \left( \begin{array}{cc}
           0 & k_+ \\
           1 & 0
       \end{array} \right).
\end{equation}
This simple 2$\times$2 Hamiltonian may help to understand the topological nature without too many calculations. Here, the coupling is non-reciprocal and is parametrized by the complex ``wave vector'' $k_+=k_x+ik_y$, with $k_x = |k_+| \cos \left(\theta + \pi/2\right)$ and $k_y = |k_+| \sin \left(\theta + \pi/2\right)$ illustrated in Fig.~\ref{epswitch}(a). It yields the eigenvalues $E_{\pm}(k_+) = \pm \sqrt{k_+}$ with EP located at $k_x = k_y = 0$ and the right and left  eigenvectors $|\Phi_{\pm}\rangle= \left(1,\pm 1/\sqrt{k_+}\right)^{T}/2$ and $\langle\Psi_{\pm}|= \left(1,\pm \sqrt{k_+}\right)$.  Since the eigenvalues are multivalued, in the dependence on the parameters $k_x$ and $k_y$ in Fig.~\ref{epswitch}(a), $\text{Re}E_{\pm}$ locates at the Riemann surface of two different branches \cite{heiss2012physics,bergholtz2021exceptional}. As the parameter $k_+$ evolves counterclockwise on a circle, as shown in Fig.~\ref{epswitch}(a), the two eigenvalues $E_{\pm}$ evolve along the red or blue curves, respectively. After encircling EPs once,  $E_{\pm}$ switches from one Riemann surface to the other. Correspondingly, the eigenvector also swaps from $|\Phi_{\pm}\rangle$ to $|\Phi_{\mp}\rangle$.
 Thereby only when the parameter encircles the EPs twice, the eigenvalue can return to its initial value. Figure~\ref{epswitch}(b) shows how the eigenvalues swap and return to their initial values on the complex plane.

The swapping of eigenvalues is captured by energy vorticity \cite{shen2018topological}
\begin{equation}
\nu_{+-}(\Gamma)=-\frac{1}{2 \pi} \oint_{\Gamma} \frac{d  \arg \left[E_{+}(k_+)-E_{-}(k_+)\right]}{d k_+} d k_+= -\frac{1}{2 \pi}\oint_{\Gamma} \frac{d \left(\theta/2 + \pi/4 \right)}{d k_+} d k_+ = -\frac{1}{2},
\label{vorticity}
\end{equation}
where $\Gamma$ is a closed loop surrounding the EPs as show in Figure~\ref{epswitch}(a). The energy vorticity $-1/2$ implies both two branches evolve counterclockwise and their path merge to form a closed loop on the complex plane, as shown in Figure~\ref{epswitch}(b). This result also implies the EP can be characterized by half-integer topological charges \cite{shen2018topological}.

The adiabatic parametric evolution of the wavefunction in the Hermitian scenario along a closed loop around the degeneracy point contributes to the Berry phase \cite{berry1984quantal}.
An evolution of $k_+$ encircling non-Hermitian degeneracy point, \textit{i.e.}, EPs, also brings intriguing and novel effects \cite{mailybaev2005geometric,heiss2012physics,gao2015observation}. The accumulated geometric phase by encircling the EPs twice is
\cite{mailybaev2005geometric,zhong2018winding,ashida2020non}:
\begin{equation}
    i\oint_{2 \Gamma} d{k_+} \left<\Psi_{\pm}\right|\frac{d}{dk_+} \left|\Phi_{\pm}\right> = \frac{1}{2}\oint_{2 \Gamma} d{k_+} \frac{d \left(\theta/2 + \pi/4 \right)}{d {k_+}}=\pi,
\end{equation} which is equal to that by encircling the Dirac point once in the Hermitian counterpart [Eq.~\eqref{berry}]. This also leads to a fractional winding number $(\pi/\pi)/2 = 1/2$ \cite{ding2016emergence,yin2018geometrical}, which is confirmed experimentally \cite{zhou2018observation} and is regarded as being related to the anomalous edge mode in a non-Hermitian lattice model \cite{lee2016anomalous,zhang2020non}. We note that one may find various forms of $2\times2$ non-Hermitian Hamiltonian that were used to elucidate the topological properties in terms of EPs in literature \cite{heiss2012physics,shen2018topological}, but these properties such as the above-mentioned nonzero energy vorticity and fractional winding number only depend on how the loop in parameter space encircles a single EP, but not on the specified form of Hamiltonian.

\subsubsection{Nodal phase: non-Hermitian Weyl semimetal}
\label{Nodal_phase}

As addressed above, the EPs in the wave-vector space can be topologically characterized by a half-integer topological charge. More complicated band structures can appear in the non-Hermitian scenario. The nodal phase refers to the emergence of energy degeneracy \textit{in the reciprocal wave-vector space}. Such energy degeneracy can be points, lines, rings, and surfaces. The possible structures are not limited to these, however. For example, two rings, making an analogy to two ``keychains'', can lock together with each passing through the center of the other, which is called ``Hopf-link'' \cite{chang2017topological,yang2019non}. Even for a single degeneracy line in the reciprocal space, it can be entangled with itself, which is called ``knot'' \cite{bi2017nodal,carlstrom2019knotted}. The simplest nontrivial knot is the trefoil knot and its configuration can be obtained by tying an overhand knot on the degeneracy line and then connecting two ends. These nontrivial degeneracy structures enrich topological characterization with new topological invariants such as link number \cite{yang2019non} and knot invariant \cite{yang2020jones}.

\begin{figure}[htp]
    \centering
    \includegraphics[width=16cm]{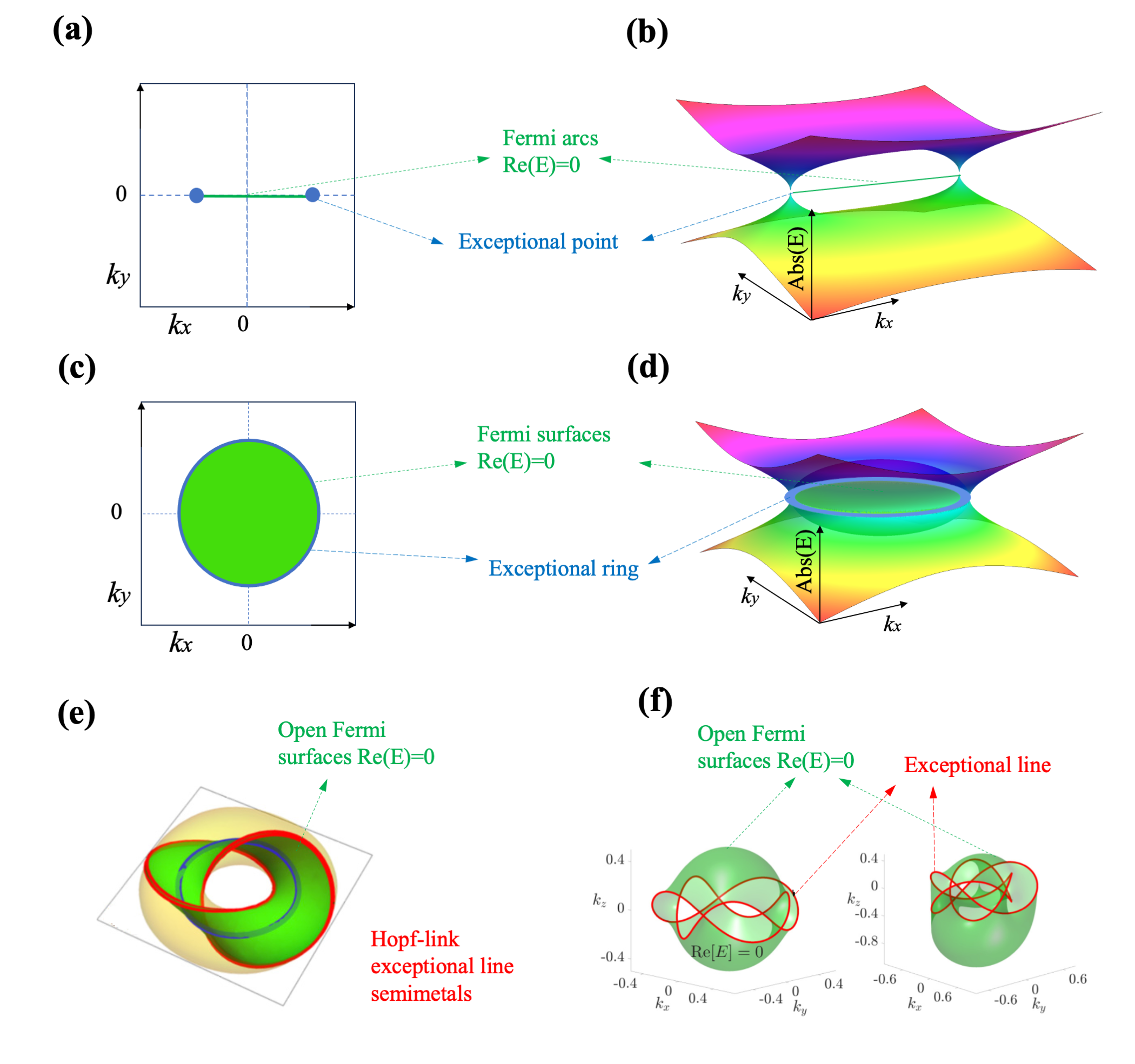}
\caption{Non-Hermitian or exceptional nodal phases. (a)-(f) compares the energy dispersion with EPs, exceptional rings, Hopf-link exceptional lines, and exceptional nodal knots. The open Fermi ``arc'' and ``surface'' are defined by vanishing real component $\text{Re}E=0$ in the dispersion, denoted by the green lines or green regions that connect the EP pairs in (a) and (b), exceptional rings in (c) and (d), Hopf-link exceptional lines in (e), and knot exceptional line in (f). Figures~(e) and (f) are, respectively, adapted with permission from Ref.~\cite{yang2019non} and Ref.~\cite{carlstrom2018exceptional}.
    \label{fig:EPnodalphase}}
\end{figure}

Similar to the Weyl points in the three-dimensional space that are stable without protection by special symmetry (refer to Sec.~\ref{BDGHamiltonian}), there are stable degeneracies in the two-dimensional non-Hermitian system that require no symmetry protection.
We exploit a two-band non-Hermitian Hamiltonian to illustrate the physics of such non-Hermitian band structures, which is parametrized in the two-dimensional momentum  ${\bf k}=(k_x,k_y)$:
\begin{equation}
H(\boldsymbol{k})=\left(\mathbf{d}_R({\bf k})+i\mathbf{d}_I({\bf k})\right)\cdot\pmb{\sigma},
	\label{eq:Hamiltonian}
\end{equation}
where the two-dimensional vectors $\mathbf{d}_R({\bf k})$ and $\mathbf{d}_I({\bf k})$ are the real functions such that the Hamiltonian is not Hermitian.
Its complex energy spectra
\begin{equation}
	E_{\pm}({\bf k})=\pm\sqrt{\mathbf{d}_R^2({\bf k})-\mathbf{d}_I^2({\bf k})+2i\mathbf{d}_R({\bf k})\cdot\mathbf{d}_I({\bf k})}
	\label{eq:Hamiltonian2}
\end{equation}
indicates that the energy degeneracies, i.e., $E=0$, appear at those momenta satisfying
\begin{subequations}\label{eq:tbm}
\begin{align}
&\boldsymbol{\mathrm{d}}_R^2({\bf k})-\boldsymbol{\mathrm{d}}_I^2({\bf k})=0,\\
&\boldsymbol{\mathrm{d}}_R({\bf k})\cdot\boldsymbol{\mathrm{d}}_I({\bf k})=0.
\end{align}
\end{subequations}
There are two equations for the two variables $(k_x,k_y)$, so the degeneracy points, saying ${\bf k}_d$, are usually isolated at the momenta ${\bf k}$-space.
The stability of such degeneracies is characterized by the energy vorticity of energy spectra \cite{shen2018topological,kawabata2019classification,hu2021knots}. With ${\rm det}[H({\bf k})]=-\left(\mathbf{d}_R^2({\bf k})-\mathbf{d}_I^2({\bf k})+2i\mathbf{d}_R({\bf k})\cdot\mathbf{d}_I({\bf k})\right)=-E^2_{\pm}({\bf k})$,
the extended topological charge (energy vorticity) of the EP at ${\bf k}_d$ is defined as \cite{shen2018topological,kawabata2019classification,yang2020jones,hu2021knots}
\begin{align}
    \nu({\bf k}_d)& =\frac{i}{2\pi}\oint_{\Gamma({\bf k}_d)}d{\bf k}\cdot \nabla_{\bf k}\ln{\rm det}\left[H({\bf k})\right] \nonumber\\
    &  = \frac{i}{2\pi}\oint_{\Gamma({\bf k}_d)}d{\bf k}\cdot \nabla_{\bf k}\ln\left(-\frac{1}{4} \left[E_{+}({\bf k})-E_{-}({\bf k})\right]\left[E_{-}({\bf k})-E_{+}({\bf k})\right]\right) \nonumber\\
    & = \frac{-1}{2 \pi} \oint_{\Gamma({\bf k}_d)}d{\bf k}\cdot \nabla_{\bf k} \left[E_{+}({\bf k})-E_{-}({\bf k})\right]+ \frac{-1}{2 \pi} \oint_{\Gamma({\bf k}_d)}d{\bf k}\cdot \nabla_{\bf k} \left[E_{-}({\bf k})-E_{+}({\bf k})\right] \nonumber\\
    & = \nu_{+-}({\Gamma({\bf k}_d)})+ \nu_{-+}({\Gamma({\bf k}_d)}), 
\end{align}
where $\Gamma({\bf k}_d)$ is a closed loop encoding ${\bf k}_d$ in the reciprocal space and $\nu_{+-}({\bf k})=\nu_{-+}({\bf k})$ [refer to Eq.~(\ref{vorticity})]. Since the sum of the topological charge of all EPs equals zero, these EPs should appear in pairs in two-dimensional wave-vector space, which is similar to paired Weyl points in three-dimensional Hermitian semimetals \cite{yan2017topological,armitage2018weyl,yang2021fermion}.
For example, for the non-Hermitian Hamiltonian (the constant $c\ne 0$)
\begin{align}\label{nonHermitianEP}
H({\bf k})=k_x\sigma_x+k_y\sigma_y+i c\sigma_y,
\end{align}
the eigenvalues $E_{\pm}=\pm \sqrt{k_x^2 + \left(k_y + i c\right)^2}$. When $k_x^2 +k_y^2 -c^2 =0$ and $k_y=0$, a pair of EPs emerge, which are located at $\left(k_x,k_y\right)=\left(\pm c, 0\right)$, as shown in Fig.~\ref{fig:EPnodalphase}(a) and (b). 

The nodal structures can be realized in both the Hermitian and non-Hermitian systems, but the symmetry is less demanding for the latter, thereby rendering the non-Hermitian nodal phases more robust to perturbations. To see this, we compare the properties of the Dirac point and the exceptional degeneracy in two dimensions.  
We emphasize that although the stability of the Dirac point in two dimensions requires symmetry, the exceptional degeneracy in the two dimensions can be realized in the absence of symmetry, as shown in the following. In the Hamiltonian  $H({\bf k})=k_x\sigma_x+k_y\sigma_y$,  the Dirac point is located at $\left(k_x,k_y\right)=\left(0, 0\right)$. The Dirac point is gapped when a perturbation term $\epsilon_3\sigma_z $ ($\epsilon_3$ is a small constant) that breaks the chiral symmetry is introduced. For the above non-Hermitian Hamiltonian $H({\bf k})=k_x\sigma_x+k_y\sigma_y+i c\sigma_y$ [\eqref{nonHermitianEP}] with $\mathbf{d}_R({\bf k})= \left(k_x,k_y,0\right)$ and $\mathbf{d}_I({\bf k})= \left(0,c,0\right)$, any perturbations are allowed such that $\mathbf{d}_R({\bf k})\rightarrow \left(k_x+\epsilon_1,k_y+\epsilon_2,\epsilon_3\right)$ and $\mathbf{d}_I({\bf k})\rightarrow \left(\epsilon_4,c+\epsilon_5,\epsilon_6\right)$, where $\epsilon_{1,\cdots,6}$ are small constants, but the non-Hermitian degeneracy still exists, i.e., they exist as the roots of the equations 
\begin{subequations}
    \begin{align}
    k_x^2 +k_y^2 -c^2 & = -2 k_x \epsilon_1 -2 k_y \epsilon_2 + 2 c \epsilon_5 -\epsilon_1^2 -\epsilon_2^2 + \epsilon_4^2 + \epsilon_5^2 + \epsilon_6^2, \\
   c k_y  & = -  \left( k_x + \epsilon_1\right)\epsilon_4 -c \epsilon_2 - k_y \epsilon_5- \epsilon_4 \epsilon_6. 
\end{align} 
\end{subequations}
Since the right side of the above equation is a small perturbation, the new non-Hermitian degeneracy can be obtained by slightly adjusting the position of the original EP. The above discussions imply that the symmetry is less demanding for the realization of the non-Hermitian nodal phase, thereby rendering them more robust to perturbations in comparison to the Hermitian nodal phase.
Such analysis is also applied to more complex nodal structures. For example, the Hopf-link semimetals have been predicted in both the Hermitian \cite{chang2017topological} and non-Hermitian systems \cite{yang2019non} and its structure can be seen in Fig.~\ref{fig:EPnodalphase}(e), while the latter one can be realized in the absence of symmetry. Another example is that fine-tuning or symmetry is required to obtain the Hermitian knot semimetals \cite{bi2017nodal}, while none of them are needed for the realization of knotted non-Hermitian metals \cite{carlstrom2019knotted}.

As addressed above, in Hamiltonian (\ref{eq:Hamiltonian}) the non-Hermitian degeneracy emerges when $\boldsymbol{\mathrm{d}}_R^2({\bf k})-\boldsymbol{\mathrm{d}}_I^2({\bf k})=0$ and $\boldsymbol{\mathrm{d}}_R({\bf k})\cdot\boldsymbol{\mathrm{d}}_I({\bf k})=0$. There exists a region with  $\boldsymbol{\mathrm{d}}_R^2({\bf k})-\boldsymbol{\mathrm{d}}_I^2({\bf k})< 0$ and $\boldsymbol{\mathrm{d}}_R({\bf k})\cdot\boldsymbol{\mathrm{d}}_I({\bf k})=0$, rendering  $\text{Re}E=0$ and $\text{Im}E\neq0$, which is called ``bulk Fermi arcs'' or ``surfaces''. Since such bulk Fermi arcs or surfaces are bounded by non-Hermitian degeneracies, their configuration depends on the structure of degeneracies \cite{kozii2017non,zhou2018observation}. For example, for Eq.~(\ref{nonHermitianEP}), the bulk Fermi arcs emerge and connect the pair of EPs ($k_x^2-c^2<0$)  as shown in Fig.~\ref{fig:EPnodalphase}(a) and (b).
To illustrate the bulk Fermi arc in the ring degeneracy, let us consider the Hamiltonian 
\begin{align}
H({\bf k})=k_x\sigma_x+k_y\sigma_y+i c\sigma_z, 
\end{align}
with the eigenvalue $E_{\pm}({\bf k})=\pm \sqrt{k_x^2+k_y^2-c^2}$ as shown in Figure~\ref{fig:EPnodalphase}(d). Now the exceptional degeneracy is a ring that is located at $k_x^2+k_y^2=c^2$, also known as the nodal ring semimetal. Within the exceptional ring, there exist bulk Fermi surfaces with ${\rm Re}E({\bf k})=0$, as plotted in Fig.~\ref{fig:EPnodalphase}(c). 
In more complicated structures, such a bulk Fermi surface can form a topological structure resembling the M{\"o}bius strip when two exceptional rings in the reciprocal space form a Hopf-link structure as shown in Fig.~\ref{fig:EPnodalphase}(e) \cite{yang2019non}. It becomes more complicated when the exceptional line forms knots as shown in Fig.~\ref{fig:EPnodalphase}(f) \cite{carlstrom2019knotted}.

Compared to the well-known surface Fermi arcs in Weyl semimetal \cite{yan2017topological,armitage2018weyl}, the bulk Fermi arcs in the non-Hermitian nodal phase also have a topological nature that lies in the winding of complex energy around EPs, \textit{i.e.}, nonzero energy vorticity \cite{kozii2017non,bergholtz2021exceptional}, which characterizes the bulk states rather than the surface ones. Experimentally, the open Fermi arc or open Fermi surface has been realized in several systems such as photonic crystal \cite{zhen2015spawning,zhou2018observation} and helical waveguide array \cite{cerjan2019experimental}, which may be detected by angle-resolved scattering measurements \cite{zhou2018observation}. The Fermi arcs that connect distinct EPs in the reciprocal space have been proposed theoretically in
several systems that involve magnon or spin \cite{mcclarty2019non,bergholtz2019non,yang2021exceptional}, while the experimental realizations of Fermi arcs and theoretical exploration of open Fermi surfaces in magnonic devices are, to the best of our knowledge, still lacking (refer to Sec.~\ref{Nodal_phase}).

\section{Exceptional points, lines, and surfaces in magnonic systems}

\label{Magnonic_EPs}

As a general physical phenomenon, EPs widely exist in many coupled systems, including the optic \cite{hodaei2014parity,el2018non,ozdemir2019parity,miri2019exceptional}, acoustic \cite{zhu2014p,jing2014pt,fleury2015invisible,liu2018unidirectional}, electronic~\cite{schindler2011experimental,choi2018observation,sakhdari2019experimental,xiao2019enhanced,stegmaier2021topological}, as well as spintronic~\cite{lee2015macroscopic,zhang2017observation,cao2019exceptional,PhysRevLett.123.237202,yuan2020steady} systems. Many properties, such as the topological property (Sec.~\ref{exceptionaltopology}) \cite{heiss2012physics,mailybaev2005geometric,gao2015observation,zhou2018observation}, the mutation of eigenfrequencies \cite{moiseyev2011non,el2018non,miri2019exceptional,ozdemir2019parity,ding2022non}, and the spontaneous symmetry breaking \cite{bender1998real,bender1999pt,bender2002complex,bender2005introduction,bender2007making}, have been discovered when the system is driven to be around or at the EPs \cite{kawabata2019symmetry,ashida2020non,bergholtz2021exceptional,ding2022non,okuma2023non}, as reviewed in Sec.~\ref{exceptional_points} below. In this context, searching for exceptional points, lines, and surfaces in magnonic systems is an attractive and significant topic. Contemporary new breakthroughs lead to new functionalities with magnons. The magnonic systems are highly tunable by diverse techniques, such as the magnetic/electric field, the thermal gradient, and even the stress, which makes the realization and manipulation of the EPs in magnonic systems fertilized with other coupled systems. In other words, the magnonic systems provide us with a powerful platform for engineering the EPs and the unique functionalities associated with them. A significant milestone was reached with the experimental realization of an exceptional surface in a four-dimensional synthetic space, achieved through the tunable properties of the cavity magnonic system \cite{PhysRevLett.123.237202}, which not only creates new prospects for high-dimensional control of non-Hermitian systems but also exerts an immediate and profound impact on the study of non-Hermitian physics in various systems. Magnons are excellent information carriers that can interact with different particles, such as electrons, phonons, photons, and other magnons. Given this, researchers are motivated to study the variation of the properties of magnons near the EPs, such as their eigenfrequencies and dissipation, which have impacts on the electrical, mechanical, or optical properties of the magnonic devices. For example, the introduction of EPs into magnonic systems strongly enhances the sensitivity of the magnonic frequency comb, achieving the recorded 32 teeth~\cite{wang2023MFC}. In addition, magnons carry spins inherited from the intrinsic spin precession direction and possess unique ``chirality''---a fixed product of the spin, propagation direction, and the surface normal that allow chiral interactions between magnons and many other quasiparticles such as phonons, other magnons, electrons, Cooper pairs, and microwave photons~\cite{yu2023chirality}. This property opens up new opportunities for the study of non-Hermitian physics based on the magnonic hybridized system, achieving many fascinating phenomena such as the magnon accumulation \cite{yu2020magnon_accumulation,yu2022giant,deng2022non,zeng2023radiation} and the non-Hermitian skin effect~\cite{deng2022non,yu2022giant,yu2020magnon_accumulation,zeng2023radiation,cai2023corner,li2023reciprocal}.

Thus far, we have addressed unified approaches to handle the non-Hermitian dynamics of magnons from the master equation (Sec.~\ref{master_equation_approach}) and Green's function (Sec.~\ref{Green_function}) approaches. In this Section, we review the magnonic realization of EPs and their potential applications aroused by the singularity in spectra and eigenvectors, as summarized in Table~\ref{table_EP_application} for typical examples. We start our discussion by elucidating the concept of exceptional degeneracies (Sec.~\ref{exceptional_points}).  We then turn to address recent theoretical studies on EPs with magnons, including several systems, for instance, the dissipatively coupled spin systems (Sec.~\ref{spin_pumping_dissipative_coupling})  and spin systems with balanced gain and loss (Sec.~\ref{EPs_magnonics}). 
We discuss several experimental paths to realize the EPs in magnonics (Sec.~\ref{Sec_magnon_PT} and \ref{Sec_EPs_cavity_magnonics}). The first technical path is to construct EPs in magnetic heterostructures (Sec.~\ref{Sec_magnon_PT}). These devices are composed of two or more adjacent ferromagnetic layers separated by such as a normal metal layer \cite{yu2020higher, Liu2019Observation, wang2020steering}. Two ferromagnetic layers with different dissipative properties strongly couple with each other via the spin transfer torque or the magnetic dipolar interaction. Precisely controlling their separations can control the coupling strength between them, such that EPs can be achieved. Another path to experimentally realize the EPs is based on cavity magnonics (Sec.~\ref{Sec_EPs_cavity_magnonics}) \cite{soykal2010strong, huebl2013high, tabuchi2015coherent, goryachev2014high, lachance2019hybrid, rao2021interferometric, rao2019analogue, wang2018bistability, rameshti2022cavity,wang2023MFC}, which possess excellent versatility and high tunability, leading the study of EPs in cavity magnonics to be feasible. Consequently, digging exotic properties around or at EPs has become one of the central tasks in cavity magnonics. We finally discuss the realization of exceptional surfaces in the magnonic system (Sec.~\ref{Exceptionalsurfacescavity}) \cite{PhysRevLett.123.237202,grigoryan2022pseudo}.

\begin{table}
   \caption{Potential applications of EPs in magnonics.} \label{table_EP_application}
  \begin{tabular}{ccc}
   \toprule[1pt]
  \makebox[0.1\textwidth][c]{Key features}&\makebox[0.5\textwidth][c]{Model and Results}&\makebox[0.2\textwidth][c]{ Functionalities} \\
        \midrule[0.5pt]
 \cellcolor[HTML]{FBE5D6}Absorption  & \begin{minipage}[m]{.5\textwidth}
      \centering\vspace*{4pt}\includegraphics[width=7cm]{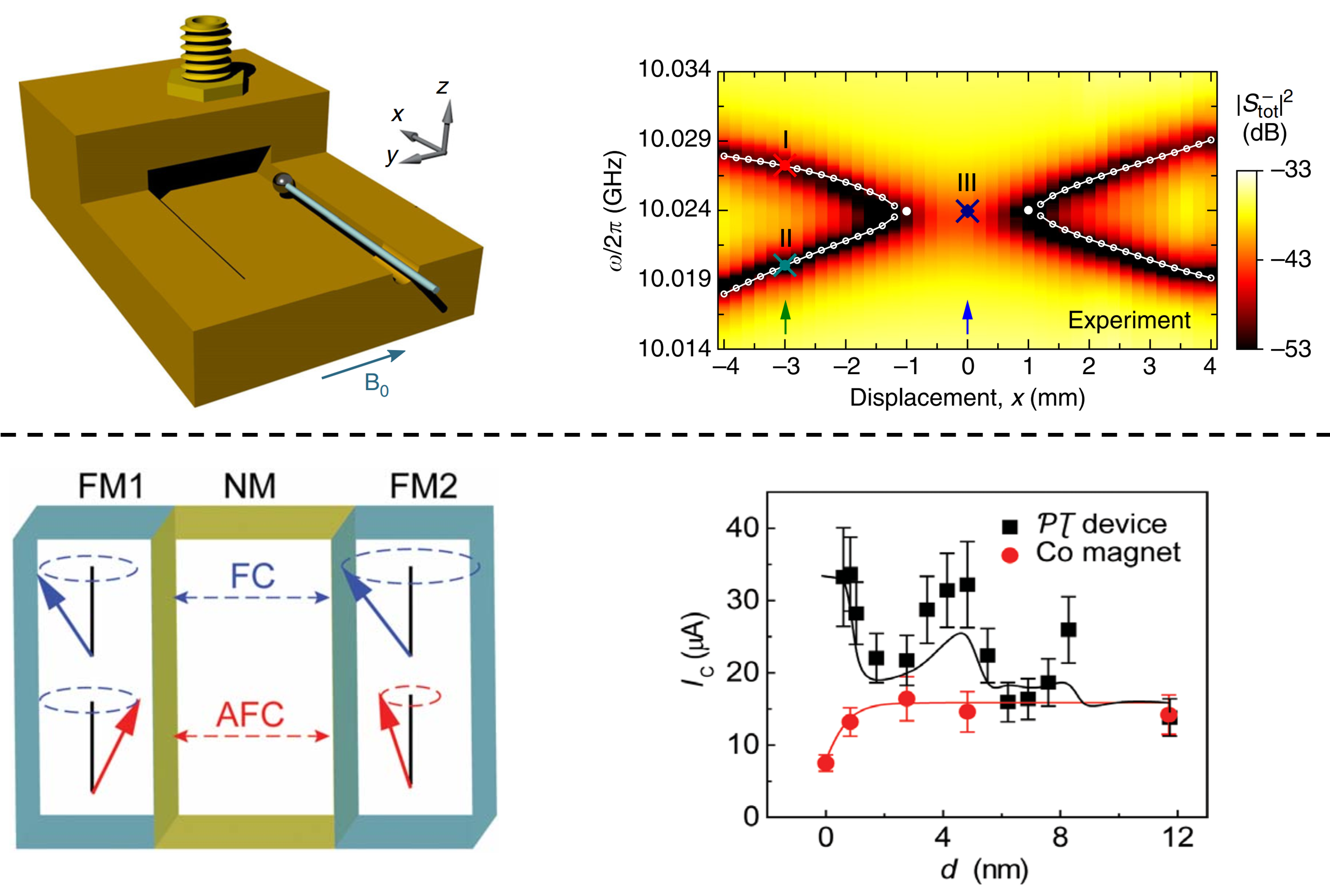}\vspace*{2pt}
    \end{minipage} & \begin{minipage}[m]{.3\textwidth}
    \begin{itemize}
    \item Polaritonic coherent perfect absorption~\cite{zhang2017observation}
    \item Enhanced magnon current~\cite{Liu2019Observation}
    \end{itemize}
    \end{minipage}
    \\
        \midrule[0.5pt]
    \cellcolor[HTML]{DEEBF7}Amplification &  \begin{minipage}[m]{.4\textwidth}
      \centering\vspace*{5pt}
      	\includegraphics[width=7.0cm]{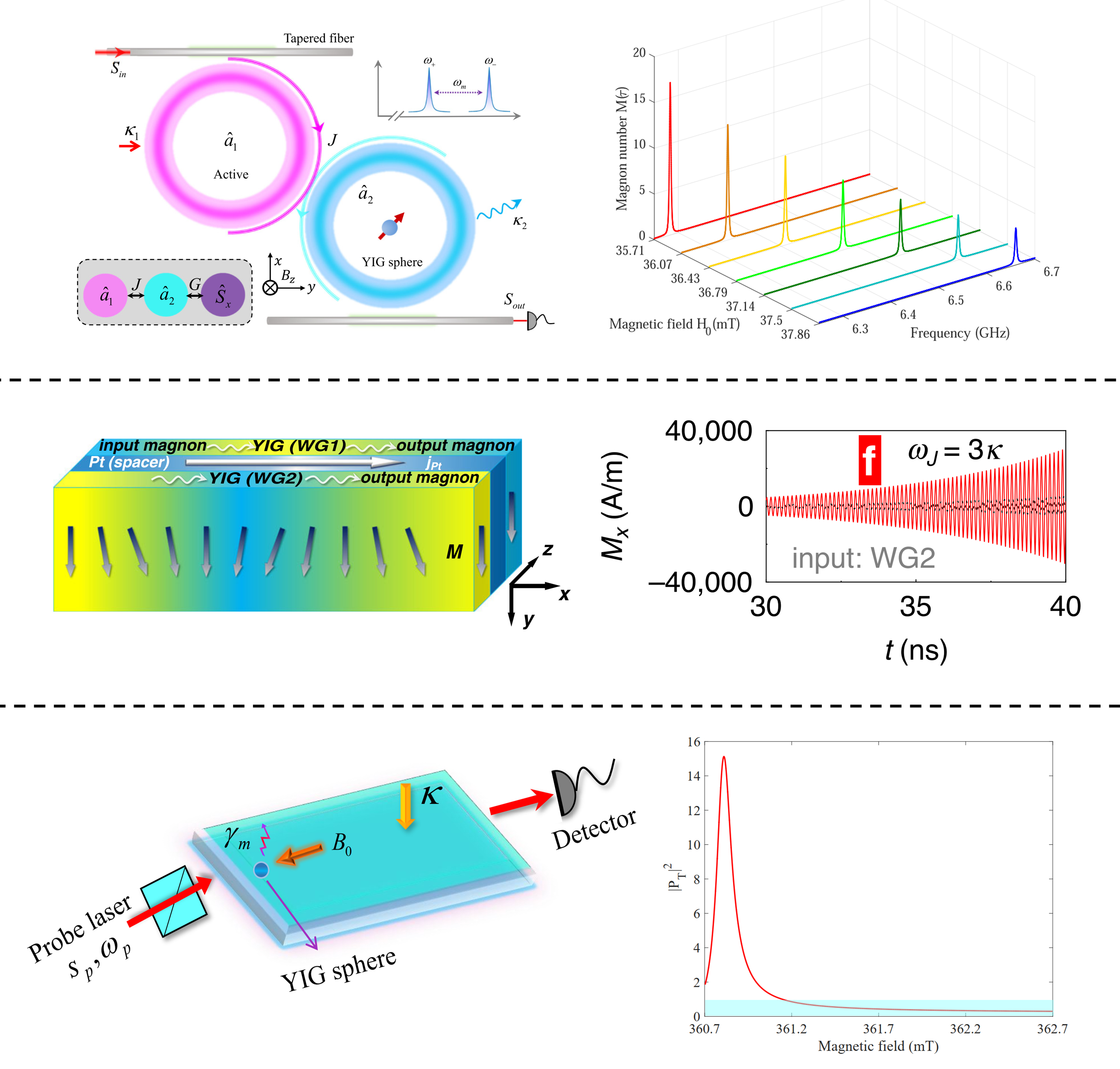}\vspace*{2pt}
    \end{minipage} & \begin{minipage}[m]{.3\textwidth}
    \begin{itemize}
    \item Stimulated emitted magnon number \cite{wang2022pt}
    \item Enhancement of magnonic permeability~\cite{wang2020steering}
    \item Magnon amplification~\cite{wang2020steering}
    \item Transmission enhancement ~\cite{wang2018magnon}
    \end{itemize}
    \end{minipage}  \\
        \midrule[0.5pt]
     \cellcolor[HTML]{E2F0D9}Sensitivity &   \begin{minipage}[m]{.4\textwidth}
      \centering\vspace*{5pt}
      	\includegraphics[width=6.8cm]{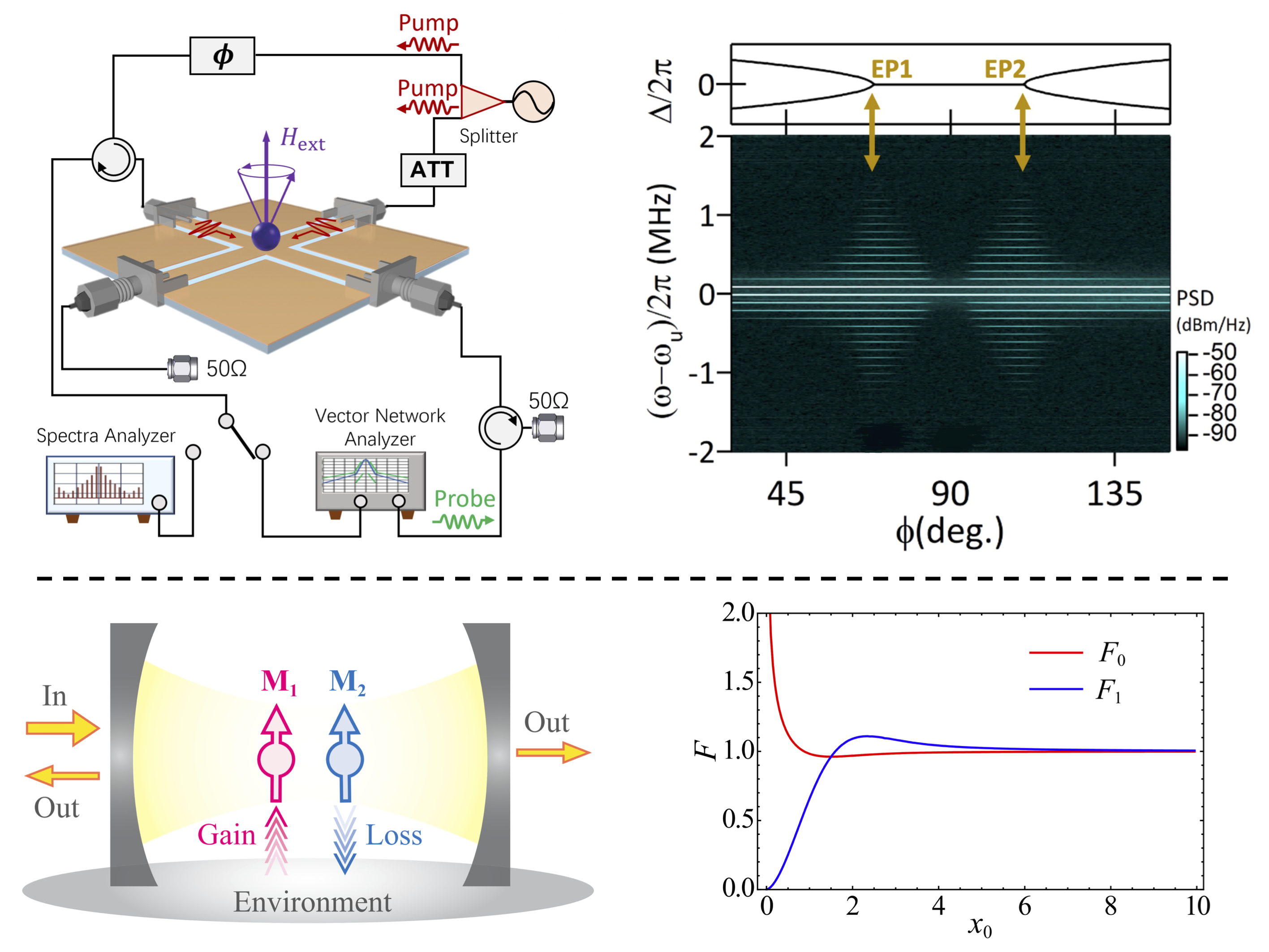}\vspace*{2pt}
    \end{minipage} & \begin{minipage}[m]{.3\textwidth}
    \begin{itemize}
    \item Giant enhancement of magnonic frequency comb \cite{rao2023unveiling,wang2023MFC}
    \item Third-order EP~\cite{cao2019exceptional}
    \item  Sensitivity approaches fetotesla~\cite{cao2019exceptional}
    \end{itemize}
    \end{minipage}  \\
   \toprule[1pt]
  \end{tabular}
\end{table}

\subsection{Exceptional points, lines, and surfaces}
\label{exceptional_points}

One of the most striking features of a non-Hermitian Hamiltonian matrix is the existence of the EPs \cite{heiss2012physics,el2018non,miri2019exceptional,ozdemir2019parity,bergholtz2021exceptional,ding2022non,okuma2023non}, with which the eigenvalues are degenerate at some isolated points in the parameter space with the same real and imaginary components, and the associated eigenvectors are coalescent to be the same or ``parallel'' \cite{kato1966perturbation,berry1998diffraction,heiss2001chirality,moiseyev2011non,heiss2012physics,brody2013biorthogonal,kunst2018biorthogonal}. For example, at the EPs, only a single left or a single right eigenvector remains for a $2\times 2$ matrix. For the general higher ranking $N\times N$ ( $N>2$) non-Hermitian Hamiltonian matrix, the $N$-fold degeneracies of the eigenvalues and the corresponding coalesce of the eigenvectors into a single one lead to $N$th-order EPs~\cite{hodaei2017enhanced,wang2019arbitrary,zhang2019higher,zhong2020hierarchical,wang2021enhanced,yu2020higher,mandal2021symmetry}.  For simplicity, unless specified we mostly focus on the second-order EPs in this review article. Not limited to a mathematical phenomenon, the EPs emerge in a variety of physical systems such as mechanics \cite{liu2016metrology,chen2020revealing,mao2020enhanced}, optics \cite{el2007theory,makris2008beam,bendix2009exponentially,guo2009observation,ruter2010observation,kottos2010broken,makris2010pt,szameit2011p,regensburger2012parity,lin2011unidirectional,hodaei2014parity,feng2014single,christensen2016parity}, acoustics \cite{zhu2014p,jing2014pt,fleury2015invisible,liu2018unidirectional}, electronic circuits \cite{schindler2011experimental,choi2018observation,sakhdari2019experimental,xiao2019enhanced,stegmaier2021topological}, as well as magnonic devices \cite{yang2018antiferromagnetism,xiao2019enhanced,Liu2019Observation,yang2020Unconventional,yuan2020steady,wang2021coherent,rao2021interferometric,wang2023floquet} by specified designs and leads to intriguing physics that is strikingly distinct from its Hermitian counterpart, as reviewed in this part.

The EPs can be treated as a special case of exceptional lines or surfaces, where the coalescence happens at a line or surface in the parameter space.

\subsubsection{EPs}
\label{exceptionalpointanalysis}

To appreciate the features of the EPs in a non-Hermitian system, we address in detail the coupled system described by a $2\times 2$ matrix
\begin{align}
    \hat{H}/\hbar=(\hat{p}^{\dagger},\hat{m}^{\dagger})\left(\begin{array}{cc}
    \omega_p-i\kappa_p&g_a\\
    g_b&\omega_m-i\kappa_m
    \end{array}\right)\left(\begin{array}{c}
         \hat{p}  \\
         \hat{m}
    \end{array}\right)=(\hat{p}^{\dagger},\hat{m}^{\dagger}){\cal H}_M\left(\begin{array}{c}
         \hat{p}  \\
         \hat{m}
    \end{array}\right),
    \label{eq:2times2}
\end{align}
where $\omega_p$ and $\omega_m$ are, respectively, the frequencies of the two different bosonic quasiparticles $\hat{p}$ and $\hat{m}$, such as microwave photons and magnons, with their dissipation or gain rates $\{\kappa_p, \kappa_m\}$ indicated by the positive or negative sign, and $g_a$ and $g_b$ represent their mutual coupling that is not necessarily conjugated to each other, e.g., the dissipative coupling $g_a=g_b^*=i{\rm Re}(g)$ \cite{heinrich2003dynamic,liu2016metrology,asenjo2017exponential,chang2018colloquium,zhang2019theory,zhang2020subradiant,harder2021coherent,rao2021interferometric,rameshti2022cavity,zou2022prb,zeng2023radiation}  or chiral coupling $\{g_b=0,g_a\ne 0\}$ \cite{sollner2015deterministic,lodahl2017chiral,yu2019chiral,yu2020chiral,yu2020magnon_accumulation,yu2023chirality}. The matrix ${\cal H}_M\ne {\cal H}^{\dagger}_M$ for such a system is generally non-Hermitian. More complicated cases can be constructed accordingly in terms of a higher-rank matrix.

A Hermitian matrix can be diagonalized with conjugated right and left eigenvectors. For a non-Hermitian matrix, the right $\left|\Phi\right>$ and left $\left<\Psi\right|$ eigenvectors, however,  are not conjugated to each other in general, but their relationship may be governed by the special symmetries of the non-Hermitian Hamiltonian \cite{bender1999pt,ahmed2003c,ahmed2003pseudo,bender2005introduction,brody2013biorthogonal,kunst2018biorthogonal,kawabata2019}. We may expect the orthonormal condition for different states $\{i,j\}$ via $\left<\Psi_i|\Phi_j\right>=\delta_{ij}$ in principle, but at the EPs they are ill-defined.
In terms of the left $\left(a^L_{\pm},b^L_{\pm}\right)$ and right $\left(a^R_{\pm},b^R_{\pm}\right)^T$ eigenvectors, the non-Hermitian matrix ${\cal H}_M$ is diagonalized as
\begin{equation}\label{epmatrix}
 \left(\begin{array}{cc}
    a_{+}^L & b_{+}^L \\
    a_{-}^L & b_{-}^L
\end{array}  \right)  {\cal H}_M \left(\begin{array}{cc}
    a_{+}^R & a_{-}^R \\
    b_{+}^R  & b_{-}^R
\end{array} \right) = \left( \begin{array}{cc}
\omega_+ & 0 \\
0 & \omega_{-}
\end{array}\right).
\end{equation}
The two eigenfrequencies
\begin{equation}\label{NHEPeigenvalue}
   \omega_{\pm}=\omega_0-i\kappa_0\pm \sqrt{\left(\Delta\omega -i\Delta\kappa\right)^2+g_ag_b},
\end{equation}
where $\omega_0=({\omega_p+\omega_m})/{2}$ and $\kappa_0=({\kappa_p+\kappa_m})/{2}$ are the averaged frequency and damping rates, respectively, and their difference  $\Delta\omega=({\omega_p-\omega_m})/{2}$ and $\Delta\kappa=({\kappa_p-\kappa_m})/{2}$. 
After normalization the associated left eigenvectors
\begin{equation}\label{NHEPlefteigenvector}
  \left(a_{\pm}^L,b_{\pm}^L \right)= \left(\Delta\omega-i\Delta\kappa \pm \sqrt{\left(\Delta\omega -i\Delta\kappa\right)^2+g_ag_b},g_a\right),
\end{equation}
and the right eigenvectors 
\begin{equation}
 \left(\begin{array}{c}
    a_{\pm}^R \\
    b_{\pm}^R
\end{array}\right)=\pm\frac{1}{ 2g_a  \sqrt{\left(\Delta\omega -i\Delta\kappa\right)^2+g_ag_b} }
\left(\begin{array}{c}
g_a \\
-\left(\Delta\omega-i\Delta\kappa\right) \pm \sqrt{\left(\Delta\omega -i\Delta\kappa\right)^2+g_ag_b}
\end{array}\right).
\label{NHEPrighteigenvector}
\end{equation} 
Substituting Eq.~\eqref{epmatrix} into  Hamiltonian \eqref{eq:2times2}, we arrive at
\begin{align}
    \hat{H}/\hbar=(\hat{p}^{\dagger},\hat{m}^{\dagger})\left(\begin{array}{cc}
    a_{+}^R & a_{-}^R \\
    b_{+}^R  & b_{-}^R
\end{array} \right)
  \left( \begin{array}{cc}
\omega_+ & 0 \\
0 & \omega_{-}
\end{array}\right)
  \left(\begin{array}{cc}
    a_{+}^L & b_{+}^L \\
    a_{-}^L & b_{-}^L
\end{array}  \right)
\left(\begin{array}{c}
         \hat{p}  \\
         \hat{m}
    \end{array}\right).
\end{align}
Thereby the hybridized modes for two eigenvalues $\omega_{\pm}$ are 
\begin{align}
\hat{h}_{\pm}=\left(\Delta\omega-i\Delta\kappa \pm \sqrt{\left(\Delta\omega -i\Delta\kappa\right)^2+g_ag_b}\right)\hat{p}+  g_a \hat{m}.
\end{align}

\begin{figure}[!htp]
	\centering
	\includegraphics[width=0.78\textwidth]{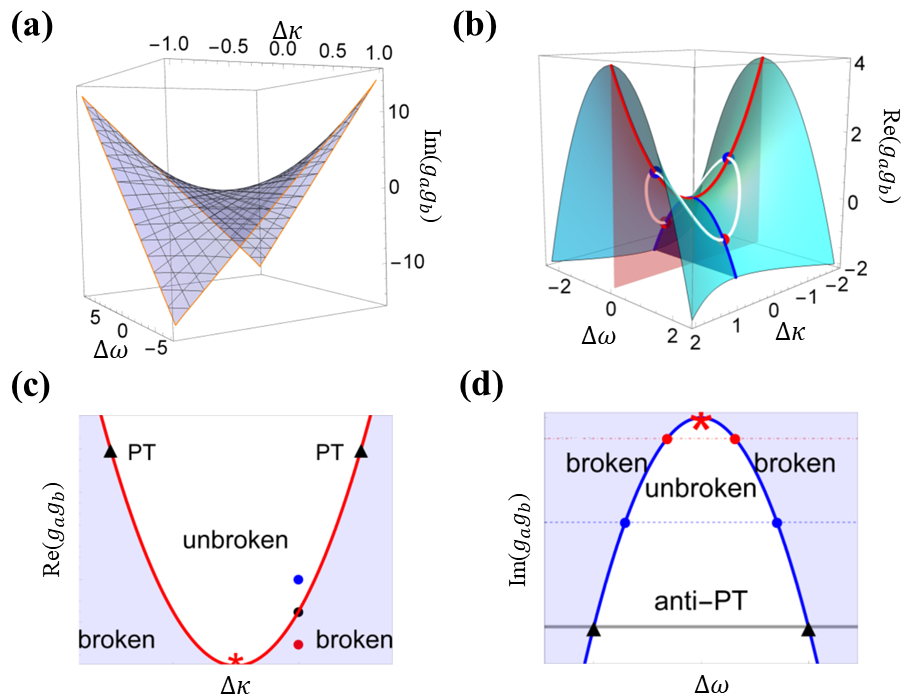}
	\caption{Conditions for the realizations of EPs. At the EPs, $\text{Im}(g_ag_b)$ [(a)] and $\text{Re}(g_ag_b)$ [(b)] depends on frequency and dissipation asymmetries $\Delta \omega$ and $\Delta \kappa$. 
 The red and blue lines on the surface of (b) correspond to the non-Hermitian matrix preserving the $\mathcal{P} \mathcal{T}$ [(c)] and anti-$\mathcal{P} \mathcal{T}$ [(d)] symmetries, respectively.  This figure is adapted with permission from Ref.~\cite{grigoryan2022pseudo}.  }
	\label{fig:epcondition}
\end{figure}

The degeneracy of the modes is governed by the square root $\sqrt{\left(\Delta\omega -i\Delta\kappa\right)^2+g_ag_b}$, which enters Eqs.~\eqref{NHEPeigenvalue}, \eqref{NHEPlefteigenvector}, and \eqref{NHEPrighteigenvector}. Such a term is always real in the Hermitian case and leads to level repulsion or band anticrossing between two modes  \cite{lidzey1998strong,verhagen2012quantum,bhoi2019abnormal,xu2020floquet,rao2023unveiling}. However, this square root term can be complex in the non-Hermitian counterparts if  $r>1$ in $\sqrt{1-r^2}$. Particularly, when this term becomes zero, the two eigenvalues become the same, and the eigenvectors coalesce into one, which implies that the non-Hermitian matrix is defective. This singularity in non-Hermitian physics is dubbed EPs in the parameter space. The EPs emerge when \cite{grigoryan2022pseudo} 
\begin{subequations}
\begin{align}
\label{universalEPRe}      &\left(\Delta\omega\right)^2-\left(\Delta\kappa\right)^2+\text{Re}(g_ag_b)=0, \\
&-2\Delta\omega\Delta\kappa+\text{Im}(g_ag_b)=0 .
\label{universalEPIm}
\end{align}
\end{subequations}
From conditions \eqref{universalEPRe} and \eqref{universalEPIm}, the solutions of $\text{Re}(g_a g_b)$ and $\text{Im}(g_a g_b)$ on $\Delta \omega$ and $\Delta \kappa$, as shown in Fig.~\ref{fig:epcondition}(a) and (b), construct surfaces. The emergence of EPs via meeting the above conditions has been realized by many experimental designs in different systems, and some typical examples are reviewed in the following. 

\textbf{EPs by tuning dissipation and Hermitian coupling}.---We start with a simple and conventional case: the coupling is Hermitian with $g_a = g^*_b$ and the system possesses the damping as usual. With condition \eqref{universalEPIm}, $\Delta\omega=0$ or $\Delta \kappa=0$ for the EPs. 
\begin{itemize}
\item When 
$\Delta \omega = 0$,  the difference of the damping of the two modes follows Eq.~\eqref{universalEPRe}: $-\left(\Delta\kappa\right)^2 + |g_a|^2=0$. So the EPs locate at $\Delta \kappa=\pm |g_a|$, where the frequency of the two modes is degenerate to be
$\omega_0-i\kappa_0$
with the coalescence eigenvector  $\left(a_{\pm}^R,a_{\pm}^R\right)^T\propto \left(g_a, i\Delta \kappa\right)$.
\item On the other hand, $\Delta \kappa=0$ can be excluded for the EPs since 
$\left(\Delta \omega\right)^2 + |g_a|^2 =0$ cannot be established. 
\end{itemize}

By tuning the dissipation rates of two modes under the Hermitian coupling, there are many realizations of the EPs in distinct systems, such as the coupled waveguides \cite{guo2009observation}, the whispering-gallery resonators \cite{jing2014pt}, the metamaterials crystal \cite{liu2018unidirectional}, as well as the magnetic layers \cite{Liu2019Observation}.

\textbf{EPs by balanced gain and loss in the $\mathcal{PT}$-symmetric system}.---We go on assuming the coupling is Hermitian with $g_a = g^*_b$, but change one of the mode's damping to be negative or the ``gain''. The above discussion still holds since it only relies on the specific damping difference $\Delta\kappa=|g_a|$ but not the sign of damping.

However, a stable system with coherent coupling appears to require that the gain and loss are balanced, i.e., the averaged $\kappa_0 = 0$.
The gain has been realized in many physical systems by various methods, as summarized in Table.~\ref{tab:gain} for the typical realizations \cite{shi2016accessing,zhang2017observation,hodaei2017enhanced,sakhdari2019experimental,xiao2019enhanced,tang2020exceptional,wang2021coherent}. 
With balanced gain and loss 
 the non-Hermitian matrix reads 
\begin{equation}
    {\cal H}_M =\left(\begin{array}{cc}
    \omega_0 - i \kappa&g_a\\
    g^*_a& \omega_0 + i  \kappa
    \end{array}\right).
    \label{matrix_PT}
\end{equation}
Such a matrix is unique since it is governed by the $\mathcal{PT}$ symmetry, i.e., $[\mathcal{PT},\hat{H}]=0$. Here, $\mathcal{P}$ is the parity operator
\begin{equation}
\mathcal{P}=\left(\begin{array}{cc}
	0&1\\
	1&0
	\end{array}\right),
\end{equation}
and  $\mathcal{T}$ is the time-reversal operator that changes the matrix element to be the complex conjugation. Such a system is identified as $\mathcal{P}\mathcal{T}$-symmetric
system.

\begin{table}\caption{Realizations of gain in typical systems. These figures are taken from Refs.~\cite{shi2016accessing,hodaei2017enhanced,zhang2017observation,sakhdari2019experimental,xiao2019enhanced,tang2020exceptional,wang2021coherent,rao2023meterscale}.}
  \label{tab:gain}

\begin{tabular}{|c|c|c|c|}
\hline
\hspace{-0.53cm}
\begin{minipage}[p]{0.25\textwidth}

\begin{tabular}{p{1\textwidth}}
\\
 \multicolumn{1}{c}{  \textbf{Acoustics}}\\ \\ 

\multicolumn{1}{c}{ \includegraphics[width=0.85\textwidth]{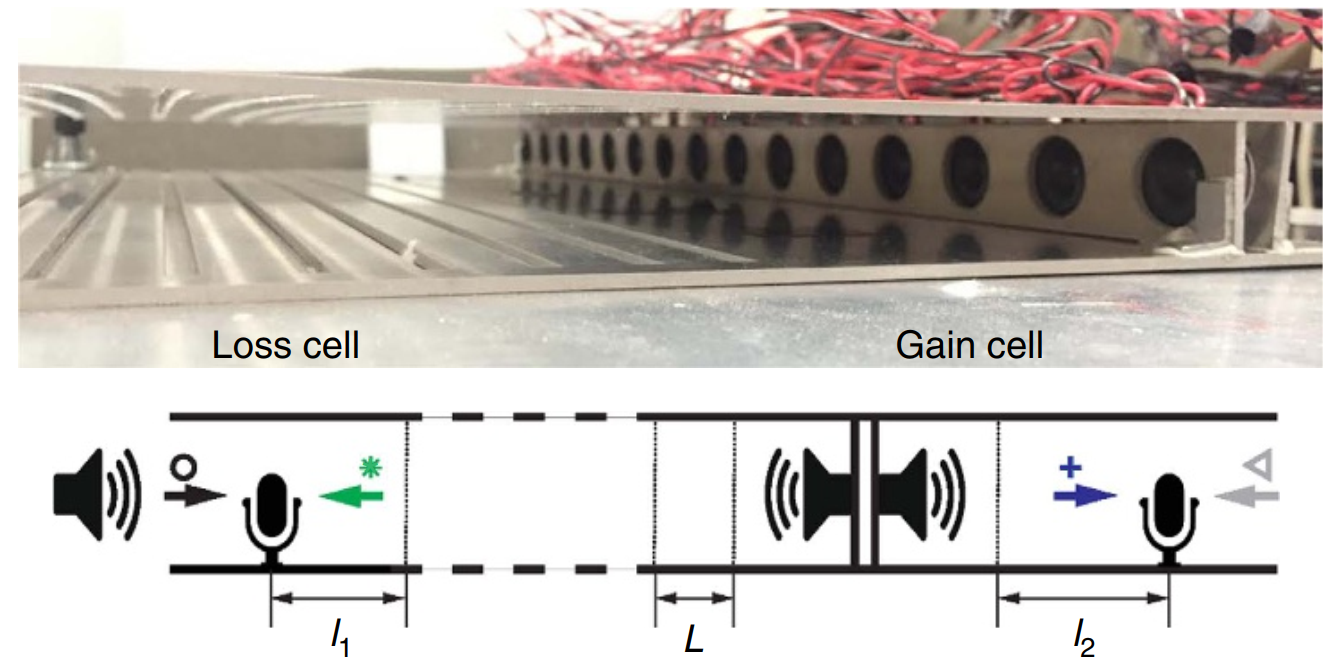}}\\
 \multicolumn{1}{c}{\scriptsize Gain from Speaker \cite{shi2016accessing}} \\
 \\
 \multicolumn{1}{c}{ \includegraphics[width=0.72\textwidth]{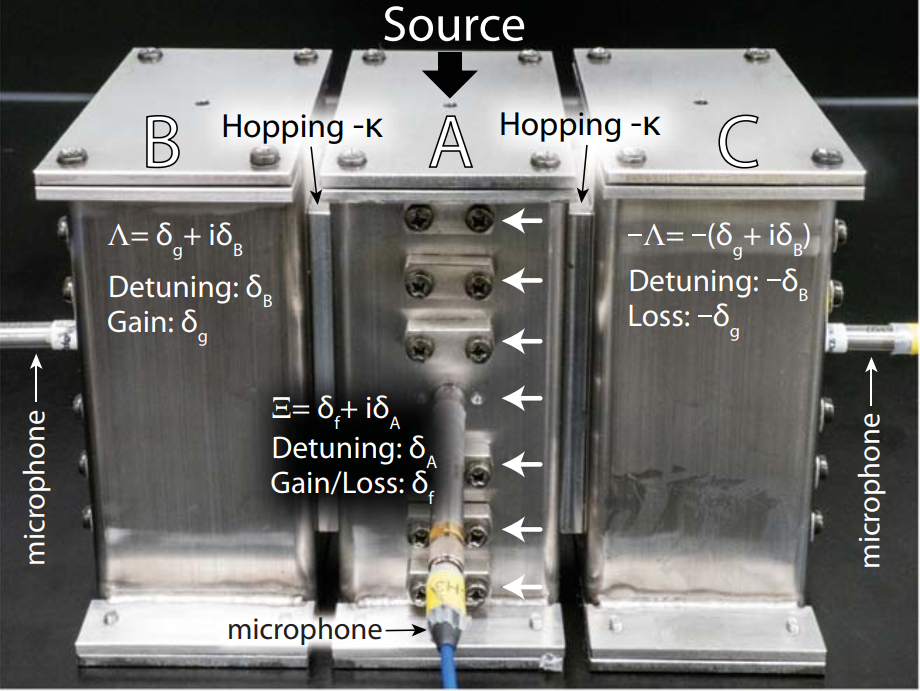}}\\
 \multicolumn{1}{c}{\scriptsize Gain from Microphone Input \cite{tang2020exceptional}}
\end{tabular}
 \end{minipage}
 &
\hspace{-0.53cm}
\begin{minipage}[p]{0.25\textwidth}

\begin{tabular}{p{1\textwidth}}
\\
 \multicolumn{1}{c}{\textbf{Electronics}}\\ \\

\multicolumn{1}{c}{ \includegraphics[width=0.72\textwidth]{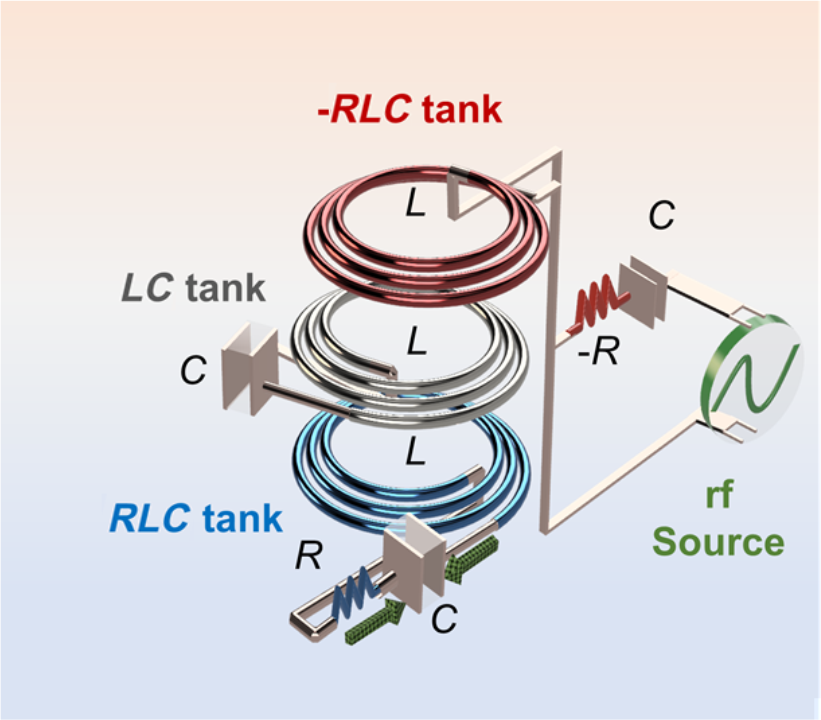}}\\
 \multicolumn{1}{c}{\scriptsize Gain from RF Power Supply \cite{sakhdari2019experimental}} \\
 \\
 \multicolumn{1}{c}{ \includegraphics[width=0.83\textwidth]{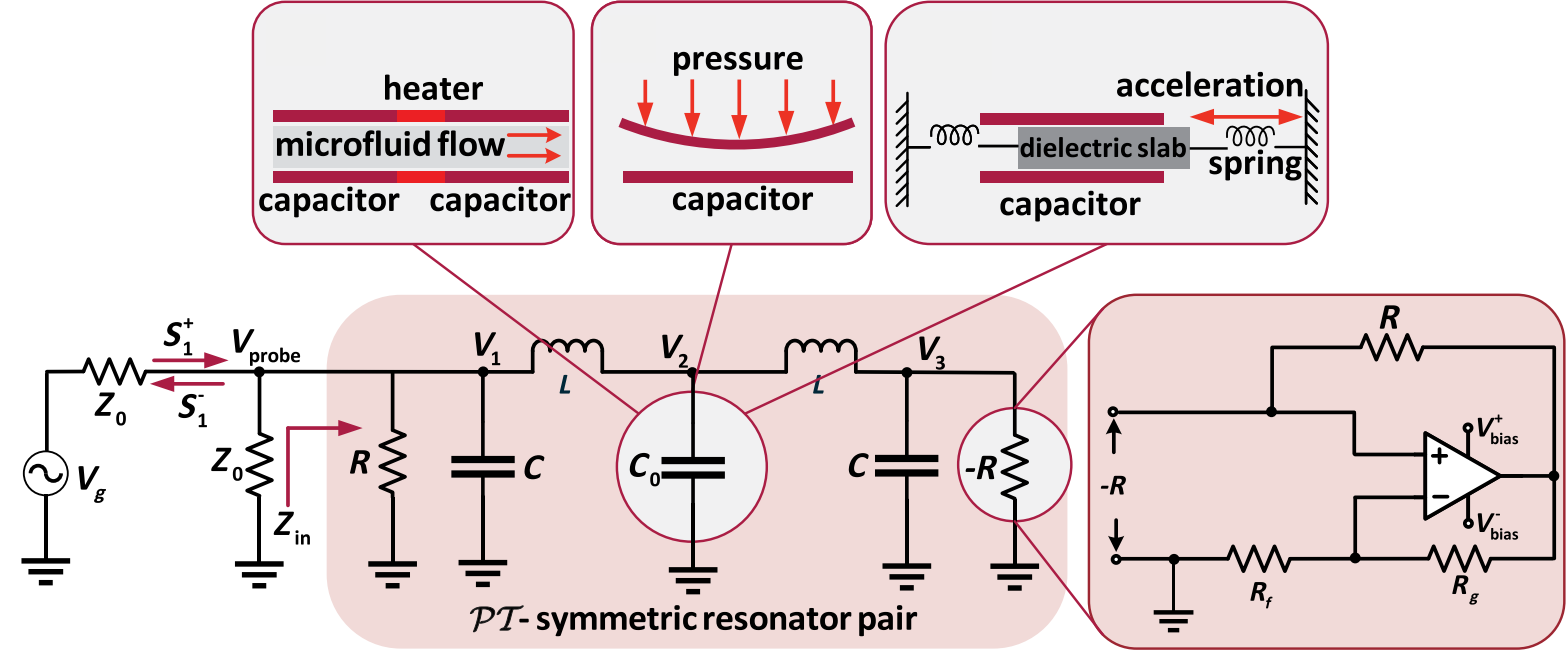}}\\
 \multicolumn{1}{c}{\scriptsize Gain from RF Power Supply \cite{xiao2019enhanced}}
\end{tabular}
 \end{minipage}

  &
\hspace{-0.53cm}
 \begin{minipage}[p]{0.25\textwidth}

\begin{tabular}{p{1\textwidth}}
\\
 \multicolumn{1}{c}{ \textbf{Optics}}\\ \\

\multicolumn{1}{c}{ \includegraphics[width=0.85\textwidth]{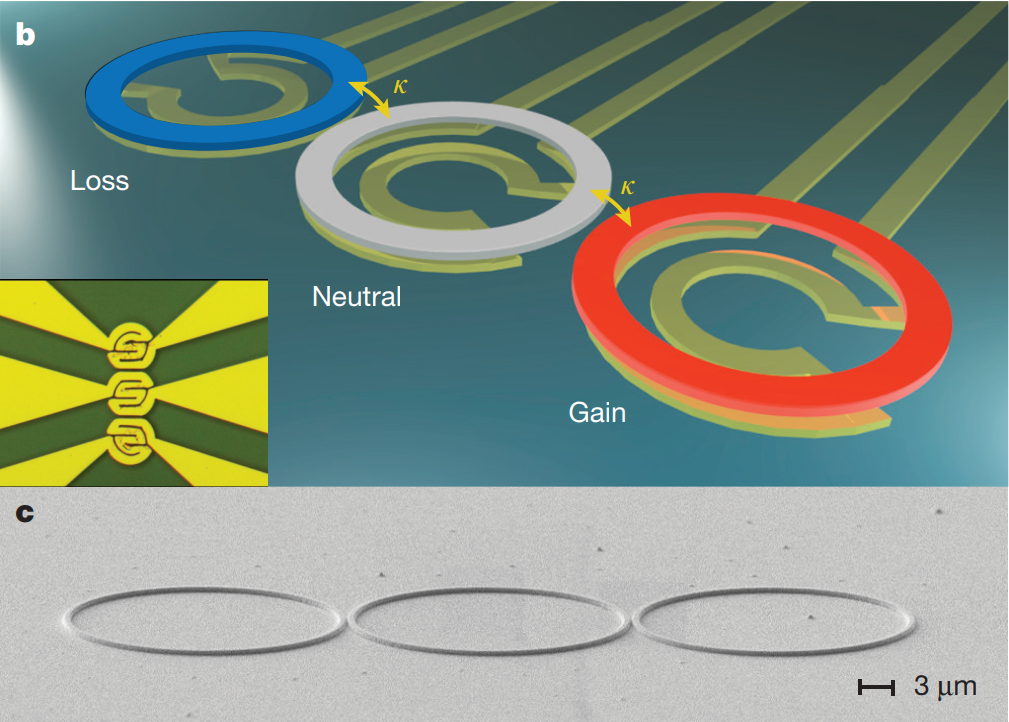}}\\
 \multicolumn{1}{c}{\scriptsize Gain from the Laser \cite{hodaei2017enhanced}} \\
 \\
 \multicolumn{1}{c}{ \includegraphics[width=0.76\textwidth]{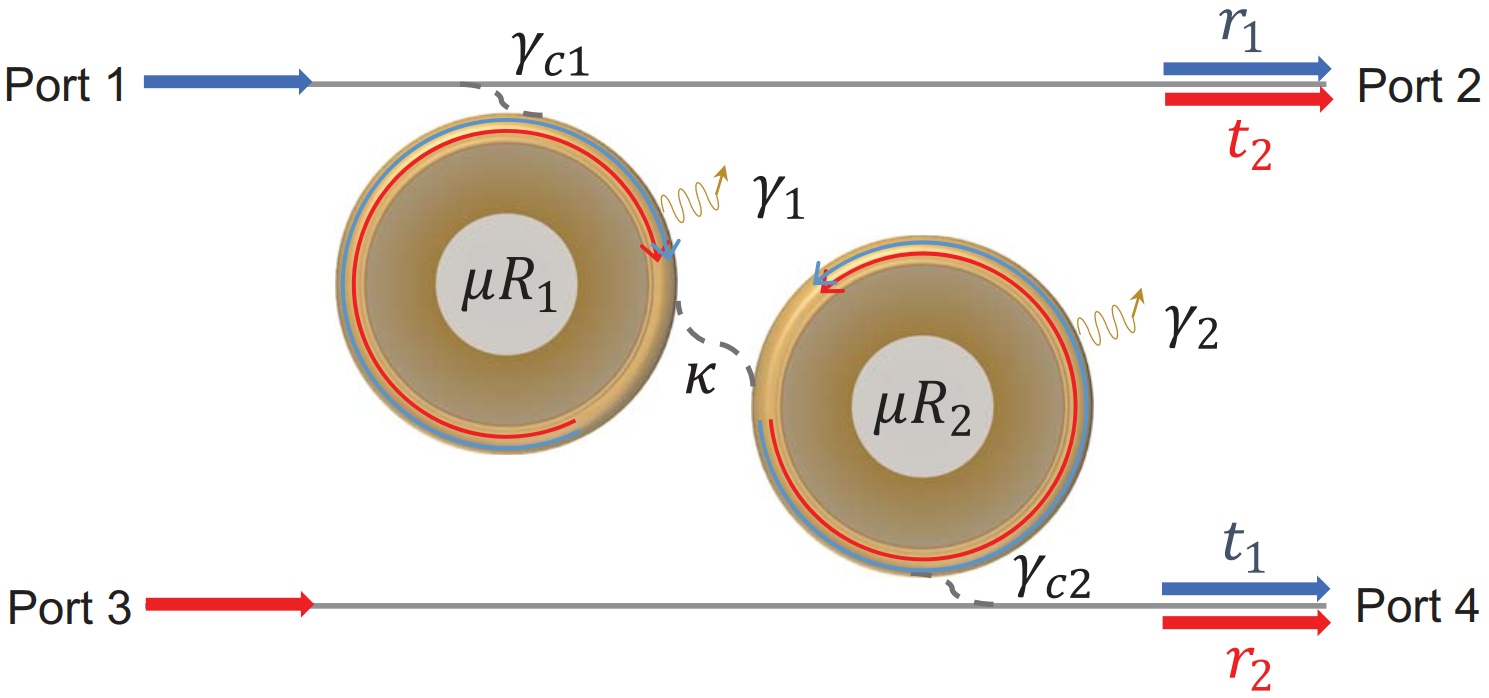}}\\
 \multicolumn{1}{c}{\scriptsize Gain from the Laser \cite{wang2021coherent}}
\end{tabular}
 \end{minipage}

   &
\hspace{-0.53cm}
\begin{minipage}[p]{0.25\textwidth}

\begin{tabular}{p{1\textwidth}}
\\
 \multicolumn{1}{c}{ \textbf{Magnonics}}\\ \\

\multicolumn{1}{c}{ \includegraphics[width=0.76\textwidth]{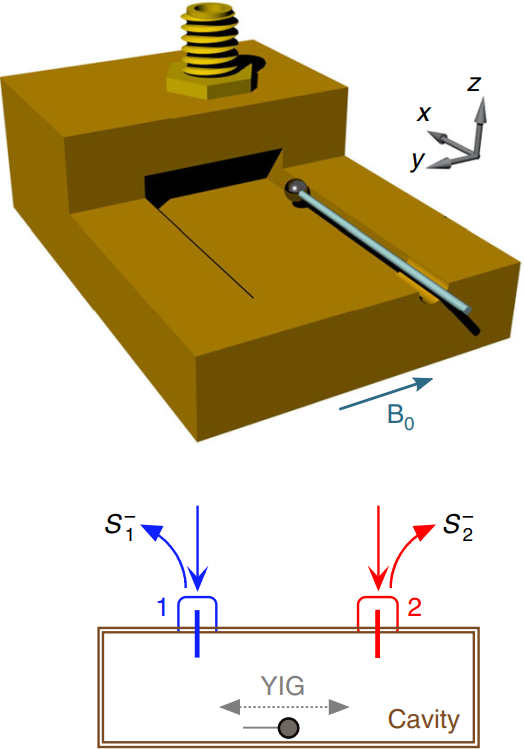}}\\
 \multicolumn{1}{c}{\scriptsize Gain from Microwave Input} \\
  \multicolumn{1}{c}{\scriptsize via Cavity Port \cite{zhang2017observation,rao2023meterscale}} \\
\end{tabular}
 \end{minipage}\\
 \hline
\end{tabular}
\end{table}

With the $\mathcal{PT}$ symmetry, 
\begin{equation} \label{eq:PTsymmetry}
    {\cal H}_M \mathcal{P} \mathcal{T} \left| \Phi \right> = \mathcal{P} \mathcal{T} {\cal H}_M \left|\Phi \right> = \mathcal{P} \mathcal{T} \omega \left|\Phi \right> = \omega^{*} \mathcal{P} \mathcal{T}\left|\Phi \right>,
\end{equation}
i.e., $\mathcal{P} \mathcal{T}\left|\Phi \right>$ is the eigenvector with eigenvalue $\omega^{*}$.
The eigenvalues of Eq.~\eqref{matrix_PT} 
\begin{equation}
   \omega_{\pm}=\omega_0\pm \sqrt{-\kappa^2+|g_a|^2}
\end{equation}
are real when the Hermitian coupling dominates the gain or loss with $|g_a|^2>\kappa^2$
and the separate 
 right eigenvectors
\begin{equation}
 \left(\begin{array}{c}
    a_{\pm}^R \\
    b_{\pm}^R
\end{array}\right)\propto
\left(\begin{array}{c}
g_a \\
i\kappa \pm \sqrt{-\kappa^2+|g_a|^2}
\end{array}\right)
\end{equation}
 preserve the  $\mathcal{PT}$ symmetry since $\mathcal{PT}\left|\Phi\right> \propto \left|\Phi\right>$, thanks to Eq.~\eqref{eq:PTsymmetry}. 
Further, when the gain or loss dominates with $|g_a|^2<\kappa^2$, the eigenvalues become complex conjugated pair, and thereby the eigenvectors enter the $\mathcal{PT}$-broken regime. Considering this mutation, the $\mathcal{PT}$-unbroken and $\mathcal{PT}$-broken regime is separated by EPs at $|g_a|^2=\kappa^2$ for the eigenvectors plotted in Fig.~\ref{fig:epcondition}(c).

On the other hand, with only loss involved some systems can also exhibit  $\mathcal{PT}$-symmetry breaking phenomena~\cite{guo2009observation,ozdemir2019parity,miri2019exceptional}. For example, a non-Hermitian matrix with degenerate frequencies $\omega_0$ and asymmetric loss $\kappa_p \neq \kappa_m$
\begin{equation}
    \hat{H}_{\mathcal{PT}-{\rm passive}} =\left(\begin{array}{cc}
    \omega_0 - i \kappa_p&g_a\\
    g^*_a& \omega_0 - i  \kappa_m
    \end{array}\right) = \frac{-i \left( \kappa_p + \kappa_m \right)}{2} + \left(\begin{array}{cc}
    \omega_0 -\frac{i\left( \kappa_p - \kappa_m \right)}{2}&g_a\\
    g^*_a& \omega_0  + \frac{i\left( \kappa_p - \kappa_m \right)}{2}
    \end{array}\right)
    \label{matrix_PT_passive}
\end{equation} can be divided into an average loss term and $\mathcal{PT}$-symmetric term, which governs the evolution (e.g., in optical systems~\cite{guo2009observation,ozdemir2019parity,miri2019exceptional})
\begin{equation}
    i \frac{\partial |\Psi(z)\rangle}{\partial z} =  \hat{H}_{\mathcal{PT}-{\rm passive}} |\Psi(z)\rangle,
\end{equation} where $|\Psi\rangle$ is the wavefunction parametrized by $z$. Performing a gauge transformation $|\Psi\rangle =  e^{{-( \kappa_p + \kappa_m )z}/{2} }|\Phi\rangle$ , we obtain that $|\Phi(z)\rangle$ is governed by $\mathcal{PT}$-symmetric Hamiltonian
\begin{equation}
    i \frac{\partial |\Phi(z)\rangle}{\partial z} =  \left(\begin{array}{cc}
    \omega_0 -\frac{i\left( \kappa_p - \kappa_m \right)}{2}&g_a\\
    g^*_a& \omega_0  + \frac{i\left( \kappa_p - \kappa_m \right)}{2}
    \end{array}\right) |\Phi(z)\rangle.
\end{equation}
Accordingly, such a dissipative system holding EPs is identified as a passive $\mathcal{P}\mathcal{T}$-symmetric system.

Tuning the relative magnitude of the Hermitian coupling and the balanced gain and loss in a coupled system with the $\mathcal{PT}$-symmetry to realize the EPs is the most-studied case theoretically and experimentally, ranging from optics \cite{regensburger2012parity,hodaei2017enhanced,el2018non,ozdemir2019parity,miri2019exceptional}, metamaterials \cite{lin2011unidirectional,liu2018unidirectional,liu2019willis,zhang2023diffusion}, quantum spin models \cite{lenke2021high}, and electronic circuits \cite{schindler2011experimental,sakhdari2019experimental,xiao2019enhanced,stegmaier2021topological}. In magnetic system, Lee \textit{et al.} \cite{lee2015macroscopic} considered two coupled classic macroscopic magnetic structures with the same frequency, and one of them is naturally dissipated while the other one is driven by external force with equal effective gain, a design similar to the system with $\mathcal{PT}$ symmetry, as reviewed in Sec.~\ref{Sec_magnon_PT}. Similar designs are further extended to trilayer ferromagnetic heterostructure \cite{yu2020higher,Liu2019Observation,wang2023floquet}, one-dimensional spin-torque oscillator arrays~\cite{flebus2020non,gunnink2022nonlinear} and two-dimensional van der Waals magnets \cite{li2022multitude}. Experimentally, the realization of EPs by tuning the gain and loss is observed in cavity-magnon system \cite{zhang2017observation,bhoi2017robust,wang2018magnon,cao2019exceptional,PhysRevLett.123.237202} and coupled magnonic waveguides or magnetic layer \cite{Liu2019Observation,wang2020steering,wang2021enhanced} (refer to Sec.~\ref{Sec_EPs_cavity_magnonics}).

The balanced gain and loss is not the necessary condition to realize the $\mathcal{PT}$-symmetry, which often requires fine-tuning for a system. Several alternative strategies were proposed to realize $\mathcal{PT}$ symmetry without real gain, including e.g., the nonlinear wave mixing processes~\cite{miri2016nonlinearity}, energy accumulation~\cite{mirmoosa2019time,li2020parity}, and parametric pumping~\cite{koutserimpas2018parametric}. In particular, effective ``gain'' can be realized via time modulation of the system~\cite{mirmoosa2019time,li2020parity,koutserimpas2018parametric}.

These cases are of fundamental importance since they emphasize the importance of  $\mathcal{PT}$-symmetry in the generalization of quantum mechanics \cite{bender1998real,bender1999pt,bender2002complex,bender2007making}, the role of symmetry breaking \cite{bender2005introduction,el2018non}, and reveal many exotic physical phenomena \cite{miri2019exceptional,ashida2020non,bergholtz2021exceptional}, such as double refraction \cite{makris2008beam}, single-mode laser \cite{feng2014single}, and orbital angular momentum microlaser \cite{miao2016orbital,zhang2020tunable}.

\textbf{EPs by competition of dissipative coupling and frequency asymmetry}.---Different from the Hermitian coupling, between two objects there exists a different interaction that is referred to as the ``dissipative coupling'', namely a coupling that is purely imaginary and contains no coherent component, i.e., $\text{Re}(g_a)=\text{Re}(g_b)=0$. With both the coherent and dissipative couplings, $g_a \ne g^*_b$, e.g., the chiral coupling with one of them being zero \cite{sollner2015deterministic,lodahl2017chiral,yu2019chiral,yu2020chiral,yu2020magnon_accumulation,yu2023chirality}.  With pure dissipative coupling, $\text{Im}\left(g_ag_b\right)=0$.
Thus to meet conditions \eqref{universalEPRe} and \eqref{universalEPIm}, $\Delta\kappa = 0$, i.e., the two modes have the same dissipation rates.
 Taking the average frequency as the frequency reference, the non-Hermitian matrix
\begin{equation}
    {\cal H}_M =\left(\begin{array}{cc}
     \Delta \omega -i \kappa_0&g_a\\
    g_a&  - \Delta \omega-i \kappa_0
    \end{array}\right),
    \label{eq:anPT}
\end{equation}
with eigenvalues 
\begin{equation}
   \omega_{\pm}=-i\kappa_0 \pm \sqrt{\left(\Delta\omega\right)^2-\left|g_a\right|^2}.
\end{equation}
 The EPs take place at $\left(\Delta \omega\right)^2 - \left|g_a\right|^2 =0$ with the coalescence of right eigenvectors $\left(a_{\pm}^R,a_{\pm}^R\right)^T\propto \left(g_a, -\Delta \omega\right)$, which can be realized by tuning the frequency asymmetry and the strength of the dissipative coupling.

The matrix Eq.~(\ref{eq:anPT}) preserves the anti-$ \mathcal{P} \mathcal{T}$ symmetry, \textit{viz.} $\mathcal{PT}{\cal H}_M \mathcal{PT} =  -{\cal H}_M $. Following the same step as Eq.~\eqref{eq:PTsymmetry}, we arrive at 
\begin{equation} \label{eq:anPTsymmetry}
    {\cal H}_M \mathcal{P} \mathcal{T} \left| \Phi \right> = -\mathcal{P} \mathcal{T} {\cal H}_M \left|\Phi \right> = -\mathcal{P} \mathcal{T} \omega \left|\Phi \right> = -\omega^{*} \mathcal{P} \mathcal{T}\left|\Phi \right>.
\end{equation}
The eigenvalues also experience mutation when across the EPs, but different from the case with $ \mathcal{P} \mathcal{T}$ symmetry, the region by $\left(\Delta\omega\right)^2-\left|g_a\right|^2<0$ is the $ \mathcal{P} \mathcal{T}$-unbroken regime since $\omega = -\omega^*$, while the $ \mathcal{P} \mathcal{T}$-broken regime locates at $\left(\Delta\omega\right)^2-\left|g_a\right|^2>0$, as shown in Fig.~\ref{fig:epcondition}(d).

In magnetic system, the anti-$ \mathcal{P} \mathcal{T}$ symmetry is realized experimentally in cavity magnonic system \cite{zhao2020observation,yang2020Unconventional,nair2021enhanced}, synthetic antiferromagnets \cite{sui2022emergent} and hybrid system \cite{bhoi2019abnormal}, which we refer to Sec.~\ref{EPs_magnonics} for details. One potential application in a magnonic system possessing anti-$ \mathcal{P} \mathcal{T}$ symmetry is the highly enhanced spin-wave excitation on the exceptional line, as proposed in two dissipatively coupled ferromagnet layers mediated by normal-metal layer \cite{pan2023imbalanced}.

\textbf{General case}.---The cases addressed above focus on the realization of the EPs via adjusting the parameters in Eq.~\eqref{universalEPRe} when the condition \eqref{universalEPIm} is presumed, where $\text{Im}(g_ag_b)$ is taken to be zero. However, one can also realize the EPs with wide parameter choice from conditions \eqref{universalEPRe} and \eqref{universalEPIm}.
 Such general realizations of EPs are proposed theoretically in coupled magnetization or LCR circuit with microwaves in a microwave cavity \cite{grigoryan2018synchronized,grigoryan2022pseudo,bhoi2019abnormal},  and two-port driven cavity magnon-polariton \cite{boventer2019steering,boventer2020control}.

\subsubsection{Unique features of EPs}
\label{UniqueEP}

The EPs have attracted tremendous attention not only because they act as spectral singularities, but also because they show exotic features and functionalities. 
The emergent unique features often root in the existence of the square root in the eigenvalues and eigenvectors. Here we address several typical cases that are described by a single parameter. We focus on how the properties are varied when the parameter passes across the EPs,  the topological properties when encircling the EPs (Sec.~\ref{exceptionaltopology}), and the sensitive response when applying a perturbation in the vicinity of EPs.

\textbf{Mutations across EPs}.---To address the mutations across the EPs, we choose three typical non-Hermitian systems, i.e., systems with anti-$ \mathcal{P} \mathcal{T}$ symmetry, with $ \mathcal{P} \mathcal{T}$ symmetry, and with chiral coupling, as summarized in the first column of Table~\ref{uniqueness of EPs}.
We focus on the dependence of the eigenvalue and eigenvector on a single parameter $r$ when passing across the EPs, with which the eigenvalues for these three systems are given by $\pm\sqrt{r^2-1}$, $\pm\sqrt{1-r^2}$, and  $\pm\sqrt{r}$, respectively.
When passing across the EPs, the eigenvalues can change from the real value to the complex pairs. We accordingly divide in the second column of Table~\ref{uniqueness of EPs} the region with real eigenvalue and complex pairs in terms of white and green, which denote the region below and beyond the EPs, respectively. For a non-Hermitian Hamiltonian, the condition of all the eigenvalues being real does not necessarily request the balanced gain and loss. For example, with dissipative coupling, all eigenvalues in the Hamiltonian matrix can be real in typical examples with anti-${\cal PT}$ symmetry or chiral coupling, as summarized in Table~\ref{uniqueness of EPs}.
The appearance of the real eigenvalue for a non-Hermitian Hamiltonian implies that the states are stable, and can survive in the time evolution. On the other hand, the complex pair of eigenvalues implies that the lasing (dissipation) mode with a positive (negative) imaginary frequency with the amplitudes increasing (decreasing) during the time evolution \cite{peng2014loss,brandstetter2014reversing,wong2016lasing}.

The lasing state evolves with the amplification of amplitude, 
but the lasing process appears to be limited by the nonlinearity or an induced phase transition \cite{lee2015macroscopic,yang2018antiferromagnetism,deng2023exceptional}. 
For example, Lee \textit{et al.} studied the time evolution of the magnetization in the coupled ferromagnetic bilayers in a $ \mathcal{PT}$ symmetry broken regime and found the amplification of amplitude for the lasing mode is suppressed by magnon nonlinear interaction \cite{lee2015macroscopic}.
Yang \textit{et al.} proposed the magnetizations in the ferromagnetic bilayers in $ \mathcal{PT}$ symmetry broken phase evolves to antiferromagnetic skyrmion phase \cite{yang2018antiferromagnetism}.
A similar phase transition from the ferromagnetic to antiferromagnetic phase was found in a similar ferromagnetic bilayer in the $\mathcal{PT}$-symmetry broken regime, which can be understood as the magnetization in the ``gain'' layer that tends to be reversed and the ``loss'' layer that recovers to equilibrium orientation \cite{deng2023exceptional}. 
By introducing easy-plane anisotropy, Deng \textit{et al.} found that there exists another regime, similar to $\mathcal{PT}$-symmetry broken regime, where a large amplitude of magnetization precession can survive for a long time \cite{deng2023exceptional}.

\begin{table}
   \caption{Mutations across the EPs. The region with real eigenvalue and complex pairs are labeled with white and green. The boundary between different regions corresponds to the location of EPs.} \label{uniqueness of EPs}
  \begin{tabular}{cccc}
   \toprule[1pt]
  \makebox[0.15\textwidth][c]{Hamiltonian}&\makebox[0.25\textwidth][c]{Eigenvalue}&\makebox[0.25\textwidth][c]{Eigenvector}& \makebox[0.25\textwidth][c]{Entanglement}\\
    \midrule[0.5pt] \begin{minipage}[m]{.15\textwidth} 
   \begin{equation}
      {\cal H}_{Anti-\mathcal{P} \mathcal{T}} = \left( \begin{array}{cc}
           r & 1 \\
           -1 & -r
       \end{array} \right) \nonumber
   \end{equation}  \end{minipage}  & \begin{minipage}[m]{.15\textwidth}
      \centering\vspace*{4pt}
      \includegraphics[width=3cm]{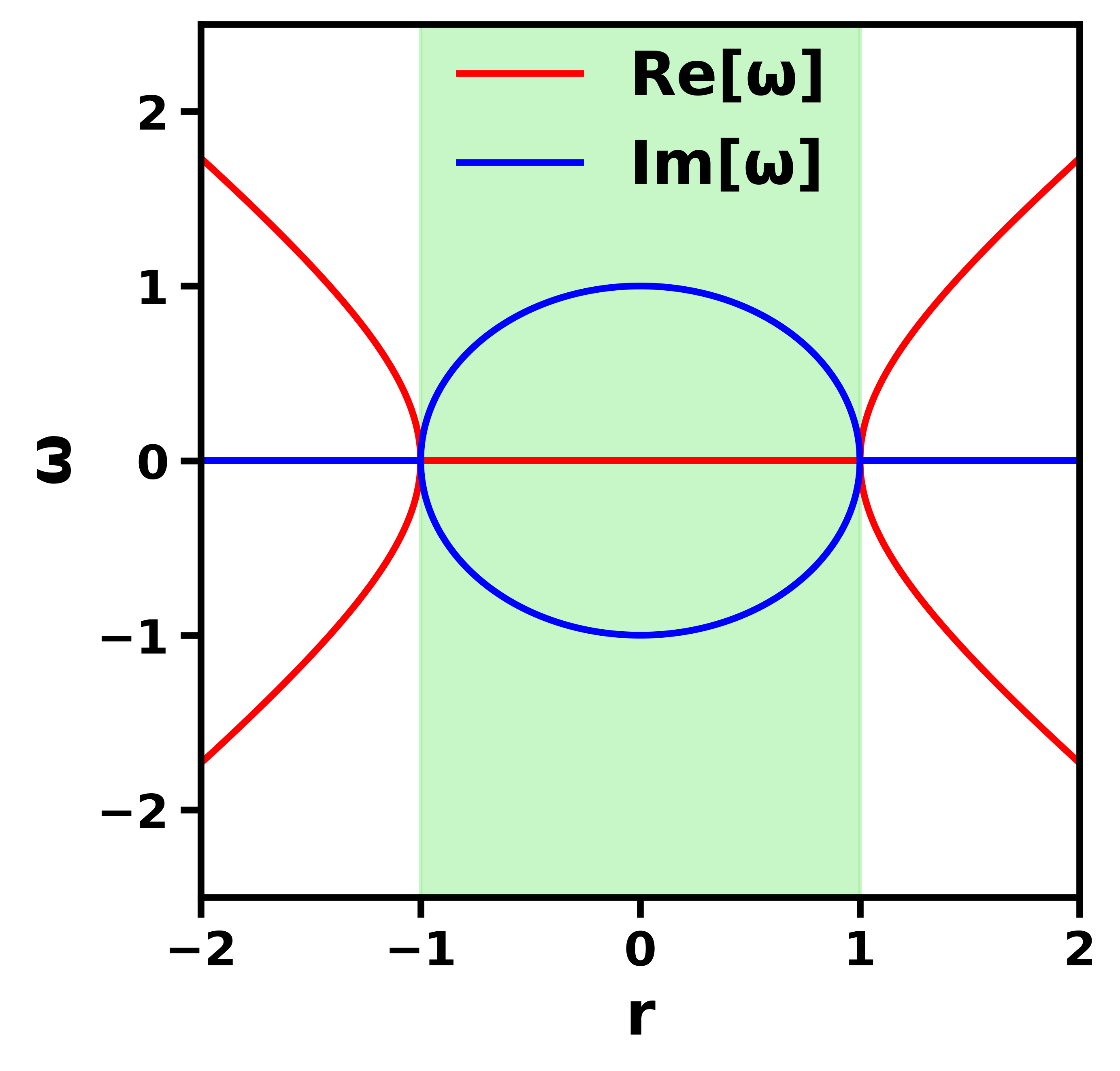}\vspace*{2pt}
    \end{minipage} & \begin{minipage}[m]{.15\textwidth}
    \centering\vspace*{4pt}
    \includegraphics[width=3.4cm]{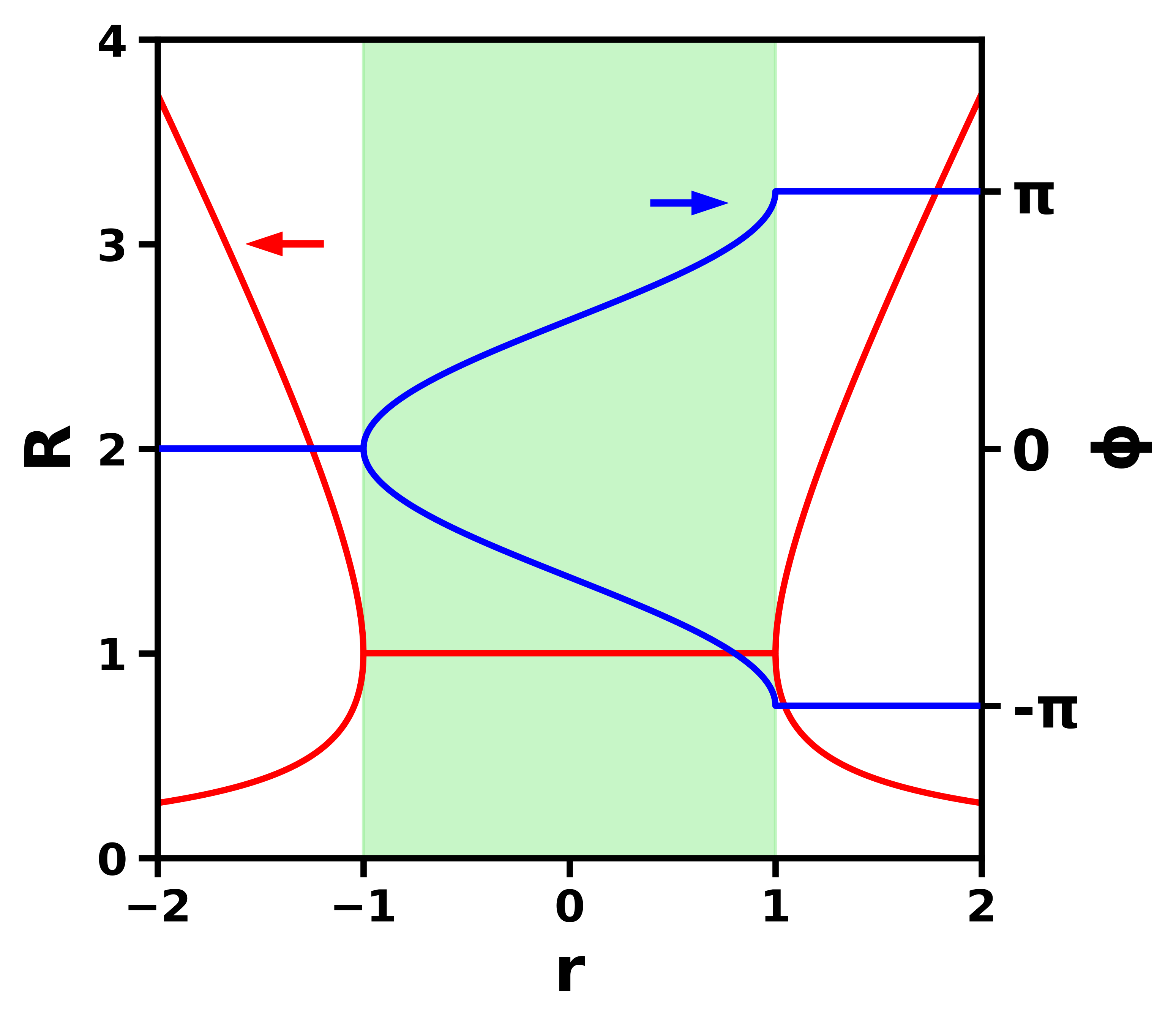}\vspace*{2pt}
    \end{minipage} & \begin{minipage}[m]{.15\textwidth}
    \centering\vspace*{4pt}
    \includegraphics[width=3.2cm]{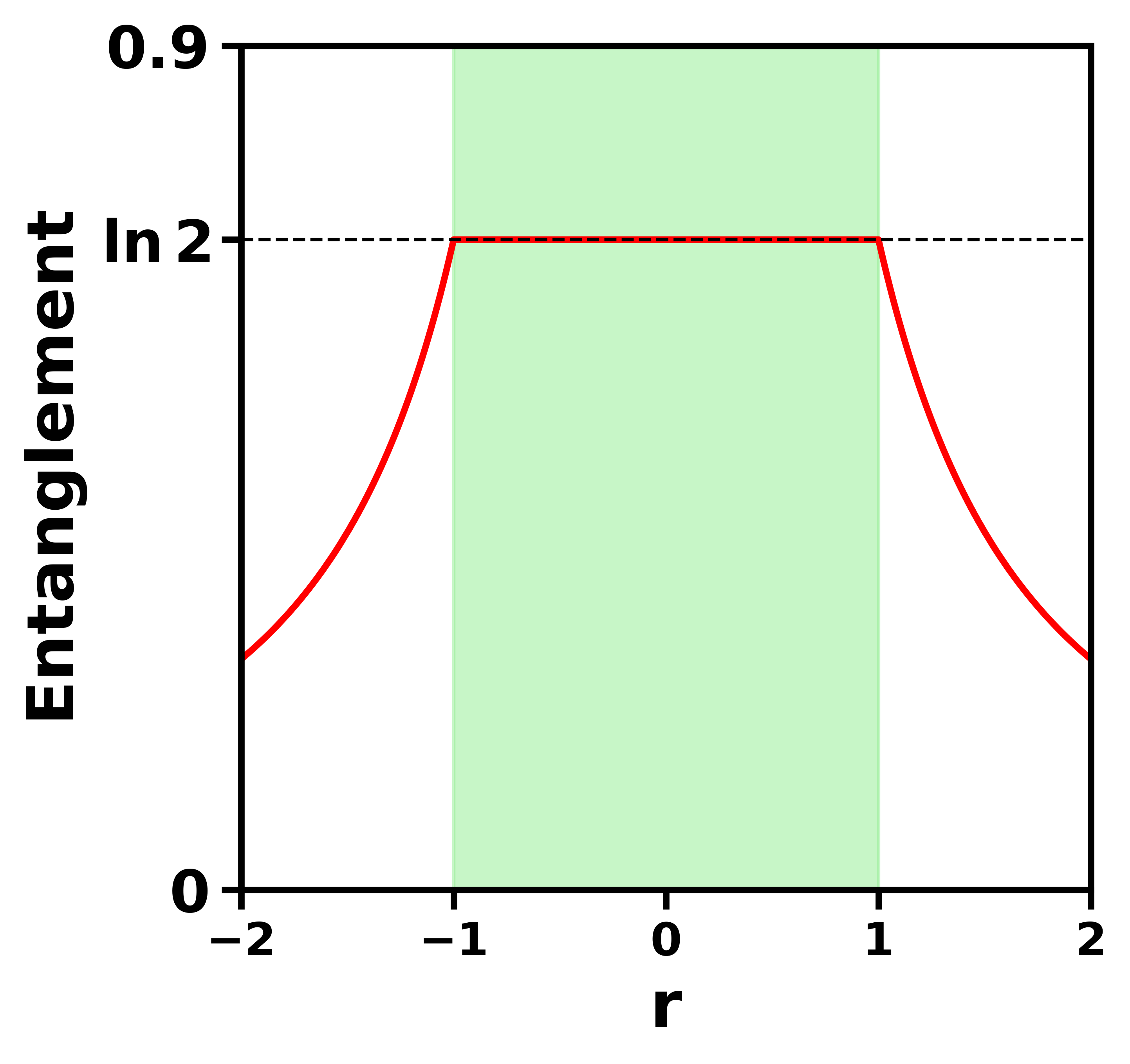}\vspace*{2pt}
    \end{minipage}\\
    \midrule[0.5pt]
\begin{minipage}[m]{.15\textwidth} 
   \begin{equation}
      {\cal H}_{\mathcal{P} \mathcal{T}} = \left( \begin{array}{cc}
           ir & 1 \\
           1 & -ir
       \end{array} \right) \nonumber
   \end{equation}  \end{minipage}  & \begin{minipage}[m]{.15\textwidth}
    \centering\vspace*{4pt}
    \includegraphics[width=3cm]{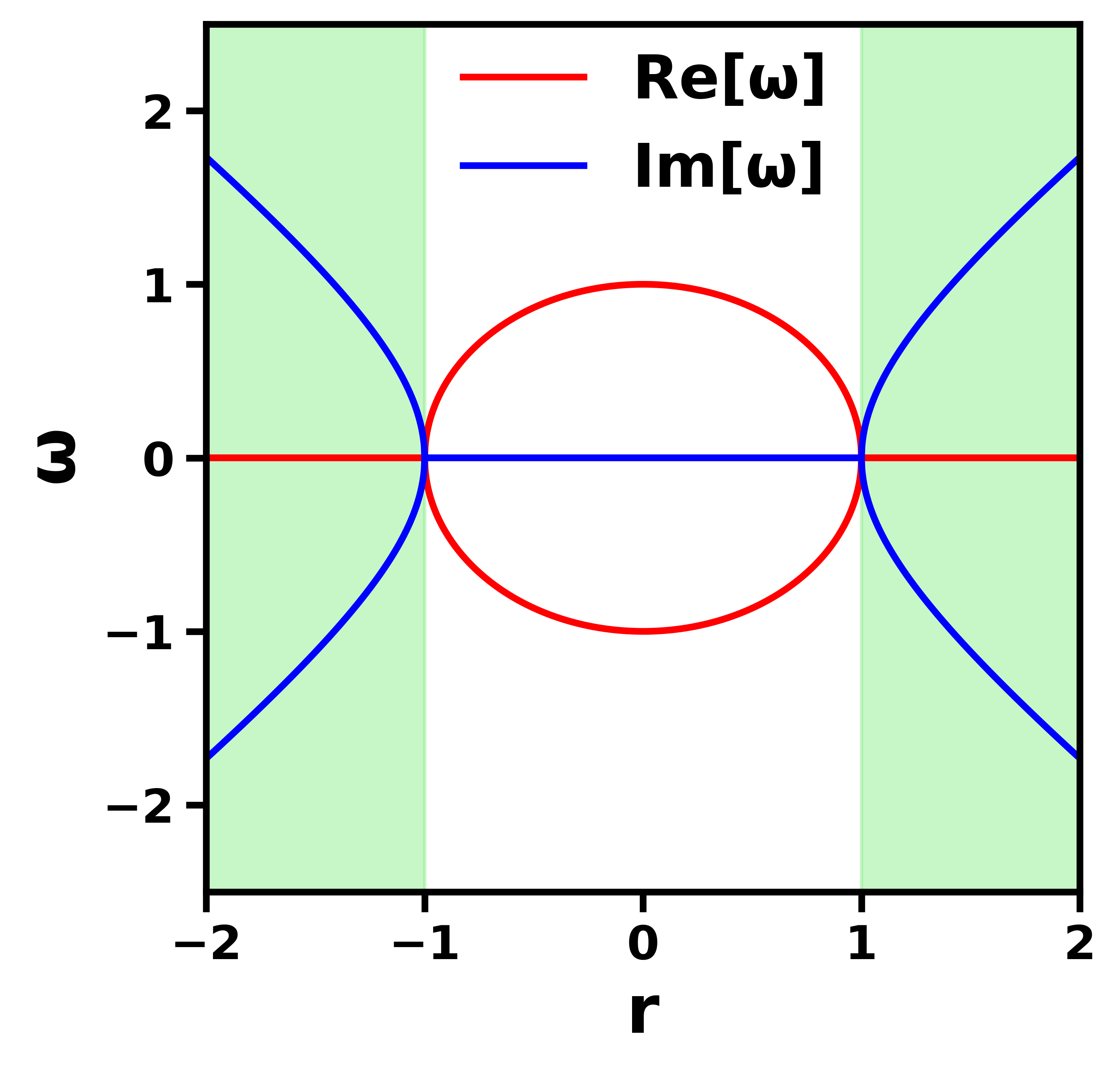}\vspace*{2pt}
    \end{minipage} & \begin{minipage}[m]{.15\textwidth}
    \centering\vspace*{4pt}
    \includegraphics[width=3.4cm]{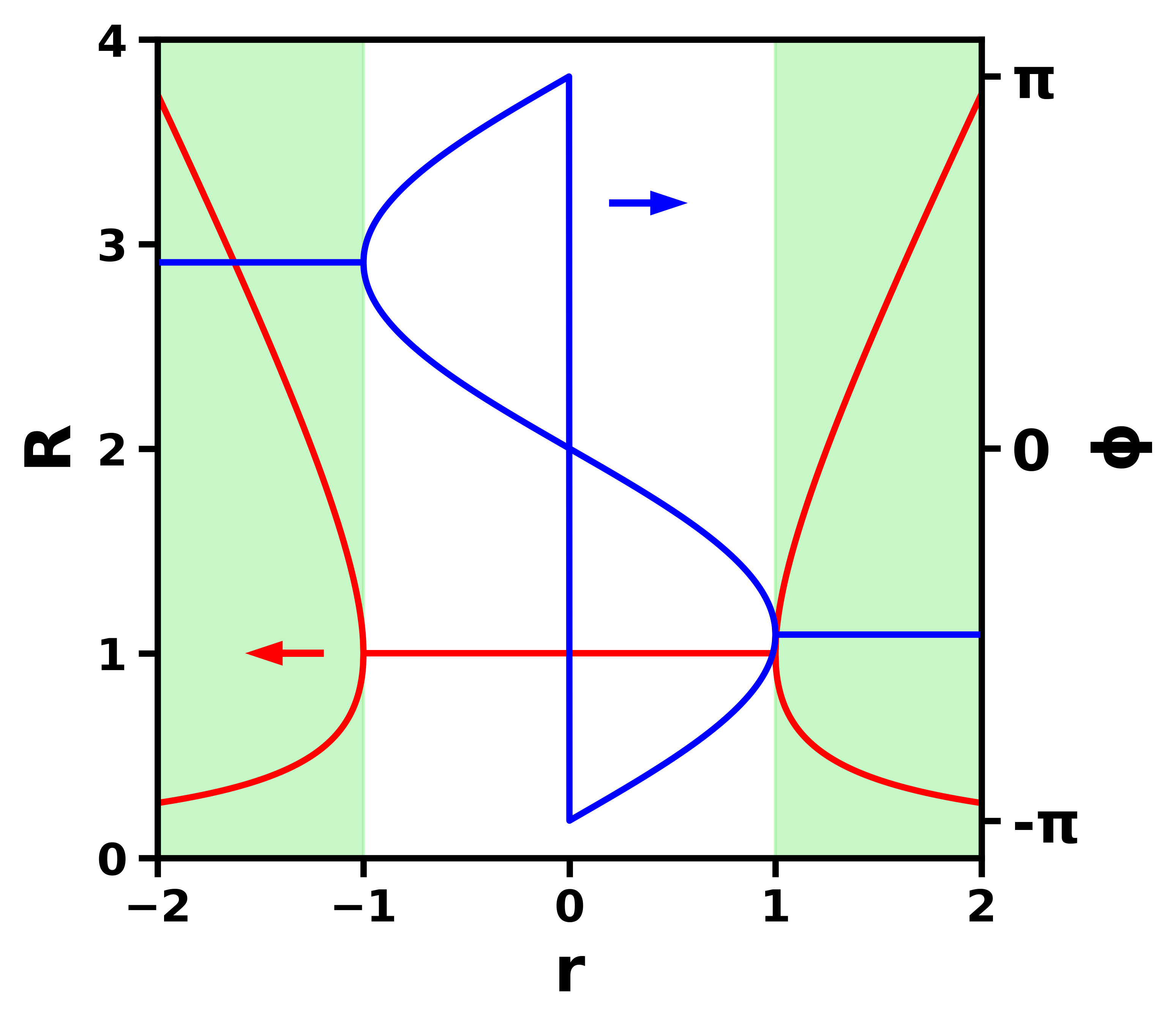}\vspace*{2pt}
    \end{minipage} & \begin{minipage}[m]{.15\textwidth}
    \centering\vspace*{4pt}
    \includegraphics[width=3.2cm]{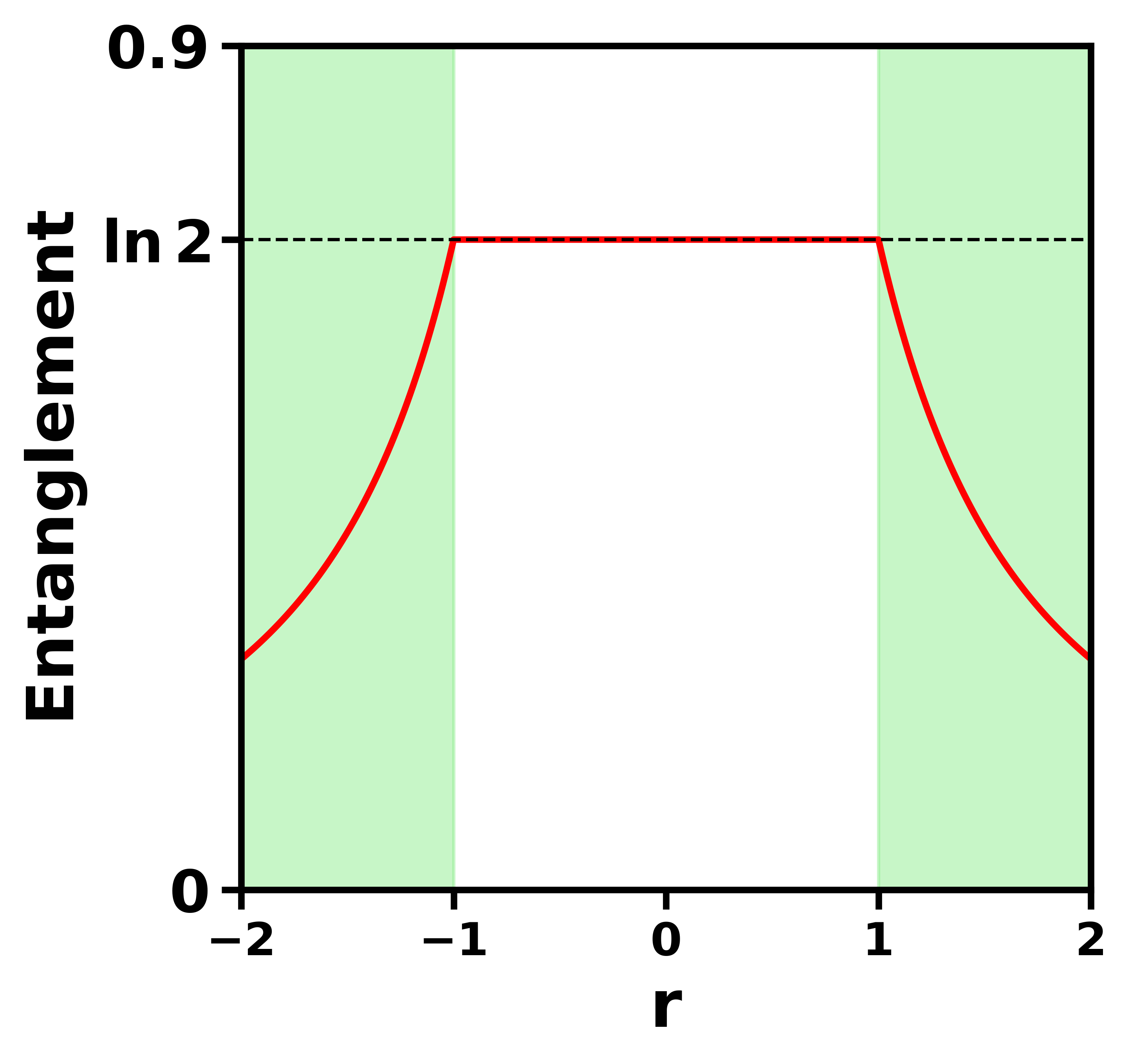}\vspace*{2pt}
    \end{minipage}
                     \\
    \midrule[0.5pt]
\begin{minipage}[m]{.15\textwidth} 
   \begin{equation}
      {\cal H}_{chiral} = \left( \begin{array}{cc}
           0 & 1 \\
           r & 0
       \end{array} \right) \nonumber
   \end{equation}  \end{minipage}  & \begin{minipage}[m]{.15\textwidth}
      \centering\vspace*{4pt}
    \includegraphics[width=3cm]{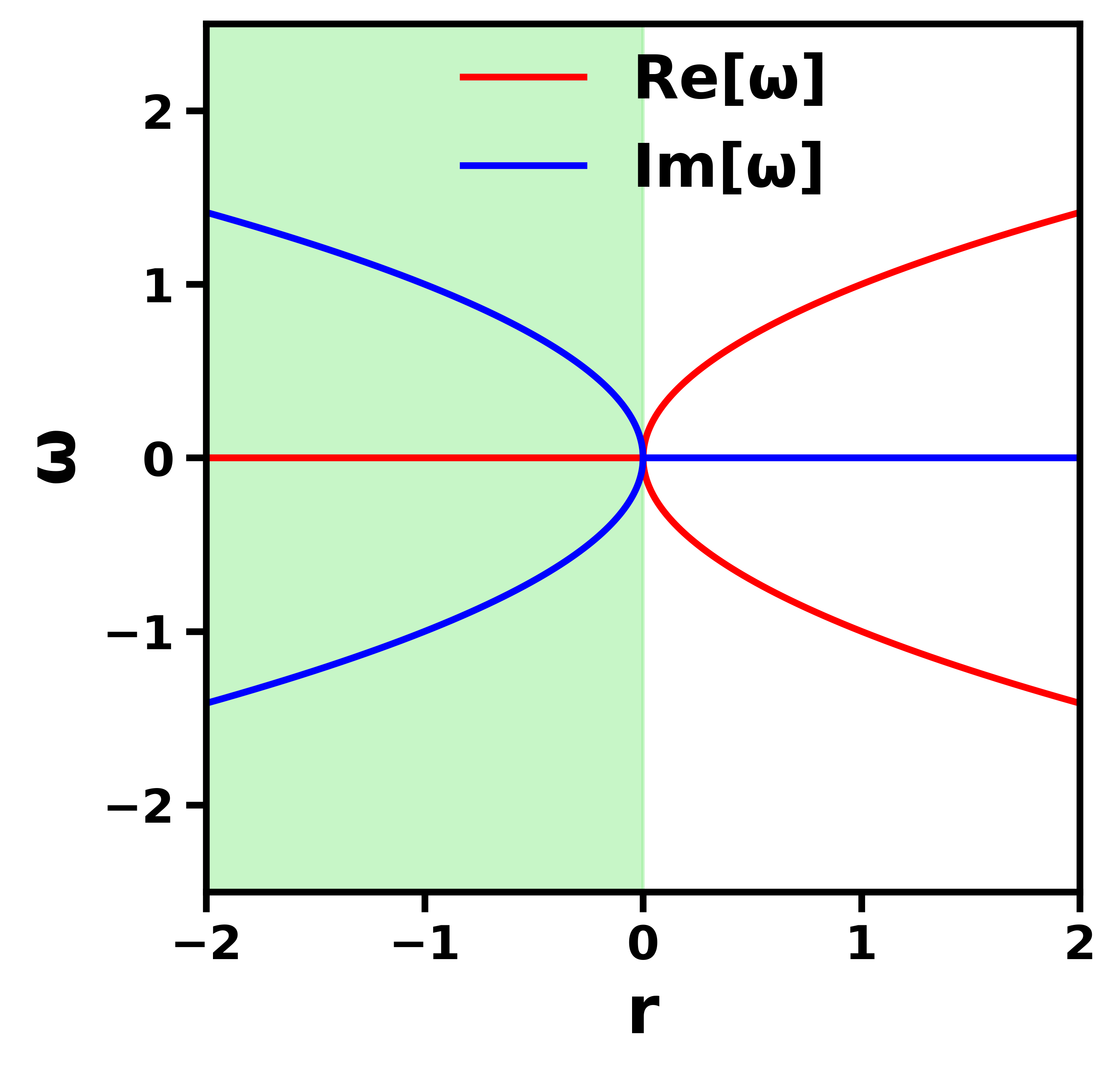}\vspace*{2pt}
    \end{minipage} & \begin{minipage}[m]{.15\textwidth}
    \centering\vspace*{4pt}  	\includegraphics[width=3.4cm]{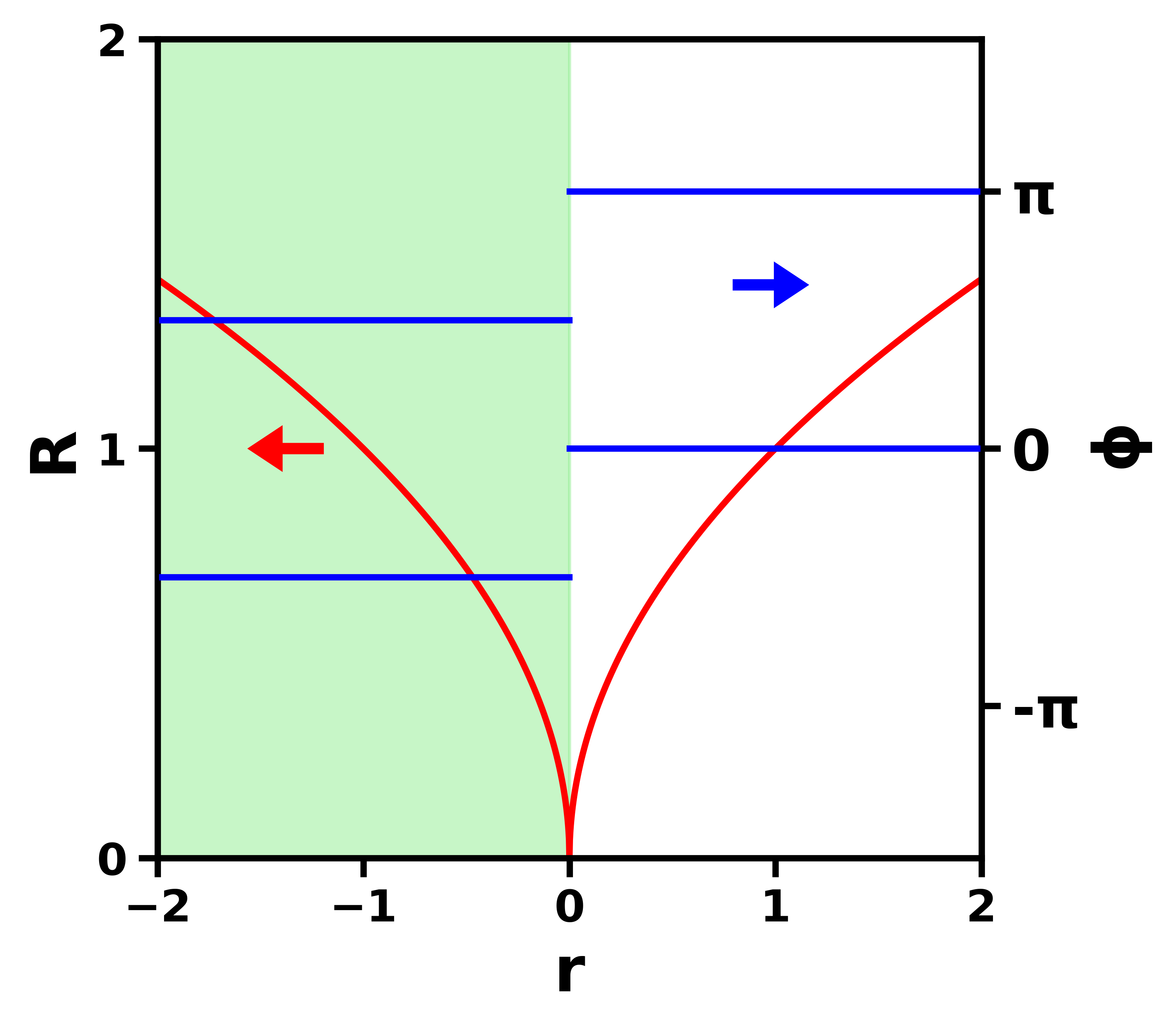}\vspace*{2pt}
    \end{minipage} & \begin{minipage}[m]{.15\textwidth}
    \centering\vspace*{4pt}
    \includegraphics[width=3.2cm]{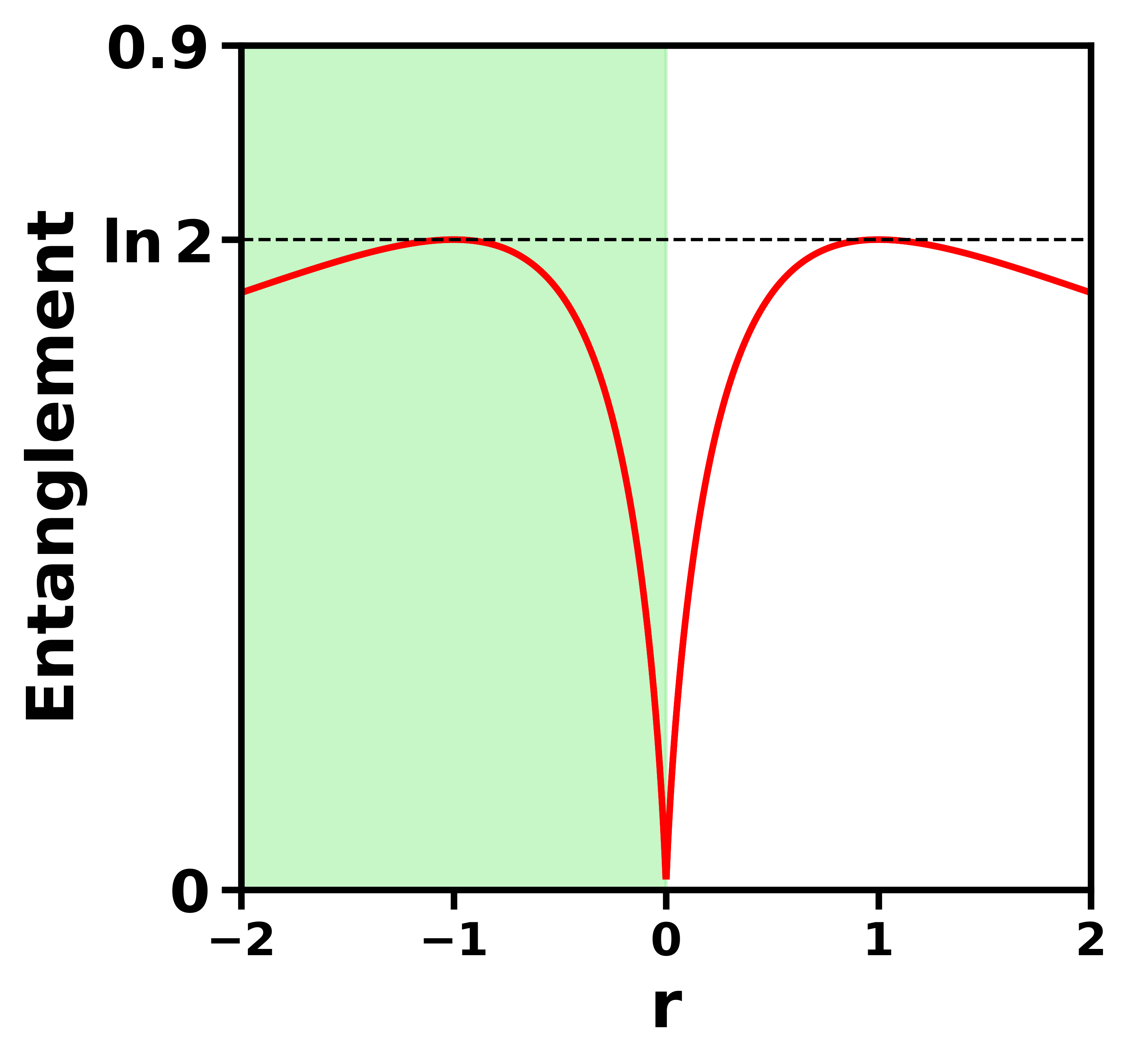}\vspace*{2pt}
    \end{minipage}
                     \\
   \toprule[1pt]
  \end{tabular}
\end{table}

Not limited to the eigenvalue mutations addressed above, the eigenvector also experiences mutations. We write the right eigenvector as $\left|\Phi\right> \propto (1,R e^{i \phi})^{T}$, where $R$ and $\phi$ are the relative amplitude and phase between the two quasiparticles. Given that the eigenvectors are not physical quantities since they are not gauge-invariant, we focus on the information (e.g., the entanglement entropy) encoded within the eigenstate---a physical quantity. Indeed, the EPs have proven 
to be valuable in quantum information processing and in the generation of entangled states \cite{yuan2020steady,lachance2020entanglement,lee2022exceptional,zou2022prb}, as also detailed in the follow-up Sec.~\ref{spin_pumping_dissipative_coupling} on the quantum regime for the spin excitation \cite{kamra2019antiferromagnetic,candido2020predicted,zou2022prb,zou2022aspects,zou2020prb}. Understanding the entanglement behavior across the EPs may be useful even for magnonics since the community is paying attention to the quantum regime \cite{zeng2019quantum,yu2021optical,li2022interaction,yuan2022quantum,zou2022aspects,rameshti2022cavity,xu2023quantum}.

We look into the quantum regime with only one particle population  $\hat{p}^{\dagger} \hat{p} + \hat{m}^{\dagger} \hat{m}$ = 1. For coupled magnon--microwave-photon system, this can be realized at a cryogenic temperature~{\rm mK}~\cite{tabuchi2015coherent}.  Consider the particle number operator $\hat{N} = \hat{p}^{\dagger} \hat{p} + \hat{m}^{\dagger} \hat{m}$, $[\hat{N},\hat{H}]=0$, so the total particle number is conserved. Consequently, if the system starts in a state like $\ket{01}$, it will continue to evolve within the subspace formed by $\ket{01}$ and $\ket{10}$.
The Fock states $\left|01\right>$ and $\left|10\right>$ form a complete basis and the normalized eigenvector 
\begin{align}
\left|\Phi\right> = \frac{1}{\sqrt{1+R^2}} \left(\left|01\right> +R e^{i \phi} \left|10\right>\right).
\label{wavefunction_NH}
\end{align}
Since this pure state cannot be written as a product state, these two particles are entangled. The quantity that can measure the degree of quantum entanglement is the entropy of entanglement, defined by \cite{horodecki1994quantum,schumacher1995quantum,horodecki2009quantum}
\begin{equation}
    S_p = S_m = -{\rm Tr}_m(\rho_p \ln \rho_p ) = -{\rm Tr}_p(\rho_m \ln \rho_m ),
\end{equation}
where $\rho_p  = {\rm Tr}_m(\rho_{p,m})$ and $\rho_m  = {\rm Tr}_p(\rho_{p,m})$ are the reduced density operator for particles $\hat{p}$ and $\hat{m}$,
and $\rho_{p,m} = \left|\Phi\right>\left<\Phi \right|$ is the density operator for the hybridized quasiparticle. Correspondingly, 
\begin{equation}
    S_p = -  \frac{1}{1+R^2} \ln\left(\frac{1}{1+R^2} \right)  -  \frac{R^2}{1+R^2} \left(\frac{R^2}{1+R^2} \right) 
\end{equation}
is solely determined by $R$.
We show the entanglement entropy as a function of parameter $r$ in the last column of Table~\ref{uniqueness of EPs}. 
We address the different features of the system with anti-$\mathcal{PT}$ symmetry, $\mathcal{PT}$ symmetry, and chiral coupling as follows.

\begin{itemize}
\item 1) anti-$\mathcal{PT}$ symmetry:\\
As shown in the first row of Table~\ref{uniqueness of EPs} for the Hamiltonian that possesses the anti-$ \mathcal{P} \mathcal{T}$-symmetry, the relative amplitude $R$ in Eq.~(\ref{wavefunction_NH}) is asymmetrically distributed when $|r|>1$, while the relative phase is fixed to be 0 or $\pi$. 
In the green region beyond the EPs, where the eigenvalues become a complex pair, $\left|\Phi_{1,2}\right> =  \left(\left|01\right> + e^{i \phi_{1,2}} \left|10\right>\right)/\sqrt{2}$ have the same amplitudes $R=1$ for the two states $\left|01\right>$ and $\left|10\right>$ but with parameter-dependent phase differences $\phi_{1,2}$.  Such states differ from the well-known Bell states up to a phase factor \cite{clauser1978bell,brunner2014bell}.
The entanglement entropy reaches $\ln 2$ which is the maximum degree of entanglement for bi-particle, namely the Bell state when the eigenvalue becomes the complex pair \cite{yuan2020steady,yang2021bistability,zou2022prb}.
Further, since in the complex pair the mode with positive imaginary energy is a lasing mode, i.e., its amplitude increases with time, it survives during the time evolution and the system is spontaneously entangled \cite{yuan2020steady,zou2022prb}. We note that ``lasing'' in this context means an exponential increase in the amplitude of one mode, but does not imply an exponential increase in the particle count as in optics.  According to Yuan \textit{et al.}~\cite{yuan2020steady} in the study of steady Bell state generation in the coupled magnon-photon system,  the particle number during the time evolution remains unchanged and the entanglement keeps increasing until it reaches the maximum entanglements $\ln 2$~\cite{yuan2020steady}. This is in contrast to the case with real eigenvalues, where the entanglement entropy typically oscillates due to the Rabi oscillation \cite{dudin2012observation,tabuchi2015coherent}. 
\item 2) $\mathcal{PT}$ symmetry:\\
Compared to the former case with anti-$ \mathcal{P} \mathcal{T}$-symmetry, when the Hamiltonian takes the form that respects the $ \mathcal{P} \mathcal{T}$-symmetry the eigenvalues become a complex pair in the green region, as shown in the second row of Table~\ref{uniqueness of EPs}, while the relative amplitude $R\neq 1$ for the two states and the relative phase is fixed to $\pm \pi/2$. Such asymmetrically distributed amplitude between two states implies the information tends to accumulate at one mode~\cite{ruter2010observation,hodaei2014parity,grigoryan2022pseudo}. At the EPs, the eigenvector $\left|\Phi\right> \propto (1,\pm i)^{T}$ exhibits the chirality that resembles the Jones vector for circularly polarized light \cite{guenther2015modern}, i.e., the left- or right-handedness defined for light with electric fields rotating around the propagation direction anticlockwise or clockwise. It is exploited for chiral propagation and directional lasing \cite{peng2016chiral,ashida2018full,miri2019exceptional}.
\item 3) Chiral coupling:\\
An extreme example is shown in the third row of Table~\ref{uniqueness of EPs}, where the coupling is chiral with $|r|\ne 1$ in the off-diagonal terms. The dependence of eigenvalues and eigenvectors on parameter $r$ becomes more complex and the maximum entangled state can be found in the regimes both below and beyond the EPs. 
\end{itemize}

From these Hamiltonian with different structures, we can conclude that the eigenvalue, eigenvector, and their related physical quantity experience mutation when parameters pass through the EPs. However, how they change depends on the form of a non-Hermitian matrix. Meanwhile, in most cases, the eigenvalue becomes a complex pair beyond the EPs, which leads to the mode selectivity in the time evolution, \textit{i.e.}, the lasing mode survives. Because of this flexibility, diversity, and mode selectivity, tuning the parameter across the EPs brings many interesting properties such as unidirectional invisibility \cite{lin2011unidirectional,feng2013experimental,fruchart2021non}, a stable entangled state \cite{yuan2020steady,zou2022prb,lee2022exceptional}, and single-mode lasing \cite{peng2014loss,brandstetter2014reversing,wong2016lasing}.

\begin{figure}[!htp]
	\centering
	\includegraphics[width=0.8\textwidth]{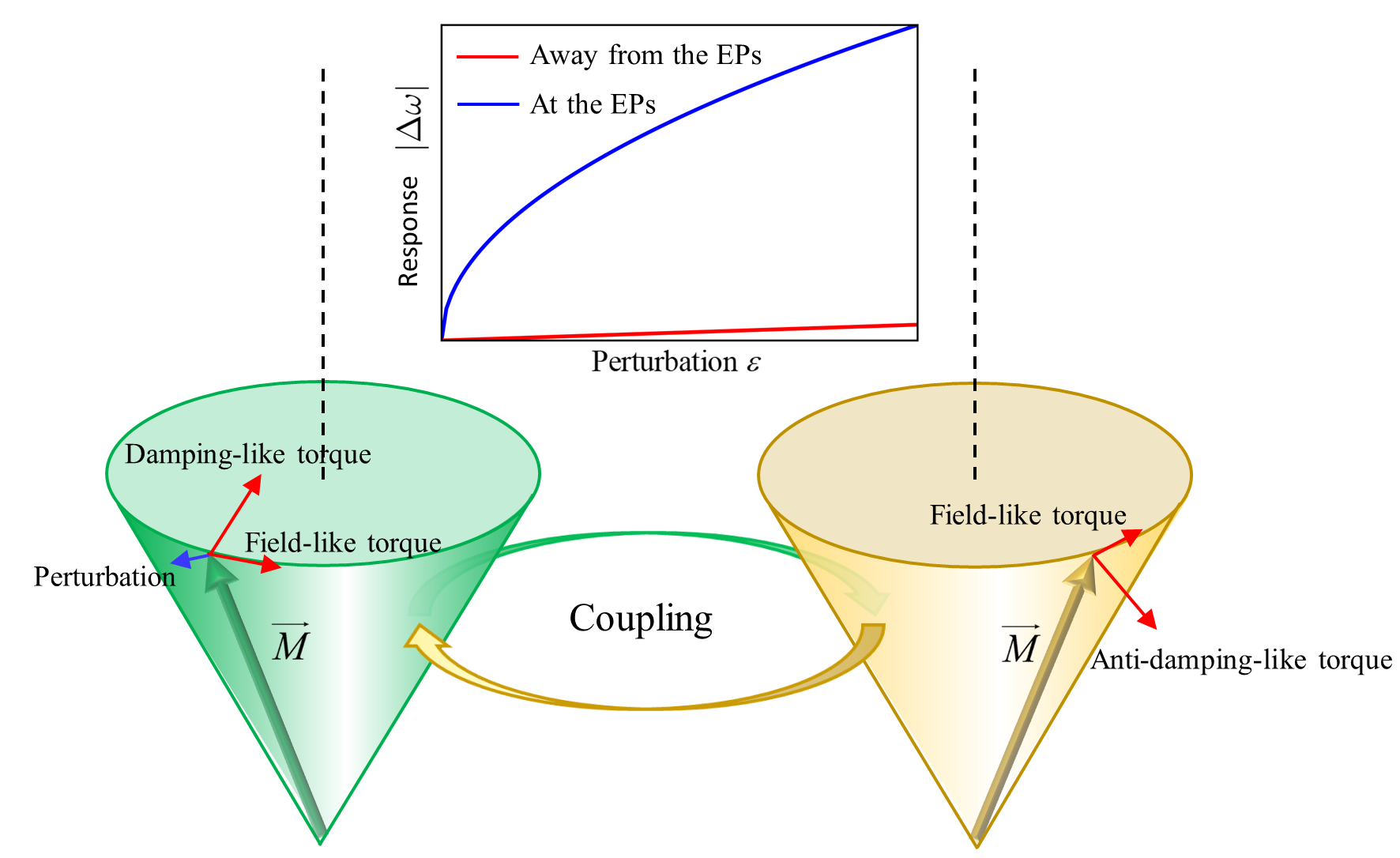}
	\caption{Improved sensitivity at the EPs in a system composed of coupled magnetizations, which obeys the $\mathcal{PT}$-symmetry with the on-site ``loss'' and ``gain'' originating from the damping and anti-damping like torques. A manifestation of improved sensitivity is the enhanced response of eigenvalue variation $\Delta \omega$ to perturbation $\epsilon$ at the EPs (blue curve) compared to that away from the EPs (red curve).}
	\label{fig:sensitivity}
\end{figure}

\textbf{Unprecedented Sensitivity}.---Many studies in the quantum sensing community demonstrated the much-improved sensitivity of a dissipative system at the EPs \cite{wiersig2014enhancing,liu2016metrology,hodaei2017enhanced,lai2019observation,cao2019exceptional,wang2021enhanced}. As one manifestation, the eigenvalue varies sensitively when applying a perturbation in the vicinity of the EPs. Experimentally, introducing a perturbation $\epsilon$ to  the $ \mathcal{P} \mathcal{T}$-symmetric Hamiltonian [Eq.~\eqref{matrix_PT}] has been realized in various experimental platforms \cite{liu2016metrology,wang2021enhanced}, which typically possess the form  
\begin{equation}
	H=\left(\begin{array}{cc}
		ir+\epsilon&s\\
		s&-ir
	\end{array}\right).
	\label{eq:NHhamiltonian}
\end{equation}
An intuitive example that satisfies the above Hamiltonian may be two coupled magnetic moments in Fig.~\ref{fig:sensitivity} that are exerted by the damping and anti-damping like torques, respectively, which may act as ``loss'' and ``gain'' for the magnetization precession. 
The eigenvalues of the two collective modes read
\begin{equation}\label{eq:eigenvalue}
    \omega_{\pm}= \frac{1}{2}\left({\epsilon\pm \sqrt{\epsilon^2-4\left(r^2-s^2-i\epsilon r \right)}}\right).
\end{equation}
When $r^2\neq s^2$, i.e., away from the EPs, the variation of the eigenvalue depends linearly on the perturbation $\Delta \omega\approx \epsilon/2 \pm ir\epsilon/\left(2\sqrt{s^2-r^2}\right)\propto \epsilon$; while at EPs with $r^2 = s^2$, the change of the eigenvalue $\Delta \omega \propto  \epsilon/2 \pm 2 \sqrt{ir\epsilon}$ is dominated by the $\epsilon^{{1}/{2}}$-term. Such nonlinear response near the EPs has been confirmed in the experiment of two YIG spheres embedded in a cavity, where the associated Hamiltonian possesses anti-$\mathcal{PT}$ symmetry \cite{zhao2020observation}. Figure~\ref{fig:sensitivity} plots the dramatic different sensitivity at and away from the EPs, where $\epsilon^{{1}/{2}}$  is much more sensitive than $\epsilon$ with small perturbations. 
Such behavior is more significant around the high-order EPs since the variation of eigenvalue follows $\epsilon^{{1}/{n}}$ with $n$ denoting the order of EPs. The improved sensitivity is now widely used for such as enhanced frequency splitting between modes \cite{wiersig2014enhancing}, the photon-photon interaction \cite{liu2016metrology}, and spontaneous emission of emitters \cite{lin2016enhanced}. Recently,  Duine \textit{et.al.} predicted in synthetic antiferromagnetic
spin-torque oscillators that the system can exhibit an EP with suitable parameters and a small perturbation at the EP can greatly affect the power of spin-torque oscillators \cite{duine2023non}.

Another manifestation of sensitivity is the dramatically enhanced coupling between modes at the EPs. A typical example may be the enhanced photon-photon interaction as observed in a composite model including an active cavity (with a gain rate $\kappa$), a passive cavity (with a loss rate $\gamma$), and a mechanical resonator \cite{liu2016metrology}. The dynamics of the coupled system can be described by a non-Hermitian Hamiltonian:
\begin{align}
\hat{H}=(\omega-i\kappa)\hat{a}^\dagger \hat{a}+(\omega+i\gamma)\hat{c}^\dagger \hat{c}+J(\hat{a}^\dagger \hat{c}+\hat{c}^\dagger \hat{a})+\omega_m \hat{b}^\dagger \hat{b}+g \hat{c}^\dagger \hat{c}(\hat{b}^\dagger+\hat{b}),
	\label{HH}
\end{align}
in which $\hat{a}$, $\hat{c}$, and $\hat{b}$ represent the active cavity, passive cavity, and mechanical phonon operators. Under the condition with $\kappa=\gamma$ and $J>\kappa$, the system respects the $\mathcal{PT}$-symmetry.  An effective coupling between the collective modes of photons and the mechanical phonons can be achieved with strength
\begin{align}
	g_{\rm eff}=\frac{g(\kappa+\sqrt{\kappa^2-J^2})}{2\sqrt{\kappa^2-J^2}}.
\end{align}
As $J\rightarrow\kappa$, $g_{\rm eff}\rightarrow\infty$. This significantly enhances the rate of coherent energy transfer between different physical entities, which would greatly boost its sensitivity compared to conventional optomechanical sensors.

\subsubsection{Exceptional lines and surfaces}

The EPs refer to energy degeneracies at isolated points, while the exceptional lines or surfaces imply the energy degeneracies at a line or on a whole surface. Since an independent parameter corresponds to a degree of freedom, we can treat it as a dimension \textit{in the parameter space} when it can be continuously changed. For a 2$\times$2 non-Hermitian matrix such as Eq.~\eqref{eq:2times2}, introducing the additional parameters that retain the  EPs, which is equivalent to increasing independent degrees of freedom in the parameter space, may lead to the emergence of exceptional lines or surfaces in high-dimensional parameter space \cite{heiss2012physics,xu2017weyl,zhou2018observation,carlstrom2018exceptional,yang2019non,budich2019symmetry,PhysRevLett.123.237202,bergholtz2019non,grigoryan2022pseudo,liu2022experimental}. Here we take a $\mathcal{PT}$-symmetric Hamiltonian $\hat{H} =  \left(\begin{array}{cc}
     i \kappa _0 & k \\
     k & -i \kappa _0 
    \end{array}\right)$
as an example~\cite{meade2008photonic,zhen2015spawning}, 
where $k$ is the magnitude of the wave vector ${\bf k}=k_x\hat{\bf x}$ in the one dimension, with eigenvalues $E_{\pm} = \pm \sqrt{k^2 - \kappa _0^2}$. The EPs appear at two isolated points $k=k_x = \pm \kappa _0$. We now introduce an independent parameter $k_y$ in the wave vector ${\bf k}=(k_x,k_y)$, which retains the EPs when $k_y=0$. An exceptional ring with $k_x^2 + k_y^2 = \kappa _0^2$ emerges in the reciprocal space.

The EPs share many similarities with the conventional energy degenerate points in the Hermitian systems.  It is predicted that the non-Hermitian perturbation, which introduces two independent parameters in the complex number space, can induce a deformation from Dirac or Weyl point to exceptional lines \cite{szameit2011p,bergholtz2019non,zhang2022universal}. This is later confirmed in photonic crystal experimentally, where an exceptional ring is observed by engineering the non-Hermitian radiation of coupled dipole modes \cite{zhen2015spawning}.

In magnonics, a three-dimensional exceptional surface is realized in the system with a magnetic YIG sphere coupled to the cavity, where the position in two directions and the strength and direction of the applied magnetic field constitutes a four-dimensional parameter space \cite{PhysRevLett.123.237202}. As addressed by 2$\times$2 non-Hermitian matrix such as Eq.~\eqref{eq:2times2} at the beginning of this section, Grigoryan and Xia proposed the exceptional surface appears in a high-dimensional parameter space constituted by frequency asymmetry, dissipation asymmetry, and coupling as shown in Fig.~\ref{fig:epcondition}(b) \cite{grigoryan2022pseudo}.

\subsection{EPs in dissipatively coupled spins}
\label{spin_pumping_dissipative_coupling}

\begin{figure}[!htp]
	\centering
	\includegraphics[width=0.98\columnwidth]{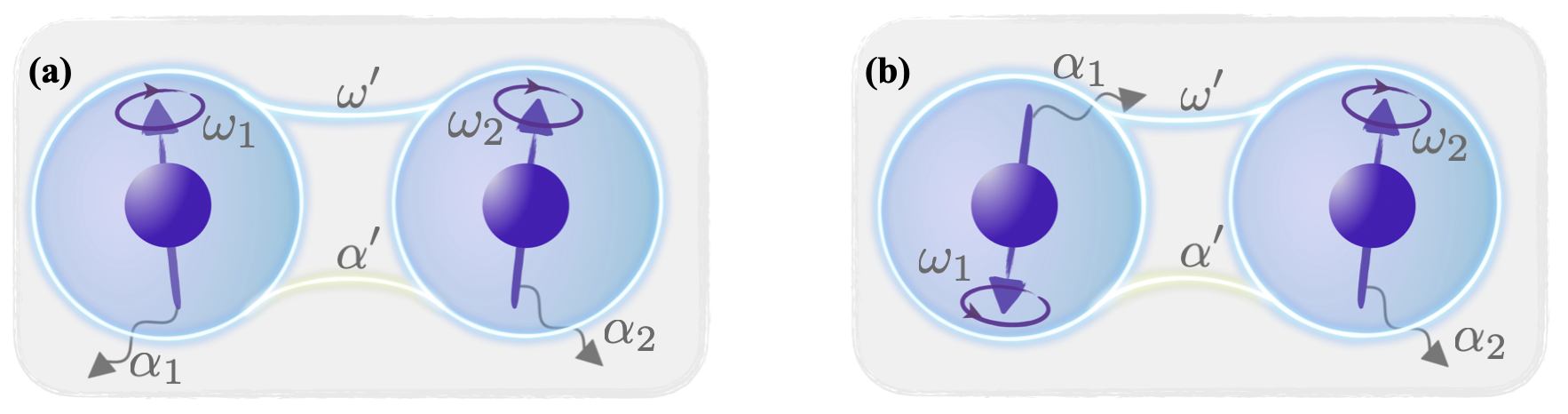}
	\caption{Schematic representation of two (a) ferromagnetically and (b) antiferromagnetically coupled spins, denoted by $\vb S_1$ and $\vb S_2$, precessing in their respective effective frequency $\omega_i$ with effective local  damping $\alpha_i$. The spins couple through both coherent interaction $\omega^\prime$ and dissipative coupling $\alpha^\prime$. }
	\label{fig:411}
\end{figure}

We start with the simplest case proposed by Tserkovnyak \cite{tserkovnyak2020} as shown in Fig.~\ref{fig:411}, where only two classical spins are considered. The generalization to a spin chain and thus EPs in momentum space appears to be straightforward~\cite{tserkovnyak2020}. These classical spins can be regarded as the simplification of magnetic moment in the ferromagnetic layer, while the coupling between them can be mediated by the normal-metal spacer. Typically, this coupling includes coherent coupling, which offers a real coupling strength and leads to a reactive coupling effect, and dissipative coupling due to damping torque caused by spin pumping~\cite{tserkovnyak2005nonlocal}.

\textbf{Ferromagnetic alignment.}   When the two classical spins are coupled ferromagnetically as shown in Fig.~\ref{fig:411}(a), the linearized  equations of motion for the Larmor precession of two spins can be cast in the following Schr\"{o}dinger-like form: 
\begin{align}
i\frac{d}{dt}\left(\begin{array}{c}
\hat{\mathscr{L}}_1\\
\hat{\mathscr{L}}_2
\end{array}\right)=\mathscr{H}\left(\begin{array}{c}
\hat{\mathscr{L}}_1\\
\hat{\mathscr{L}}_2
\end{array}\right),
\end{align}\label{Eq:twospin}
where $\hat{\mathscr{L}}_l\equiv(\hat{S}_l^x+i\hat{S}_l^y)/\sqrt{2S_l}$ obey the canonical algebra $i\{\hat{\mathscr{L}}_l,\hat{\mathscr{L}}_l^*\}=1$. Here, $\hat{S}^i_l$ with $i=\{x,y,z\}$ and $l=\{1, 2\}$ stands for the three components of the $l$-th spin, and $S_l$ is the magnitude of spins. With local applied magnetic field $b_l$ along the $\hat{\bf z}$-direction $\omega_1\equiv b_1 +JS_2$ and $\omega_2\equiv b_2 +JS_1$ are effective frequencies when coupled to each other coherently with strength $J$. 
The effective Hamiltonian matrix takes the form of~\cite{tserkovnyak2020}
\begin{equation}
    \mathscr{H}=(\gamma_-  - i\alpha_-  )\omega_+{\sigma}_z - ( \gamma^\prime -  i\alpha')\omega_+{\sigma}_x, \label{eq:411}
\end{equation}
where terms proportional to $\hat{\sigma}_0\equiv I_{2\times 2}$ are disregarded since they do not affect the following analysis. Here, $\omega_+\equiv (\omega_1+\omega_2)/2$ is the symmetrized spin frequency. $\gamma_-\equiv \omega_-/\omega_+$ is the  normalized detuning with $\omega_-=(\omega_1-\omega_2)/2$ being the asymmetry in the local frequencies. $\alpha_-=(\alpha_1-\alpha_2)/2$ is the damping asymmetry, and $\alpha^\prime$ is the dissipative coupling between the two spins.  $\gamma^\prime\equiv \omega^\prime/\omega_+$ is the normalized coherent coupling with $\omega^\prime\equiv J\sqrt{S_1S_2}$. In the derivation of the above Hamiltonian, one has assumed the smallness of the damping parameters, $\alpha_l\ll 1$, and of the interspin detuning and coupling, $\{\omega_-, \omega^\prime \}\ll \omega_+$.

Upon diagonalization of the Hamiltonian~\eqref{eq:411}, we obtain the eigenvalues
\(  \lambda_\pm =\pm \sqrt{ (\gamma_--i\alpha_-)^2 +(\gamma^\prime -i\alpha^\prime)^2  }. 
\label{eigenvalues}
\)
The EPs occur when the diagonalization of the Hamiltonian breaks down, which takes place when $\lambda_\pm=0$, while the individual terms under the square root are not both zero. In this scenario, $\mathscr{H}\neq 0$, while $\mathscr{H}^2= 0$, thereby confirming that the matrix cannot be diagonalized. This EP condition translates into 
\(  \gamma_--i\alpha_-=\pm i(\gamma^\prime -i\alpha^\prime)\neq 0.  \)
We address two fundamental scenarios.

First, when the coupling between two spins is purely reactive, the dissipative coupling $\alpha^\prime=0$. Further, when the resonance condition $\omega_1=\omega_2$ is assumed, $\gamma_-=0$. Hence, the eigenvalues (\ref{eigenvalues}) reduce to $\lambda_\pm=\pm \sqrt{\gamma^{\prime 2}-\alpha_-^2 }$. The EPs occur when the damping asymmetry $\alpha_-$ and the normalized coherent coupling $\gamma^\prime$ have the same magnitude, i.e., $\alpha_-=\pm \gamma^\prime$. It is evident that the system is in the $\mathcal{PT}$-exact regime when the coherent coupling dominates, $|\gamma^\prime|>|\alpha_-|$. In this case, the eigenvalues are real, suggesting two oscillating modes. In contrast, when the damping asymmetry is relatively large compared to the coherent coupling, $|\alpha_-|>|\gamma^{\prime}|$, we are in the $\mathcal{PT}$-broken regime. Here, the eigenvalues are purely imaginary, and the spin on one site is more damped compared to the other (depending on the sign of $\alpha_-$).

Second, when the coupling between two spins is purely dissipative,  the coherent coupling $\gamma^\prime=0$. Further, we require the local dissipation to be symmetric with $\alpha_-=0$. Thereby, the eigenvalues (\ref{eigenvalues}) reduce to $\lambda_\pm = \pm \sqrt{\gamma_-^2- \alpha^{\prime 2} }$. When the detuning $\gamma_-$ dominates over the dissipative coupling, i.e., $|\gamma_-| >|\alpha^\prime|$, the system is in the $\mathcal{PT}$-exact regime. Conversely, the $\mathcal{PT}$ symmetry is broken when the dissipative coupling is large, i.e., $|\alpha^\prime|>|\gamma_-|$. In the latter case, the eigenvalues are purely imaginary, with the antisymmetric mode more damped relative to the symmetric one. Additionally, both modes are perfectly synchronized in regard to their real frequency components, which is referred to as level attraction~\cite{bernier2018level,harder2018level}, as opposed to the usual level repulsion observed in a hybridized Hermitian system. The EPs are achieved when $\alpha^\prime = \pm\gamma_-$ in this scenario, a natural occurrence in a magnetic bilayer system with a diffusive normal-meta spacer~\cite{tserkovnyak2005nonlocal,heinrich2003dynamic,tserkovnyak2003dynamic}.

\textbf{Antiferromagnetic alignment.} We consider two antiferromagnetically ($J<0$) coupled spins with equal magnitudes as shown in Fig.~\ref{fig:411}(b), $\vb S_i\approx (-1)^i S\hat{\bf z}$, for sites $i=\{1,2\}$. 
Let us suppose the local magnetic fields to be $\vb b_i=(-1)^i b_i \hat{\bf z}$, where $b_i=(-1)^i b+K$. Here, $K\geq 0$ represents the easy-axis anisotropy, and $b$ is the applied uniform magnetic field that is collinear with the anisotropy axis. We define the canonical transverse coordinates as $\hat{\mathscr{L}}_j\equiv (-1)^j (\hat{S}_j^x +i\hat{S}^y_j)/\sqrt{2S}$, which satisfy $i\{ \hat{\mathscr{L}}_j, \hat{\mathscr L}_j^* \}\approx (-1)^j$. By using these coordinates, the linearized equations of motion of the two spins can be cast into the form of Eq.~\eqref{Eq:twospin} with the following Hamiltonian matrix
\( \mathscr H=i(\alpha \omega^\prime-\alpha^\prime \kappa) {\sigma}_x +(i\omega^\prime {\sigma}_y -\kappa {\sigma}_z  ),   \)
where we have dropped the constant part and assumed the smallness of all damping parameters as before.  Here, $\alpha\equiv g+\alpha^\prime$ is the effective  local damping, consisting of the intrinsic local damping $g$ (assumed to be equal for the two sites) and the dissipative coupling $\alpha^\prime$ due to spin pumping. $\kappa\equiv K+\omega^\prime$ is the local Larmor frequency consisting of the anisotropy $K$ and the  exchange coupling $\omega^\prime =|J|S$. Thus we have the constraints $\alpha\geq \alpha^\prime$ and $\kappa\geq \omega^\prime$. To locate the EP, we set $\mathscr H^2=0$, giving rise to the condition 
\(  \alpha \omega^\prime -\alpha^\prime \kappa =\pm \sqrt{\kappa^2 -\omega^{\prime 2}},     \)
which, in terms of the intrinsic local damping $g$, can be rewritten as 
\( g\omega^\prime =\alpha^\prime K \pm \sqrt{K(K+2\omega^\prime)}.  \)
Assuming a weak dissipative coupling $\alpha^\prime\ll 1$ and a dominant exchange interaction $\omega^\prime \gg K$, we obtain $g\approx \sqrt{2K/\omega^\prime} =\omega/\omega^\prime \ll 1$ with $\omega=\sqrt{2K\omega^\prime}$ being  the intrinsic antiferromagnetic resonance frequency. This  is consistent with the small value of the damping parameter $g$.

We observe that two antiferromagnetically coupled spins naturally lead to an EP when the Hamiltonian is solely determined by the exchange coupling, with the Hamiltonian taking the form of
\( \mathscr H_{\text{AF}}=\omega^\prime (i\hat{\sigma}_y-\hat{\sigma}_z),  \)
which is not diagonalizable.
However, in the ferromagnetic case, one needs to introduce damping and additional fields to achieve EPs. This is evident from the Hamiltonian taking the form of
\( {\mathscr{H}}_{\text{F}}=\omega^\prime (1-\hat{\sigma}_x),  \)
in the presence of only ferromagnetic exchange. 
To gain insights into such behavior, we decompose the small-angle dynamics into symmetric and antisymmetric components, given by $\mathscr{L}\pm=(\mathscr{L}_1\pm \mathscr{L}_2 )/2$. 
In both scenarios, there exists a zero mode $\mathscr{L}_+$, corresponding to a reorientation of the overall order parameter. In the ferromagnetic case, the distortion of the order parameter triggers its small-angle precession with frequency $2\omega^\prime$, while in the antiferromagnetic case, it results in an unbounded growth of $\mathscr{L}_+$. This suggests that the antiferromagnetic EP leads to a breakdown of the linearized treatment. In the context of spin dynamics,  it is simply the precession of the N\'{e}el order parameter in the plane perpendicular to the distortion $\mathscr{L}_-$ with frequency $\propto \omega^\prime \mathscr{L}_-$.

\begin{figure}[!htp]
	\centering
	\includegraphics[width=16.3cm]{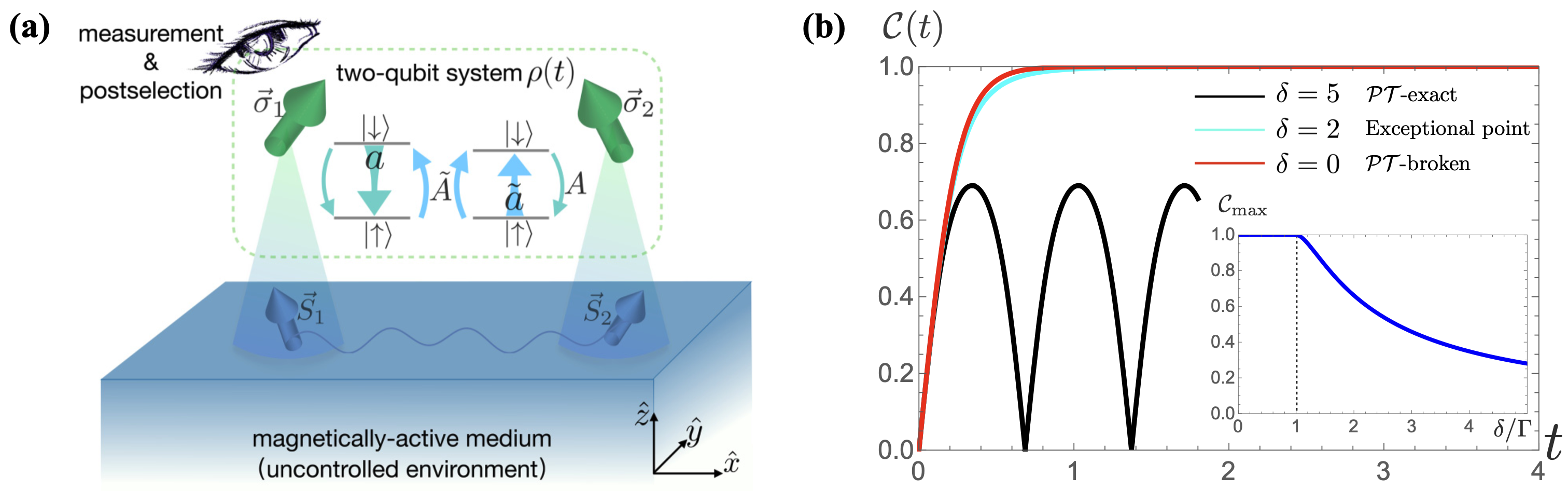}
	\caption{(a)  A system composed of two spin qubits is coupled
with a magnetic environment, which induces local relaxation
$\{a,\tilde{a}\}$, mediates dissipative couplings $\{A,\tilde{A}\}$, as well as coherent
couplings between two qubits. The two qubits may achieve a
stable entangled state with large enough $A$ and $\tilde{A}$, and even
a Bell state with the help of measurement and postselection.  (b) Concurrence of qubits as a function of time with initial
state $\ket{\uparrow\downarrow}$ under continuous measurements and post selections. We set $\Gamma=2.$ The black curve $\delta=5$ is in the $\mathcal{P}\mathcal{T}$-exact regime, where entanglement oscillates and its maximal value is less than 1. At the
exceptional point (cyan curve), there is no oscillation and its maximal
value is 1.  In the $\mathcal{P}\mathcal{T}$-broken regime (red curve), entanglement is $\mathcal{C}(t)=\tanh 2\Gamma t$. The inset shows the maximal concurrence as a function
of $\delta/\Gamma$. The figure is reproduced with permission from Ref.~\cite{zou2022prb}.  }
\label{fig:3351}
\end{figure}

\textbf{EPs and quantum spins}.---Here,  we derive the non-Hermitian dynamics of two quantum spins, i.e., spin qubits, coupled to magnons with appropriate post-selections~\cite{candido2020predicted,kamra2019antiferromagnetic,zou2022prb,zou2022aspects} by using the master equation approach introduced in Sec.~\ref{master_equation_approach}, see Fig.~\ref{fig:3351}(a). Specifically, we demonstrate that the magnon-mediated dissipative couplings between the qubits result in the EP in the dynamics.
This EP enables the generation of a Bell state, which can play a critical role in quantum information processing.

Let us consider an illustrative model consisting of two spin qubits weakly coupled to a magnet, with Hamiltonian
$ \hat{H}=\hat{H}_{\text{S}}+ \hat{H}_{\text{E}}+\hat{H}_{\text{SE}}$.
Here, $\hat{H}_{\text{S}}=-( \Delta_1\hat{\sigma}_1^z+\Delta_2\hat{\sigma}_2^z)/2$ is the Hamiltonian for the system with two qubits subjected to two local magnetic fields $\Delta_1$ and $\Delta_2$, respectively, along the $\hat{\bf z}$-direction, $\hat{H}_{\text{E}}$ is an unspecified  Hamiltonian of the medium as an environment for the system, and $\hat{H}_{\text{SE}}=\lambda (\hat{\pmb \sigma}_1\cdot \hat{\bf S}_1  + \hat{\pmb \sigma}_2\cdot \hat{\bf S}_2 )$ 
describes the system-environment interaction with coupling strength $\lambda$, where $\hat{\pmb \sigma}_i$ stands for the Pauli matrice of the $i$-th qubit, and $\hat{\bf S}_i$ represents local spin density operator within the medium.
Without loss of generality,  we assume $\Delta_1\geq \Delta_2\geq 0$. We focus on an axially-symmetric environment $\hat{H}_{\text{E}}$ in spin space, while a generalization is straightforward. It would also be interesting to generalize the physics to the dipolar coupling between the qubit and the medium \cite{fukami2021opportunities}.

Similar to the example that we discussed in Sec. \ref{bilinear_coupling}, we can derive the Lindblad master equation of the density matrix of the two-qubit system based on Born-Markov approximation: 
\(  \dv{}{t}\hat{\rho} = -i\big[\hat{H}_{\text{S}}+\hat{H}_{\text{eff}} ,\hat{\rho} \big]  - \mathcal{L}[\hat{\rho}].  \label{master} \)
Here, $\hat{H}_{\text{eff}}$ refers to the medium-induced effective coherent coupling between qubits, which leads to  the unitary evolution of the system. However, we will disregard this factor in the ensuing discussion as we aim to investigate the impact of induced dissipative couplings. It is worth noting that the coherent interactions can also be derived using the Schrieffer-Wolff transformation~\cite{trifunovic2012long,trifunovic2013long,flebus2018quantum,flebus2019entangling,neuman2020nanomagnonic,fukami2021opportunities}.
$\mathcal{L}[\hat{\rho}]$ is the  dissipative Lindbladian expanded in the usual form:
 $  \mathcal{L}[\hat{\rho}]= \sum_{nm}h_{nm}  \big(  \hat{\mathcal{O}}^\dagger_m \hat{\mathcal{O}}_n \hat{\rho} + \hat{\rho} \hat{\mathcal{O}}^\dagger_m \hat{\mathcal{O}}_n -2 \hat{\mathcal{O}}_n \hat{\rho} \hat{\mathcal{O}}^\dagger_m   \big),$
where the coefficient matrix $h$ is Hermitian and positive-semidefinite, and  $\hat{\mathcal{O}}\!\! =\!\! ( \hat{\sigma}_1^- , \hat{\sigma}_2^- , \hat{\sigma}_1^+ , \hat{\sigma}_2^+  )$ comprises qubit operators. We have neglected pure-dephasing effects, which in practice
may be mitigated by dynamic decoupling~\cite{viola1999dynamical,khodjasteh2005fault,franco2014preserving}.
In the dissipative Lindblad part, $h$ is block diagonal due to the axial symmetry, given by
\(  h=\mqty(\tilde{a} & \tilde{A} \\ \tilde{A}^* & \tilde{a}) \oplus  \mqty(a & A^* \\ A & a).  \)
Here, all parameters are given by Green's functions of the magnetic medium $a=i\lambda^2 G^>_{\hat{S}_1^+ \hat{S}_1^-}(\Delta)/2, A=i\lambda^2 G^>_{\hat{S}_1^+\hat{S}_2^-}(\Delta)/2 $ and $\tilde{a} = e^{-\beta \Delta} a$ and $\tilde{A} = e^{-\beta \Delta} A$, where $\beta = 1/(k_B T)$ and $\Delta \equiv (\Delta_1+\Delta_2)/2$. The greater and lesser Green's functions follow the conventional definitions: 
$ G^>_{\hat{X},\hat{Y}}(t)\equiv -i\langle \hat{X}(t)\hat{Y}\rangle, G^<_{\hat{X},\hat{Y}}(t)\equiv -i\langle \hat{Y}\hat{X}(t)\rangle,$
and the Fourier transformation is given by $G(\omega)=\int dt\; G(t) e^{i\omega t}$. $a$ and $\tilde{a}$ are associated with local decay and the reverse process. They govern the local relaxation of individual qubits, giving rise to the relaxation time $T_1$ and contributing to the decoherence time $T_2$ of a single qubit. In contrast, $A$ and $\tilde{A}$ are related to cooperative decay and the reverse process involving both qubits and are referred to as dissipative couplings, which depend on the distance between the two qubits.
The thermodynamic stability of the magnetic medium imposes
 $   a \geq |A|$ and $\tilde{a} \geq |\tilde{A}| $,
which ensures the matrix $h$ is positive-semidefinite.

 When the Lindblad master equation governing a quantum system does not involve quantum jumps, the quantum dynamics can be described using an effective non-Hermitian Hamiltonian. In our case, by subjecting the two qubits to continuous measurements of the absolute value of their total spin $z$ component $\hat{\underline{\sigma}}^z=\hat{\sigma}_1^z+\hat{\sigma}_2^z$  and subsequently conditioning the postselection on zero outcomes, we can effectively 
forbid all quantum jump processes. As a possible low-temperature implementation of the proposed post-selection scheme, we may post-select on the absence of any emitted magnons.
  In the basis $\{\ket{\uparrow\downarrow}, \ket{\downarrow\uparrow}\}$, the non-Hermitian Hamiltonian is~\cite{brody2012mixed,bender2007making} 
\( \hat{\mathcal{H}}_{\text{eff}}= \mqty[ -\delta-i(a+\Tilde{a}) &  -i\Gamma \\ -i\Gamma &  \delta -i(a+\Tilde{a})  ],  \)
where $\Gamma\equiv A+\Tilde{A}$ is the dissipative coupling and $\delta=(\Delta_1-\Delta_2)/2$. It is clear that there are three different regimes: $\mathcal{PT}$ symmetry broken regime with $\delta<\Gamma$, the exceptional point with $\delta=\Gamma$, and $\mathcal{PT}$-exact regime with $\delta>\Gamma$.

In the $\mathcal{PT}$-broken regime, the eigenvalues of this Hamiltonian are purely imaginary. Assuming the initial state is $\ket{\uparrow\downarrow}$ and $\delta=0$, the entanglement between the two qubits is $\mathcal{C}(t)=\tanh (2\Gamma t)$. The final state is maximally entangled, as shown in Fig.~\ref{fig:3351}(b).
At the EP,  $\hat{\mathcal{H}}_{\text{eff}}$ is non diagonalizable, since two eigenstates
coalesce into $\left( \ket{\uparrow\downarrow}+i\ket{\downarrow\uparrow} \right)/\sqrt{2}$. The two qubits will gradually evolve into this sole state where they are maximally entangled. For example, 
starting with a trivial state $\ket{\uparrow \downarrow}$, the  
concurrence 
$\mathcal{C}(t)=  2\Gamma t  \sqrt{1+\Gamma^2 t^2}  /( 1+2\Gamma^2 t^2 ), $
algebraically approaching 1, see Fig.~\ref{fig:3351}(b).
In the $\mathcal{PT}$-exact regime, two eigenenergies have nonzero real parts. The entanglement will oscillate with frequency $2\sqrt{\delta^2-\Gamma^2}$, as shown in Fig.~\ref{fig:3351}(b). The maximal entanglement one can achieve is 
$   \mathcal{C}_{\text{max}}(\eta)=  \sqrt{2-  1/\eta^2}/\eta,  $ with $\eta=\delta/\Gamma$, which is less than 1.
Notably, the second derivative of $\mathcal{C}_{\text{max}}$ is discontinuous across the EP ($\eta=1$), reflecting a  phase transition [see Fig.~\ref{fig:3351}(b)].

\subsection{EPs by balanced gain and loss of spin dynamics}

\label{EPs_magnonics}

\subsubsection{Theoretical proposals}
\label{EPs_magnonics_theoretical}

\begin{figure}[!htp]
\centering
\includegraphics[width=1\textwidth]{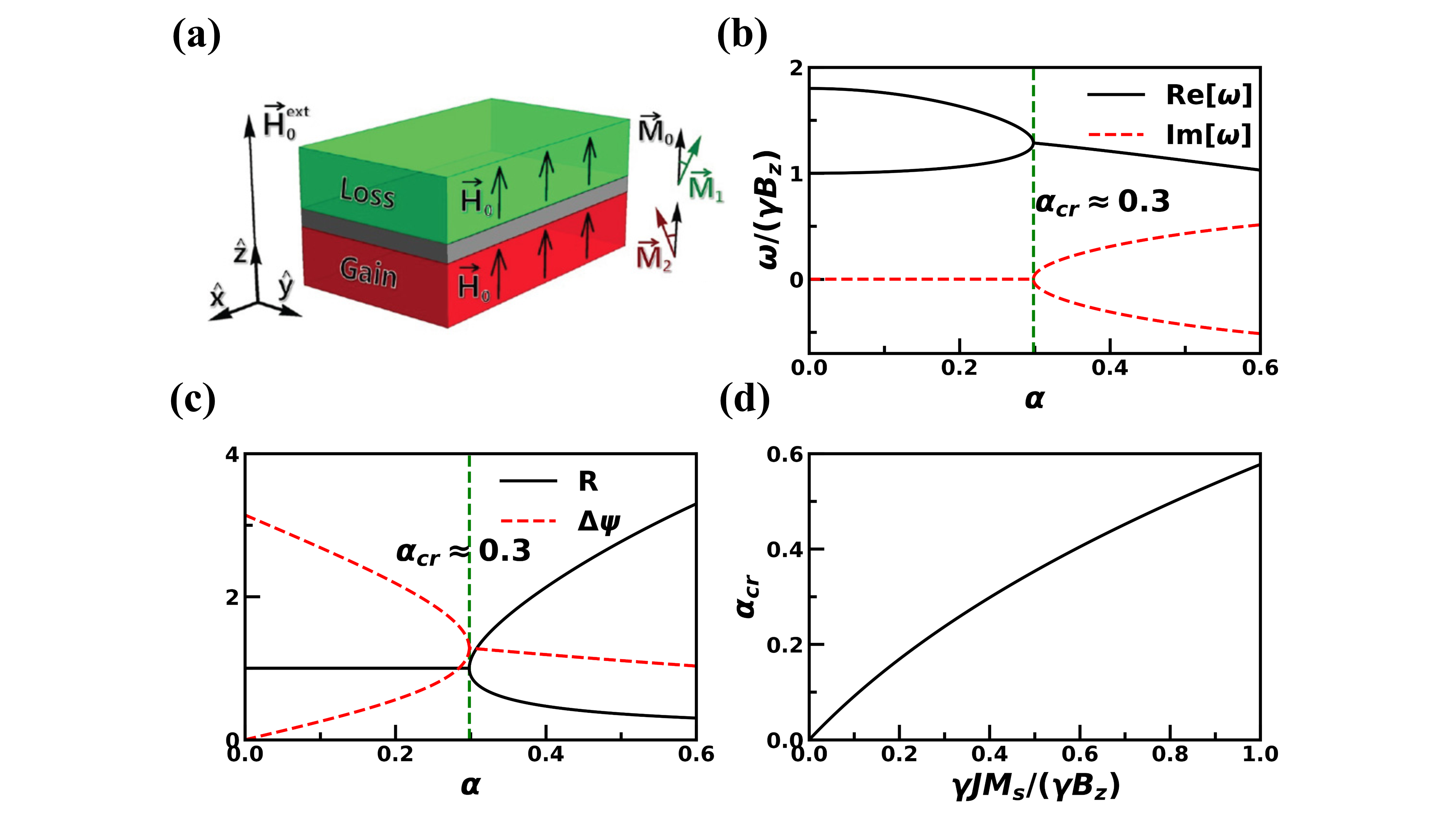}
\caption{Realization of EPs in coupled magnetic structure by balanced gain and loss. (a) is the configuration. (b) and (c) plot the mutation of complex energy $\omega$, amplitude ratio $R$, and phase difference $\Delta \psi $ of the two modes around the EP. $\alpha_{cr}\approx 0.3$, denoted by the green dashed line in (b) and (c), implies the EP. (d) shows the dependence of critical damping coefficient $\alpha_{cr}$ on $\gamma M_s J/\gamma B_z $. The figures are adapted with permission from Ref.~\cite{lee2015macroscopic}.   }
	\label{fig:magnonep}
\end{figure}

\textbf{Second-order EPs}.---Let us start with a simplified model to address how the EPs are realized in  the collective dynamics of macro spins and how the magnonic functionality is improved by the EPs. 
The model of two macroscopic magnetic structures with balanced gain and loss that are coupled with ferromagnetic exchange constant $J>0$ \cite{PhysRevB.91.094416} is shown in Fig.~\ref{fig:magnonep}(a).  Similar non-Hermitian magnetic models were extensively studied later in Refs.~\cite{lee2015macroscopic,yang2018antiferromagnetism,yu2020higher,tserkovnyak2020,PT_bilayer}. 

The macroscopic magnetic structures extended in the $y$-$z$ plane are applied by the external field ${\bf B}=B_z\hat{\bf z}$ along the $\hat{\bf z}$-direction, with the free energy
\begin{equation}\label{Hamil}
  F = - {\bf M}_1\cdot {\bf B} - {\bf M}_2\cdot {\bf B} - J  {\bf M}_1 \cdot {\bf M}_2,
\end{equation}
where ${\bf M}_{1}$ and ${\bf M}_{2}$ are the magnetization of two magnetic layers that are coupled with ferromagnetic exchange coupling $J>0$. 
The magnetization dynamics are governed by the LLG equation \cite{gilbert2004phenomenological,landau1992theory}
\begin{equation}
\label{LLGtwolayer}
\begin{aligned}
 \frac{d{\bf M}_{1}}{dt} &=-\gamma  {\bf M}_{1} \times (B_z\hat{\bf z} + J  {\bf M}_{2})  + \frac{\alpha }{\left|{\bf M}_{1}\right|} {\bf M}_{1} \times \frac{d{\bf M}_{1}}{dt},  \\
 \frac{d{\bf M}_{2}}{dt}  &=-\gamma  {\bf M}_{2} \times (B_z\hat{\bf z} + J  {\bf M}_{1})  - \frac{\alpha }{\left|{\bf M}_{2}\right|}  {\bf M}_{2} \times \frac{d{\bf M}_{2}}{dt}.
\end{aligned}
\end{equation}
The positive Gilbert damping coefficient $\alpha$ describes the intrinsic damping of magnetization, while a negative one implies the amplification or ``gain'', which may be realized effectively by such as parametric driving \cite{demokritov2006bose,lee2015macroscopic,bracher2017parallel}, spin transfer torque \cite{slavin2005current,apalkov2005slonczewski,galda2016parity,galda2018parity}, and optomagnonic interaction \cite{zhang2016cavity,cao2022negative,rameshti2022cavity}.

In the linear regime, the magnetization 
\begin{equation}
    {\bf M}_{1,2}(t) \approx M_s\hat{\bf z}+ \left({m}_{1,2x}\hat{\bf x}+{m}_{1,2y}\hat{\bf y}\right)e^{-i\omega t},
\end{equation}
where $ M_s = \left|{\bf M}_{1}\right|=\left|{\bf M}_{2}\right|$ is the saturated magnetization and $\left|{m}_{1,2x,y}\right|\ll M_s$. With ${m}_{1,2}^{-}={m}_{1,2x}-i{m}_{1,2y}$, the LLG equation (\ref{LLGtwolayer}) reduces to  
\begin{equation}
    \omega \left(
         \begin{array}{c}
           m_{1}^{-} \\
           m_{2}^{-} \\
         \end{array}
       \right)  = \frac{\gamma}{1+\alpha^2}\left(
        \begin{array}{cc}
        B_z + M_s J -i \alpha \left( B_z + M_s J\right) & -M_s J + i \alpha M_s J \\
        -M_s J -i \alpha M_s J &  B_z + M_s J + i \alpha \left(B_z + M_s J\right) \\
       \end{array}
        \right)\left(
         \begin{array}{c}
           m_{1}^{-} \\
           m_{2}^{-} \\
         \end{array}
       \right),
\end{equation}
with the eigenvalues and eigenvectors for the two branches
\begin{subequations}
\begin{align}
\label{eq:maneticEP}
  \omega_{\pm} &= \frac{\gamma}{1+\alpha^2}\left({ B_z + M_s J \pm \sqrt{(M_s J)^2 -\alpha^2  ( B_z^2 + 2 B_z M_s J)}}\right),\\
  \left(
\begin{array}{cc}
  m_{1}^{-} \\
  m_{2}^{-}
\end{array}\right)_{\pm}&=\left(
        \begin{array}{c}
        \frac{i}{(1+i\alpha)M_s J}\left(\alpha(B_z +M_s J)\pm i \sqrt{(M_s J)^2 -\alpha^2  ( B_z^2 + 2 B_z M_sJ)}\right) \\
                       1 \\
        \end{array}
        \right).
\end{align}
\end{subequations}  

According to Eq.~\eqref{eq:maneticEP}, the EP appears with the specific Gilbert damping coefficient $\alpha_{\rm cr} =M_s J/\sqrt{B_z^2 + 2B_zM_sJ} $. The eigenvalue is purely real when $\alpha < \alpha_{\rm cr}$, which, on the other hand, contains the imaginary component when  $\alpha > \alpha_{\rm cr}$. As shown in Fig.~\ref{fig:magnonep}(b) with a specific external field $B_z=2.5M_s J   $, the two eigenvalues merge at $\alpha = \alpha_{\rm cr}\approx 0.3$. The amplitude ratio $R=\left|{m_1^{-}}/{m_2^{-}}\right|_{\pm}$  and the phase difference $\Delta \psi = {{\rm Arg}(m_1^{-})}/{{\rm Arg}(m_2^{-})}$ of the two modes experience a sudden change around the EP, as shown in Fig.~\ref{fig:magnonep}(c).  The two eigenmodes share the same amplitude when $\alpha <
\alpha_{\rm cr}$ as in  Fig.~\ref{fig:magnonep}(c), implying an equal distribution of energy. On the other hand, when $\alpha >
\alpha_{\rm cr}$, the relative amplitude becomes asymmetric with a fixed phase difference, suggesting that the energy tends to accumulate in one of the two eigenmodes. 
With an increase of ratio $M_s J/B_z$, $\alpha_{\rm cr}$ for the EP increases as in Fig.~\ref{fig:magnonep}(d). It also implies that the condition for the emergence of EP becomes a specific line in the two-dimensional parameter space constituted by  $M_s J$ and $B_z$. By analogy, it can be envisioned that the exceptional lines and surfaces can emerge in higher dimensional parameter space when more freedoms are introduced such as the frequency asymmetry.

A similar design is also proposed in the system with two magnetic layers embedded with a metallic Pt layer, where the injected current in the Pt layer exerts opposite torques for two adjacent magnetic layers~\cite{wang2020steering}. For such a $\mathcal{PT}$-symmetric system, Wang~\textit{et.al.} found many unique functionalities near the EPs such as enhanced magnetic susceptibility, magnon trapping, and magnon enhancement~\cite{wang2020steering}.

\begin{figure}[!htp]
	\centering
	\includegraphics[width=0.96\textwidth]{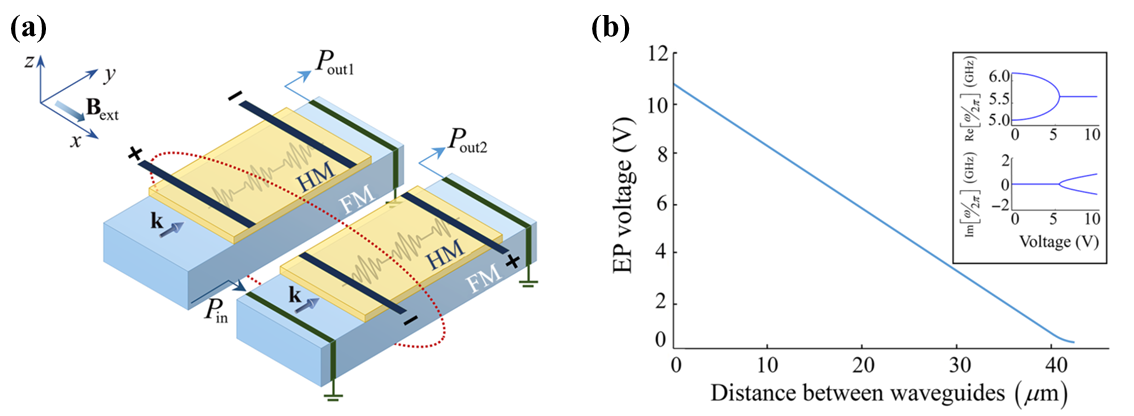}
	\caption{Realization of EPs in planar $\mathcal{PT}$-symmetric heterostructures of coupled ferromagnetic--heavy-metal layer. (a) is the configuration. (b) shows the dependence of critical voltage for the emergence of EP on the distance between two ferromagnetic layers. The inset in (b) shows the frequency variation near the EP as s function of voltage. The figure is reproduced with permission from Ref.~\cite{PT_bilayer}.} 
	\label{fig:planar}
\end{figure}

Different from the indirect coupling between two ferromagnetic layers mediated by a metallic spacer, Temnaya \textit{et al.} studied a direct coupling by the magneto-dipolar interaction in two identical planar ferromagnetic layers and found that the EPs can be realized by injected spin current from thin layers of heavy metal ~\cite{PT_bilayer}, as shown in Fig.~\ref{fig:planar}(a). The magnon in the ferromagnetic layer is dissipated governed by the Gilbert damping, while the injected spin current can compensate or enhance this dissipation via tuning the sign of the voltage in the heavy metal layer, thus with the opposite voltage for the two heavy-metal layers achieving  controllable gain or loss for the ferromagnetic layer. Let us denote one magnetization of the ferromagnetic layer as ``${\bf M}_{1}$'' and the other one  ``${\bf M}_{2}$''. The magnetic dynamics of the coupled ferromagnetic layers are governed by 
\begin{equation}\label{planarLLG}
\begin{aligned}
 \frac{d{\bf M}_{1}}{dt} &=-\mu_0 \gamma  {\bf M}_{1} \times \mathbf{H}_{\mathrm{eff},1}  + \frac{\alpha }{\left|{\bf M}_{1}\right|} {\bf M}_{1} \times \frac{d{\bf M}_{1}}{dt} + 
 \frac{\sigma J_s}{\left|{\bf M}_{1}\right|}\left({\bf M}_{1} \times\left({\bf M}_{1}\times{{\bf{p}}_1}\right)\right),  \\
 \frac{d{\bf M}_{2}}{dt} &=-\mu_0 \gamma  {\bf M}_{2} \times \mathbf{H}_{\mathrm{eff},2}  + \frac{\alpha }{\left|{\bf M}_{2}\right|} {\bf M}_{2} \times \frac{d{\bf M}_{2}}{dt} + 
 \frac{\sigma J_s}{\left|{\bf M}_{2}\right|}\left({\bf M}_{2} \times\left({\bf M}_{2}\times{{\bf{p}}_2}\right)\right),
\end{aligned}
\end{equation}
where $\mu_0$ is the vacuum magnetic permeability and $\mathbf{H}_{\mathrm{eff,1}}$ and $\mathbf{H}_{\mathrm{eff,2}}$ are, respectively, the effective field for ${\bf M}_{1}$ and ${\bf M}_{2}$, which is dominated by the external field and mutual magnetic dipolar field. The third term phenomenologically describes the extra gain or loss introduced by the injected spin current $J_s$ with ${\bf p}$ denoting the polarization direction of electron spin and $\sigma$ characterizing the efficiency of spin transfer torque. As shown in Fig.~\ref{fig:planar}(a), the balanced extra gain or loss by the injected spin current is controlled by applying the opposite voltage on the heavy-metal layer.

Similar to the previous analysis, the EPs emerge in the competition between the relative gain and loss of two ferromagnetic layers, and their coupling strength.
At a fixed distance, the EPs emerge at a certain voltage accompanied by a sudden change of eigenvalue in the parameter space. An increase in distance causes weaker magneto-dipolar coupling, leading to a small required voltage for the emergence of EP, as shown in Fig.~\ref{fig:planar}(b). In addition, the non-contact magneto-dipolar coupling and the injected spin current can be modulated by the distance between ferromagnetic layers and applied voltage, respectively, providing a feasible way of producing and modulating EPs experimentally. 

The EP  is also proposed to be realized in coupled magnetic waveguides mediated by a nanoscale conductive layer, where the injected current exerts balanced gain and loss for the magnetic waveguides \cite{wang2023floquet}, in which configuration Wang \textit{et al.} found that the required current for the realization of EP is lower when using a time-periodic current that with a constant current \cite{wang2023floquet}.

\begin{figure}[!htp]
	\centering
	\includegraphics[width=0.98\textwidth]{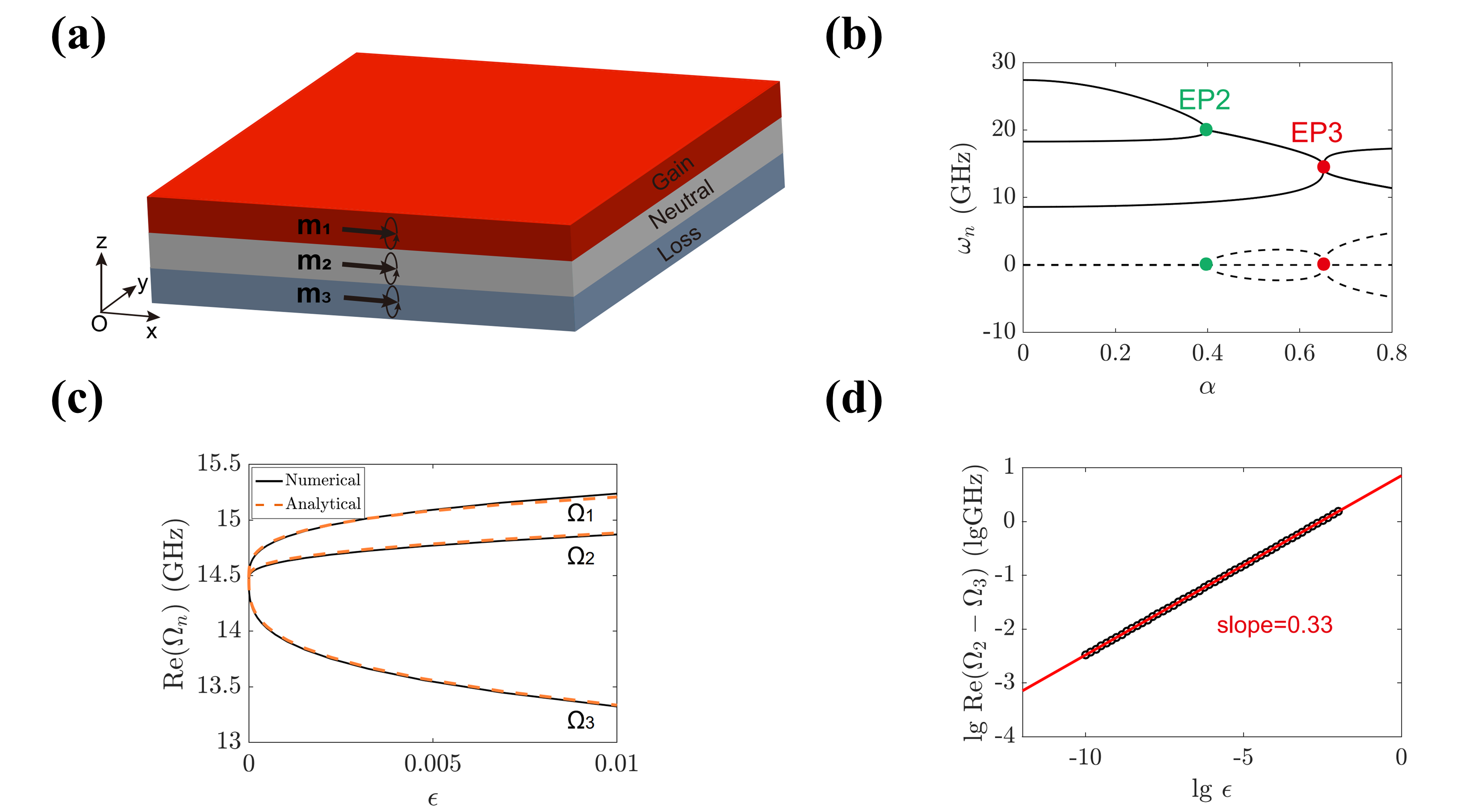}
	\caption{Realization of high-order EPs in a ferromagnetic trilayer. (a) is the configuration, where the red, gray, and blue layers imply the ferromagnetic film with a gain of rate $\alpha$, without net dissipation, and with a loss of the same rate $\alpha$. (b) The dependence of frequency spectra on the gain-loss parameter $\alpha$, where ``EP2'' and ``EP3'' represent the second-order and third-order EPs, respectively. (c) shows the dependence of the ferromagnetic resonance frequency on the perturbation $\epsilon$. (d) depicts the frequency splitting on the logarithmic scale around the third-order EPs. The figure is reproduced with permission from Ref.~\cite{yu2020higher}.}
	\label{fig:EPrro}
\end{figure}

\textbf{High-order EPs}.---As addressed above, the second-order EPs exist with two coupled ferromagnetic layers or macro spins with balanced gain and loss. Yu \textit{et al.} predicted the third-order EPs that can be realized in a three-layer structure composed of a ferromagnetic layer without net damping (``neutral'') sandwiched by two ferromagnetic layers with gain and loss of the same magnitude governed by parameter $\alpha$~\cite{yu2020higher}, as depicted in Fig.~\ref{fig:EPrro}.
The dependence of the three eigenvalues $\omega_n$ with the mode indexes $\{n=1,2,3\}$ on the gain and loss parameter $\alpha$ is shown in Fig.~\ref{fig:EPrro}(b), where the second-order EPs (``EP2'') and third-order EPs (``EP3'') exist with different $\alpha$ when choosing suitable exchange-coupling strength and external magnetic field. Similar to the merging of two modes at EP2, EP3 implies that three modes share the same eigenvalue and eigenvectors.

As addressed in Sec.~\ref{UniqueEP}, one application of the EPs is the enhanced sensitivity when a perturbation is implemented in the vicinity of EPs. Here for the ferromagnetic trilayer, a small perturbation $\epsilon$ can be induced by an external magnetic field. As shown in 
Fig.~\ref{fig:EPrro}(c), the dependence of ferromagnetic resonance frequency $\Omega_n$ on $\epsilon$ in the vicinity of ``EP3'' exhibits non-linear behavior, different from the conventional linear response. The frequency splitting can be fitted by a logarithmic function $\text{Re}\left(\omega_2-\omega_3\right) \propto \epsilon^{1/3}$, corresponding to the third-order EPs [Fig.~\ref{fig:EPrro}(d)], i.e., a superior response by small perturbations.

\subsubsection{Observation of EPs in passive $\mathcal{PT}$-symmetric magnonic devices} 
\label{Sec_magnon_PT}

\begin{figure}[!htp]
	\centering
	\includegraphics[width=0.86\textwidth]{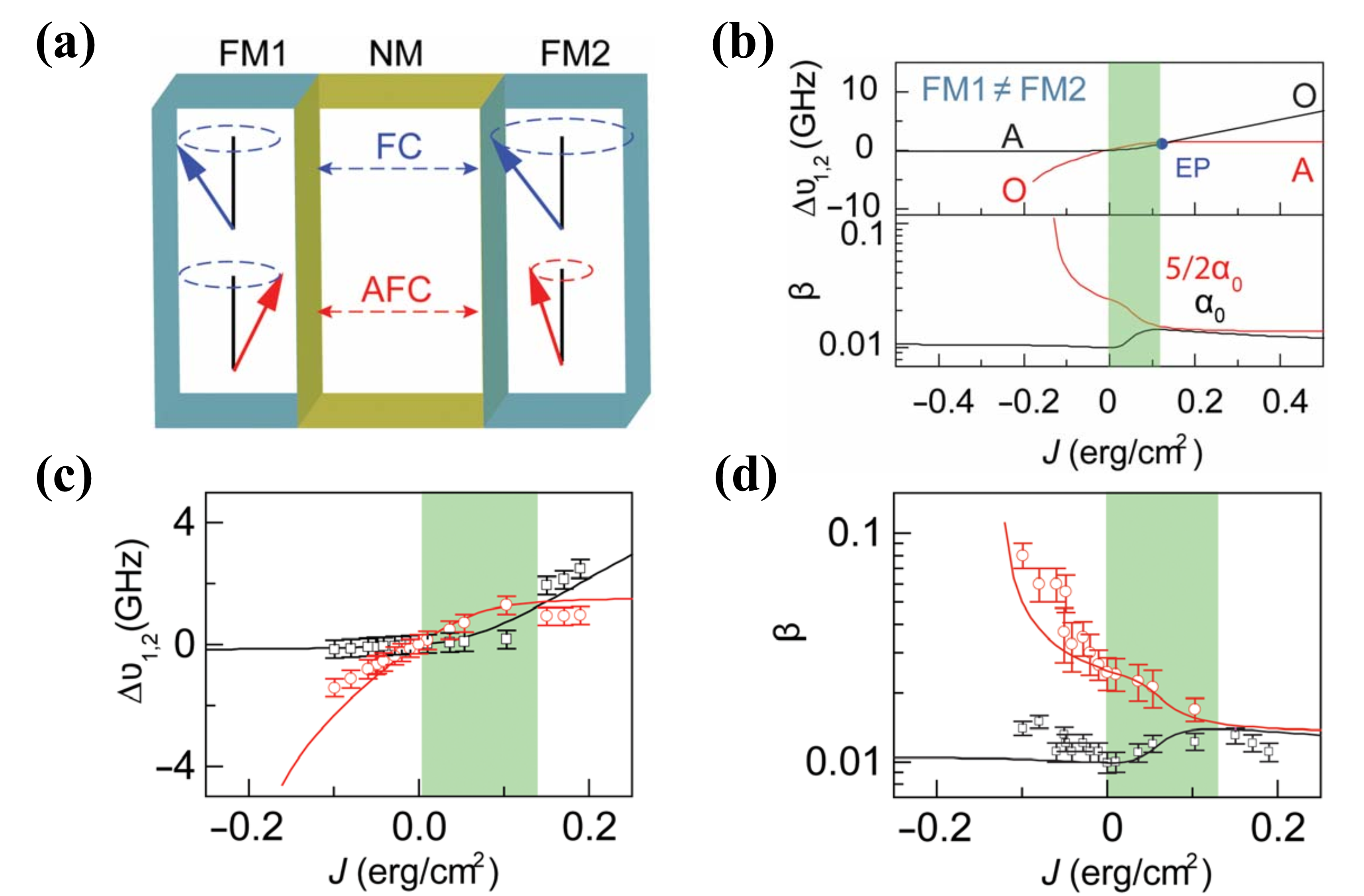}
	\caption{(a) A sandwiched magnetic trilayer consisting of two coupled FM layers and a nonmagnetic layer.  (b) Frequency differences between two eigenmodes and the corresponding frequencies of each FM layer without couplings, where $\alpha_1=\alpha_0$ and $\alpha_2=5/2\alpha_0$ are the Gilbert damping rates of the Co and NiFe layers, respectively. (c) and (d) Experimental results of the frequency changes and damping constants of two eigenmodes. The figures are reproduced with permission from Ref.~\cite{Liu2019Observation}.} 
	\label{fig:EPE}
\end{figure}

As mentioned above, the heterostructures composed of two ferromagnetic metal (FM) layers and one normal metal (NM) spacer are the choices to investigate non-Hermitian topological magnonics \cite{yu2020higher}. Despite the intensive studies among theories \cite{lee2015macroscopic,yang2018antiferromagnetism,wang2021enhanced,yu2020higher,PT_bilayer,li2022multitude,deng2023exceptional,duine2023non,wang2023floquet}, an inconvenient fact is a lag in experimental research. Recently, Liu \textit{et al.} \cite{Liu2019Observation} made efforts to achieve $\mathcal{PT}$-symmetry in a magnonic device of sandwiched structure, as shown in Fig. \ref{fig:EPE}(a). In their device, two FM layers are separated by a Pt layer of thickness $d$. The coupling between two FM layers  is governed by the Ruderman-Kittel-Kasuya-Yosida (RKKY) exchange interaction, parametrized by the constant $J$. As $d$ increases,  $J$ decreases and alternates between the ferromagnetic and antiferromagnetic couplings \cite{Liu2019Observation}, which affects the effective magnetic field and the damping torque of each FM layer (due to the spin pumping effect), such that the experiment observes the variation of both the resonant frequencies and damping rates of two FM layers with changing $d$. Based on such tunability, the EPs via the passive  $\mathcal{PT}$-symmetry is experimentally possible to be achieved.

Besides precisely controlling the coupling strength, the Gilbert damping rates of two FM layers need to differ to achieve the $\mathcal{PT}$-symmetry. Accordingly, the two FM layers are chosen to be different materials with cobalt (30 nm) and permalloy (5 nm). Their Gilbert damping rates $\alpha_0({\rm Co})=0.0103$ and $\alpha_0({\rm NiFe})=0.0250$. With two macro spins, there are two eigenmodes, i.e., the acoustic and optical modes with low and high frequencies, which repel each other and form an anti-crossing  when $J$ changes. The frequency differences between the eigenmodes and the individual frequencies of each FM layer are defined as $\Delta\nu_{1,2}=\nu_{1,2}-\omega_{1,2}$, where $\nu_{1,2}$ are the frequencies of the acoustic and optical modes, $\omega_{1,2}$ are the individual frequencies of two FM layer. The frequency differences and the damping constants $\beta$ of these two eigenmodes cross each other at $J>0$, as shown in Fig.\ref{fig:EPE}(b). Thereby, the authors addressed that EP and the $\mathcal{PT}$ symmetric transition is theoretically allowed in their device.

To experimentally test the expectation, the authors fabricate a series of devices with the thickness of the Pt layer varying from 0 to 12 nm.
An external in-plane magnetic field $B=700$~G is applied.
Figure~\ref{fig:EPE}(c)-(d) shows a comparison between the experimental data and theoretical calculations. As $J$ decreases, the eigenfrequencies of acoustic and optical modes get close. Fitted by the Kittel formula $\nu=\mu_0\gamma\sqrt{H_{\rm{eff}}(H_{\rm{eff}}+ M_{\rm{eff}})}$ at a fixed magnetic field of $700$~G, where  $H_{\rm{eff}}=H_0+H_{\rm{ex}}$, and $M_{\rm{eff}}=J/(H_{\rm{ex}}d)$ are the effective field, and effective magnetization, respectively, the two eigenmodes cross twice at $J=0$ and $0.13$ ~$\mathrm{erg/cm^2}$. The authors define the area between the two intersections as the $\mathcal{PT}$ symmetry regime, while the other areas as the $\mathcal{PT}$-symmetry broken phase. At the second intersection with $J=0.13$~$\mathrm{erg/cm^2}$, the damping rates of two eigenmodes are equal, which appears to be evidence of EPs. This experimental finding may inspire further exploration of the exotic properties around EPs in magnonic devices.

\subsection{Observation of EPs in cavity magnonic systems}

\label{Sec_EPs_cavity_magnonics}

Coupled magnon-photon system in ``cavity magnonics'' \cite{soykal2010strong,huebl2013high,tabuchi2015coherent,bai2017cavity,goryachev2014high,li2018magnon,lachance2019hybrid,bhoi2020roadmap,cao2015exchange,boventer2018complex,yu2019prediction,grigoryan2018synchronized,rao2019analogue,rao2021interferometric,li2022coherent,rameshti2022cavity,yuan2023periodic} is an alternative choice to achieve the EPs.
A cavity magnonic system involves discrete photon modes in a microwave cavity and magnons in a bulk magnet. They can strongly couple to the photons in a microwave cavity via the magnetic dipole-dipole interaction. The coupling strength is proportional to the square root of the spin number of the magnet such that it is enhanced by increasing the material volume. In addition, changing the overlap coefficient of the two modes can also tune the photon-magnon coupling strength.

Both the cavity photon and magnon can be treated as harmonic oscillators, with annihilation operators denoted by $\hat{p}$ and $\hat{m}$, respectively.
The Hamiltonian for bare cavity photon $\hat{H}_c=\hbar\widetilde{\omega}_c(\hat{p}\hat{p}^\dagger+1/2)$ and magnon  $\hat{H}_m=\hbar\widetilde{\omega}_m(\hat{m}\hat{m}^\dagger+1/2)$, 
where $\tilde{\omega}_c=\omega_c-i(\kappa+\beta)$ and $\tilde{\omega}_m=\omega_m-i(\gamma+\alpha)$ are the complex frequencies of the photon and magnon modes with their intrinsic (external) damping rates $\beta$ ($\kappa$) and $\alpha$ ($\gamma$). The magnetic dipolar interaction produces a coherent photon-magnon coupling $\hat{H}_{int}=\hbar J(\hat{p}+\hat{p}^\dagger)(\hat{m}+\hat{m}^\dagger)$ with a coupling strength $J$. Additional effects appear 
when the cavity mode and magnon interact with the same bath that supports traveling waves, which offer a common reservoir for both modes with additional dissipative rates $\kappa$ and $\gamma$. 
An indirect ``dissipative'' coupling $i\Gamma= i\sqrt{\kappa\gamma}$ is induced due to the cooperative radiation of two modes into the common reservoir \cite{yao2019microscopic, rao2020interactions, wang2019nonreciprocity}. Including both the coherent and dissipative couplings, the dynamics of a cavity magnonic system is governed by \cite{rameshti2022cavity}
\begin{align}
i\frac{d}{dt}\left(\begin{array}{c}
\hat{p}\\
\hat{m}
\end{array}\right)=
\hbar\left(\begin{array}{cc}
\omega_c-i(\kappa+\beta) &J+i\Gamma\\
J+i\Gamma &\omega_m-i(\gamma+\alpha)	\end{array}\right)\left(\begin{array}{c}
\hat{p}\\\hat{m}
\end{array}\right).
\label{HMCM}
\end{align}
The two eigenvalues 
\begin{equation}
	\tilde{\omega}_{\pm}=\omega_0-i\kappa_0\pm\frac{1}{2}\sqrt{(\Delta\omega-i\Delta\kappa)^2+4(J+i\Gamma)^2},
	\label{EGV}
\end{equation}
where $\omega_0=(\omega_c+\omega_m)/2$, $\kappa_0=(\kappa+\beta+\gamma+\alpha)/2$, $\Delta\omega=\omega_c-\omega_m$, and $\Delta\kappa=(\kappa+\beta)-(\gamma+\alpha)$. When the square root in Eq.~(\ref{EGV}) becomes zero, the two eigenvalues and eigenfunctions of the coupled photon-magnon system coalesce into one, indicating the emergence of the EP.

\subsubsection{Topological properties around EPs in cavity magnonic systems}

\begin{figure}[!htp]
           \centering
	\includegraphics[width=\textwidth]{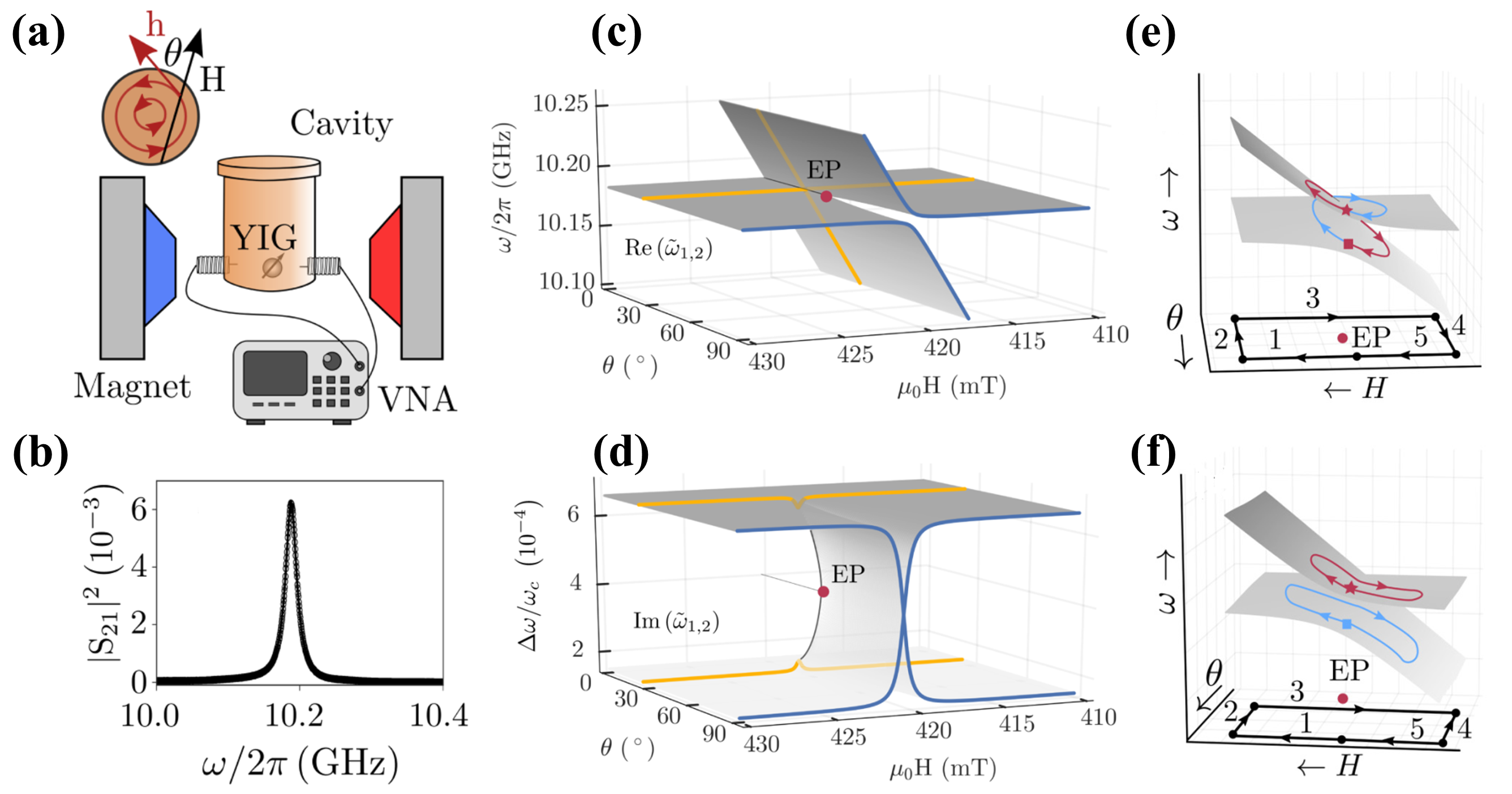}
	\caption{(a) Experiment setup with an \textit{in situ} control of the photon-magnon coupling strength by changing the direction of the static magnetic field with an angle $\theta$ with respect to the local microwave field. (b) Transmission spectrum of microwaves for the TM$_{011}$ mode. (c) and (d) Calculated frequencies and line widths of two eigenmodes at different magnetic-field strengths and directions. The blue and yellow curves show the measured mode frequencies and linewidths when $\theta>\theta_{\rm EP}$ and $\theta<\theta_{\rm EP}$, respectively, where $\theta_{\rm EP}$ is the critical angle for the realization of EPs. (e) and (f) Path taken to encircle or avoid encircling the EPs. The figure is reproduced with permission from Ref.~\cite{PhysRevB.95.214411}.}
\label{TPEP}
\end{figure}

As addressed in Sec.~\ref{exceptionaltopology}, the topological mode switching near the EP means that encircling an EP in a three-dimensional parameter space enables the dynamic switching of the eigenmodes of the system. Such a topological property was first demonstrated in cavity magnonics by Harder \textit{et al.}~\cite{PhysRevB.95.214411}. In the experiment, a yttrium iron garnet (YIG) sphere of diameter 0.3-mm was placed at the outer and bottom edge of a cylindrical microwave cavity made of oxygen-free copper (diameter=25~cm, height=33~cm), as sketched in Fig.~\ref{TPEP}(a). The TM$_{011}$ cavity mode is exploited in the experiment, with resonant frequency $\omega_c/(2\pi)=10.2$~GHz and quality factor $Q=660$ [Fig.~\ref{TPEP}(b)]. The frequency of the magnon mode, on the other hand, can be tuned by the static magnetic field $H$ following a linear dependence, i.e., $\omega_m=\gamma(H+H_A)$, where $H_A$ represents the anisotropy field.

When the damping rates of the cavity photon and magnon modes are fixed, a method for achieving the EPs is to continuously tune the coupling strength $J$. According to Eq.~\eqref{EGV}, the EP can be realized as long as this cavity magnonic system possesses $\mathcal{PT}$ symmetry, i.e., $J=\Delta\kappa/2$ and $\Delta\omega=0$. Via changing the relative direction $\theta$ between the local microwave and static magnetic fields, the coupling strength $J=J_M|\sin\theta|$ \cite{PhysRevB.95.214411}. Figure~\ref{TPEP}(c) and (d) address the calculated eigenfrequencies [real part of Eq.~(\ref{EGV})] and linewidths [imaginary part of Eq.~(\ref{EGV})] at different magnetic-field directions and strengths. The solid curves are experimental data measured at $\theta=90^\circ$ and $23^\circ$, corresponding to the strong and weak coupling cases, respectively. It is noted that according to the calculation, the EP occurs at $\theta=33^\circ$, at which the eigenfrequencies of two modes merge with each other.

To demonstrate the geometric mode switching in this cavity magnonic system, the coupling strength is  tuned to encircle the EP in the $\theta$-$H$ parameter space \cite{PhysRevB.95.214411}. Two different evolution paths are addressed in Fig.~\ref{TPEP}(e). Along the evolution path, the mode initially at low (high) frequencies can continuously evolve to high (low) frequencies after a closed path around the EP, i.e., a mode switching. Such switching cannot be achieved when the EP is not encircled as plotted in Fig.~\ref{TPEP}(f). Because of this mutation, one eigenstate of the photon-magnon system can be adiabatically transitioned to the other without crossing the coupling gap.

\subsubsection{Coherent perfect absorption at EPs in cavity magnonics}

As a singularity in a non-Hermitian system, the EP relates to a ``phase'' transition in a coupled system and leads to novel phenomena, for instance, the spontaneous symmetry breaking across the EPs (refer to Sec.~\ref{exceptional_points}). 
In a cavity magnonic system, besides the abrupt change of the system's eigenfrequencies near EPs, the dissipation rates of eigenmodes are dramatically altered as well.
Since the dissipation rates have a significant impact on the electromagnetic scattering of the coupled photon-magnon modes, the energy absorption of the cavity magnonic system exhibits a significant change near EPs. Zhang \textit{et al.} observed the polaritonic coherent perfect absorption at the EP~\cite{zhang2017observation}.

\begin{figure}[!htp]
\centering
\includegraphics[width=\textwidth]{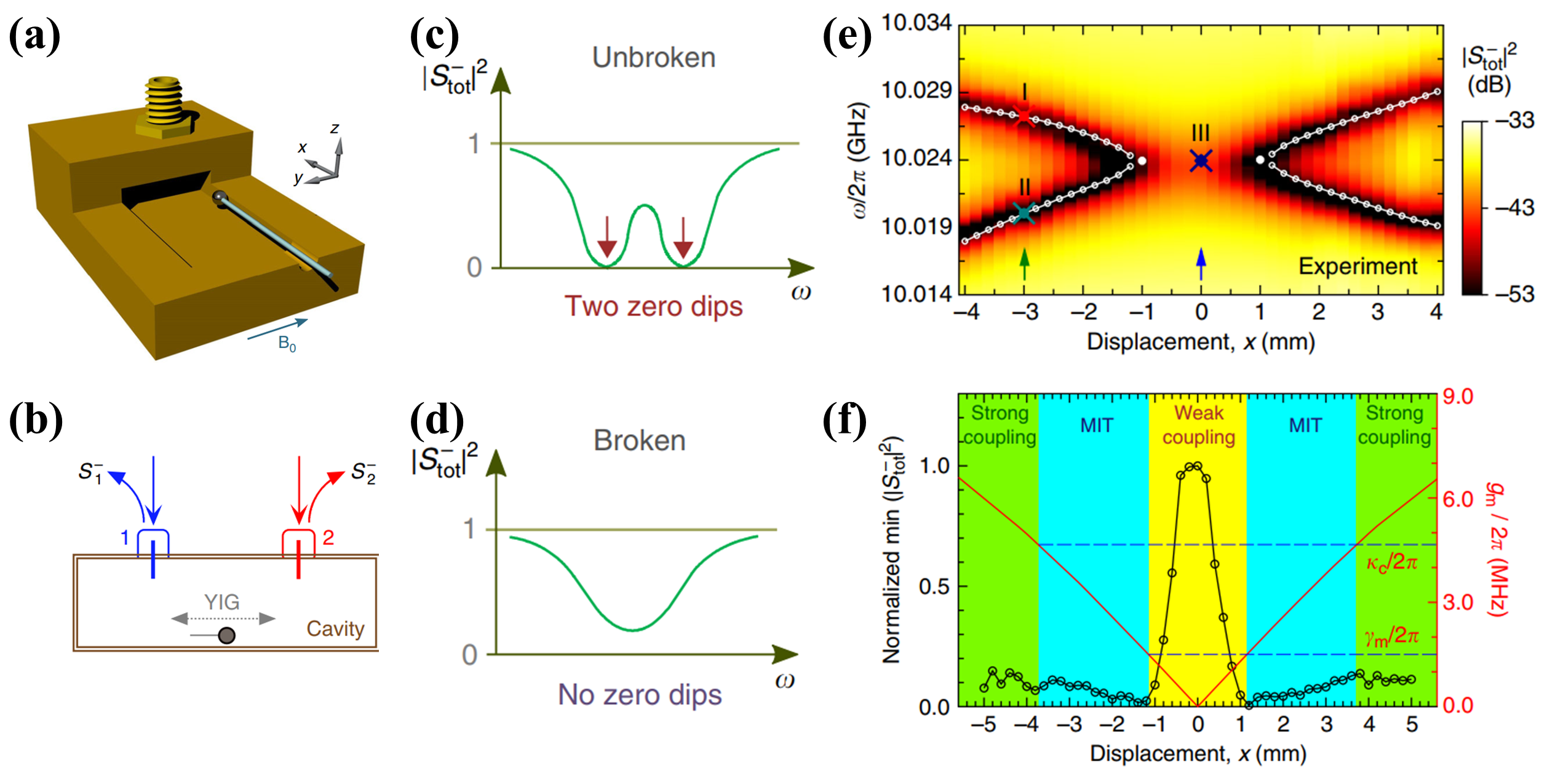}
\caption{(a) shows the cavity magnonic system, where a YIG sphere glued on a wooden rod is placed at the bottom of a rectangular cavity. 
(b) illustrates the two feedings, indicated by the blue and red arrows, and the position of the YIG sphere, which is horizontally moved to tune the photon-magnon coupling strength. (c) and (d) Total output microwave spectra when the $\mathcal{PT}$-symmetry is unbroken and broken. (e) Measured total microwave output spectra \textit{vs}. the position of the YIG sphere in the cavity.  The white circles indicate the two photon-magnon hybridized modes that merge with each other at $\pm1$ mm, corresponding to the EPs in the system. 
(f) Different magnon-photon coupling scenarios, including the weak coupling, the MIT, and the strong coupling regimes, indicated by different colors. The EPs determine the phase transition boundary between the weak coupling and the MIT regime. The black circles represent the minimal microwave output spectra measured at different positions of the YIG sphere. The red solid curve indicates the magnon-photon coupling strengths. The figure is reproduced with permission from \cite{zhang2017observation}.} 
\label{EPCPA}
\end{figure}

In the experiment~\cite{zhang2017observation}, a YIG sphere was placed near the bottom of a rectangular microwave cavity as shown in Fig.~\ref{EPCPA}(a). A wooden rod attached to the YIG sphere is used to precisely control its position in the cavity and the photon-magnon coupling strength $g_m$ as well, noting $g_m=J$ in Eq.~(\ref{EGV}).
To achieve the EP in this system, the authors tuned the damping rates of the cavity modes, while the damping rate of the magnon mode $\gamma_m$ is not affected. For the cavity modes, besides the intrinsic damping rate $\kappa_{int}$ that arises from the material and surface roughness, the cavity has two external damping rates $\kappa_{1,2}$ because of the photon leakage at two ports. These two external damping rates can be precisely controlled by adjusting the lengths of the pins.

Further, to construct an effective Hamiltonian with $\mathcal{PT}$-symmetry, the authors input two coherent microwave tones into the cavity from the two ports to drive the coupled photon-magnon system into a steady state.
The authors treat the two input tones as effective gains and phenomenologically model their system as
\begin{equation}
    \hat{H}_{\rm CPA}/\hbar=[\omega_c+i(\kappa_1+\kappa_2-\kappa_{int})]\hat{p}^\dagger\hat{p}+(\omega_m-i\gamma_m)\hat{m}^\dagger\hat{m}+g_m(\hat{p}^\dagger\hat{m}+\hat{p}\hat{m}^\dagger).
\end{equation}
When $\omega_c=\omega_m$ and $\kappa_1+\kappa_2-\kappa_{int}=\gamma_m$ are satisfied by tuning the magnetic field and the lengths of pins, respectively, a non-Hermitian Hamiltonian with the $\mathcal{PT}$-symmetry can be achieved. The two eigenfrequencies $\omega_\pm=\omega_0\pm\sqrt{g^2_m-\gamma^2_m}$ of the system become real as long as $g_m>\gamma_m$. Furthermore, by setting the power ratio as $\kappa_1/\kappa_2=1.23$ and the phase difference as $\Delta\phi=0$ in their experiment, coherent perfect absorption of electromagnetic waves is achieved: in the measured total output spectrum $|S_{tot}^-|^2=|S_1^-|^2+|S_2^-|^2$, two zero dips, at which there is nearly no microwave leakage, indicate that all feeding microwave photons from the two ports are absorbed by the system.

On the other hand, in this coupled system the EPs can be achieved by tuning the photon-magnon coupling strength $g_m$. Because of the field profile of TE$_{102}$ mode, $g_m$ almost linearly decreases from 5.5~MHz to zero when the YIG sphere is moved from $x=\pm 4$~mm to the center.
The total microwave output spectra measured at each YIG position are plotted in Fig.~\ref{EPCPA}(e). The two eigenmodes merge at $x=\pm1$~mm, indicating the emergence of EPs. 
Between the two EPs, only one absorption dip occurs, which is much smaller in magnitude than that of two dips outside the EPs. It implies that microwave absorption is suppressed when the modes coalesce, and EPs imply a transition boundary. This transition was explained as the $\mathcal{PT}$-symmetry breaking. The authors argued that, between the two EPs, the eigenfrequencies evolve from purely real numbers to complex pairs, such that additional energy dissipation is produced and breaks the balance between the effective gain and loss.

In addition, these two EPs indicate the boundary between the magnetic induce transparency (MIT) ($\kappa_{int}+\kappa_1+\kappa_2>g_m>\gamma_m$) and weak-coupling ($\kappa_{int}+\kappa_1+\kappa_2,\gamma_m>g_m$) regimes, as indicated in Fig.~\ref{EPCPA}(f). The black circles show the minimum of $|S_{tot}^-|^2$ as a function of the position of the YIG sphere. Two abrupt changes of $|S_{tot}^-|^2$ at $x=\pm1$~mm correspond to the two EPs. This work combines three intriguing phenomena, i.e., coherent perfect absorption, EPs, and $\mathcal{PT}$-symmetry breaking, in a single system, which demonstrates the potential of hybridized magnonic devices in the exploration of non-Hermitian physics.

\subsubsection{EPs and singularities in a dissipatively coupled  photon-magnon system}

\begin{figure}[!htp]
\centering
\includegraphics[width=\textwidth]{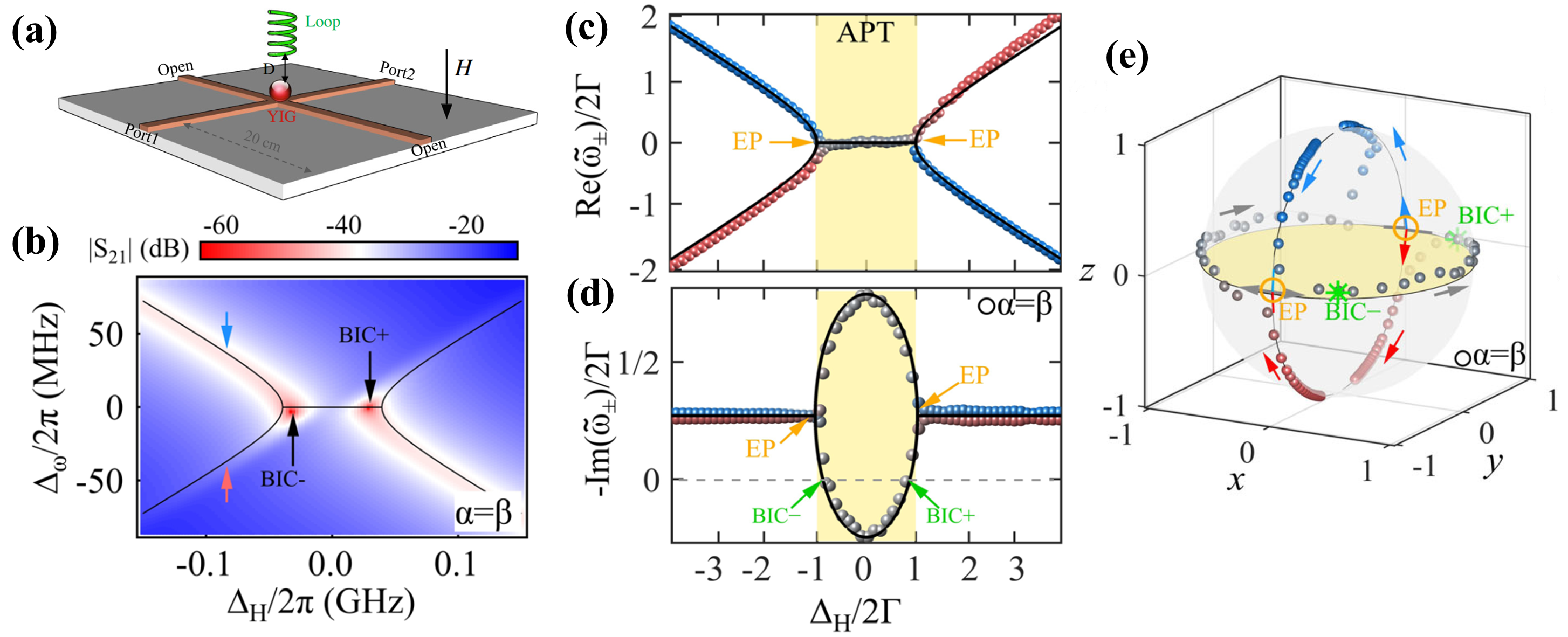}
\caption{(a) A YIG sphere biased by a perpendicular magnetic field $H$ is placed at the center of the cross cavity. A loop antenna above the YIG sphere can control the damping rate of the magnon mode. (b) Measured transmission spectra as a function of the frequency detuning $\Delta_\omega=\omega-\omega_{\rm ref}$ and the field detuning $\Delta_H$ at the condition of $\alpha=\beta$. The two red spots indicate two unconventional BICs, whose transmission amplitude approaches zero. The black curves represent the calculated eigenfrequencies. (c) and (d) The real and imaginary parts of the eigenvalues at the condition $\alpha=\beta$. The blue and red parts are dominated, respectively, by the cavity and magnon modes. The green arrows in (d) indicate the zero imaginary parts of an eigenmode, implying the unconventional BICs. 
When $|\Delta_H/(2\Gamma)|<1$, the system lies in the anti-$\mathcal{PT}$ symmetry phase, as shaded by the yellow color. (e) The Bloch sphere is constructed from the two eigenvectors of the system when $\alpha=\beta$. The orange circles and the green stars represent the EPs and two unconventional BICs. The equator of the Bloch sphere with $\theta=\pi/2$ indicates the anti-$\mathcal{PT}$ symmetry phase. The figure is reproduced with permission from Ref.~\cite{yang2020Unconventional}.}
\label{DEP}
\end{figure}

The ``level attraction'' that exhibits the coalescence of real components and repulsion of imaginary components of eigenfrequencies of a coupled magnon-photon system was theoretically proposed by  Grigoryan \textit{et al.}~\cite{grigoryan2018synchronized}.
The experimental observation was first reported in a microwave optomechanical circuit \cite{PhysRevA.98.023841}. Shortly after it, a similar phenomenon was found in a cavity magnonic system and attributed to the cavity Lenz effect \cite{harder2018level} [see Fig.~\ref{DEP}(a) and (b) for similar spectra]. These motivate efforts to explore the physical origin and the potential functionalities of level attraction in cavity magnonic devices \cite{yang2019control, yu2019prediction, rao2019level, bhoi2019abnormal, boventer2019steering, yao2019microscopic, proskurin2019microscopic, rao2020interactions}. 
One mechanism that can produce level attraction is the multi-channel interference in a cavity magnonic system, which was experimentally demonstrated in Ref.~\cite{rao2021interferometric}. Another mechanism is the dissipative photon-magnon coupling mediated by a high-damping mode or photon reservoirs, such as traveling waves in cavity magnonics systems~\cite{ yu2019prediction, yao2019microscopic, rao2020interactions, wang2019nonreciprocity}. The coupling constant is real in the coherent photon-magnon coupling case but is purely imaginary number $i\Gamma= i\sqrt{\kappa\gamma}$ for the dissipative photon-magnon coupling, noting $\kappa$ and $\gamma$ are defined in Eq.~\eqref{HMCM}. This imaginary coupling strength leads, respectively, to the coalescence of the real parts and the repulsion of the imaginary parts of eigenfrequencies of the system [see Fig.~\ref{DEP}(c) and (d)]. 

Level attraction is a typical non-Hermitian phenomenon, which makes searching for the EPs in a purely dissipative cavity magnonic system possible. However, to reach such a goal, a paradox must be properly solved. In order to achieve an observable dissipative photon-magnon coupling, $\Gamma=\sqrt{\kappa\gamma}$ should be as large as possible, i.e., it requires either $\kappa$ or $\gamma$ to be sufficiently large, noting $\kappa$ and $\gamma$ are the photon and magnon radiation rates via the traveling waves. Enhancing $\gamma$ in experiments appears to be a challenge, but adjusting the design of microwave cavities can produce a large radiation damping rate $\kappa$. The band-stop cavity such as the planar cross-cavity shown in Fig.~\ref{DEP}(a) can support a large $\kappa$ that is several orders in magnitude larger than $\gamma$. Placing a YIG sphere in the center of the cross cavity enables an observable dissipative photon-magnon coupling. However, this huge $\kappa$ also renders a nonzero square root in Eq. (\ref{EGV}), indicating the difficulty to achieve the EPs in a dissipative cavity magnonic system. To obtain a similar phenomenology to the EPs, a trade-off is needed. Yang \textit{et al.} reduces the full Hamiltonian (\ref{HMCM}) to an effective one that specifically describes the anti-resonances in the coupled system \cite{yang2020Unconventional}, i.e., disregarding the photon background radiation $\kappa$ in Eq.~(\ref{HMCM}). 
The reduced Hamiltonian in a rotating frame with respect to the reference frequency $\omega_{\rm ref}=(\omega_m+\omega_c)/2$ reads 
\begin{equation}
\hat{H}_{\rm rot}/\hbar=\left(\begin{array}{cc}
		-\Delta_H/2-i\beta &i\Gamma\\
		i\Gamma &\Delta_H/2-i(\alpha+\gamma)
	\end{array}\right),
	\label{RH}
	\end{equation}
where $\Delta_H=\omega_m-\omega_c$ is the field detuning tuned by an external magnetic field. From this effective Hamiltonian \eqref{RH}, two pseudo-EPs are expected to occur at $\Delta\omega=\pm\Gamma$, when $\beta=\alpha+\gamma$.

Figure~\ref{DEP}(a) shows the experimental setup for studying the EPs in a dissipative cavity magnonic device. A loop antenna is used to enhance the damping rate $\gamma$ of the YIG sphere to meet the criteria $\beta=\alpha+\gamma$ for EPs. The transmission spectra of the system measured at different field detuning are plotted in Fig. \ref{DEP}(b), where 
the frequencies and line widths of two anti-resonances are extracted by curve fitting. The real and imaginary parts of the eigenvalues calculated from the reduced Hamiltonian (\ref{RH}) can well reproduce the measured results, as shown by the black solid lines in Fig.~\ref{DEP}(c) and (d). At $\Delta\omega=\pm\Gamma$, the frequencies of two anti-resonances coalesce, indicating the occurrence of EPs. By contrast, the linewidths of two anti-resonances bifurcate to the upper and lower branches. When $\Delta\omega=\pm\Gamma\sqrt{1-(\beta/\Gamma)^2}$, the lower branch can reach zero, as indicated by green arrows in Fig. \ref{DEP} (d), such that two  singularities occur, which is referred to as unconventional bound state in the continuum (BIC). 
Back to the transmission spectra Fig.~\ref{DEP}(b), there indeed exist two ultra-sharp dips symmetrically occurring with respect to $\Delta_H=0$, as shown by the red spots. These two unconventional BICs are very sensitive to the damping rates of the dissipative cavity magnonic system.

The two EPs define the boundaries between the phases preserving and breaking the anti-$\mathcal{PT}$ symmetry. 
Mathematically, a Hamiltonian with an anti-$\mathcal{PT}$ symmetry is obtained by multiplying the $\mathcal{PT}$ symmetric Hamiltonian by $i$, such that $\{\hat{H}^{({\rm APT})}, \mathcal{PT}\}=0$. The authors calculated the eigenvector evolution in this system with the anti-$\mathcal{PT}$ symmetry ~\cite{yang2020Unconventional}. From Eq.~(\ref{RH}), the two eigenvectors 
\begin{equation}
    |\lambda_\pm\rangle=\left(\begin{array}{cc}
		v_{1\pm}\\
		v_{2\pm}
	\end{array}\right)=\left(\begin{array}{cc}
		i\Gamma\\
	\frac{\Delta_H}{2}-i\frac{\alpha-\beta}{2}\pm\sqrt{\left(\frac{\Delta_H}{2}-i\frac{\alpha-\beta}{2}\right)^2-\Gamma^2}
	\end{array}\right).
\end{equation}
By using the notation $v_{1,2}=|v_{1,2}|e^{i\phi_{1,2}}$, a Bloch sphere when $\alpha=\beta$ can be constructed from the two eigenvectors as follows. The relative intensity and phase difference of the two eigenvectors is defined, respectively, as $\theta=2\arctan(|v_2|/|v_1|)$ and $\phi=\phi_1-\phi_2$, which represent the polar and azimuthal angles in the Bloch sphere. Mapping $(\theta, \phi)$ of the calculated eigenvectors at different detuning, a Bloch sphere is constructed [see Fig. \ref{DEP}(e)], where the north and south poles represent the uncoupled cavity mode $|\hat{p}\rangle$ and magnon mode $|\hat{m}\rangle$, respectively. The arrows indicate the direction of increasing $\Delta_H$. Two eigenvectors coalesce at $\Delta_H=\pm2\Gamma$, i.e., two EPs on the equator, and one of them turns into BICs at $\Delta_H=\pm\Gamma\sqrt{1-(\beta/\Gamma)^2}$. From the Bloch sphere, it indicates that in the anti-$\mathcal{PT}$ symmetry-preserved phase, both eigenvectors locate at the ``equator modes'' with equal contributions from cavity and magnon modes. As addressed by the authors, these equator modes are nontrivial and may find applications in quantum information processing.

\subsubsection{Giant enhancement of magnonic frequency combs by EPs}

So far, researchers have exploited two different paths to realize the EPs and investigate the functionality associated with the EPs in magnonics. 
The first one is to fabricate the magnetic heterostructures \cite{PT_bilayer}. A significant challenge in this path is to achieve high-quality ferromagnet/normal metal interfaces and proper film thickness. The other path is based on the cavity magnonics by engineering either the photon-magnon coupling strength or the individual dissipation. Compared to the former path, the technical obstacle in the second path is smaller, since the design and fabrication of a centimeter-scale microwave cavity is easy to implement. So we have witnessed a rapid development of the experimental study on non-Hermitian physics in cavity magnonics in recent years~\cite{zhang2017observation,PhysRevLett.123.237202,wang2020steering,yang2020Unconventional}. Even so, there still exist difficulties in the precise control of the photon-magnon coupling inside the cavity. An alternative technical path that can get rid of the restrictions is in high demand.

\begin{figure}[!htp]
\centering
\includegraphics[width=\textwidth]{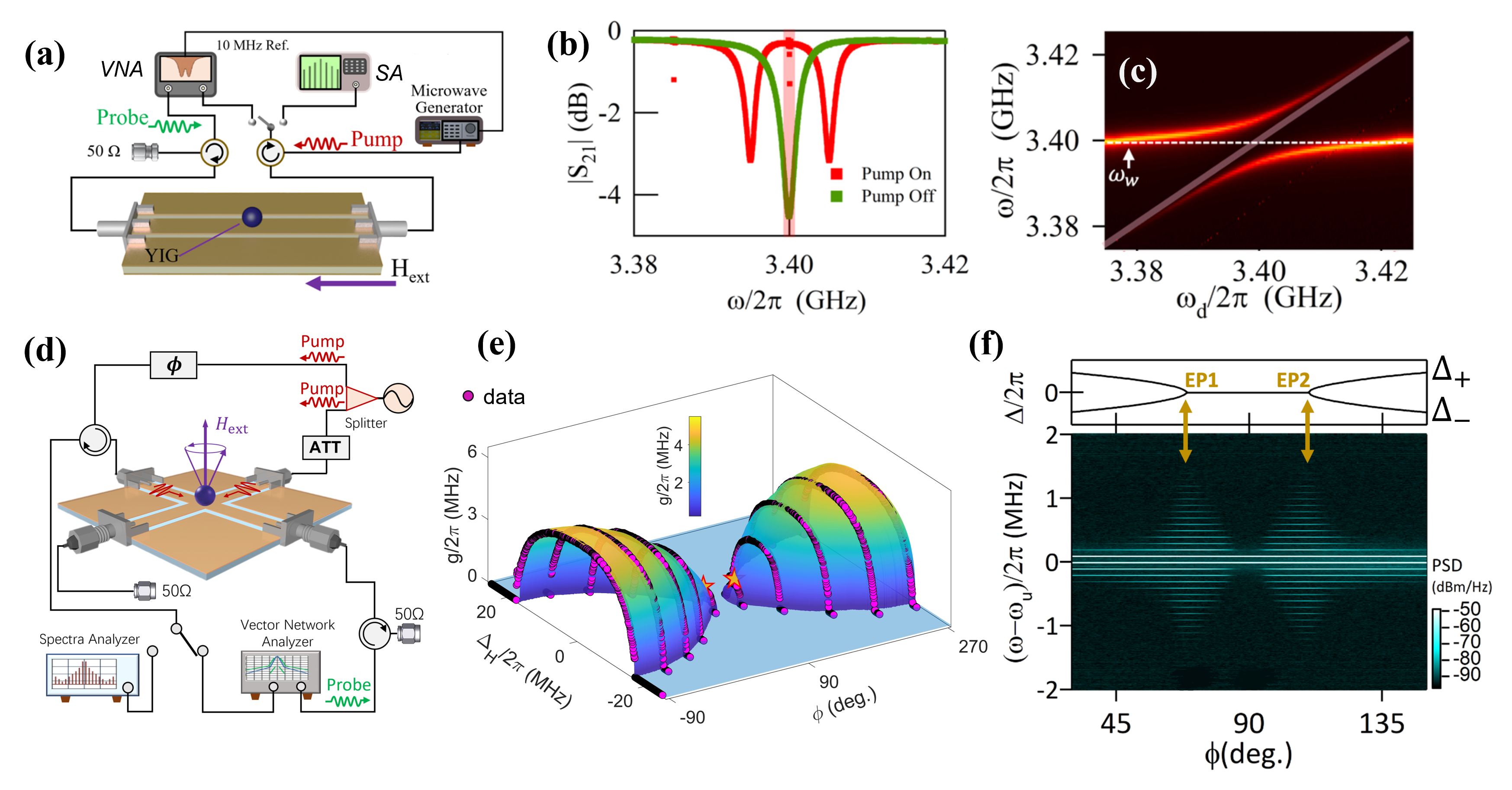}
\caption{(a) Experimental setup with loading a YIG sphere to the coplanar waveguide. Continuous pumps and probes of microwaves excite and detect the magnon modes in the YIG sphere.
(b) Measured transmission spectra $S_{21}$ of the (2,2,0) Walker mode with or without the pump. (c) Measured transmission spectra of the coupled PIM and Walker modes at different pump frequencies. The frequency of the Walker mode is fixed at $3.4$~GHz. The pink strips in (b) and (c) indicate the pump. (d) 
Experimental setup for the realization of EPs in the driven magnetic sphere when loaded in the cross-shaped coplanar waveguide. (e) compares the measured and calculated coupling strength $g/2\pi$ between the PIM and Kittel mode in the YIG sphere. The circles are experimental data, while the colored surface is a theoretical calculation. Two stars mark the EPs. (f) Giant enhancement of the magnonic frequency combs near the two EPs. Top: Eigenfrquencies $\Delta_\pm$ of the coupled system under different polarizations of the pump field. Bottom: Radiation spectra measured at different polarizations of the pump field. The tone number of the combs dramatically increases when the system is tuned to approach the EPs. The figure is reproduced with permission from Refs.~\cite{rao2023unveiling,wang2023MFC}.}
\label{MFC_EP}
\end{figure}

A recent experimental work~\cite{rao2023unveiling} reported that the magnon modes in YIG spherical samples, namely the Walker modes~\cite{walker1957magnetostatic, gloppe2019resonant, dillon1957ferrimagnetic}, can exhibit mode-splitting behavior when the YIG sphere loaded in a waveguide is continuously driven by a strong microwave pump. Although such a mode splitting is common when a magnetic sphere is loaded in a cavity that supports discrete photon modes, this phenomenon in a waveguide without discrete modes comes from a nonlinear process since the mode splitting appears only when there exists a strong microwave drive. The experimental setup is depicted in Fig.~\ref{MFC_EP}(a), in which a YIG sphere is mounted on a coplanar waveguide. A strong microwave with a single frequency works as the ``pump'' to drive the YIG sphere, while a weak probe with frequency resolution across a wide frequency window (``probe'') transmits through the YIG sphere and then is collected by a vector network analyzer (VNA) to measure the transmission spectrum of magnon modes. Figure~\ref{MFC_EP}(b) compares the two transmission spectra without and with the pump: without the pump, the signal is simply a resonant dip representing the (2,2,0) Walker mode~\cite{walker1957magnetostatic, gloppe2019resonant, dillon1957ferrimagnetic} of the YIG sphere; with the pump, the resonant dip splits into two dips, exhibiting as a pump-induced splitting behavior. This splitting behavior shows a nontrivial dependence on either the magnetic field or the pump frequency. As shown in Fig.~\ref{MFC_EP}(c), the (2,2,0) Walker mode is fixed at 3.4 GHz, while the pump frequency is continuously swept through the (2,2,0) Walker mode. An anti-crossing behavior appears in the measured transmission mapping, which may act as important experimental evidence for the coherent coupling effect between \textit{two} resonant modes, but there is no known resonant mode besides the (2,2,0) Walker mode in this narrow observation band. Further, the unknown mode strictly follows the pump frequency and is independent of the external magnetic field, such that we can easily control its coupling with the desired Walker mode by tuning the pump frequency $\omega_d$, as addressed in Fig.~\ref{MFC_EP}(c). In addition, the experiment finds the coupling strength $g$ between two modes can be precisely controlled by the pump power $P$, following a relation of $g\propto P^{1/4}$~\cite{rao2023unveiling, wang2023MFC}.

In fact, the mode splitting driven by an external field is a common phenomenon and can be produced by several different physical mechanisms. For example, the Autler-Townes splitting \cite{autler1955stark} appears when a two-level system is driven strongly by the AC field, which induces fast oscillations between the ground and excited states. The associated splitting gap of the excited state is proportional to $P^{1/2}$ and, when driven resonantly, the splitting is centered around the frequency of the excited state~\cite{autler1955stark}. This mechanism explains the experiment by Xu \textit{et al.}, in which additional mode splittings when driven by an alternating magnetic field of MHz are observed in the cavity-magnonic system with the splitting gap proportional to the drive field strength~\cite{xu2020floquet}. Recently, Li \textit{et al.} reported the mode splitting of the ferromagnetic resonance when a nanomagnet is driven into a \textit{deeply} nonlinear regime, which is attributed to the spin nutation effect~\cite{li2019nutation}, i.e., an additional precession of the magnetization around the saturated magnetization that precesses around the static field.  These possible origins, as addressed by the authors, cannot explain their observed anti-crossing behavior, as evidenced by the dependence of the splitting gap on the pump power (i.e., $g\propto P^{1/4}$) and the applied driven power that is not sufficient to drive the system into a deep nonlinear regime. Therefore, they phenomenologically proposed that a resonant mode is excited by the pump signal, but the microscopic understanding of this assumption remains wanting and calls for future studies.

Unlike the Walker modes in the YIG sphere, this resonant mode strictly follows the pump frequency [Fig.~\ref{MFC_EP}(c)] and can only exist when a weak magnetic field is applied to the sample. When the external magnetic field becomes strong, this mode disappears, regardless of the magnitude of the pump power. These features indicate that this mode is not a Walker mode but rather relates to the nonlinear magnetization state of the YIG sphere. Based on these facts, the authors attribute this mode to the cooperative precession of unsaturated spins in the YIG sphere, appearing when the sphere is driven by the pump signal.

Unlike the normal magnon modes, e.g., the Walker modes, the PIM exhibits excellent tunability by the pump.  This property facilitates the realization of the EPs between the PIM and desired Walker modes via matching the coupling constants and the dissipation difference.
Compared with conventional methods, the new method based on the coupling of the PIM and Walker modes merely needs to control the pump signal, such as its frequency and power, and hence greatly simplifies the experimental operations.

In addition, when the system is simultaneously driven by the pump and probe, the coupling between the Walker mode and PIM can exhibit nonlinear behavior. 
This nonlinear coupling enables the frequency conversion and hence produces the magnonic frequency combs~\cite{rao2023unveiling, wang2021magnonic, hula2022spin, wang2023MFC, xiong2023magnonic}. As an intriguing singularity in the coupled systems, the EPs have been found to exhibit a significant enhancement of sensitivity to the disturbance. Using the EPs to amplify the sensitivity in a coupled system with the Kittel modes and PIM may greatly enhance the magnonic frequency comb generation. Recently, this idea has been experimentally tested in Ref.~\cite{wang2023MFC}. The authors constructed a coupled system comprising a PIM and the (1,1,0) Walker mode or Kittel mode, and then precisely tune the coupling strength between the PIM and Kittel mode by tuning the driving power when varying its polarization. The authors fabricate a cross-shaped coplanar waveguide with two orthogonal arms to regulate the polarization of the pump field [Fig.~\ref{MFC_EP}(d)]. By adjusting the relative phase $\phi$ of the polarizations between two coherent pump microwaves that are, respectively, input from two orthogonal arms of the cross-shaped coplanar waveguide, the polarization of the pump field can be tuned from the right-handedness to the left-handedness. As a consequence, by the angular momentum conservation between magnons, the coupling strength between the PIM and Kittel mode can be precisely controlled.

Figure~\ref{MFC_EP}(e) shows the evolution of the coupling constant $g/(2\pi)$ in the parameter space constructed by $\phi$ and the frequency difference $\Delta_H$ between the Kittel mode and PIM. The colored surface is given by the numerical calculation, which nicely supports the experimental data with the purple circles. When $g$ is adjusted to meet the criterion $g=\delta_-/2$, where $\delta_-$ is the damping difference between the two modes, the EPs occur at $\Delta_H=0$  [refer to Eq.~\eqref{NHEPeigenvalue}], which is marked by the two stars in Fig.~\ref{MFC_EP}(e). Further, when the coupled system is simultaneously driven by the pump and probe, the coupling strength $g$ between the PIM and Kittel mode oscillates periodically in time, instead of a constant under a single-tone drive. Such a nonlinear process reaches its maximum near the EPs, leading to the giant enhancement of the magnonic frequency combs, as shown in Fig.~\ref{MFC_EP}(f). In the measurement, the Kittel mode is fixed at the pump frequency, and the detuning between the pump and probe microwaves is set to be 0.1~MHz. When the polarization of the pump field is tuned by $\phi$, the authors observe the dramatic enhancement of the tone number of the combs from several to more than 32 near the two EPs [Fig.~\ref{MFC_EP}(f)]. This work establishes a strong link between the non-Hermitian topology and the frequency combs and provides a new avenue for frequency multiplication by taking advantage of the EPs. Although it is demonstrated in the microwave regime, it has the potential to be applied in other frequency ranges.

\subsection{Exceptional surfaces in cavity magnonics}
\label{Exceptionalsurfacescavity}

\begin{figure}[!htp]
	\centering
	\includegraphics[width=\textwidth]{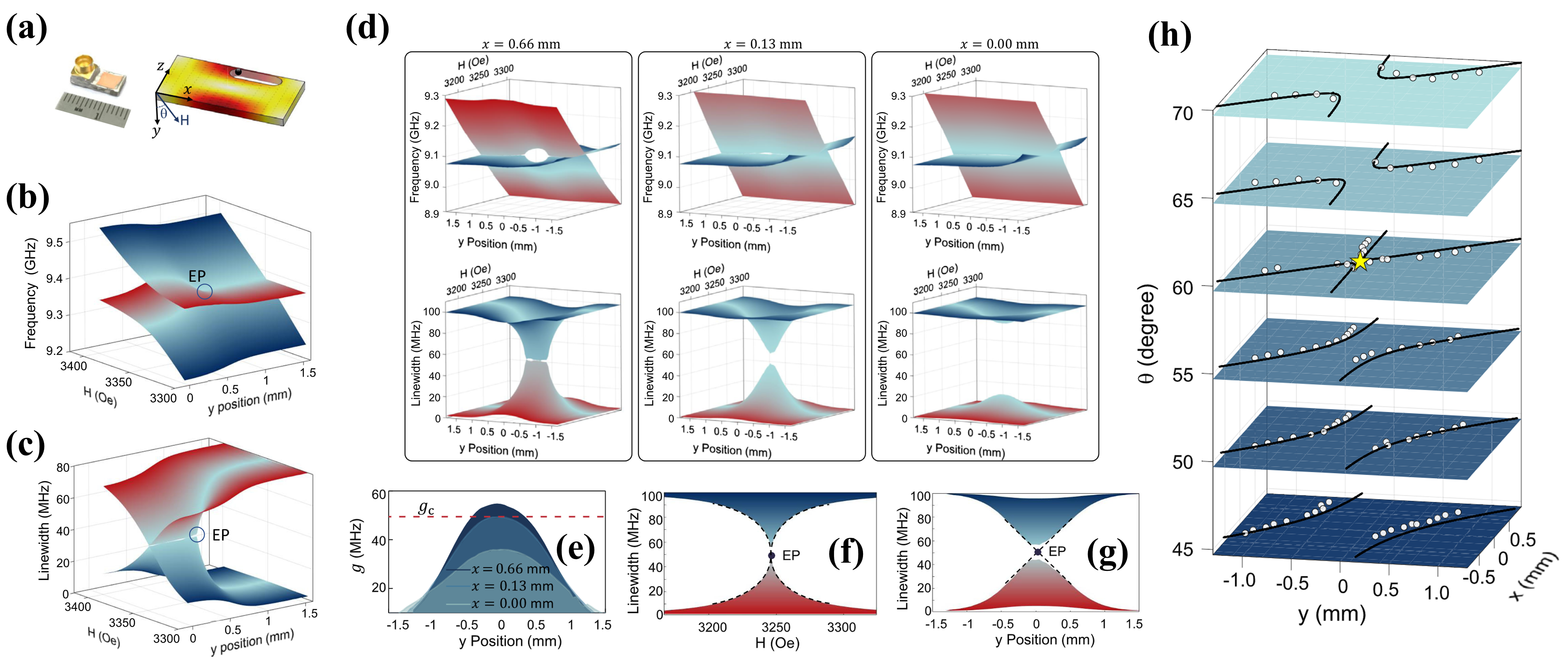}
	\caption{(a) Schematic of the microwave cavity with simulated magnetic fields. (b) and (c) show, respectively, the measured Riemann surfaces for the real (resonant frequency) and imaginary (resonant linewidth) parts of the eigenfrequency. (d)  show the dependence of the measured Riemann surfaces for the real (top) and imaginary (bottom) parts of the eigenfrequencies on the position $x=0.66$, $0.13$, and $0.00$~mm of the YIG sphere. The EP coalescence is observed at $x=0.13$~mm when $g=g_c$. (e)  The coupling strength $g$ as a function of $y$ at different $x$ locations. (f) and (g) Cross-sectional views of the case with $x=0.13$~mm. (h) Slices of exceptional surface in the 3D ($x$,$y$,$\theta$) parameter space. The figure is reproduced with permission from Ref.~\cite{PhysRevLett.123.237202}.} 
	\label{fig:ZXFEPSF}
\end{figure}

The EPs are isolated points in the parameter space, as discussed above. Bringing these ideas to higher dimensions may help to provide chances to expand EPs to exceptional surfaces, i.e., a collection of EPs on surfaces~\cite{PhysRevLett.123.237202,grigoryan2022pseudo}. It requires additional degrees of freedom and flexible tunability, thus posing a significant challenge for the experimental demonstration. A cavity magnonic system can be easily manipulated by either adjusting the geometric boundaries of microwave cavities or tuning the magnetic field on the magnet. This prominent tunability provides an ideal solution for the experimental realization of exceptional surfaces. The experimental setup~\cite{PhysRevLett.123.237202} consists of a microwave cavity and a YIG sphere of 400 $\mu$m that is placed inside the cavity, as shown in Fig.~\ref{fig:ZXFEPSF}(a). The cavity is fabricated from a printed circuit board with a high dielectric constant. The cavity mode volume is much smaller than that of a conventional three-dimensional resonant cavity, making it very suitable for the realization of strong photon-magnon coupling. A small rod attached to the YIG sphere is used to control the position of the YIG sphere in the cavity such that the coherent photon-magnon coupling strength $g$ is precisely controlled. When $g$ matches a critical coupling strength $g_c$, i.e., $g=g_c=\Delta\kappa/2$, the EP in this coupled system is realized according to Eq.~(\ref{EGV}).

To realize an exceptional surface, Zhang \textit{et al.} introduces multiple degrees of freedom to manipulate the EPs in a four-dimensional synthetic space, including the magnitude of the bias field ${\bf H}$, the direction $\theta$ of ${\bf H}$ with respect to the $\hat{\bf y}$-direction, the position of YIG sphere in both the $x$ and $y$ directions [see Fig.~\ref{fig:ZXFEPSF}(d)]. When $x=0.66$~mm, two EPs are widely separated. As $x$ decreases, they approach and coalesce at $x=0.13$~mm. When $x$ keeps decreasing, no EP can be observed (e.g. at $x=0$) and the real parts of two eigenfrequencies always cross each other at diabolic points, while their imaginary parts are separated. Comparing $g$ at different $x$ positions can help to find the relation between coupling strength $g$ and $g_c$, as displayed in Fig.~\ref{fig:ZXFEPSF}(e). It is worth noting that the curve of coupling strength crosses with the line of $g_c$ twice at $x=0.66$~mm, crosses only once at $x=0.13$~mm, and has no intersections at $x=0.00$~mm. These results show a good correspondence with Fig.~\ref{fig:ZXFEPSF}(d). When an EP pair coalesces, no signature of mode coupling can be observed in the Riemann surface for the real part of the eigenfrequency. However, in the imaginary part, the singularity condition associated with this EP can be easily observed. At this point, this coalesced EP behaves as a linear and square-root dependence respectively [Fig.~\ref{fig:ZXFEPSF}(f) and (g)].

Although it appears that the EP coalescence only applies via tuning the $y$ position of the YIG sphere, by changing the magnetic-field direction $\theta$, the coalescence of EP pairs can be realized by tuning the $x$ positions as well. Figure~\ref{fig:ZXFEPSF}(h) plots the distribution of EPs in a 4D synthetic space defined by $x$, $y$, $\theta$ and $H$. Here the hidden dimension $H$ is fixed at 3237~Oe for ensuring the zero-detuning condition. White dots in Figure~\ref{fig:ZXFEPSF}(h) are the EPs calculated from experimental data, while the solid lines are theoretical calculations from extrapolated data. They are consistent with each other. These white dots depict the outline of the exceptional surface in the 4D synthetic space. By varying $\theta$, the overall amplitude of the saddle surface of $g$ is tuned. At a critical angle $\theta_c$, the saddle surface intersects the plane at the saddle point, suggesting the coalescence of EPs in both $x$ and $y$ dimensions. This work points out a novel direction for creating high-dimensional EPs. Four-dimensional exceptional volumes or even higher-order EP assembles can be achieved by introducing higher-order synthetic dimensions, laying the groundwork for magnonic non-Hermitian physics, and paving a new avenue for magnon-based signal processing.

\begin{figure}[!htp]
	\centering
	\includegraphics[width=\textwidth]{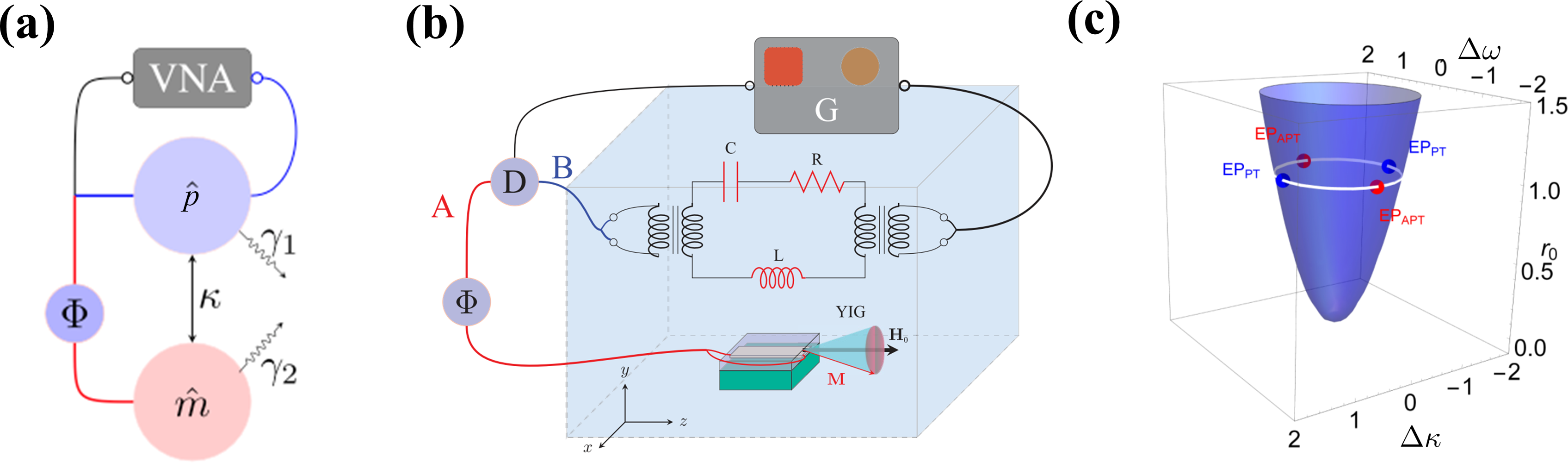}
	\caption{Proposal of exceptional surface in cavity magnonics via microwave modulation. In (a) the magnon-photon coupling in the cavity is modulated by additional phase $\Phi$. (b) is the proposal of the microwave modulation that realizes (a). (c) plots the exceptional surface, on which the blue and red dots correspond to the EPs with $\mathcal{PT}$ and anti-$\mathcal{PT}$symmetry breakings, respectively. Figures (a) and (c) are adapted with permission from Ref.~\cite{grigoryan2022pseudo}. Figure (b) is adapted with permission from Ref.~\cite{grigoryan2018synchronized}.} 
	\label{fig:EPSurface}
\end{figure}

The experiments above exploit the frequency and dissipation asymmetries $\Delta\omega$ and $\Delta\kappa$ between the magnon and microwave modes in Eq.~\eqref{HMCM} to realize the exceptional surface. An alternative choice is to use the competition between the dissipative and coherent couplings, as addressed in Sec.~\ref{exceptionalpointanalysis}. Grigoryan \textit{et al.} \cite{grigoryan2018synchronized,grigoryan2022pseudo} propose the very first mechanism to realize the dissipative coupling in cavity magnonic system: Beyond the conventional magnon-cavity coupling, they introduce a phase shift ``$\phi$'' in the feedback microwaves driven by the magnetization dynamics in the cavity, as shown in Fig.~\ref{fig:EPSurface}(a). 
In Fig.~\ref{fig:EPSurface}(b), the magnetization precession induces a driving voltage in the LCR circuit that models the electromagnetic field in the cavity, while the induced current is modulated by a phase shift ``$\phi$''   and amplitude $\delta_0$ that affects the magnetization precession~\cite{bai2015spin,wolz2020introducing,grigoryan2018synchronized,grigoryan2022pseudo}. Recent experiments exploit such a design and realize a regime transition for the cavity photon-magnon states from conventional ``level repulsion'' to ``level attraction'' by tuning the phase shift $\phi=\pi$~\cite{boventer2019steering,boventer2020control}. Such a system is governed by the Hamiltonian
\begin{equation}\label{eq:chiralcavity}
   \hat{H} = \left(\begin{array}{cc}
       \omega_p-i\kappa_p & g_a \\
     g_a\left(1+ \delta_0 e^{i\phi}\right)   &   \omega_m-i\kappa_m
   \end{array}    \right),
\end{equation}
where $\omega_p$ and $\omega_m$ are the frequencies of photon and magnon modes with damping rates $\kappa_p$ and $\kappa_m$, respectively, $g_a$ is the coupling constant, and the dissipative component is contained in $\delta_0 e^{i\phi}$ \cite{wirthmann2010direct,cao2013spintronic}. Equation.~\eqref{eq:chiralcavity} is equivalent to Eq.~\eqref{eq:2times2} when replacing $g_a\left(1+ \delta_0 e^{i\phi}\right) = g_b$ (refer to Sec.~\ref{exceptional_points}). 
The exceptional saddle surface  is achievable by the frequency asymmetries $\Delta\omega$, dissipation asymmetries $\Delta\kappa$, and coupling between two modes ~\cite{grigoryan2022pseudo}.  Figure \ref{fig:EPSurface}(c) plots the condition $\left(\Delta\omega\right)^2+\left(\Delta\kappa\right)^2=r_0$ as a new exceptional surface and reveals features associated with the $\mathcal{PT}$- and anti-$\mathcal{PT}$-symmetric breakings.

\section{Exceptional nodal phases of magnons}
\label{exceptional_nodal_phases_magnons}	
  
As addressed in above Sec.~\ref{Magnonic_EPs}, the EPs, exceptional lines and exceptional surfaces have been realized in various magnonic systems \cite{yang2018antiferromagnetism,xiao2019enhanced,PhysRevLett.123.237202,Liu2019Observation,yang2020Unconventional,wang2021coherent,nair2021enhanced}, exhibiting their potential advantages and functionalities in exploring the non-Hermitian physics due to the flexible way to engineer the dissipation in these devices. 
Nevertheless, we postpone the review of the efforts for the realization of EPs in the reciprocal wave-vector space, which were already proposed in the magnetism community~\cite{bergholtz2019non,mcclarty2019non,yang2021exceptional,cayao2023bulk}. These realizations are special since they correspond to the degeneracies of the energy bands, i.e., (exceptional) nodal phases. As reviewed in Sec.~\ref{Nodal_phase}, the exceptional nodal phases in the non-Hermitian scenario may lead to the non-Hermitian bulk Fermi arcs in magnonic systems \cite{bergholtz2019non,mcclarty2019non,wang2022exceptional,yang2021exceptional,cayao2023bulk}. In this section, we first review the proposal of the non-Hermitian nodal phase originating from the self-energy corrections from the proximity ferromagnet to the electronic surface states \cite{bergholtz2019non} and address the similar realization in pure magnonic system originating from the contribution of the Dzyaloshinskii-Moriya interaction to the magnon scattering in the framework of the Green's function approach \cite{mcclarty2019non} (for the approach, refer to Sec.~\ref{Green_function}). 
Then we address a new type of spin liquid with EPs in the reciprocal space in a strongly correlated spin model when coupled to the environment in the framework of the master-equation approach \cite{yang2021exceptional} (for the approach, refer to Sec.~\ref{master_equation_approach}). Finally, we highlight the presence of unique exceptional nodal phases of magnons in a van der Waals ferromagnetic bilayer \cite{li2022multitude}, where a multitude of EPs covers extended portions of the first Brillouin zone (Sec.~\ref{NHNPmultiple}).

\subsection{Non-Hermitian Weyl physics}
\label{NHNPbylead}

\begin{figure}[!htp]
\centering
\includegraphics[width=13.8cm]{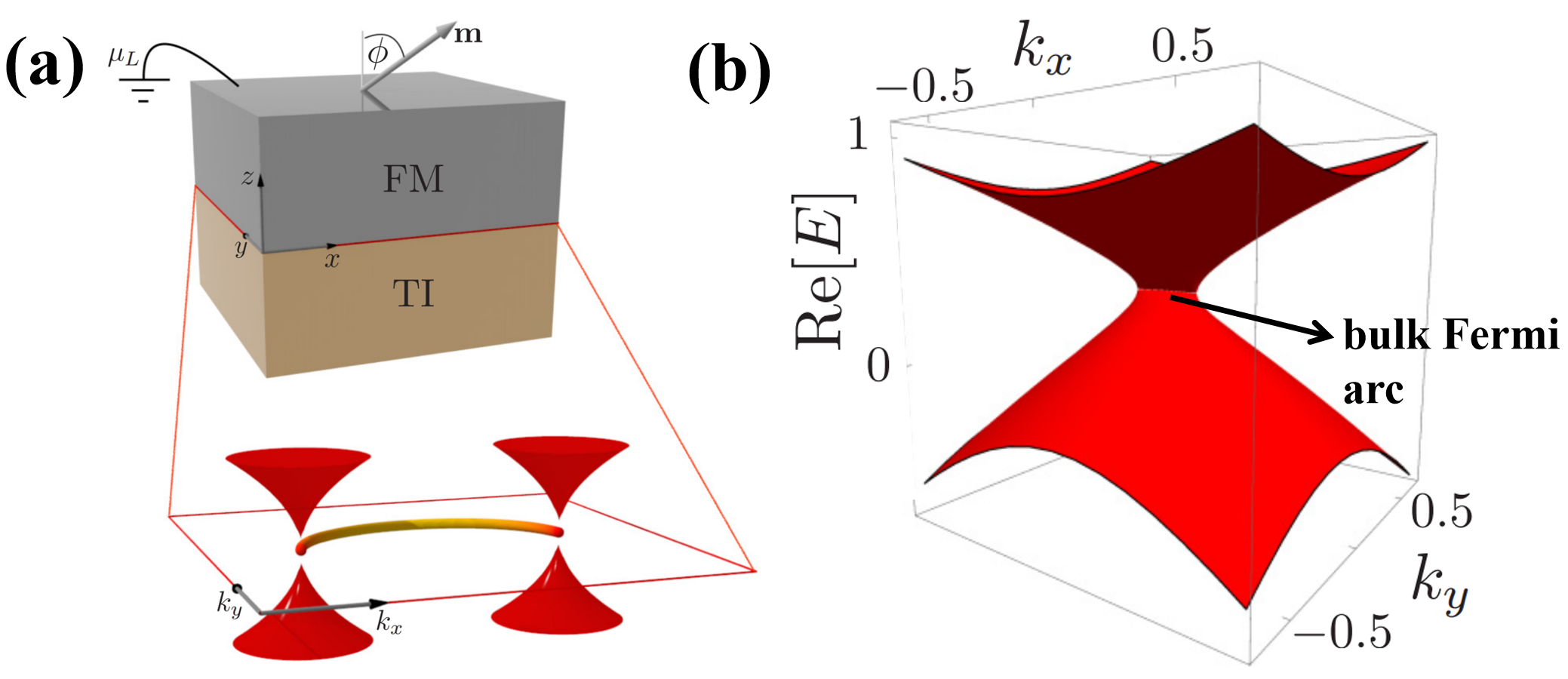}
  \caption{Non-Hermitian Weyl physics in topological insulator$|$ferromagnet junctions. (a) is the configuration in which the surface states of a 3D topological insulator couple via the hopping of electrons to the ferromagnetic metal with magnetization $\boldsymbol{\mathrm{m}}$ in the direction $\phi$. (b) describe the spectrum of a non-Hermitian Weyl phase with EPs connected by the bulk Fermi arc. The figure is reproduced with permission from Ref.~\cite{bergholtz2019non}.} 
\label{fig:NHW1}
\end{figure}

One strategy to realize the non-Hermitian Weyl semimetal involves introducing a non-Hermitian perturbation on the Hermitian degeneracy point as mentioned in Sec.~\ref{Nodal_phase}. Following this strategy, Bergholtz and Budich 
explored a heterostructure consisting of three-dimensional (3D) topological insulator and ferromagnetic metal~\cite{bergholtz2019non}, as shown in Fig.~\ref{fig:NHW1}(a). The surface states of 3D topological insulators behave as a two-dimensional electron gas governed by the Rashba-like spin-orbit coupling \cite{hasan2010colloquium,qi2011topological}. Due to the non-Hermitian self-energy corrections arising from the coupling with the ferromagnetic lead, the quasiparticle lifetime becomes finite. This correction results in a transition from gapless surface states into a non-Hermitian nodal phase with a pair of EPs in the reciprocal space, as detailed below \cite{bergholtz2019non}.

Among the heterostructure, the surface states of a 3D topological insulator at the interface ($x$-$y$ plane) are described by a two-band model \cite{qi2011topological}
	\begin{align}
		H_{\rm{TI}} (\boldsymbol{k})=\lambda(k_y\sigma_x-k_x\sigma_y),
		\label{3D_TI}
	\end{align}
where the spin of electrons is locked to their momentum.
Here ${\pmb \sigma}=\{\sigma_x,\sigma_y,\sigma_z\}$ is the Pauli matrix in the spin spaces, and
	$\lambda$ characterizes the group velocity of the fermion with a linear dispersion $\pm \lambda k$. This Hamiltonian acts as a Rashba-like spin-orbit interaction in the spintronics study \cite{sinova2004universal,galitski2013spin,bercioux2015quantum}. In the ferromagnetic metal, the itinerant electron interacts with the local magnetization ${\bf m}$ via the Zeeman interaction, described by the Hamiltonian
\begin{align}
H_F(\boldsymbol{k})=-2t\left[\cos(k_x)+\cos(k_y)\right]\sigma_0-2t_z\cos(k_z)\sigma_0-\mu_F\sigma_0+\boldsymbol{\mathrm{m\cdot\sigma}},
\label{eq:itinerant}
\end{align}
where $t$ is the hopping amplitude in the $x$-$y$ plane, $t_z$ is the out-of-plane hopping parameter along the $z$-direction, and $\mu_F$ is the chemical potential.

Due to the interfacial tunneling, the electrons at the interface acquire the surface self-energy mediated by the electrons in the ferromagnet, which breaks the time-reversal symmetry and alters the topological surface state to be a non-Hermitian Weyl phase without specific symmetries.
The surface retarded self-energy
\begin{align}
\Sigma_F^r(\omega)=V_{\rm SF}G_F^r(\omega)V_{\rm SF}^{\dagger}
\label{surfaceselfenergy}
\end{align}
depends on the spin since
the retarded Green's function $G_F^r(\omega)$ of the electrons in the ferromagnet  
is spin-dependent~\cite{ryndyk2009green}, where $V_{\rm SF}$ is the spin-independent hopping amplitude between the surface states of the topological insulator and the ferromagnet. Inclusion of the surface retarded self-energy $\Sigma_F^r(\omega = 0)$ [Eq.~(\ref{surfaceselfenergy})] at the chemical potential into the Hamiltonian (\ref{3D_TI}), the effective Hamiltonian for the surface states becomes
\begin{align}
{H}_{\rm eff}&=\lambda(k_y\sigma_x-k_x\sigma_y)+\Sigma_F^r(\omega = 0)-B\sigma_z\equiv \epsilon_0+\boldsymbol{\mathrm{d}}(\boldsymbol{k})\cdot\boldsymbol{\sigma},
\label{effective_Hamiltonian_Weyl}
\end{align}
where the out-of-plane magnetic field $B$ breaks the time-reversal symmetry, $\epsilon_0$ is a complex number and $\boldsymbol{\mathrm{d}}(\boldsymbol{k})=\boldsymbol{\mathrm{d}}_R(\boldsymbol{k})+i\boldsymbol{\mathrm{d}}_I(\boldsymbol{k})$ are two-dimensional complex vectors with real components $\boldsymbol{\mathrm{d}}_R(\boldsymbol{k})$ and $\boldsymbol{\mathrm{d}}_I(\boldsymbol{k})$. $\boldsymbol{\mathrm{d}}_R(\boldsymbol{k})$ is contributed by the Rashba-like spin-orbit coupling of surface state and the real part of the self-energy correction, and  $\boldsymbol{\mathrm{d}}_I(\boldsymbol{k})$ comes from the imaginary part of the self-energy correction.
We note the in-plane wave vector ${\bf k}={k}_x\hat{\bf x}+k_y\hat{\bf y}$, so the Hamiltonian (\ref{surfaceselfenergy}) is similar to the general form (\ref{eq:Hamiltonian}) addressed in Sec.~\ref{Nodal_phase}.

  Some properties of the effective Hamiltonian (\ref{effective_Hamiltonian_Weyl}) have been reviewed in Sec.~\ref{Nodal_phase}. Its complex spectrum $E_{\pm}(\boldsymbol{k})=\epsilon_0\pm\sqrt{\boldsymbol{\mathrm{d}}_R^2(\boldsymbol{k})-\boldsymbol{\mathrm{d}}_I^2(\boldsymbol{k})+2i\boldsymbol{\mathrm{d}}_R(\boldsymbol{k})\cdot\boldsymbol{\mathrm{d}}_I(\boldsymbol{k})}$. Two EPs at $E(\boldsymbol{k})=\epsilon_0$ are isolated in the reciprocal space when $\boldsymbol{\mathrm{d}}_R^2(\boldsymbol{k})-\boldsymbol{\mathrm{d}}_I^2(\boldsymbol{k})=0$ and $\boldsymbol{\mathrm{d}}_R(\boldsymbol{k})\cdot\boldsymbol{\mathrm{d}}_I(\boldsymbol{k})=0$, as shown in Figure~\ref{fig:NHW1}(a) and (b). Between the pair of EPs, there exist ``bulk Fermi arcs'' with $\boldsymbol{\mathrm{d}}_R^2({\bf k})-\boldsymbol{\mathrm{d}}_I^2({\bf k})< 0$ and $\boldsymbol{\mathrm{d}}_R({\bf k})\cdot\boldsymbol{\mathrm{d}}_I({\bf k})=0$, \textit{i.e.}, $\text{Re}E=0$ and $\text{Im}E=0$,  as addressed in Sec.~\ref{Nodal_phase}.

  The coupled 3D topological insulator and ferromagnetic metal hold the tunability by several degrees of freedom such as the magnetic field $B$ as well as the magnitude and direction of magnetization in the ferromagnetic lead. The bulk Fermi arcs can be detected by angle-resolved
photoemission spectroscopy experiments \cite{nagai2020dmft}. The topological-insulator$|$ferromagnetic heterostructure may provide an experimentally feasible and tunable platform for studying and observing such distinct phase~\cite{bergholtz2019non,bergholtz2021exceptional,song2023observation}. Similar design concepts have been explored in other systems such as coupled Dirac semimetal and superconductor \cite{jana2021emergence} as well as coupled superconductor and ferromagnetic lead \cite{cayao2023bulk}.

\subsection{Non-Hermitian nodal phase by magnon interactions}
\label{NHNPbymagnon}

Without choosing the heterostructure~\cite{bergholtz2019non}, McClarty \textit{et al.} predicted the non-Hermitian nodal phase of magnons in a single ferromagnet \cite{mcclarty2019non} in the presence of the Dzyaloshinskii-Moriya interaction. They considered a mechanism of spontaneous magnon decay in a spin-1/2 ferromagnet on the honeycomb lattice. Spontaneous decay is a fundamental characteristic of magnon when the nonlinear interaction is sufficiently strong~\cite{zhitomirsky2013colloquium}, which induces magnon scattering and thereby affects the magnon lifetime.  For such a system, the Hamiltonian 
	\begin{equation}
		\hat{H}=-\sum\limits_{n=1}^3\sum\limits_{\langle i,j\rangle_n}J_n(\hat{\boldsymbol{S}}_i\cdot\hat{\boldsymbol{S}}_j)+D\sum\limits_{\langle i,j\rangle_2}\nu_{ij}\hat{\boldsymbol{z}}\cdot(\hat{\boldsymbol{S}}_i\times\hat{\boldsymbol{S}}_j),
		\label{eq:NHM}
	\end{equation}
     where $J_n$ is the $n$-th neighboring ferromagnetic couplings between spin $\hat{\boldsymbol{S}}$. The last term in Eq.~\eqref{eq:NHM} is the Dzyaloshinskii-Moriya interaction with the index $\nu_{ij}=-1$ ($+1$) for the clockwise (counterclockwise) hopping with respect to the normal $\hat{\boldsymbol{z}}$ axis, from the site $i$ to $j$, which induces spontaneous magnon decay in the nonlinear magnetization dynamics.

\begin{figure}[!htp]
\centering
\includegraphics[width=0.85\textwidth]{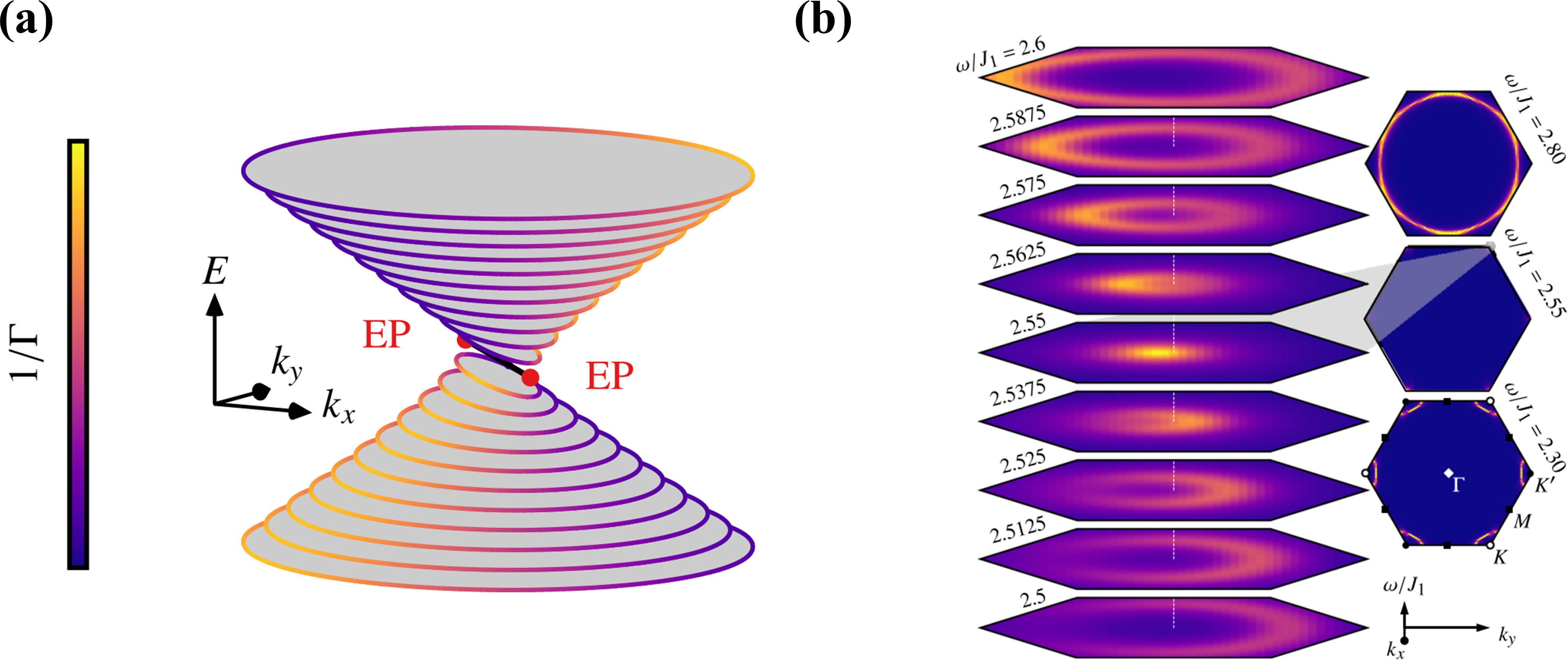}
\caption{Prediction of bulk Fermi  arc between a pair of EPs in the magnon spectra of the honeycomb lattice model. (a) is the constant slices of the real part  $E_{\boldsymbol{k}}$ of the mode spectra, where the inverse of the imaginary part $\Gamma_{\boldsymbol{k}}$  is denoted by the color bar. (b) show the spectral function $A(\boldsymbol{k},\omega)$ of different energies around the EPs. These figures are reproduced with permission from Ref.~\cite{mcclarty2019non}.}
\label{fig:NHM}
\end{figure}

Retaining the linear order of the Holstein-Primakoff transformation~\cite{holstein1940field,auerbach2012interacting} in terms of the boson operator $\hat{a}_{\boldsymbol{k}\alpha/\beta}$, where ${\boldsymbol{k}}$ is the wave vector  and $\{\alpha,\beta\}$ denote the sublattices, leads to the Hamiltonian of the form
\begin{equation}\label{BDGFermiarc}
\hat{H}(\boldsymbol{k})=\sum\limits_{\boldsymbol{k}}\sum\limits_{\alpha\beta}\left(A_{\boldsymbol{k}}^{\alpha\beta}\hat{a}_{\boldsymbol{k}\alpha}^{\dagger}\hat{a}_{\boldsymbol{k}\beta}+\frac{1}{2}\left(B_{\boldsymbol{k}}^{\alpha\beta}\hat{a}_{\boldsymbol{k}\alpha}^{\dagger}\hat{a}_{-\boldsymbol{k}\beta}^{\dagger}+\rm{H.c.}\right)\right).
\end{equation}
Here, $A_{\boldsymbol{k}}$ and $B_{\boldsymbol{k}}$ represent the spin-conserving and nonconserving interactions, both of which depend on the detail of the model. Equation~\eqref{BDGFermiarc} is a BdG-type Hamiltonian as addressed in Sec.~\ref{BDGHamiltonian}. The magnon spectrum in the linear regime is given by diagonalizing the matrix~\cite{blaizot1986quantum}
\begin{equation}
\sigma_zM_{\boldsymbol{k}}\equiv
\left(\begin{array}{cc}
A_{\boldsymbol{k}}&B_{\boldsymbol{k}}\\
-B_{-\boldsymbol{k}}^{\ast}&-A_{-\boldsymbol{k}}^{\ast}
\end{array}\right)
\end{equation}
with the matrix $M_{\boldsymbol{k}}$ and $\sigma_z={\rm diag}(1,-1)$. Under specified parameters, such spectrum possesses a linear Dirac point in the Brillouin zone \cite{mcclarty2019non}. 
The authors showed that in the linear regime, the Dzyaloshinskii-Moriya interaction does not strongly affect the magnon dispersion. But entering the nonlinear regime, the Dzyaloshinskii-Moriya interaction contributes a retarded magnon self-energy \cite{negele2018quantum,blaizot1986quantum}    near the band degeneracy point $\omega_0$
\begin{equation}
         \Sigma(\boldsymbol{k},\omega) \approx  \Sigma'(\boldsymbol{k},\omega_0) +  \Sigma''(\boldsymbol{k},\omega_0)
     \end{equation}
to the retarded Green function of magnons 
\begin{equation}
	G(\boldsymbol{k},\omega)\equiv\Big\{(\omega+i0^+)\sigma_z-[M_{\boldsymbol{k}}+\Sigma(\boldsymbol{k},\omega)]\Big\}^{-1}.
\end{equation}
Here the real part $\Sigma'(\boldsymbol{k},\omega_0)$ renormalizes the magnon spectra, and the imaginary part $\Sigma''(\boldsymbol{k},\omega_0)$ leads to spontaneous decay of magnons. In the vicinity of the band degeneracy point, such non-Hermitian corrections lead to a two-band non-Hermitian model
 \begin{align}
{H}_{\rm eff}&=\lambda(k_x\sigma_x+k_y\sigma_y) - i \left( a_0 + \boldsymbol{a}({\bf k})\cdot\boldsymbol{\sigma}\right),
\label{effective_Hamiltonian_Weyl_twoband}
\end{align}
 which is similar to Eq.~\eqref{effective_Hamiltonian_Weyl}.
Here $\lambda$ is the group velocity, and the constants $a_0$ and $\boldsymbol{a}({\bf k})$ contribute to the non-Hermitian perturbation. The real part $E_{\boldsymbol{k}}$ of the spectra from Eq.~\eqref{effective_Hamiltonian_Weyl_twoband} and the corresponding inverse of the imaginary part ${\Gamma}_{\boldsymbol{k}}$ denoted by the color bar are shown in Fig.~\ref{fig:NHM}(a), where a pair of EPs emerge in the reciprocal space. Such a nodal phase is similar to the case of applying a non-Hermitian perturbation on the band degeneracy point as addressed in Sec.~\ref{Nodal_phase}. These two EPs possess the same real part and there is a bulk Fermi arc that connects them. Compared to the Fermi arcs with zero real energy, the only difference is that this bulk arc is shifted by a finite real energy.

 Figure~\ref{fig:NHM}(a) also exhibits the anisotropic characteristic of magnon lifetime, in which the magnon modes below the EPs with the shortest lifetime are skewed to one side in the reciprocal space but located on the other side when they are above the EPs.
The calculated spectrum in Fig.~\ref{fig:NHM}(a) is further confirmed by the spectral function as shown in Fig.~\ref{fig:NHM}(b), which reveals the density of state at certain momentum $\boldsymbol{k}$ and energy $\omega$. The spectral function can be detected by angle-resolved photoemission spectroscopy~\cite{lv2019angle}, providing a way of experimental verification of bulk Fermi arc.
	
In Sec.~\ref{NHNPbylead} and \ref{NHNPbymagnon}, the quasiparticle renormalization is captured by the energy shift and finite lifetime in the framework of the Green-function approach~\cite{kozii2017non,bergholtz2019non,mcclarty2019non,jiang2021optical}.  Such non-Hermitian perturbation from the self-energy correction to the quasiparticle spectra renders the band degenerate points such as the Dirac or Weyl points split into a pair of EPs that is connected by the bulk Fermi arc, similar to the mechanism addressed in Refs.~\cite{zhou2018observation,yang2021exceptional} (refer to Sec.~\ref{Nodal_phase}). A similar approach is used to model the bulk Fermi arc in several other systems such as heavy-fermion system~\cite{nagai2020dmft}, Dirac fermion system~\cite{papaj2019nodal}, superconductors coupled with ferromagnetic lead~\cite{cayao2023bulk} and photonic system~\cite{zhou2018observation,wang2021simulating}. A different approach to describe the non-Hermitian perturbation lies in the framework of the master-equation approach, which is used to deal with the excitation of the spin liquid in Ref.~\cite{yang2021exceptional}, which is one nontrivial ground state of the strongly correlated quantum spins ~\cite{balents2010spin,savary2016quantum,zhou2017quantum}. Recently, using the Lindblad master equation, Yang \textit{et al.} proposed an ``exceptional spin liquid'' with several EPs in the reciprocal space when the strongly correlated quantum spins on the Kitaev honeycomb lattice couple to the environment that introduces an effective complex effective coupling between spins~\cite{yang2021exceptional}. They find that these EPs are not only connected by bulk Fermi arcs with $\text{Re}E=0$ but also the i-Fermi arcs with $\text{Im}E=0$, which are thereby robust to perturbations.

Experimentally, the bulk Fermi arc has been observed in periodic photonic crystals by angle-resolved scattering measurements~\cite{zhou2018observation}, and can also be detected in principle by angle-resolved photoemission
spectroscopy~\cite{nagai2020dmft} or interferometric measurements~\cite{wang2021simulating}. Such measurements may be extended to detect the bulk Fermi arcs for magnonic systems as well. A recent study demonstrated that the optical conductivity in the topological phase with non-Hermitian Fermi arc exhibits a different response to the optical field distinguished from other topological phases~\cite{jiang2021optical}.

\subsection{A multitude of EPs in reciprocal space}
\label{NHNPmultiple}

\begin{figure}[!htp]
		\centering
		\includegraphics[width=16cm]{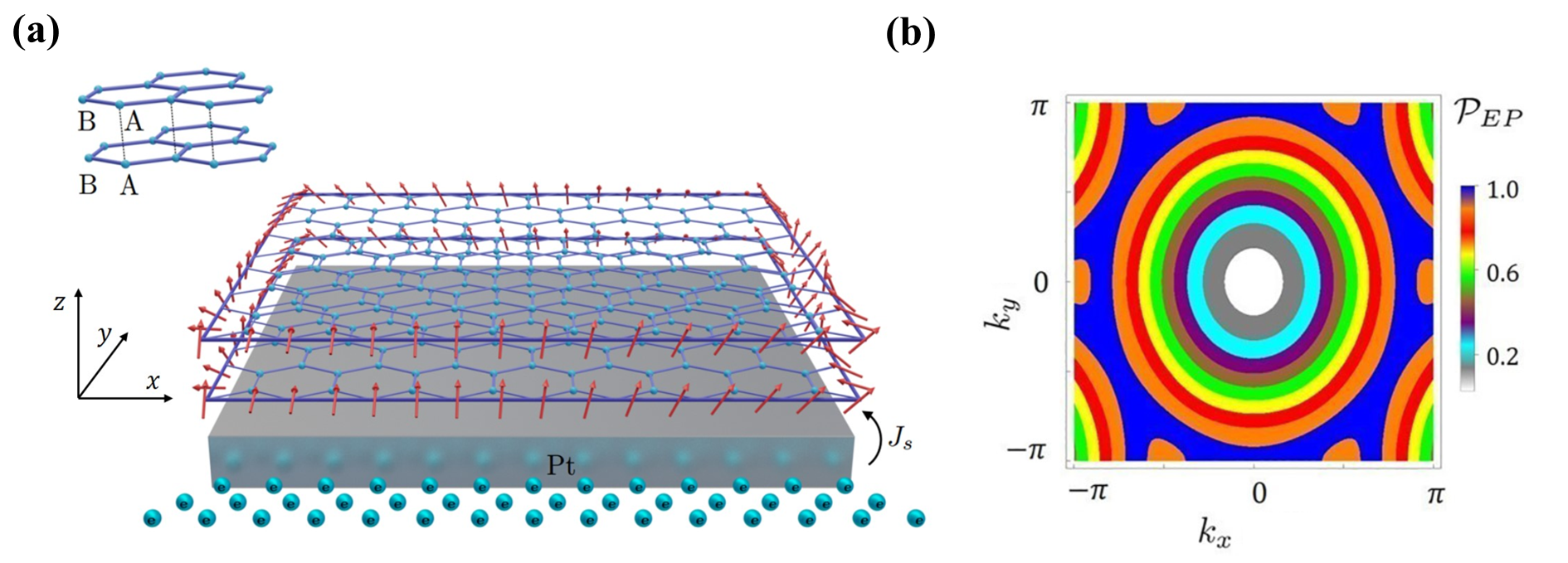}
		\caption{Prediction of a multitude of EPs in AA-stacked van der Waals ferromagnetic bilayer. (a) is the configuration, where the spin current is injected to one layer from the adjacent Pt layer. (b) shows the overlap of the eigenvectors $\mathcal{P}_{E P} \equiv\left|\left\langle\phi_1^R | \phi_2^R\right\rangle\right|^2 $ for the two lowest bands, where the regions occupied by the EPs are denoted by blue color.
The figure is reproduced with permission from Ref.~\cite{li2022multitude}.} 
		\label{fig:multiEP}
	\end{figure}

No matter whether the non-Hermitian degeneracies are EPs, exceptional lines and surfaces, as well as Hopf links and knots, they often have counterparts in the Hermitian nodal phases. Moreover, in the $n$-dimensional reciprocal space, the degeneracies typically are of $(n-1)$-dimension or lesser. For example, zero-dimensional EPs emerge in one-dimensional reciprocal space or the one-dimensional exceptional lines appear in two-dimensional reciprocal space. Such isolated degeneracies may not affect remarkably the conductivity-related bulk properties since they require the integral over the whole Brillouin zone, while the phase space around the exceptional degeneracies appears to be quite small. By contrast, Li \textit{et al.} proposed the emergence of a multitude of EPs in a bilayer of AA-stacked ferromagnetic honeycomb lattice [Fig.~\ref{fig:multiEP}(a)], where the distribution of the non-Hermitian degeneracy have the same dimension with the reciprocal space [Fig.~\ref{fig:multiEP}(b)] \cite{li2022multitude}, and to the best of our knowledge, it has no counterpart in other systems.

 Similar to the design of $\mathcal{PT}$-symmetric systems \cite{regensburger2012parity,lee2015macroscopic,hodaei2017enhanced,el2018non,ozdemir2019parity,miri2019exceptional,wang2023floquet} as reviewed in Sec.~\ref{exceptionalpointanalysis}, Li \textit{et al}. proposed the realization of the balanced gain and loss for a coupled magnetic bilayer by the injected current from an adjacent metal (Pt) substrate   \cite{li2022multitude}.
 The magnetization dynamics are governed by the effective magnetic field $\mathbf{h}_{\text{eff }}$ involving the intralayer and interlayer couplings and the intrinsic dissipation (gain) denoted by minus $\alpha$ (positive $\alpha$):
\begin{equation}
\begin{aligned}
& \dot{\mathbf{m}}_{i,j}^{(l=1)}=-\gamma \mathbf{m}_{i,j}^{(l=1)} \times \mathbf{h}_{{i,j}, \text { eff }}^{(l=1)}-\alpha \mathbf{m}_{i,j}^{(l=1)} \times \dot{\mathbf{m}}_{i,j}^{(l=1)}, \\
& \dot{\mathbf{m}}_{i,j}^{(l=2)}=-\gamma \mathbf{m}_{i,j}^{(l=2)} \times \mathbf{h}_{{i,j}, \text { eff }}^{(l=2)}+\alpha \mathbf{m}_{i,j}^{(l=2)} \times \dot{\mathbf{m}}_{i,j}^{(l=2)},
\end{aligned}
\end{equation}
where $l=\{1,2\}$ denotes the bottom and top layers and $\{i,j\}$ represents two sublattices of the honeycomb lattice. Here, the positive $\alpha$ originates from the injected spin current from the adjacent Pt layer. With balanced gain and loss, such a system possesses $\mathcal{PT}$-symmetry, where the interlayer coupling is Hermitian as addressed in Sec.~\ref{exceptional_points}. Considering that the eigenvectors are coalescent at the EPs, the authors introduce $\mathcal{P}_{E P} \equiv\left|\langle \phi_1^R | \phi_2^R\rangle\right|^2 $ to denote the region of EPs for the two lowest bands $\{1,2\}$, which becomes parallel for the right eigenvector at the EPs. As shown in Fig.~\ref{fig:multiEP}(b), the EPs cover a large area in the Brillouin zone with suitable parameters $\alpha$, implying the EPs may affect significantly the macroscopic transport property for the magnons~\cite{li2022multitude}.

\section{Topological edge state \textit{vs.} non-Hermitian skin effect in magnonic systems}

\label{Non_Hermitian_skin_effect_magnon}

Above we have addressed several aspects of the non-Hermitian topology associated with the exceptional energy degeneracies in the parameter or wave-vector space without resorting to the nontrivial collective modes, which, however, can be governed by the wavefunction topology (Sec.~\ref{SSH_model_summary}) or non-Hermitian spectral topology as well (Sec.~\ref{Nonhermitian_skin_effect}) \cite{hasan2010colloquium,moore2010birth,qi2011topological,shen2012topological,bernevig2013topological,breunig2022opportunities}. In this section, we start with the magnonic non-Hermitian SSH model as a generalization of the Hermitian one \cite{su1979solitons,heeger1988solitons,fradkin1983phase,li2014topological,lieu2018topological,yao2018edge,obana2019topological} in Sec.~\ref{SSH_model_summary}  by emphasizing the unique functionalities with the topological characterization (Sec.~\ref{magnonics_SSH}). We then focus on the realization of the non-Hermitian skin effect in magnonic devices~\cite{deng2022non,yu2022giant,yu2020magnon_accumulation,zeng2023radiation,cai2023corner,li2023reciprocal} when the conventional bulk-boundary correspondence breaks down. We emphasize the role of the long-range chiral interaction that can be realized in magnonic systems (Sec.~\ref{Chiral_interaction}), which displays several nontrivial features compared to the short-range chiral interaction in the Hatano-Nelson model as stated in Sec.~\ref{Nonhermitian_skin_effect}. We finally emphasize the recent progress of the non-Hermitian skin effect in a higher dimension with a magnonic realization in van der Waals ferromagnetic honeycomb lattice and magnetic array with generalization of topological characterization (Sec.~\ref{high_dimension_skin_effect}).

\subsection{Magnonic non-Hermitian SSH model}
\label{magnonics_SSH}

As addressed in Sec.~\ref{SSH_model_summary}, there are two types of non-Hermitian SSH models. The type II non-Hermitian SSH model allows the realization of the non-Hermitian skin effect in the presence of the non-reciprocal coupling (Fig.~\ref{fig:SSH1}), which breaks the conventional bulk-boundary correspondence~\cite{yao2018edge}. Nevertheless, for type I non-Hermitian SSH model~\cite{flebus2020non}, the non-Hermiticity is introduced via the alternative gain and loss in neighboring sites, which exhibits the conventional topological edge state but not the non-Hermitian skin effect with the GBZ coinciding with the Brillouin zone. In this case, conventional bulk-boundary correspondence still holds, and the topological invariant refers to the Berry phase or winding number defined on the Brillouin zone.

\begin{figure}[!htp]
\centering
\includegraphics[width=0.98\textwidth]{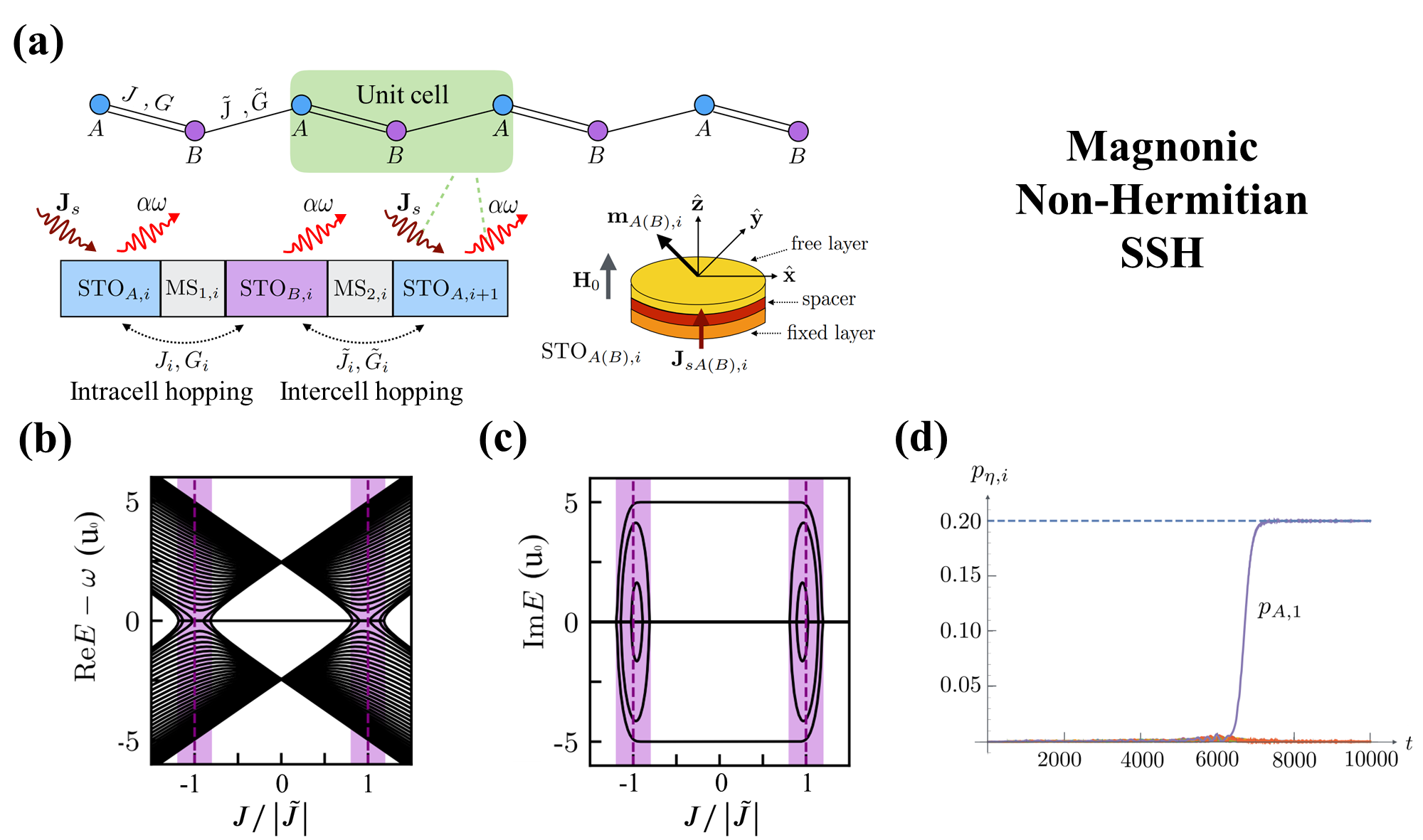}
\caption{Magnonic realization of the non-Hermitian SSH model in terms of an array of STOs that are spaced by the metallic layers. In (a) the metallic spacers mediate the RKKY exchange coupling, $J$ and $\tilde{J}$, and spin-pumping induced dissipative coupling, $G$ and $\tilde{G}$, between neighboring STOs.  (b) and (c) plot the numerically calculated energy spectrum for $N=40$ unit cells as a function of $J/\left|\tilde{J}\right|$ under the $\mathcal{PT}$ symmetry, where the topological modes with Re$E=0$ lie within the gap. In (d), only the magnetization at the end of the array is driven to its threshold by the injected spin current, acting as the lasing topological edge mode. The figures are adapted with permission from Ref.~\cite{flebus2020non}.}
\label{fig:SSH2}
\end{figure}

Recently, Flebus \textit{et al.} proposed a magnonic realization of the type I non-Hermitian SSH model in terms of an array of spin-torque oscillators (STOs) \cite{flebus2020non,gunnink2022nonlinear}. 
The authors predicted the topological magnonic lasing edge modes, which can be excited by the spin current injection \cite{flebus2020non,gunnink2022nonlinear}. Each STO consists of a trilayer composed of a free ferromagnet, a metallic spacer, and  a ferromagnet with pinned magnetization, as shown in Fig.~\ref{fig:SSH2}(a), where the injected spin current is polarized by the pinned magnetization and thus exerts a torque on the misaligned magnetization in the free layer and induce oscillations \cite{tsoi1998excitation,tsoi2000generation,kiselev2003microwave,krivorotov2005time}. Different STOs are mediated by the metallic layer (``MS'') and coupled by the RKKY exchange interaction and spin pumping \cite{kaka2005mutual,litvinenko2021analog,tiberkevich2009phase,li2011coupled}, which can even achieve
mutual synchronization when their coupling exceeds their frequency mismatch \cite{sani2013mutually,houshang2016spin,romera2018vowel}.
An STO itself is a non-Hermitian system since there exists competition between an effective ``gain'' due to the spin-current injection \cite{slonczewski1996current,berger1996emission,tserkovnyak2002enhanced} and the ``loss'' by the dissipation or Gilbert damping. The metallic layer between two neighboring STOs mediates the non-Hermitian coupling including the dissipative component due to the spin pumping. Via tuning the length of metallic spacers, the distinct intracell and intercell couplings can be achieved, making a close analogy to the non-Hermitian SSH model (Sec.~\ref{SSH_model_summary}) \cite{fradkin1983phase,li2014topological,lieu2018topological,obana2019topological}. 

 The STOs in the array are divided into $A$ and $B$ sublattices when they are spaced by two different metallic spacers, \textit{viz}. ``MS$_1$'' and ``MS$_2$'' in Fig.~\ref{fig:SSH1}(a).
 The magnetization in the $A$ sublattice is exerted by the torque from the spin current $J_s$, while no spin current is injected into the sublattice $B$; on the other hand, the magnetization for both the $A$ and $B$ sublattices dissipates with the same rates $\alpha\omega$, parametrized by the Gilbert damping coefficient $\alpha$. The RKKY exchange interaction contains the intracell and intercell Hermitian couplings $J$ and $\tilde{J}$, while the spin pumping across the metallic spaces mediates the intracell and intercell dissipative couplings $G$ and $\tilde{G}$ (refer to Sec.~\ref{spin_pumping_dissipative_coupling}). Thus, the Hamiltonian for such an array of STOs, in the reciprocal space, reads
\begin{equation}
\hat{H}(k)=id_0(k)\sigma_0+{\bf d}( k)\cdot{\pmb \sigma},
\end{equation}
where $\sigma_0$ is the identity matrix and ${\pmb \sigma}$ are the Pauli matrices. $d_0(k)=(J_s-2\alpha\omega)/2$ is contributed by the on-site spin torque and damping, and the coupling between STOs governs the complex vector
	\begin{equation}
		{\boldsymbol{d}}({k})=\left(\begin{array}{c}
			-J+i\omega G-(\tilde{J}-i\omega\tilde{G})\cos k\\
			-(\tilde{J}-i\omega\tilde{G})\sin k\\
			iJ_s/2
		\end{array}\right).
	\end{equation}
    The Bloch Hamiltonian preserves the $\mathcal{PT}$ symmetry, as demonstrated in Sec.~\ref{exceptional_points}, when the gain by the injected spin current balances exactly the intrinsic dissipation of one unit cell, \textit{viz}. $J_s=2\alpha\omega$, and the spin pumping between two neighboring STOs vanishes $G=\tilde{G}=0$. The numerical calculation for the frequency spectrum with $N=40$ unit cells in terms of real part Re$E-\omega$ and imaginary part Im$E$ as a function of $J/\left|\tilde{J}\right|$, as plotted in Figs.~\ref{fig:SSH1}(b) and (c), contains a real line gap when $\left| |J/\tilde{J}|-1\right|>\left|\alpha \omega/\tilde{J}\right|$, corresponding to the area outside the purple region in Fig.~\ref{fig:SSH1}(b), and topological edge modes when $\left|J/\tilde{J}\right|<1$, i.e., the area between the purple lines. These topological edge states have zero real frequency but nonzero imaginary frequency Im$E = \pm \alpha \omega $, corresponding to a lasing ($\alpha\omega$) or lossy ($-\alpha\omega$) magnonic edge mode. Impressively, the authors showed that in the existence of dissipative coupling $G\approx\tilde{G}$, which breaks the $\mathcal{PT}$-symmetry, the edge states can survive when they are much smaller than the damping rate $\alpha$~\cite{flebus2020non,gunnink2022nonlinear}.

 The positive imaginary frequency of the edge lasing mode implies the magnon population grows exponentially with time. The numerical calculation for the dynamics of the array of STOs confirms that the edge mode with magnetization direction along the $\hat{\bf z}$-direction can be excited and driven into auto-oscillation by spin current, as shown in Fig.~\ref{fig:SSH1}(d), while the other bulk modes remain inactive. The edge states appear to be conveniently identified from the spatial distribution of wavefunction since the bulk modes are located within the array. Similar topological protected lasing modes have been found in photonics \cite{zhao2018topological,parto2018edge,harari2018topological,bahari2017nonreciprocal,bandres2018topological}, active matter \cite{sone2020exceptional} and light-matter hybrid system \cite{comaron2020non}, which are of great interest due to the single-mode lasing and their robustness to back-scattering and disorder.
The non-Hermitian band topology in the magnonic systems is far from explored, in comparison to the extensive studies of EPs topology as stated in Sec.~\ref{exceptional_points} and \ref{EPs_magnonics}.

\subsection{Non-Hermitian skin effect in long-range chirally coupled magnets}
\label{Chiral_interaction}
 
The topological edge state as stated above corresponds to a topological invariant that is characterized by the bulk property, which is known as the bulk-boundary correspondence \cite{hasan2010colloquium,qi2011topological,shen2012topological,bernevig2013topological,asboth2016short,yan2017topological,hasan2021weyl}.  But in the non-Hermitian systems there also exists an exotic topological property that is governed by the energy spectral topology \cite{zhang2020,okuma2020} as stated in Sec.~\ref{Nonhermitian_skin_effect}, which leads to one of the most remarkable features of the non-Hermitian system, \textit{viz}. the non-Hermitian skin effect with a macroscopic number of eigenstates piling up at one boundary of the system \cite{okuma2020,zhang2020correspondence,bergholtz2021exceptional,zhang2022review}.

The most important ingredient for the non-Hermitian skin effect appears to be the chiral coupling, also known as the asymmetric coupling or non-reciprocal coupling, which is already realized in many systems such as the photonic lattices \cite{weidemann2020topological}, the electric circuits \cite{hofmann2020reciprocal,helbig2020generalized,zou2021observation,zhang2022observation2}, the mechanic circuits \cite{ghatak2020observation} and the cold-atom systems \cite{li2020topological,liang2022dynamic}, in which the non-Hermitian skin effect is successfully observed. We note that all these  systems mentioned above possess only the \textit{short-range} chiral coupling. Although not necessarily limited to the nearest or the second nearest neighbors, and so on, the short-range chiral coupling can span over several interacting objects as long as the range is much shorter than the size of the system.

Another system that possesses the chiral coupling is the emitters, such as atoms or magnets, that are mediated by traveling waves of open cavities \cite{asenjo2017exponential,chang2018colloquium,yu2020chiral}. Since the propagating waves may have small dissipation,  the effective coupling between emitters is of long range, even approaching the range of the system's size. These systems are of great interest since interference by radiation of traveling waves gives rise to collective modes including the superradiant and subradiant states \cite{dicke1954coherence,asenjo2017exponential,zhang2019theory,dinc2020multidimensional,zhang2020subradiant}, induced by constructive interference (destructive interference) with faster (suppressed) decay rates than that of the individual emitter, which has the potential application in the coherent photon storage \cite{asenjo2017exponential,ferioli2021storage}, the excitation transfer \cite{moreno2019subradiance,ballantine2020subradiance}, and the many-body entangled states for quantum computation \cite{paulisch2016universal,hammerer2010quantum}.
The chiral coupling has been realized experimentally in these systems \cite{lodahl2017chiral,bliokh2015spin,chervy2018room,coles2016chirality}, but the non-Hermitian skin effect of collective modes is never observed, to the best of our knowledge, which implies distinct properties from those with short-range chiral coupling.

Recently, Yu and coauthors demonstrated that the chirality alone in the long-range coupled magnets only drives some modes with the skin tendency, while most modes are located at the bulk \cite{yu2020chiral,yu2020magnon_accumulation,yu2022giant}.  However, the combination of chirality and a considerable dissipation of traveling waves drives strongly all the modes to one edge, i.e., leading to the non-Hermitian skin effect \cite{yu2022giant,zeng2023radiation,cai2023corner}. 
They further demonstrated that the non-Hermitian skin effect of such long-range coupled systems
has the same topological origin that is depicted by spectral winding number as stated in Sec.~\ref{Nonhermitian_skin_effect}. However, they hold contrast responses to a point defect due to the nature of long-range coupling \cite{zeng2023radiation} and thereby may provide an experimentally stable platform for exciting and detecting the skin modes. In the two dimensions, they predicted the edge and corner accumulations of magnonic states and proposed the double winding indexes to characterize them~\cite{cai2023corner}.

\textbf{An overview}.---Before diving into the detailed mechanisms, we first provide an overview of the non-Hermitian skin effect in different systems with short-range and long-range couplings and compare their similarities and differences in terms of condition, topological origin, and response to a point defect, as summarized in Table~\ref{fig:SK1co}. We emphasize the new features endowed by the long-range coupling.

\begin{table}
   \caption{Comparison of the non-Hermitian skin effect in different systems with short-range and long-range couplings in terms of the condition, topological origin, and response to point defect.} \label{fig:SK1co}
  \begin{tabular}{ccc}
   \toprule[1pt]
  \makebox[0.11\textwidth][c]{Model}&\makebox[0.30\textwidth][c]{System with short-range coupling}&\makebox[0.50\textwidth][c]{System with long-range coupling}\\
    \midrule[0.5pt] \begin{minipage}[m]{.11\textwidth}
       Schematic \\ diagram
  \end{minipage}  & \begin{minipage}[m]{0.3\textwidth}
      \centering\vspace*{4pt}  	\includegraphics[width=4.0cm]{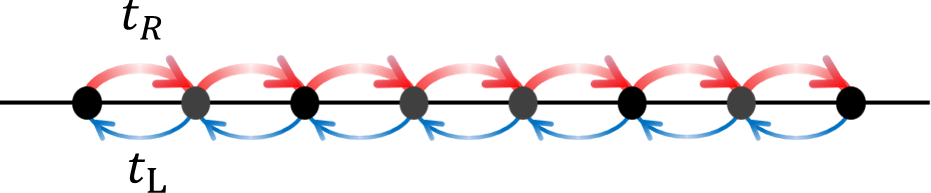}\vspace*{2pt}
      \begin{equation}
\hat{H}=\sum\limits_{i}\left(t_L\hat{c}_i^{\dagger}\hat{c}_{i+1}+t_R\hat{c}_i\hat{c}_{i+1}^{\dagger}\right) \nonumber
\end{equation}
    \end{minipage} & \begin{minipage}[m]{.5\textwidth}
    \centering\vspace*{4pt}
    \includegraphics[width=5.5cm]{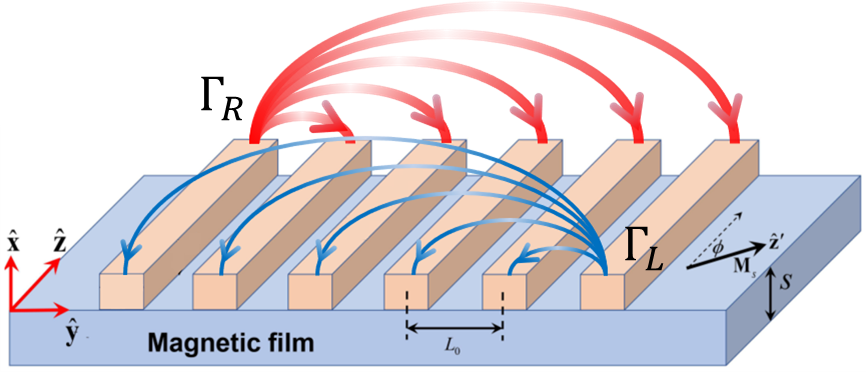}\vspace*{2pt}
 \begin{equation}
 \begin{aligned}
      \hat{H}_{\mathrm{eff}}/\hbar & = \left(\omega_{{\rm K}} - \frac{\Gamma_R+\Gamma_L}{2}\right)\sum_l \hat{\beta}_{l}^{\dagger}\hat{\beta}_{l} \\
    &   -i \Gamma_L \sum_{l < l^{\prime}} e^{ik_0\left|R_l-R_{l^{\prime}}\right|}\hat{\beta}_{l}^{\dagger}\hat{\beta}_{l^{\prime}}\\
     &   -i \Gamma_R \sum_{l > l^{\prime}} e^{ik_0\left|R_l-R_{l^{\prime}}\right|}\hat{\beta}_{l}^{\dagger}\hat{\beta}_{l^{\prime}} \nonumber
 \end{aligned}
\end{equation}
    \end{minipage}
    \\
        \midrule[0.5pt] \begin{minipage}[m]{.11\textwidth}
      Condition for\\ skin effect
  \end{minipage}  & \begin{minipage}[m]{0.3\textwidth}
      \centering\vspace*{4pt}
      	\includegraphics[width=3.3cm]{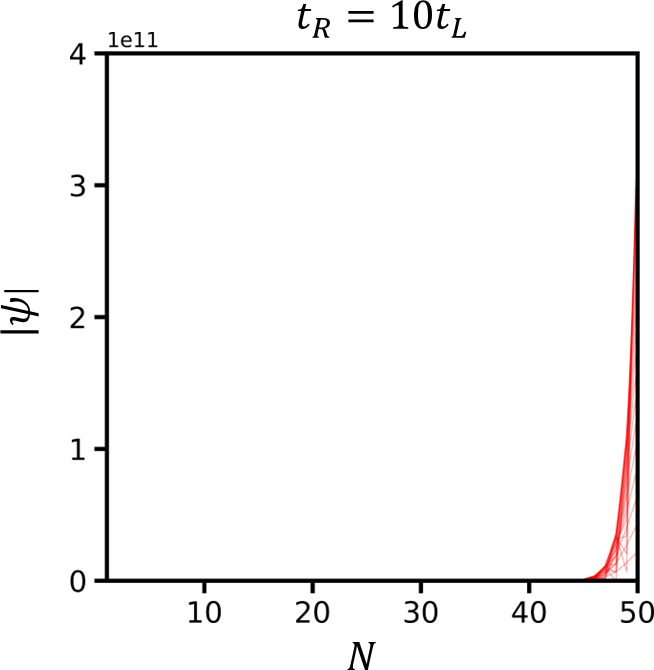}\vspace*{2pt}
    \end{minipage} & \begin{minipage}[m]{.5\textwidth}
    \centering\vspace*{4pt}
    \includegraphics[width=7.1cm]{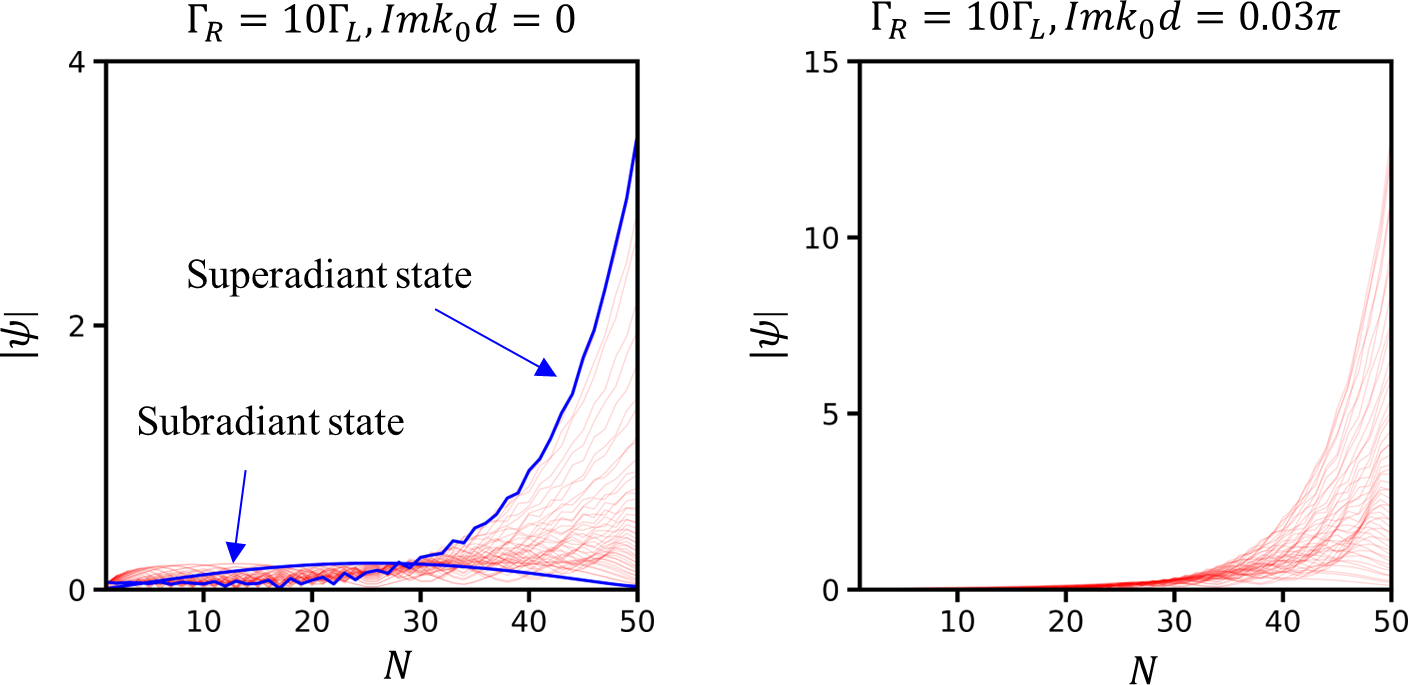}\vspace*{2pt}
    \end{minipage}
                     \\
    \midrule[0.5pt] \begin{minipage}[m]{.11\textwidth}
      Origin for\\ skin effect
  \end{minipage}  & \begin{minipage}[m]{0.3\textwidth}
      \centering\vspace*{4pt}
      	\includegraphics[width=3.3cm]{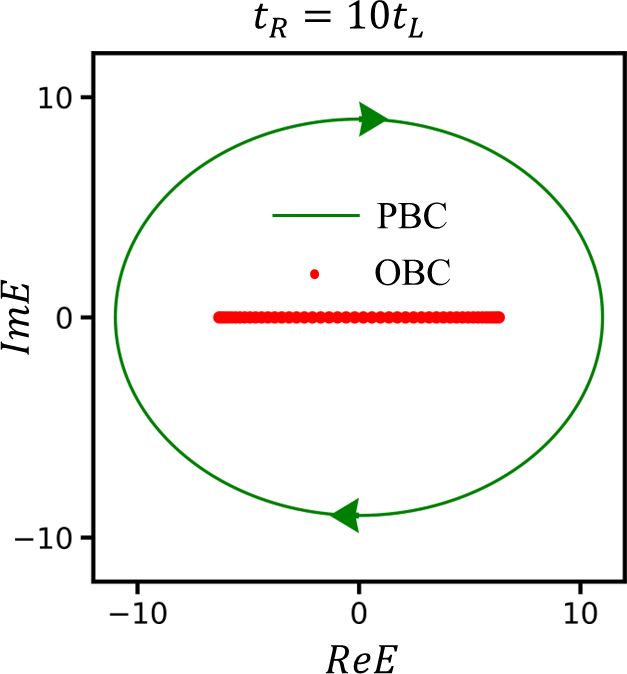}\vspace*{2pt}
    \end{minipage} & \begin{minipage}[m]{.5\textwidth}
    \centering\vspace*{4pt}
    \includegraphics[width=7.1cm]{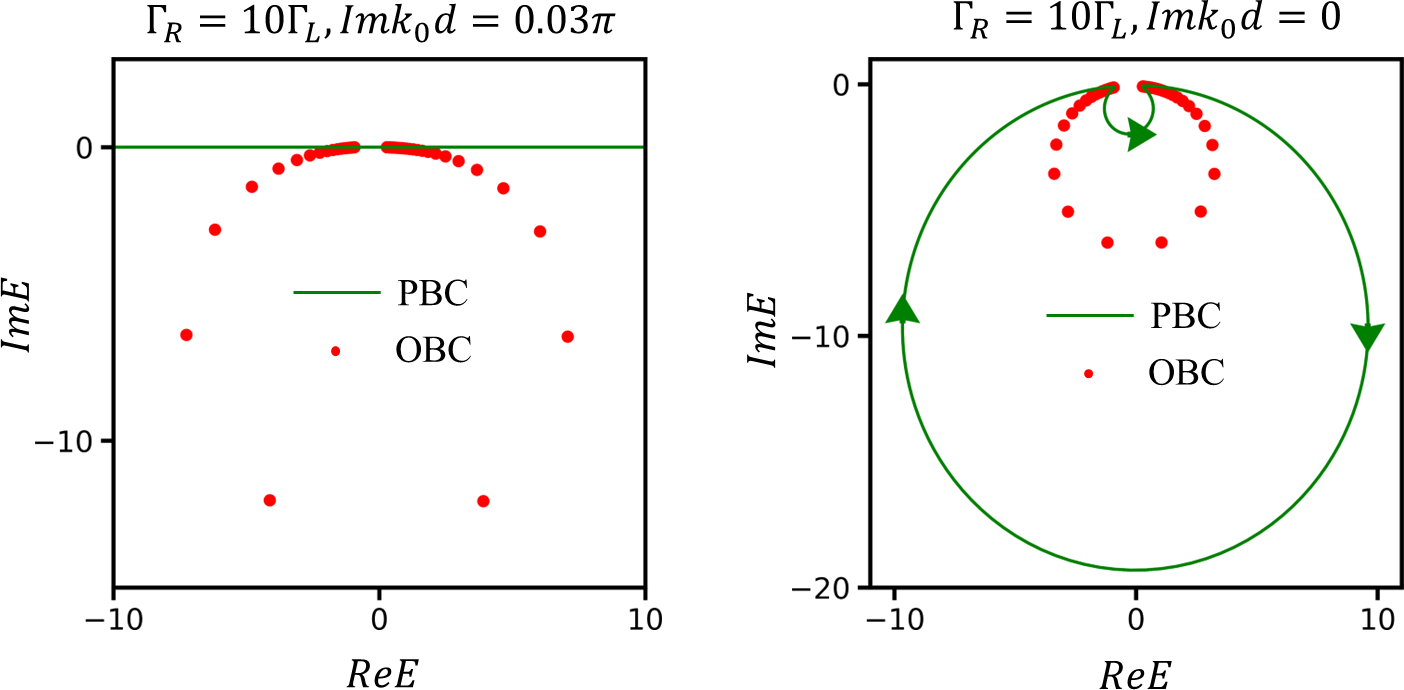}\vspace*{2pt}
                     \end{minipage}
                     \\
        \midrule[0.5pt] \begin{minipage}[m]{.11\textwidth}
      Response to \\ point defect
  \end{minipage}  & \begin{minipage}[m]{0.3\textwidth}
      \centering\vspace*{4pt}
      	\includegraphics[width=3.3cm]{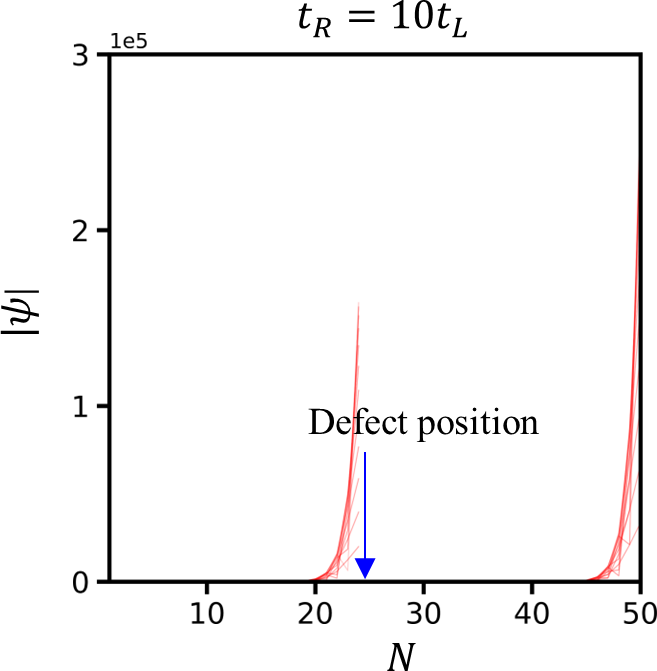}\vspace*{2pt}
    \end{minipage} & \begin{minipage}[m]{.5\textwidth}
                      \centering\vspace*{4pt}
      	\includegraphics[width=7.1cm]{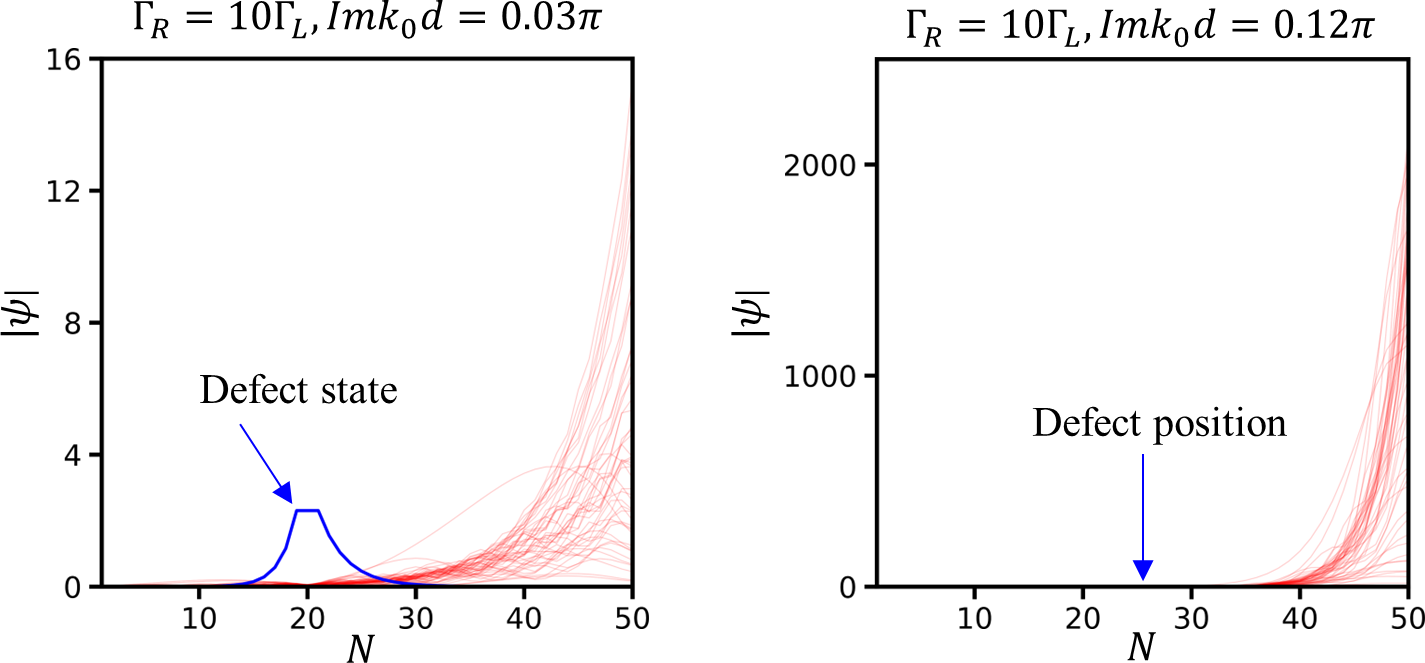}\vspace*{0pt}
                     \end{minipage}
                     \\
   \toprule[1pt]
  \end{tabular}
\end{table}

In Table~\ref{fig:SK1co}, the Hatano-Nelson model is a typical system with short-range and asymmetric hopping $t_R$ from the left to right and $t_L$ from the right to left \cite{hatano1996localization}, as addressed in Sec.~\ref{Nonhermitian_skin_effect}. While a system with long-range coupling is illustrated by the coupled magnetic nanowires array in proximity to a magnetic film, in which the long-range coupling is mediated by the spin waves of the magnetic film \cite{yu2022giant}.
In the latter case, the range 
 of coupling depends on the attenuation length of spin waves, and thereby a magnetic film with  attenuation parametrized by $\text{Im}k_0d$ allows for a long but finite-range coupling between magnetic nanowires.
The long-range coupling can also be chiral with $\Gamma_L\ne \Gamma_R$, where $\Gamma_L$ denotes the coupling directed from right to left and $\Gamma_R$ from left to right.

With the chirality $t_R=10t_L$ in the Hatano-Nelson model, all the states accumulate at the right boundary. 
While with the chirality $\Gamma_R=10\Gamma_L$ alone without attenuation ${\rm Im}k_0d=0$ in the system, the long-range coupling only drives a few states to the boundary. The combination of chirality and attenuation of the mediator leads to the non-Hermitian skin effect for such systems. Though the  conditions for realizing the non-Hermitian skin effect are distinct in both systems, they share the same topological origin: the non-Hermitian skin effect emerges only when the spectra under the PBC, as shown by the green color, encircle the  spectra under the OBC, as denoted by the red dots on the complex plane (refer to the third row in Table~\ref{fig:SK1co}).

The localization of non-Hermitian skin modes is fragile to point defects such as vacancy in the Hatano-Nelson model, as calculated in the fourth row in Table~\ref{fig:SK1co}, since by a defect the system is divided into two independent subsystems.
By contrast, the non-Hermitian skin modes in the system with long-range coupling are robust to defects, manifesting as a large amplitude at the right boundary for all of the states \cite{zeng2023radiation} except for the appearance of the defect state at the defect position \cite{qi2018defect,malzard2015topologically,stegmaier2021topological}.

\subsubsection{Long-range chiral interaction between magnets}
\label{waveguides}

 The magnetization of the collective mode of ferromagnets, i.e. magnons, rotates counterclockwise instead of clockwise around the external magnetic field. 
 So their handedness is fixed and the time-reversal symmetry is broken. 
 They tend to couple chirally with the chiral vector fields that are characterized by the generalized spin-orbit interaction, i.e., a locking of the propagation direction and their vector rotation direction \cite{yu2023chirality}. Chiral interaction between magnons of  magnetic wires or spheres has been discovered widely in the experiments when they couple with other traveling modes, such as the spin waves \cite{au2012resonant,yu2019chiral,yu2020magnon,wang2021nonreciprocal} in films, waveguide microwaves \cite{yu2020chiral,zhang2020broadband}, surface acoustic waves \cite{yu2020nonreciprocal,yamamoto2020non}, and Cooper pairs \cite{yu2022efficient,borst2023observation,zhou2023gating} to name a few.

\begin{figure}[!htp]
\centering
 \includegraphics[width=0.98\textwidth]{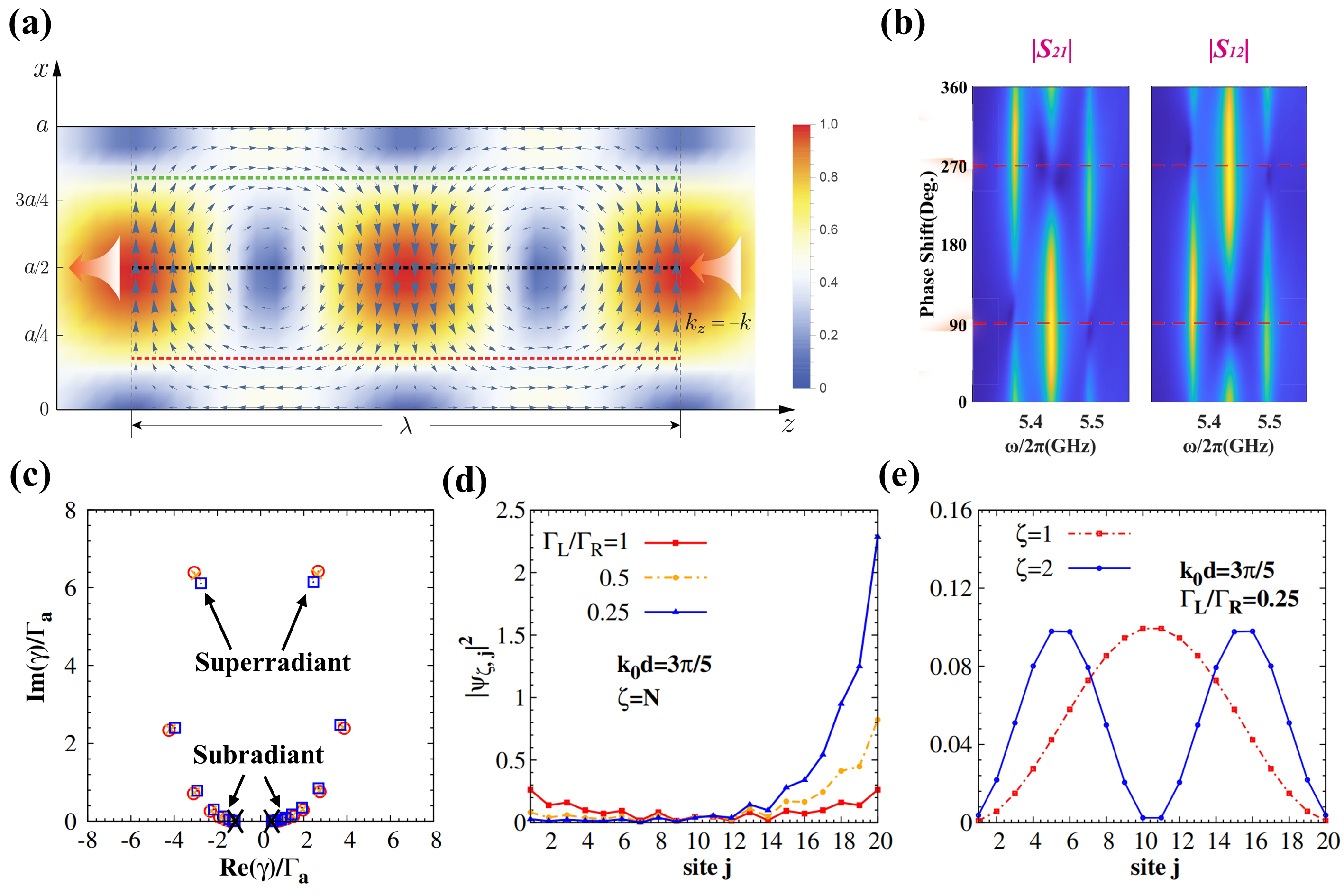}
	\caption{Origin of chirality for a system with long-range coupling and the effect of chirality on the collective magnon modes. (a) shows the spatial distribution of the magnetic field of the $TE_{10}$ mode with the direction and magnitude indicated by the arrow. (b) shows the dependence of transmission mapping $\left|S_{21}\right|$ and $\left|S_{12}\right|$ on the phase shift of the microwaves input from the two arms of the cross-shaped cavity. (c) Normalized frequencies on the complex plane, where the superradiant (subradiant) states are marked. (d) shows the dependence of the spatial distribution of superradiant states on the chirality and (e) shows the first two subradiant states. Figure~(a) is reproduced with permission from Ref.~\cite{yu2020chiral}. Figure~(b) is adapted with permission from Ref.~\cite{zhong2022controlling}. Figures~(c)-(e) are adapted with permission from Ref.~\cite{yu2020magnon_accumulation}.} 
	\label{fig:chirality}
\end{figure}

For example, when the magnetic spheres are loaded in a microwave waveguide at special planes, they can couple chirally with the microwaves, i.e., it depends on the propagation directions of microwaves \cite{yu2020chiral,yu2020magnon_accumulation}. The total Hamiltonian for such a hybridized system typically has a general form \cite{yu2020chiral,yu2020magnon_accumulation}
\begin{equation}
    \hat{H}/\hbar=\sum_l\omega_{{\rm K}} \hat{\beta}_{l}^{\dagger}\hat{\beta}_{l}+\sum_{k}\omega_{k}\hat{\alpha}^{\dagger}_{k}\hat{\alpha}_{k}+\sum_l\sum_{k}\left(g_{k}e^{-ikR_l}\hat{\beta}_{l}\hat{\alpha}^{\dagger}_{k}+g_{k}e^{ikR_l}\hat{\beta}_{l}^{\dagger}\hat{\alpha}_{k}\right),
\end{equation}
where $\omega_{{\rm K}}$ and $\hat{\beta}_{l}^{\dagger}$  are the frequency and creation operator of the Kittel magnon mode in the magnetic spheres and $R_l$ denotes the locations of the $l$-th magnetic sphere, while  $\omega_{k}$ and $\hat{\alpha}_{k}^{\dagger}$  are the frequency and creation operator of microwave photons in the waveguide. The third term describes the conversion between magnon and photon by the dipolar interaction between magnetization and microwave magnetic field parametrized by coupling constant $g_k$, which is chiral when $\left|g_k\right|\neq\left|g_{-k}\right|$ \cite{yu2020chiral,yu2020magnon_accumulation}. Mediated by the microwave photons,  the magnons in different spheres interact chirally with each other in a long-range \cite{yu2020chiral,yu2020magnon_accumulation}, as illustrated in Fig.~\ref{fig:chirality}(a). The chirality of microwaves can be inferred as follows. For the ${\rm TE}_{10}$ mode in Fig.~\ref{fig:chirality}(a), the gray arrows represent a snapshot of the magnetic field distribution. Consider the propagation of microwaves on the two special planes guided by the red and green lines. When the waves propagate at the green line from left to right, 
the arrows rotate clockwise, while they rotate counterclockwise when the waves travel from right to left, i.e., the propagation direction is locked to the circular polarization of the magnetic field. When we load an ensemble of magnetic spheres saturated along the $\hat{\bf y}$-direction over the red line (green line) in the waveguide, the Kittel magnon in any sphere only couples with the waves propagating to the left (right) by the matched chirality.  The emitted photon by one sphere can be absorbed by another which mediates a chiral coupling between magnons. Further, considering the negligible dissipation of photon propagation, the coupling length is sufficiently long compared to the length of the array of magnetic spheres.

Recently, Zhong \textit{et al.} observed the chiral coupling effect between Kittel magnons in the YIG magnetic sphere and microwave photons in a cross-shaped cavity, where the field polarizations change as well at different positions  \cite{zhong2022controlling}. They  observed the nonreciprocal transmission property of the coupling system by observing different transmission mapping $|S_{21}|$ and $|S_{12}|$, as shown in  Fig.~\ref{fig:chirality}(b).

 When loaded with an array of $N$ magnetic spheres in the microwave waveguide, the microwave photon mediates a long-range interaction between the Kittel magnons in different magnetic spheres. The motion of magnons is then collective, governed by the effective Hamiltonian \cite{yu2020chiral,yu2020magnon_accumulation}
 \begin{equation}\label{effectiveinteraction}
     \hat{H}_{\mathrm{eff}}/\hbar  = \left(\omega_{{\rm K}} - \frac{\Gamma_R+\Gamma_L}{2}\right)\sum_l \hat{\beta}_{l}^{\dagger}\hat{\beta}_{l} -i \Gamma_L \sum_{l < l^{\prime}} e^{ik_0\left|R_l-R_{l^{\prime}}\right|}\hat{\beta}_{l}^{\dagger}\hat{\beta}_{l^{\prime}} -i \Gamma_R \sum_{l > l^{\prime}} e^{ik_0\left|R_l-R_{l^{\prime}}\right|}\hat{\beta}_{l}^{\dagger}\hat{\beta}_{l^{\prime}},
 \end{equation}
 where $\Gamma_L$ and $\Gamma_R$ describe the coupling between magnon directed from right to left and from left to right, respectively, which can be nonreciprocal with different magnitudes  $|\Gamma_L|\ne |\Gamma_R|$ depending on the coupling $g_k$ between magnetic spheres and waveguide \cite{yu2020chiral,yu2020magnon_accumulation}. Here $k_0$ is the corresponding resonant wavevector of photon modes for magnon frequency.
In comparison to the Hatano-Nelson model (\ref{Hatano_Nelson}), where the chiral coupling is sufficient for realizing the non-Hermitian skin effect \cite{hatano1996localization}, the non-Hermitian skin effect may be expected since the energy tends to accumulate at one boundary. However, the studies revealed more complicated features when the interaction is of infinite long-range \cite{yu2020chiral,yu2020magnon_accumulation}.

 The frequency $\gamma$ of $N$ collective modes, divided by the spontaneous radiation rate $\Gamma_a=(\Gamma_L+\Gamma_R)/2$, is shown in Fig.~\ref{fig:chirality}(c), where the superradiant states have a decay rate much larger than $\Gamma_a$ and the subradiant states have a decay rate much smaller \cite{asenjo2017exponential,chang2018colloquium,zhang2019theory}. We introduce a band index $\zeta=\{1,2,..., N\}$ according to the increased decay rates.  $\zeta=1,2$ labels the first two subradiant states and $\zeta=N$ labels the most superradiant state.
 The spatial distribution of the typical superradiant state and first two subradiant states is shown in Fig.~\ref{fig:chirality}(d) and (e), respectively. The subradiant states extend over the bulk, forming the standing waves, and are rarely affected by the chirality $\Gamma_L/\Gamma_R = 0.25$. While the superradiant state possesses a large amplitude at the boundary without chirality, and this skewness becomes more significant with increased chirality, as shown in Fig.~\ref{fig:chirality}(d). However, the skewness of the superradiant state to the boundary does not originate from the non-Hermitian skin effect, since it typically requires almost all of the eigenstates to pile up to the boundary as stated in Sec.~\ref{Nonhermitian_skin_effect}.

Although the chirality is sufficient for inducing the non-Hermitian skin effect in the Hatano-Nelson model \cite{hatano1996localization},  it only drives skin tendency in the system with long-range coupling. Such distinction motivates our recent studies on the condition of the non-Hermitian skin effect for the system with long-range coupling and unique functionalities \cite{yu2022giant,zeng2023radiation}.

\subsubsection{Non-Hermitian skin effect driven by long-range chiral interaction}
\label{magnetic_wires}

\begin{figure}[!htp]
	\centering
	\includegraphics[width=0.95\textwidth]{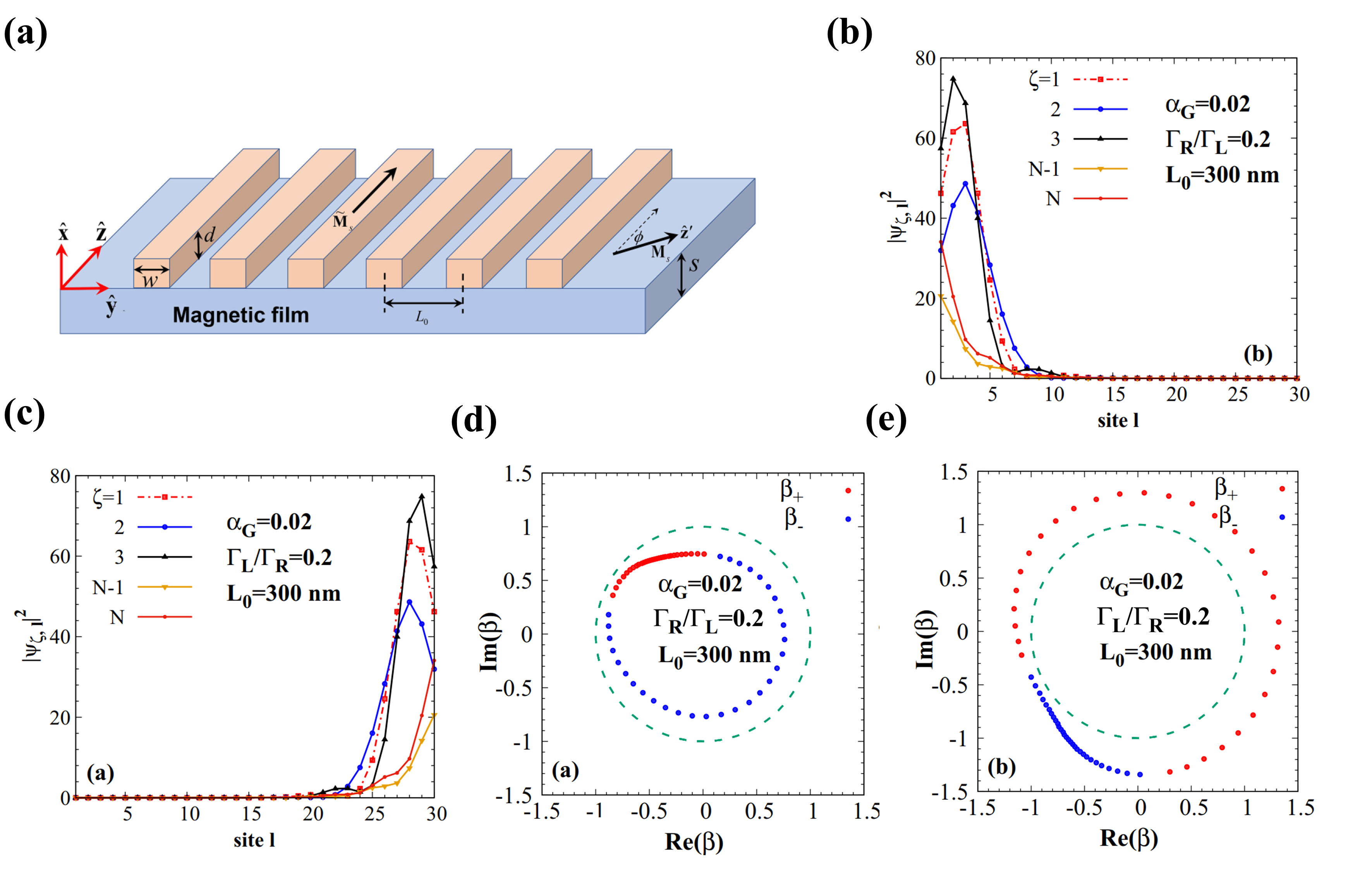}
	\caption{Prediction of the magnonic non-Hermitian skin effect in a periodic magnetic nanowire array deposited on the thin magnetic film. (a) is the configuration. In the presence of dissipation of spin wave in the film parametrized by $\alpha_G=0.02$, all the magnonic states of the subsystem of wires accumulate at the left boundary with chirality $\Gamma_R/\Gamma_L=0.2$ [(b)] and  right boundary with chirality $\Gamma_L/\Gamma_R=0.2$ [(c)]. For the corresponding complex wavevector,  $\left|\beta\right|<1$ with chirality $\Gamma_R/\Gamma_L=0.2$ [(c)] and $\left|\beta\right|>1$ with chirality $\Gamma_L/\Gamma_R=0.2$ [(e)], respectively. These figures are adapted with permission from Ref.~\cite{yu2022giant}.} 
	\label{fig:magnonicSK}
\end{figure}

It should be noted that the major difference between the Hatano-Nelson model \cite{hatano1996localization} and the system with a long-range coupling \cite{asenjo2017exponential,chang2018colloquium,zhang2019theory,yu2020chiral,yu2020magnon_accumulation} is the distinct coupling range. The microwaves propagate in a long range, but the spin waves propagate at a shorter distance. 
From this point of view, we consider a periodic array of $N$ magnetic nanowires fabricated on a thin magnetic film, as illustrated in Fig.~\ref{fig:magnonicSK}(a), where the interaction between the Kittel magnon in the wires with the spin waves in the film is chiral \cite{yu2019chiral,yu2019chiral_2,chen2019excitation,yu2020magnon_accumulation,yu2020chiral,yu2020magnon,yu2022giant} and the coupling range is limited by attenuation of traveling spin waves in the magnetic film.
The magnetic nanowires are separated by a large distance, which allows disregarding their mutual dipolar interaction.

\textbf{Microscopic model}.---The saturated magnetization ${\bf M}_s$ of the magnetic film is biased by the applied magnetic field ${\bf H}_{\rm{app}}$, while the saturated magnetization $\tilde{\bf M}_s$ of the wire is pinned along the wire by shape anisotropy, which is in an angle $\phi$ with respect to ${\bf H}_{\rm{app}}$.
In thin magnetic films, the magnetization precession is circular. In terms of the magnon operator $\hat{\alpha}_{k}$ in the film and 
$\hat{\beta}_{l}$ in the $l$-th magnetic wire,  the magnetization operators can be expanded by 
\cite{holstein1940field,kamra2016super}
\begin{align}
\nonumber
&\hat{M}_{x}(\mathbf{r})=\sqrt{2M_s\gamma\hbar}\sum
_{k}\left(m_{x}^{(k)}(x)e^{iky}\hat{\alpha}_{k}
+{\rm H.c.}\right),\\
&\hat{M}_{y}(\mathbf{r})=\cos\phi\sqrt{2M_s\gamma\hbar}\sum
_{k}\left(im_{x}^{(k)}(x)e^{iky}\hat{\alpha}_{k}
+{\rm H.c.}\right),\nonumber\\
&\hat{\tilde{M}}_{\alpha=\{x,y\},l}(\mathbf{r})=\sqrt{2\tilde{M}_{s}\gamma\hbar}
\left(\tilde{m}_{l,\alpha}^{\rm K}({\bf r})\hat{\beta}_{l}
+{\rm H.c.}\right),
\label{expansion}
\end{align} 
where $m^{(k)}_{x}(x)$ and $\tilde{m}_{l,\alpha}^{\rm K}({\bf r})$ represent the normalized amplitudes of the spin-wave eigenmodes in the wire and film. The coupling between the magnetic nanowires and the film is governed by the dipolar field $h_{\beta}(x,\pmb{\rho})$ from the spin wave in the film, as given by Eq.~\eqref{free_energy}, which is written as  \cite{landau1984ld,jackson1999classical}
\begin{equation}
\hat{H}_{\rm int}=-\mu_0\int_0^d dxd\pmb{\bf \rho}\tilde{M}_{l,\beta}(x,\pmb{\rho})h_{\beta}(x,\pmb{\rho}),
\label{interaction}
\end{equation}  
in the summation convention over repeated Cartesian
indices $\beta=\{x,y,z\}$.
After quantization, the total Hamiltonian for this coupled system reads
\begin{align}
\hat{H}/\hbar=\sum_l\left( \omega_{{\rm K}}-i \frac{\eta}{2}\right)\hat{m}_{l}^{\dagger}\hat{m}_{l}+\sum_{k}\left(\omega_{k}-i\frac{\eta_k}{2}\right)\hat{\alpha}^{\dagger}_{k}\hat{\alpha}_{k}+\sum_l\sum_{k}\left(g_{k}e^{-ikR_l}\hat{m}_{l}\hat{\alpha}^{\dagger}_{k}+g_{k}e^{ikR_l}\hat{m}_{l}^{\dagger}\hat{\alpha}_{k}\right),
\label{total_Hamiltonian}
\end{align}
where $\omega_{{\rm K}}$ and $\hat{m}_{l}^{\dagger}$ are the frequency and creation operator of Kittel magnon in the $l$-th nanowire and $R_l=ld$ is the location with the distance $d$ between nanowires, while $\omega_k=\mu_0 \gamma \left({H}_{\rm app} + \alpha_{\rm ex}{M}_sk^2\right)$ with the exchange stiffness $\alpha_{\rm ex}$ and $\hat{\alpha}^{\dagger}_{k}$ are the frequency and creation operator of the film magnon with wave vector $k$.
Here, $\eta=2\tilde{\alpha}_{\mathrm{G}}\omega_{\mathrm{K}}$ and $\eta_{k}=2\alpha_{\mathrm{G}}\omega_{k}$ represent the intrinsic dissipation for magnetic nanowires and film with Gilbert damping constants $\tilde{\alpha}_{\mathrm{G}}$ and $\alpha_{\mathrm{G}}$. 
The third term in Eq.~(\ref{total_Hamiltonian}) describes the conversion of magnons between films and nanowires by the dipolar interaction with coupling constant $g_k$ \cite{yu2022giant}
\begin{align}
g_{k}=F(k)m_x^{(k)*}\left(|k|+k\cos\phi\right)\left(\tilde{m}_{x}^{\rm K}+i{\rm sgn}(k)\tilde{m}^{\rm K}_{y}\right),
\end{align}
where $F(k)$ is the form factor governed by the geometry of the wire and film. The coupling constant depends on the propagation direction of the spin wave, and the angle $\phi$ between ${\bf M}_s$ and $\tilde{\bf M}_s$. For specified angles $\phi=0$ and $\pi$, the coupling between the magnetic nanowire and film $g_{-\left|k\right|}=0$ and $g_{\left|k\right|}=0$, exhibiting complete chirality. A tunable non-reciprocity can be realized by changing $\phi$.

We derive the effective interaction between magnetic nanowires in terms of the mean-field equation of motion of   $\hat{\alpha}_{k}$ and  $\hat{m}_{l}$ in the frequency domain:
\begin{equation}
    \begin{aligned}
           \left(\omega- \omega_{{\rm K}}+ i\frac{\eta}{2}\right) \hat{m}_{l}& = \sum_{k}g_{k}e^{ikR_l} \hat{\alpha}_{k}, \\
        \left(\omega- \omega_{k} +i\frac{\eta_k}{2}\right) \hat{\alpha}_{k}& = \sum_{l}g_{k}e^{-ikR_l}\hat{m}_{l},
    \end{aligned}
\end{equation}
which yields an effective interaction between magnons in magnetic nanowires
\begin{equation}
    \left(\omega-  \omega_{{\rm K}}+i \frac{\eta}{2}\right) \hat{m}_{l} = \sum_{l^{\prime}} \Gamma_{ll^{\prime}}\hat{m}_{l^{\prime}}.
\end{equation}
The effective coupling constants 
\begin{align}
\Gamma_{ll^{\prime}}(\omega)=\sum_{k}\frac{g_{k}^2  e^{ik\left(R_l-R_{l^{\prime}}\right)} }{\omega- \omega_{k} +i\frac{\eta_k}{2}}=\left\{
\begin{matrix}
-i\frac{\Gamma_L+\Gamma_R}{2},\quad~~~~~~\mathrm{when}~~~~R_{l}=R_{l^{\prime}} \\
-i\Gamma_R e^{ik_0\left(R_{l} - R_{l^{\prime}}\right)},\quad~\mathrm{when}~R_{l}>R_{l^{\prime}}\\
-i\Gamma_L e^{ik_0\left( R_{l^{\prime}}-R_{l}\right)},\quad\mathrm{when}~~R_{l}<R_{l^{\prime}}
\end{matrix}
\right.  .
\label{coupling}
\end{align}
Here, $k_0\approx \left(1+i\alpha_G/2\right)k_{\omega}$ is not purely real with $k_{\omega}=\sqrt{\left(\omega-\mu_0 \gamma {H}_{\rm app}\right)/\left(\mu_0 \gamma \alpha_{\rm ex}M_s\right)}$. 
The coupling constant $\Gamma_R=L|g_{k_{0}}|^{2}/v(k_{0})$ is mediated by waves propagating from left to right and $\Gamma_L=L|g_{-k_{0}}|^{2}/v(k_{0})$ from right to left, where $L$ and $v(k_{0})$ is the length of the film along $\hat{\bf y}$-direction and the group velocity of spin waves in the film. $|\Gamma_R|\ne |\Gamma_L|$ can be non-reciprocal with different magnitudes \cite{yu2022giant}. 
The effective Hamiltonian 
$\hat{H}_{\mathrm{eff}} = \hat{M}^{\dagger}\tilde{H}_{\mathrm{eff}} \hat{M}$ can be written in the form of a non-Hermitian matrix
\begin{equation}
\tilde{H}_{\mathrm{eff}}/\hbar=%
\begin{pmatrix}
	\omega _{\mathrm{K}}-i\frac{\eta}{2}-i\frac{\Gamma _{R}+\Gamma _{L}}{2} & 
			-i\Gamma _{L}e^{ik_0d} & -i\Gamma _{L}e^{2ik_0d} & \dots & -i\Gamma
			_{L}e^{i(N-1)k_0d} \\ 
			-i\Gamma _{R}e^{ik_0d} & \omega _{\mathrm{K}}-i\frac{\eta}{2}-i\frac{\Gamma
				_{R}+\Gamma _{L}}{2} & -i\Gamma _{L}e^{ik_0d} & \dots & -i\Gamma
			_{L}e^{i(N-2)k_0d} \\ 
			-i\Gamma _{R}e^{2ik_0d} & -i\Gamma _{R}e^{ik_0d} & \omega _{\mathrm{K}}-i\frac{\eta}{2}-i\frac{\Gamma _{R}+\Gamma _{L}}{2} & \dots & -i\Gamma
			_{L}e^{i(N-3)k_0d} \\ 
			\vdots & \vdots & \vdots & \ddots & \vdots \\ 
			-i\Gamma _{R}e^{i(N-1)k_0d} & -i\Gamma _{R}e^{i(N-2)k_0d} & -i\Gamma
			_{R}e^{i(N-3)k_0d} & \dots & \omega _{\mathrm{K}}-i\frac{\eta}{2}-i\frac{\Gamma
				_{R}+\Gamma _{L}}{2}%
		\end{pmatrix}%
		,  \label{H_matrix}
	\end{equation} 
 where $\hat{M}=\left(\hat{m}_1,\hat{m}_2,\cdots,\hat{m}_N\right)^T$.

 In terms of the biorthogonal  eigenvectors $\left<\Psi_l \right|$ and $\left|\Phi_l\right>$  of matrix \eqref{H_matrix} with state index $l\in\{1,2,...,N\}$, Eq.~(\ref{H_matrix}) should be diagonalized as 
\begin{equation}\label{longrangematrixdiagonal}
    \left<\Psi_l \right| \tilde{H}_{\mathrm{eff}}/\hbar \left|\Phi_l\right> = \omega_l.
\end{equation}
  The eigenvector can be constructed as a superposition of Bloch basis $\left|\beta_{\kappa_n}\right> = \left(\beta_{\kappa_n},\beta_{\kappa_n}^{2},...,\beta_{\kappa_n}^{N}\right)^{T}$, where $\beta _{\kappa} \equiv \exp({i \kappa d})$ is parameterized by the complex wave vector $\kappa$, i.e.,
 \begin{align}
 \left|\Phi_l\right> = \sum_{n} \alpha^{(l)}_n \left|\beta_{\kappa_n}\right>,
 \end{align}
 with the superposition coefficients $\alpha^{(l)}_n$.
For those systems with short-range chiral coupling such as the Hatano-Nelson model, the eigenvalue and eigenstates can be obtained by solving the eigenvalue equation with only two boundary conditions as stated in Sec.~\ref{Nonhermitian_skin_effect}.
However, for the long-range coupling in Eq.~\eqref{H_matrix},  the number of boundary equations is proportional to the system size, which renders it a challenge to solve the eigenvalues and eigenstates in terms of the approach introduced in Sec.~\ref{Nonhermitian_skin_effect}. Here, we introduce another approach to solving the eigenvalue problem.

The effective Hamiltonian $\hat{H}_{\mathrm{eff}}/\hbar = \sum_{l} \hat{M}^{\dagger} \left|\Phi_l\right> \omega_l \left<\Psi_l \right|  \hat{M}$.
Accordingly, in terms of the biorthogonality condition $\left<\Psi_l|\Phi_{l'}\right>=\delta_{ll'}$, $\left[\left<\Psi_l \right|  \hat{M}, \hat{M}^{\dagger} \left|\Phi_{l'}\right> \right]=\delta_{ll'}$, which leads to
\begin{equation}
\label{longrangeeigen1}
    \left[\hat{H}_{\mathrm{eff}}/{\hbar}, \hat{M}^{\dagger}  \left|\Phi_l\right>\right]=  \omega_l \hat{M}^{\dagger} \left|\Phi_l\right>.
\end{equation}
On the other hand, for the Bloch basis 
\begin{equation}
\label{eigenequation}
	 \left[\hat{H}_{\mathrm{eff}}/{\hbar}, \hat{M}^{\dagger}  \left|\beta_{\kappa}\right>\right]= \omega_{\kappa} \hat{M}^{\dagger}  \left|\beta_{\kappa}\right> + i \Gamma_R g_{\kappa} \hat{M}^{\dagger}  \left|\beta_{k_0}\right> - i \Gamma_L h_{\kappa} \hat{M}^{\dagger}  \left|\beta_{-k_0}\right>,
\end{equation}
with a complex dispersion relation 
\begin{equation}
\label{longrangeeigen}
\omega_{\kappa}=\omega_{\mathrm{K}}-i\frac{\eta}{2}
		-i\frac{\Gamma_R}{2}\frac{\beta_{\kappa}+\beta_{k_0}}{\beta_{\kappa}-\beta_{k_0}}-i\frac{\Gamma_L}{2}\frac{1+\beta_{\kappa}\beta_{k_0}}{1-\beta_{\kappa}\beta_{k_0}},
	\end{equation}
and $g_{\kappa}={\beta_{\kappa}}/({\beta_{\kappa}-\beta_{k_0}})$ and $h_{\kappa}={\left(\beta_{\kappa}\beta_{k_0}\right)^{N+1}}/({\beta_{\kappa}\beta_{k_0}-1})$ are dimensionless that describe the leakage into the modes with (complex) wave vector $\pm k_0$ via microwave photons.

The degenerate frequency 
\begin{align}
\omega_{\kappa_1}=\omega_{\kappa_2}
    \label{condition_a}
\end{align}
contains two roots $\{\beta_{\kappa_1},\beta_{\kappa_2}\}$, which allows to construct the eigenstate as a superposition $\left|\Phi_l\right>=\alpha^{(l)}_1\left|\beta_{\kappa^{(l)}_1}\right>+\alpha^{(l)}_2\left|\beta_{\kappa^{(l)}_2}\right>$ in terms of two coefficients $\alpha^{(l)}_1$ and $\alpha^{(l)}_2$. Below we drop the band index $l$. With Eq.~\eqref{eigenequation} we find
    \begin{align}
    [\hat{H}_{\mathrm{eff}}/{\hbar}, \hat{M}^{\dagger} \left( \alpha_1\left|\beta_{\kappa_1}\right>+\alpha_2\left|\beta_{\kappa_2}\right>\right)] &=\omega_{\kappa_{1,2}} \hat{M}^{\dagger}  \left( \alpha_1\left|\beta_{\kappa_1}\right>+\alpha_2\left|\beta_{\kappa_2}\right>\right) + i \Gamma_R \left( \alpha_1 g_{\kappa_1} +\alpha_2 g_{\kappa_2} \right)\hat{M}^{\dagger}  \left|\beta_{k_0}\right>  \nonumber\\
    & - i \Gamma_L \left(\alpha_1 h_{\kappa_1}+\alpha_2 h_{\kappa_2}\right) \hat{M}^{\dagger}  \left|\beta_{-k_0}\right>.
        \end{align}
$\omega_{\kappa_1}=\omega_{\kappa_2}$ is the eigenfrequency only when the last two terms vanish, i.e., 
\begin{equation}
    i \Gamma_R \left( \alpha_1 g_{\kappa_1} +\alpha_2 g_{\kappa_2} \right)\hat{M}^{\dagger}  \left|\beta_{k_0}\right> 
    - i \Gamma_L \left(\alpha_1 h_{\kappa_1}+\alpha_2 h_{\kappa_2}\right) \hat{M}^{\dagger}  \left|\beta_{-k_0}\right> = 0,
\end{equation} which for two linearly independent $\left|\beta_{k_0}\right>$ and $\left|\beta_{-k_0}\right>$ requests 
\begin{equation}
    \left(\begin{array}{cc}
       g_{\kappa_1}  & g_{\kappa_2} \\
       h_{\kappa_1}  & h_{\kappa_2}
    \end{array}\right)\left( \begin{array}{c}
         \alpha_1  \\
         \alpha_2
    \end{array}\right)=0.
\end{equation}
It has a solution when 
\begin{equation}
g_{\kappa_2}h_{\kappa_1}=g_{\kappa_1}h_{\kappa_2}.
 \label{condition_b}
\end{equation}
We solve $\kappa_1$ and $\kappa_2$ in terms of Eqs.~(\ref{condition_a}) and (\ref{condition_b}),
with which the eigenstate becomes
\begin{equation}\label{sposition}
    \left|\Phi\right> = \frac{ \beta_{\kappa_2}}{ \beta_{\kappa_2}-\beta_{k_0}} \left|\beta_{\kappa_1}\right> -\frac{ \beta_{\kappa_1}}{ \beta_{\kappa_1}-\beta_{k_0}} \left|\beta_{\kappa_2}\right>,
\end{equation}
up to a normalized constant. 
So the wave function at the site $i$ 
\begin{equation}
    \Phi^{(i)} = \frac{ \beta_{\kappa_2}}{ \beta_{\kappa_2}-\beta_{k_0}} \left(\beta_{\kappa_1}\right)^{i} -\frac{ \beta_{\kappa_1}}{ \beta_{\kappa_1}-\beta_{k_0}} \left(\beta_{\kappa_2}\right)^{i}.
\end{equation}
When $|\beta_{\kappa_{1,2}}|>1$ ($|\beta_{\kappa_{1,2}}|<1$), the amplitude of the magnonic mode increases with the increase of the site number, leading to the localization at the right (left) boundary.

The Hamiltonian \eqref{H_matrix} maps to a system with short-range coupling only when a sufficient strong attenuation is imposed by the dissipation of spin waves.
The magnonic collective modes under a combination of chirality and attenuation of spin waves are plotted in Fig.~\ref{fig:magnonicSK}(b) and (c) with the band index $\zeta=\{1,2,..., N\}$ according to the increased decay rates. 
With the attenuation of spin waves $\alpha_G=0.02$, all the states including the subradiant states $\zeta=\{1,2,3\}$ and superradiant states $\zeta=\{N-1,N\}$ accumulate at the left boundary of the array with chirality $\Gamma_R/\Gamma_L=0.2$, but become skewed at the right boundary with opposite chirality  $\Gamma_L/\Gamma_R=0.2$. We can understand such skin effect in terms of the allowed complex wave vectors $\beta_{\kappa_{1,2}}$, which  can be solved from dispersion relation Eq.~\eqref{longrangeeigen} with eigenvalues from matrix diagonalization, as labeled by $\beta_{\pm}$ in Fig.~\ref{fig:magnonicSK}(d) and (e), respectively.
The distribution of complex wavevector $\left|\beta_{\pm}\right|<1$ with chirality $\Gamma_R/\Gamma_L=0.2$ corresponds to the localization of eigenstates at the left boundary [Fig.~\ref{fig:magnonicSK}(b)], while $\left|\beta_{\pm}\right|>1$ with chirality $\Gamma_L/\Gamma_R=0.2$ explains the localization at the right boundary [Fig.~\ref{fig:magnonicSK}(c)].

\textbf{Relation to Hatano-Nelson model}.---The method addressed above is applied to systems with short-range coupling as well. We illustrate the Hatano-Nelson model (Sec.~\ref{Nonhermitian_skin_effect}) as an example with the Hamiltonian written as 
\begin{equation}
     \hat{H} = \sum_{l} \hat{C}^{\dagger} \left|\Phi_l\right> E_l \left<\Psi_l\right|  \hat{C},
     \label{HN_extension}
\end{equation}
where $\hat{H}$ is the Hamiltonian \eqref{Hatano_Nelson}, $\hat{C}^{\dagger}=\left(\hat{c}_{1}^{\dagger},\hat{c}_{2}^{\dagger},\cdots,\hat{c}_{N}^{\dagger}\right)$, and $E_l$ is the eigenvalue. With Eq.~(\ref{HN_extension}) the commutation relation 
\begin{equation}\label{EigenHatano}
    	 [\hat{H}, \hat{C}^{\dagger}  \left|\Phi_l\right>] = E_{\kappa}  \hat{C}^{\dagger}  \left|\Phi_l\right>.
\end{equation}
On the other hand, in terms of the Bloch basis, we substitute Eq.~\eqref{Hatano_Nelson} into $[\hat{H}, \hat{C}^{\dagger}  \left|\beta_{\kappa}\right>]$ and find 
\begin{equation}\label{eigeneuqation2}
	 [\hat{H}, \hat{C}^{\dagger}  \left|\beta_{\kappa}\right>] =  E_{\kappa}\hat{C}^{\dagger}  \left|\beta_{\kappa}\right> - t_L {\hat{c}_{N}^{\dagger}}\beta_{\kappa}^{N+1}- t_R \hat{c}_{1}^{\dagger},
\end{equation}
where $E_{\kappa}=t_{L} \beta_{\kappa}+t_R/\beta_{\kappa}$.

Since when degenerate  $E_{\kappa}$ contains two roots $\{\beta_{\kappa_1},\beta_{\kappa_2}\}$, the eigenmode can be constructed by a superposition of the Bloch basis $\left|\Phi_l\right>=\alpha^{(l)}_{1} \left|\beta_{\kappa^{(l)}_1}\right>+\alpha^{(l)}_{2}\left|\beta_{\kappa^{(l)}_2}\right>$ with the superposition coefficients $\alpha^{(l)}_1$ and $\alpha^{(l)}_2$. 
By dropping the band index $l$, Eq.~\eqref{eigeneuqation2} leads to 
\begin{equation}
     \left[\hat{H}, \hat{C}^{\dagger} \left(\alpha_1 \left|\beta_{\kappa_1}\right> +\alpha_2 \left|\beta_{\kappa_2}\right> \right)\right] = E_{\kappa}\hat{C}^{\dagger}\left(\alpha_1 \left|\beta_{\kappa_1}\right> +\alpha_2 \left|\beta_{\kappa_2}\right> \right) - t_L{\hat{c}_{N}^{\dagger}}\left(\alpha_1\beta_{\kappa_1}^{N+1}+\alpha_2\beta_{\kappa_2}^{N+1}\right)-t_R \hat{c}_{1}^{\dagger}\left(\alpha_1+\alpha_2\right).
\end{equation}
$E_{\kappa}$ is the eigenfrequency when $t_L{\hat{c}_{N}^{\dagger}}\left(\alpha_1\beta_{\kappa_1}^{N+1}+\alpha_2\beta_{\kappa_2}^{N+1}\right)+t_R \hat{c}_{1}^{\dagger}\left(\alpha_1+\alpha_2\right)=0$, which requests\begin{equation}
\left(\begin{array}{cc}
1 & 1\\
\beta_{\kappa_1}^{N+1} & \beta_{\kappa_2}^{N+1}
\end{array}\right)\left(\begin{array}{c}
\alpha_{1}\\
\alpha_{2}
\end{array}\right)=0,
\end{equation}
which is exactly the boundary condition  Eq.~\eqref{boundary_condition}.
The consistent eigenenergy and boundary condition with those in Sec.~\ref{Nonhermitian_skin_effect} should validate that the approach we propose is general for systems with both short-range and long-range couplings.

\textbf{Topological origin}.---The emergence of the non-Hermitian skin effect in the Hatano-Nelson model corresponds to a spectral winding as addressed in Sec.~\ref{Nonhermitian_skin_effect}, which requires a comparison of the energy spectra under the PBC and OBC on the complex plane. 
Since the boundary is important, the allowed wave vector $\beta_{\kappa}$  in the energy spectra under the PBC and OBC is very different as addressed in Sec.~\ref{Nonhermitian_skin_effect}. Specifically, under the PBC the allowed wavevector $\beta_{\kappa}$  lies in the range $\left[e^{-i\pi}, e^{i\pi}\right]$ with real wave vectors, \textit{i.e.}, they are the Bloch wave vectors under the translation symmetry.
Therefore, the spectra under the PBC are obtained by substituting the Bloch wave vectors into Eq.~\eqref{longrangeeigen}.

\begin{figure}[htp]
\centering
\includegraphics[width=14.5cm]{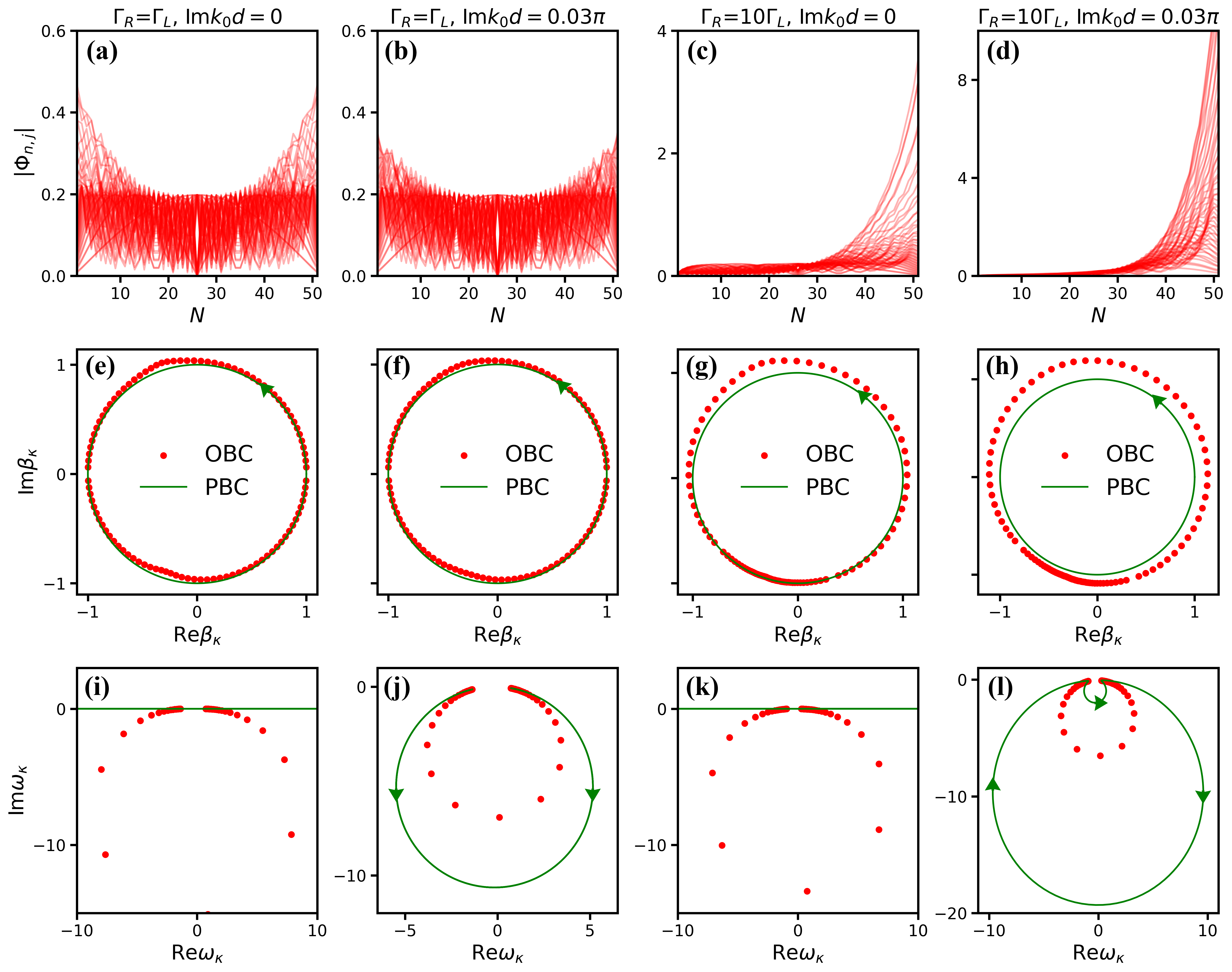}
\caption{Topological origin of non-Hermitian skin effect for the magnonic collective modes in the one-dimensional array. (a-d) plot the spatial distribution of eigenstates; (e-h) show the allowed wave vectors $\beta_{\kappa}$; (i-l) show the frequency spectra $\omega_{\kappa}$ , scaled by $(\Gamma_R+\Gamma_L)/2$. The parameters used for the calculation are addressed at the top of the figure. The allowed wavevector $\beta_{\kappa}$ and frequency $\omega_{\kappa}$ for PBC and OBC are denoted by the green lines and red dots, respectively. These figures are adapted with permission from Ref.~\cite{zeng2023radiation}.}
\label{summary}
\end{figure}

To address the topological condition for the non-Hermitian skin effect with the long-range coupling, we show the numerical results for the systems with different combinations of  chirality and  attenuation of spin waves in Fig.~\ref{summary}, with parameters for the calculation addressed at the top of the figure. The attenuation of spin waves is described by $\text{Im}k_0d$ in Eq.~\eqref{H_matrix} and a larger $\text{Im}k_0d$ corresponds to a smaller propagation length for spin waves. Figure~\ref{summary}(a)-(d) address the spatial distribution of eigenstates, and Fig.~\ref{summary}(e)-(h) and Fig.~\ref{summary}(i)-(l) compare the allowed wave vectors and energy spectra under the PBC and OBC, respectively.

As shown in Fig.~\ref{summary}(i)-(l), the energy spectra under the PBC by the green curves encircle the spectra under the OBC denoted by the red dots only when both the chirality $\Gamma_R=10\Gamma_L$ and attenuation of spin waves $\text{Im}k_0d=0.03\pi$ exist. In this case, the corresponding eigenstates as in Figs.~\ref{summary}(d) exhibit the non-Hermitian skin effect, suggesting a similar correspondence between the skin modes and spectral topology as addressed in Sec.~\ref{Nonhermitian_skin_effect}.

By the topological origin, we can explain the suppression of non-Hermitian skin effect  when an array of magnets is loaded on the ``chiral plane'' of the waveguides \cite{yu2022giant,zeng2023radiation}, where ${\rm Im}(k_0d)=0$, as addressed in Sec.~\ref{waveguides}.
Figure~\ref{summary} provides an overview of the non-Hermitian skin effect in the system with long-range coupling. 
\begin{itemize}
\item  Without chirality $\Gamma_R=\Gamma_L$ and without attenuation $\text{Im}k_0d=0$, only several eigenstates, \textit{i.e.}, the superradiant states are skewed to both boundaries, as shown in Fig.~\ref{summary}(a). The allowed wave vector $\beta_{\kappa}$ on the complex plane is almost overlapped with the unit circle, \textit{i.e.}, the Bloch wave vector, except for the superradiant states, indicating the absence of non-Hermitian skin effect, as shown in Figs.~\ref{summary}(e). These extended eigenstates imply the  spectra under the PBC do not encircle that under OBC as in Fig.~\ref{summary}(i) since under the PBC the energy is real.
\item Without chirality $\Gamma_R=\Gamma_L$ and with attenuation $\text{Im}k_0d=0.03\pi$, the skewness of superradiant states is suppressed as shown in Fig.~\ref{summary}(b), which corresponds to an overlap between $\beta_{\kappa}$ and Bloch wave vector in Fig.~\ref{summary}(f). Figure~\ref{summary}(j) exhibits no winding for the spectra under the PBC. 
\item With chirality $\Gamma_R=10\Gamma_L$ and without attenuation $\text{Im}k_0d=0$, the superradiant states develop a larger amplitude at the right boundary in Fig.~\ref{summary}(c), while most of the eigenstates remain extended over the array. Figure.~\ref{summary}(g) exhibits a small deviation of $\beta_{\kappa}$ from the Bloch wave vector, which leads to the skewness of superradiant states. However, it is still not the non-Hermitian skin effect as confirmed by Fig.~\ref{summary}(k), where the distribution of the spectra under the PBC is similar to that in Fig.~\ref{summary}(i). 
\item With chirality $\Gamma_R=10\Gamma_L$ and with attenuation $\text{Im}k_0d=0.03\pi$, all the states accumulate at the right boundary of the array as shown in Fig.~\ref{summary}(d). Figure~\ref{summary}(h) shows all $\beta_{\kappa}$ deviate strongly from the unit circle, corresponding to the increase of the amplitude of eigenstates when approaching the right boundary. In this case, the spectra under the PBC encircle those under the OBC, as shown in Fig.~\ref{summary}(l), which corresponds to the presence of a non-Hermitian skin effect.  
\end{itemize}

\subsubsection{Equilibrium thermal distribution of magnons}

One natural question concerning the pile-up of almost all the eigenstates at the edge of the wire array, as reviewed above, is whether there exists a spontaneous thermal accumulation of magnons in the wires. If so, this would imply a ``Maxwell's Demon'' that rectifies the equilibrium thermal current. Recently, Yu and coauthors addressed this issue and demonstrated that although there is a non-Hermitian skin effect, the number of magnons in every wire is the same guaranteed by the second law of thermodynamics~\cite{zeng2023radiation,distribution_Yu}. Let us consider the Kittle magnons $\{\hat{m}_1,\hat{m}_2,\cdots,\hat{m}_N\}$ in the $N$ wires that interact with the environment via the thermal noise $\{\hat{b}_1,\hat{b}_2,\cdots,\hat{b}_N\}$, while the thermal noise $\hat{c}_k$ causes the fluctuation of the magnons $\hat{a}_k$ in the substrate. The associated Langevin's equation for the hybridized system reads
  \begin{align}
    i\frac{d\hat{m}_l}{dt}&=\left(\omega_{\rm K}-i\frac{\eta}{2}\right)\hat{m}_l+\sum_k g_k e^{ikR_l}\hat{a}_k-i\sqrt{\eta}\hat{b}_l,\nonumber\\
    i\frac{d\hat{a}_k}{dt}&=\left(\omega_k-i\frac{\eta_k}{2}\right)\hat{a}_k+\sum_l g_k e^{-ikR_l}\hat{m}_l-i\sqrt{\eta_k}\hat{{c}_k}.
    \label{min1}
  \end{align} 
   Using the on-shell approximation, Eq.~(\ref{min1}) in the frequency domain reads
   \begin{align}
      \left(\omega-\tilde{H}_{\rm eff}/\hbar\right)\hat{M}=-i\sqrt{\eta}\hat{ B}- i\sum_k \frac{\sqrt{\eta_k} g_k}{\omega_{\rm K}-\omega_k+i\frac{\eta_k}{2}}Q(k)\hat{c}_k,
      \label{min7}
   \end{align}
  where $\hat{ B}=(\hat{b}_1,\hat{b}_2,\cdots,\hat{b}_N)^{T}$ and $Q(k)=(e^{ikR_1},e^{ikR_2},\cdots, e^{ikR_N}  )^{T}$. 
  The correlation-fluctuation theorem requests~\cite{clerk2010introduction}
  \begin{subequations}
    \begin{align}
    \langle\hat{{B}}(\omega)\hat{{B}}^{\dagger}(\omega')\rangle
    &=2\pi\delta(\omega-\omega')(n(\omega)+1)\overleftrightarrow{\bf I},\\
     \langle \hat{c}_k(\omega)\hat{c}_{k'}^{\dagger}(\omega')\rangle&=2\pi \delta(\omega-\omega')(n_k+1)\delta_{kk'},
    \label{min11}
   \end{align}
    \end{subequations}
   where $n(\omega)=1/\{\exp[\hbar\omega/(k_BT)]-1\}$ and $n_k=1/\{\exp[\hbar \omega_k/(k_BT)]-1\}$ are the Bose-Einstein distribution at temperature $T$. We remind that the left and right eigenvectors of $\tilde{H}_{\rm eff}/\hbar$ are ${\cal L}=(\Psi_1,\Psi_2,\cdots,\Psi_N)$ and ${\cal  R}=(\Phi_1,\Phi_2,\cdots,\Phi_N)$, which obey the orthonormal relation ${\cal  L}^{\dagger} R={R}^{\dagger}{\cal  L}=\overleftrightarrow{\bf I}$. With the eigenvalues $\omega_l$, we define $\Omega={\rm diag}(\omega-\omega_l)$, so $\omega-\tilde{H}_{\rm eff}/\hbar={\cal R}\Omega{\cal L}^{\dagger}$. 
   The equilibrium distribution of magnons is found via  
   \begin{align}
     \left\langle\hat{M}(t)\hat{M}^{\dagger}(t)\right\rangle
     ={\cal R}\left\langle\hat{A}(t)\hat{A}^{\dagger}(t)\right\rangle {\cal R}^{\dagger},
     \label{am}
 \end{align}
 where the distribution of collective eigenmodes
 \begin{align}
    \left\langle\hat{A}(t)\hat{A}^{\dagger}(t)\right\rangle
    &= \int_{-\infty}^{\infty}\int_{-\infty}^{\infty}\frac{d\omega d\omega'}{(2\pi)^2}e^{-i(\omega-\omega')t}\eta {\Omega}^{-1}{\cal L}^{\dagger}\langle\hat{B}(\omega)\hat{B}^{\dagger}(\omega')\rangle {\cal L}({\Omega}^{-1})^{\dagger}\nonumber\\
   &+\int_{-\infty}^{\infty}\int_{-\infty}^{\infty}\frac{d\omega d\omega'}{(2\pi)^2}e^{-i(\omega-\omega')t}\sum_k\sum_{k'}g_k g_{k'} \frac{\sqrt{\eta_k \eta_{k'}}\langle\hat{c}_k(\omega)\hat{c}_{k'}^{\dagger}(\omega')\rangle}{(\omega_{\rm K}- \omega_k+i\frac{\eta_k}{2})(\omega_{\rm K}- \omega_{k'}-i\frac{\eta_{k'}}{2})}\nonumber\\
   &\times {\Omega}^{-1} {\cal L}^{\dagger}{Q}(k) 
    {Q}^{\dagger}(k'){\cal L}({\Omega}^{-1})^{\dagger}.
    \label{min12}
 \end{align} 
 After performing the integrals, we obtain 
  $\left\langle\hat{A}(t)\hat{A}^{\dagger}(t)\right\rangle=\left(n(\omega_{\rm K})+1\right){\cal L}^{\dagger} \cal L$.
   Substituting into (\ref{am}) yields 
  $\left\langle\hat{M}(t)\hat{M}^{\dagger}(t)\right\rangle
      =\left(n(\omega_{\rm K})+1\right)R{\cal L}^{\dagger} \cal L {R}^{\dagger}$.
 Since ${\cal L}^{\dagger}{\cal R}=\overleftrightarrow{\bf I}$,
       $R{\cal L}^{\dagger}=R{\cal L}^{\dagger}\times (RR^{-1})=R({\cal L}^{\dagger}R) R^{-1}=\overleftrightarrow{\bf I}$
and $\cal L {R}^{\dagger}=(R{\cal L}^{\dagger})^{\dagger}=\overleftrightarrow{\bf I}$, leading to 
\begin{align}
    \left\langle\hat{M}(t)\hat{M}^{\dagger}(t)\right\rangle=\left(n(\omega_{\rm K})+1\right)\overleftrightarrow{\bf I}.
\end{align}
This demonstrates that the magnon number in every wire is the same because every wire is biased by an infinite thermal bath, i.e., the substrate.

\subsubsection{Defect states \textit{vs.} the non-Hermitian skin effect}

Although the non-Hermitian skin effect is protected by the spectral topology, it may be affected by defects since the defect may induce additional boundaries to the system with short-range coupling, as addressed in Table~\ref{fig:SK1co}.
It is an intriguing problem to explore the effect of defects on the non-Hermitian skin effect for the system with long-range coupling since the coupling range overcomes the defect range. Recently,  Zeng and Yu considered an alternative model with an array of magnetic nanowires coupled to a dielectric substrate \cite{zeng2023radiation}. They demonstrated that the non-Hermitian skin modes are robust to the defects in the system with long-range coupling, which may provide an experimental feasible system for detecting, observing, and exploring the non-Hermitian skin effect with magnons \cite{zeng2023radiation}.

\begin{figure}[ht]
\centering
\includegraphics[width=14cm]{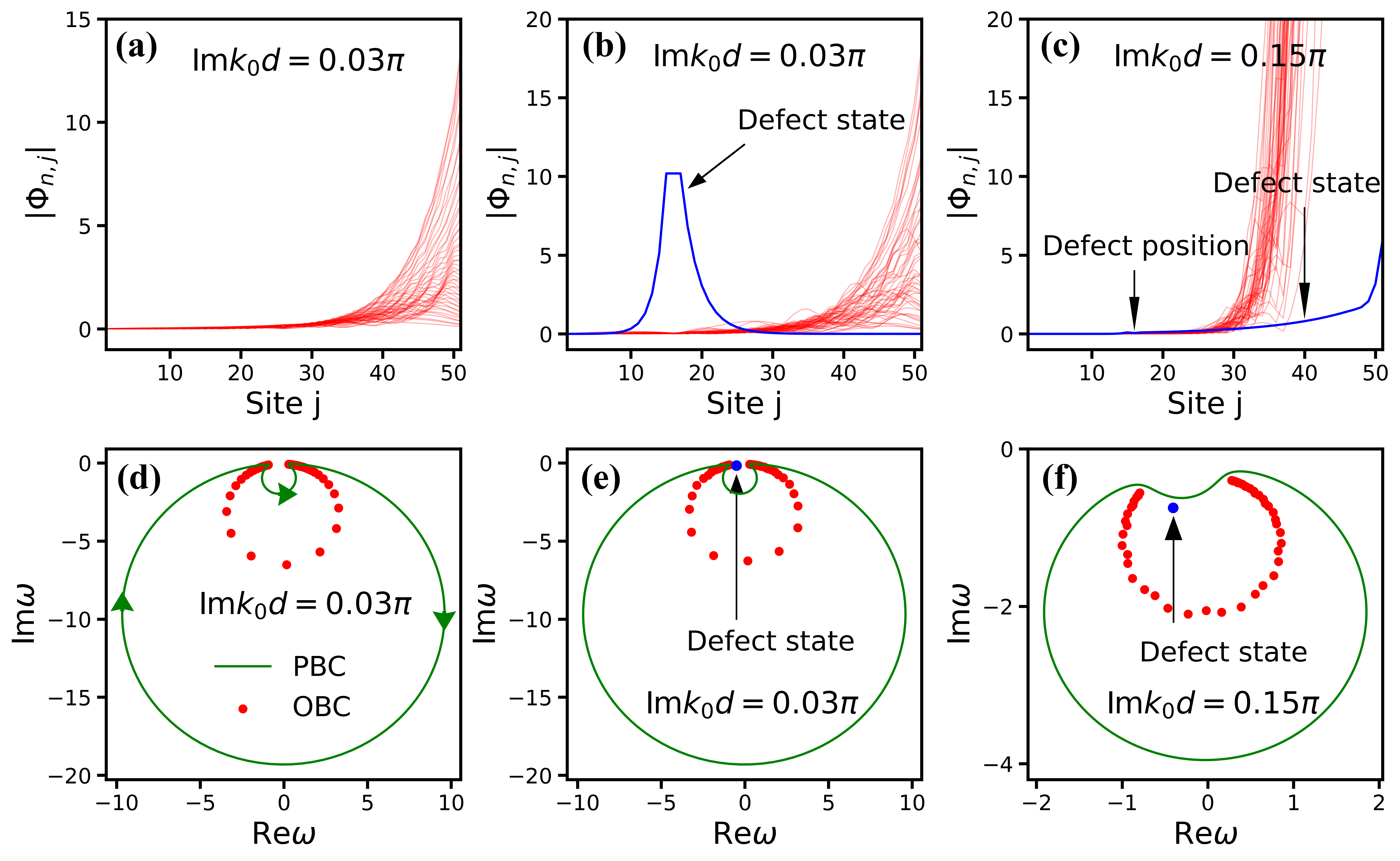}
\caption{Robustness of the non-Hermitian skin effect to the defect in the long-range coupled magnetic nanowires, governed by the chiral interaction  $\Gamma_R=10\Gamma_L$. (a) shows the spatial distribution of eigenstates with attenuation $\text{Im}k_0 d=0.03\pi$. (b) and (c) show the eigenstates when the defect is introduced with the attenuation  $\text{Im}k_0 d=0.03\pi$ [(b)] and $\text{Im}k_0 d=0.15\pi$ [(c)]. When  $\text{Im}k_0 d=0.03\pi$, (d) and (e) compare the energy spectra under the PBC  and OBC without [(d)] and with [(e)] the defect, while  (f) plots the situation with larger attenuation $\text{Im}k_0 d=0.15\pi$.
These figures are reproduced with permission from Ref.~\cite{zeng2023radiation}.}
\label{defect}
\end{figure}

The chiral pumping of surface acoustic waves  due to interference with two magnetic nanowires in proximity to the dielectric substrate  was proposed theoretically by Zhang \textit{et. al.}  \cite{zhang2020unidirectional}.  Later, Yu addressed the chiral coupling between the Kittel magnon of a single nanowire with the surface phonon in the linear regime \cite{yu2020nonreciprocal}. In the nonlinear regime, the magnetization  dynamics drive phonon frequency combs with frequency multiplication with high frequency 10~GHz \cite{cai2023acoustic}, which may overcome the technical restrictions of the electric approach for specific piezoelectric substrates. 
In terms of the chiral coupling between magnetic nanowires and dielectric substrates, we  obtain an effective interaction between the Kittel magnons in the magnetic nanowires that are mediated by the surface acoustic waves \cite{zeng2023radiation}, similar to Eq.~\eqref{effectiveinteraction}. For such a system as shown in Fig.~\ref{defect}, the defect can be introduced by the absence of nanowires in the array or the local magnon frequency shift by a locally biased magnetic field.  We use $\text{Im}k_0d$ to describe the attenuation of surface acoustic waves and focus on the chirality $\Gamma_R=10\Gamma_L$ to investigate the effect of the defect on the non-Hermitian skin effect.

\begin{figure}[ht]
\centering
\includegraphics[width=10cm]{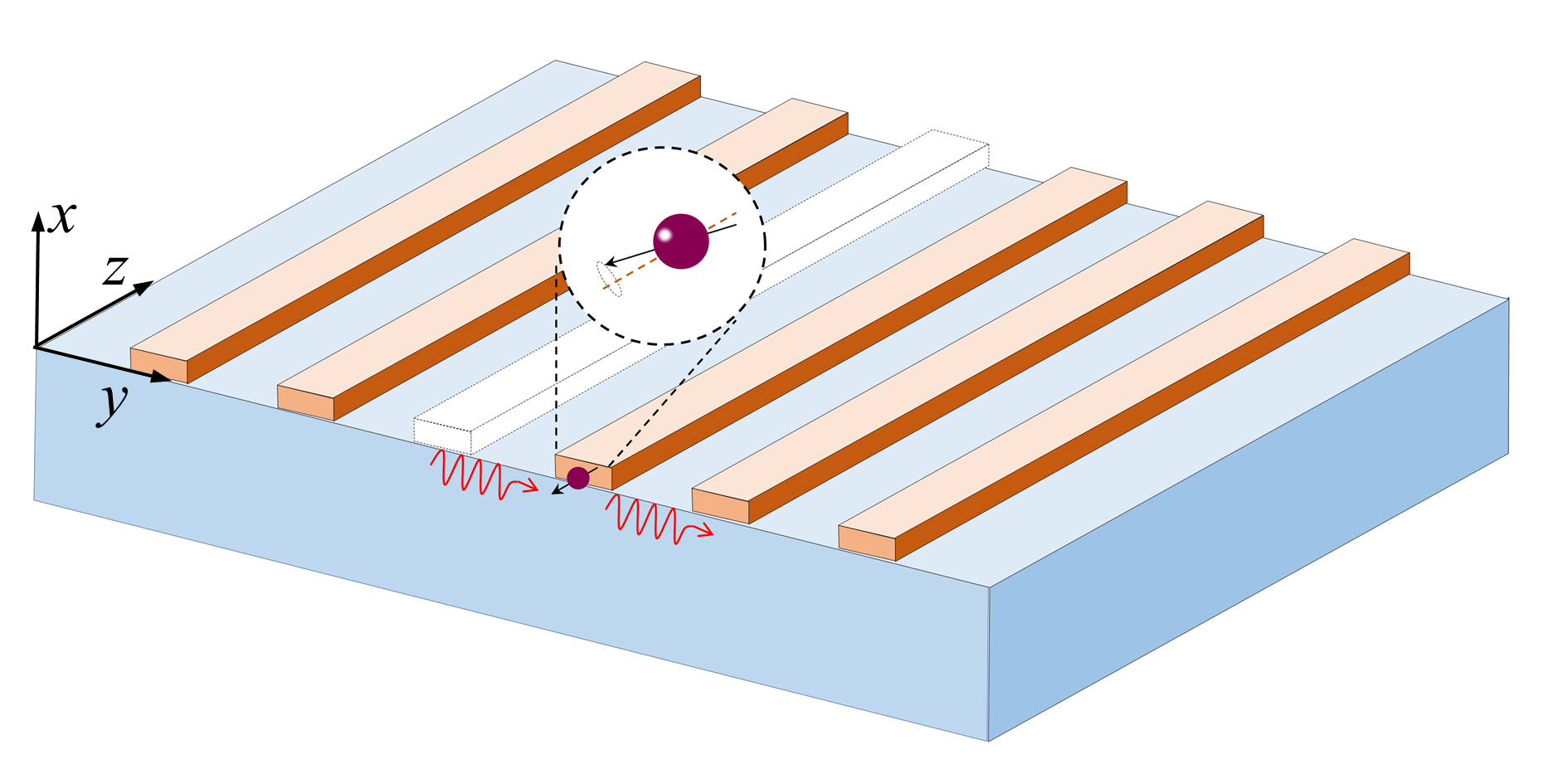}
\caption{ An array of magnetic nanowires on the dielectric substrates, where the defect can be introduced by removing a single magnetic nanowire. The figure is reproduced with permission from Ref.~\cite{zeng2023radiation}.}
\label{defectmodel}
\end{figure}

The eigenstates with chirality $\Gamma_R=10\Gamma_L$ and different attenuation of surface acoustic waves are shown in Fig.~\ref{defect}(a)-(c); the corresponding energy spectra, normalized by $(\Gamma_R+\Gamma_L)/2$, under the PBC and OBC are denoted by the green lines and red dots, respectively,  in Fig.~\ref{defect}(d)-(f).  Without a defect, Fig.~\ref{defect}(a) reproduces the same non-Hermitian skin effect as Fig.~\ref{fig:magnonicSK} and Fig.~\ref{summary}, and the spectra under the PBC encircles that under the OBC. When a defect is introduced at the left side of the magnetic array, a  defect state, which is spatially separated from other eigenstates, appears with a considerable amplitude at the defect position as shown in Fig.~\ref{defect}(b). While the other skin modes are rarely affected by the defect. From the spectra on the complex plane, as shown in Fig.~\ref{defect}(e), the frequency of the defect state just locates at the outside of the region encircled by the spectra under the PBC, demonstrating its robustness to the non-Hermitian topology. By increasing the attenuation of acoustic waves to $\text{Im}k_0d=0.15\pi$, the frequency of defect states decreases and falls into the winding circle of the spectra under the PBC as shown in Fig.~\ref{defect}(f), at which regime the defect state is also skewed to the boundary. 
This highlights again the robustness of the non-Hermitian skin effect of collective modes to the defect when the coupling mediated by phonons is of long range, which can thereby be used for producing and detecting stable magnonic skin modes.

\subsection{Non-Hermitian skin effect in higher dimensions}
\label{high_dimension_skin_effect}

The discovery of the one-dimensional non-Hermitian skin effect raises challenges and perspectives for understanding the open systems with exotic properties and phenomena \cite{ashida2020non,bergholtz2021exceptional, zhang2022review,ding2022non,lin2023topological}. It stimulates the research interest in extending the non-Hermitian skin effect from the one to higher dimensions, which yields rich and diverse manifestations of skin modes. The winding number defined in the one-dimensional system can be extended to the higher dimensions. For example, in the
non-reciprocal two-dimensional non-Hermitian systems,
the two winding numbers defined along two normal directions, i.e., a (non-vanishing) topological winding tuple $\{W_1,W_2\}$, may precisely
distinguish different edge and corner skin effects, i.e., a
precise prediction of the edge or corner on which the
modes localize, which is a straightforward generalization
of the one-dimensional winding number~\cite{kawabata2020higher,cai2023corner}. Nevertheless, the corner skin modes can still exist in higher dimensions with vanishing winding tuples as a manifestation of the higher-order
non-Hermitian skin effect~\cite{okugawa2020second}. Kawabata \textit{et al.} showed that the nonzero Wess-Zumino term leads to
the presence of higher-order corner skin modes in non-Hermitian systems~\cite{kawabata2020higher}.
Among the first and higher-order non-Hermitian skin effects, the first-order non-Hermitian skin effect is also known as the conventional non-Hermitian skin effect, which can be divided into non-reciprocal non-Hermitian skin effect and geometrically dependent non-Hermitian skin effect~\cite{zhang2022universal}. In high-dimensional systems, the boundaries can be of different shapes. The non-reciprocal non-Hermitian skin effect does not depend on the selection of open boundaries, but the geometry-dependent non-Hermitian skin effect depends~\cite{zhang2022universal}.
The rapid progress in the field also raised theoretical
challenges and urgent issues in the topological characterization of the different skin modes \cite{zhang2022review,ding2022non,lin2023topological}  The challenge arising in the characterization of the high-dimensional non-Hermitian skin effect, to the best of our knowledge, involves the following aspects. 
\begin{itemize}
\item Structure diversity. In comparison to the one-dimensional chains, there is more freedom for the configurations in the high-dimensional open systems such as the rectangle and triangle lattices.
The two-dimensional non-Hermitian skin effect has been proposed in the square \cite{kawabata2020higher,lee2019hybrid} and honeycomb \cite{li2022gain,zhu2022hybrid} lattices and other configurations \cite{zhang2022universal} as well. The extension to the three-dimensional structure has been proposed in the cubic lattice \cite{kawabata2020higher,lee2019hybrid}. 
Even for the similar non-Hermitian skin effect, its phenomenology may be very different in that the localization of states can appear at the edge or corner in the two-dimensional models \cite{zhu2022hybrid,lee2019hybrid}, and surface, hinge, or corner in the three-dimensional models \cite{lee2019hybrid}. 
\item Diverse mechanisms. In comparison to the non-Hermitian skin effect induced by the asymmetric coupling in the one-dimensional system, the non-reciprocity possesses more freedom in the two-dimensional monoatomic lattice, which can even induce the localization of bulk states at the corner  \cite{benalcazar2017quantized,lee2019hybrid}. But for the high-dimensional non-Hermitian skin effect, the global asymmetric coupling or the non-reciprocity is not indispensable. For example, even without net reciprocity in the bulk, the asymmetric coupling can still exist at the edge sites in the presence of topological edge modes that lead to the non-Hermitian skin effect, forming hybridized higher-order skin-topological modes, while the bulk modes remain extended \cite{lee2019hybrid}. A further study demonstrated the on-site gain and loss and chiral current induced by non-local flux can also drive the topological edge mode to the corner in a two-dimensional honeycomb lattice without asymmetric coupling \cite{li2022gain}. 
\item Different number scaling for the non-Hermitian skin modes. For the first-order and second-order topological mode of a two-dimensional square lattice with the system size $L\times L$, where the system length is $L$,  the number of edge states and corner states typically scales with $O(L)$ and $O(1)$ \cite{schindler2018higher,xie2021higher}.  Different from such conventional topological modes, the number of non-Hermitian skin modes can scale as $O(L^2)$ at the edge or corner. For the second-order counterpart of the non-Hermitian skin effect, it scales as $O(L)$ at the edge \cite{okugawa2020second,lin2023topological}. With a different mechanism, the higher-order hybridized skin-topological mode, i.e. the non-Hermitian skin mode of the topological edge state, scales as $O(L)$ as well at the corner for a two-dimensional square lattice \cite{lee2019hybrid}. Experimentally, the high-order non-Hermitian skin and hybridized topological skin modes have been observed in acoustic crystals \cite{zhang2021observation}, cold atoms systems \cite{li2020topological} and topolectrical circuits \cite{zou2021observation}. 
\end{itemize}

\begin{figure}[!htp]
		\centering
		\includegraphics[width=0.99\textwidth]{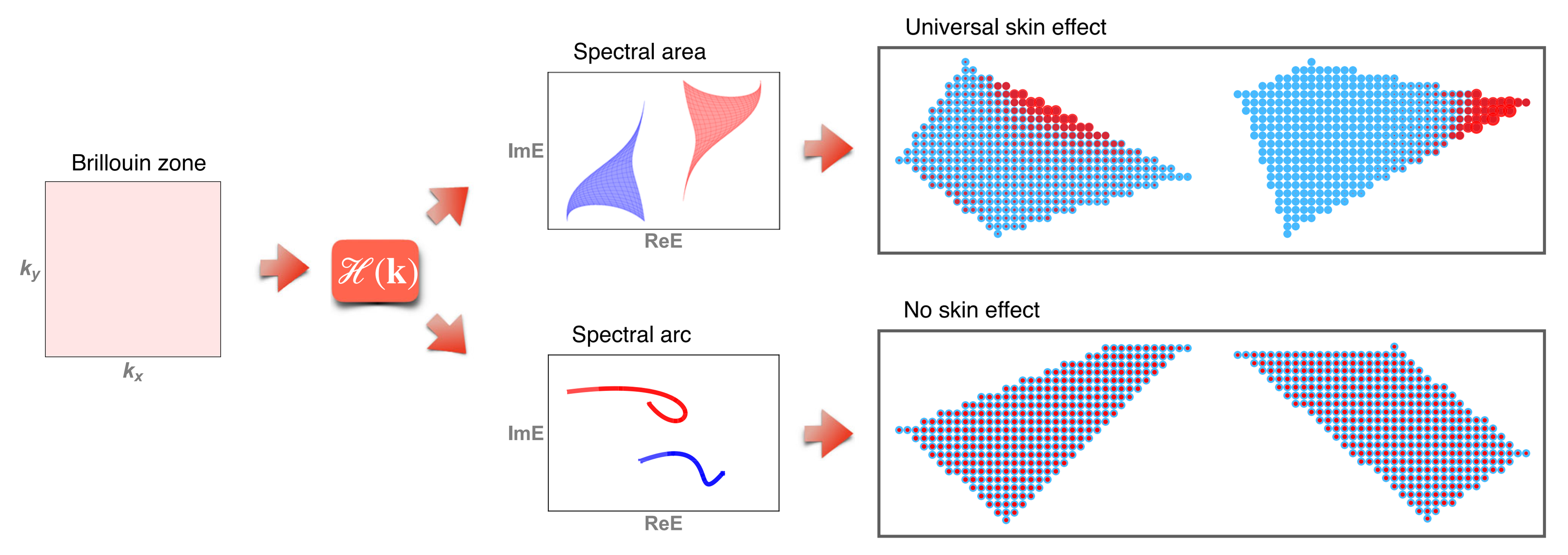}
		\caption{A ``theorem'' for characterizing the two-dimensional non-Hermitian skin effect. The energy spectra under the PBC can cover a finite area or populate as an arc, while only the former leads to a non-Hermitian skin effect with the localization of eigenstates (red region) at the boundary. The figure is reproduced with permission from Ref.~\cite{zhang2022universal}.} 
		\label{fig:SK2}
\end{figure}

Although there may exist the challenge to universally describe the condition for the non-Hermitian skin effect in high-dimensional open systems, a recent study proposed a useful ``theorem'' to describe its existence. It states that when the spectra under the PBC cover a finite area on the complex plane, the non-Hermitian skin effect appears  \cite{zhang2022universal}, as shown in Fig.~\ref{fig:SK2}. 

Let us address this issue in more detail. In the one-dimensional lattice, the energy spectra under the PBC can form a loop or arc \cite{bergholtz2019non}, as addressed in Sec.~\ref{Nonhermitian_skin_effect}. But in the higher dimension, the spectra under the PBC generally form a continuum band within an area or separated subareas on the complex plane as shown in Fig.~\ref{fig:SK2}, as well as arcs with specified Bloch Hamiltonian. 
The right panel of Fig.~\ref{fig:SK2} shows  the spatial distributions of all eigenstates \cite{zhang2022universal}, which is  calculated by 
\begin{equation}\label{eq:distributionforuniversal}
 W({\bf r})=\frac{1}{N} \sum_{n=1}^N\left|\phi_n(\boldsymbol{r})\right|^2,
\end{equation} 
where $N$ is the number of eigenstates and $\phi_n(\boldsymbol{r})$ is the $n$-th right eigenstate on the site ${\bf r}$.  $W({\bf r})$ is represented by red on the sites denoted by the blue dots. 
The localization of modes only happens when the spectra under the PBC cover a finite area.
There is no non-Hermitian skin effect with a uniform distribution of eigenstates on the sites when the spectrum under the PBC is an arc. The physical explanation may be rooted in the fact that the arc-like spectra possess high degeneracies for each energy, thus providing sufficient Bloch basis for forming standing waves when the system turns from the PBC to the OBC. Instead, when the spectra under the PBC distribute over an area with lower degeneracy, additional states that deviate from the conventional Bloch basis are required to construct the standing waves under the OBC. This is consistent with one-dimensional systems as addressed in Sec.~\ref{Nonhermitian_skin_effect}, where the non-Hermitian skin effect appears when the spectra under the PBC form a loop with decreased degeneracy compared to arcs.

Realization of the arc-like spectra in the high-dimensional non-Hermitian systems requires fine-tuning of parameters, so a randomly generated local non-Hermiticity can have a probability close to one for the two-dimensional non-Hermitian spin effect \cite{zhang2022universal}. The two-dimensional non-Hermitian skin effect is further classified into two categories: \textit{corner skin effect} and \textit{geometry-dependent skin effect} depending on the defined current functional, which can be considered as an extension of the definition of chiral current \cite{zhang2020correspondence}. A full understanding and topological characterization of the geometry-dependent skin effect appear to remain an open question to date. Interestingly, a non-Hermitian system can exhibit the non-Hermitian skin effect if stable EPs exist in the reciprocal space that ensures nonzero spectral area, thus bridging two unique phenomena in non-Hermitian systems \cite{zhang2022universal}.

\begin{figure}[!htp]
\centering
\includegraphics[width=0.92\textwidth]{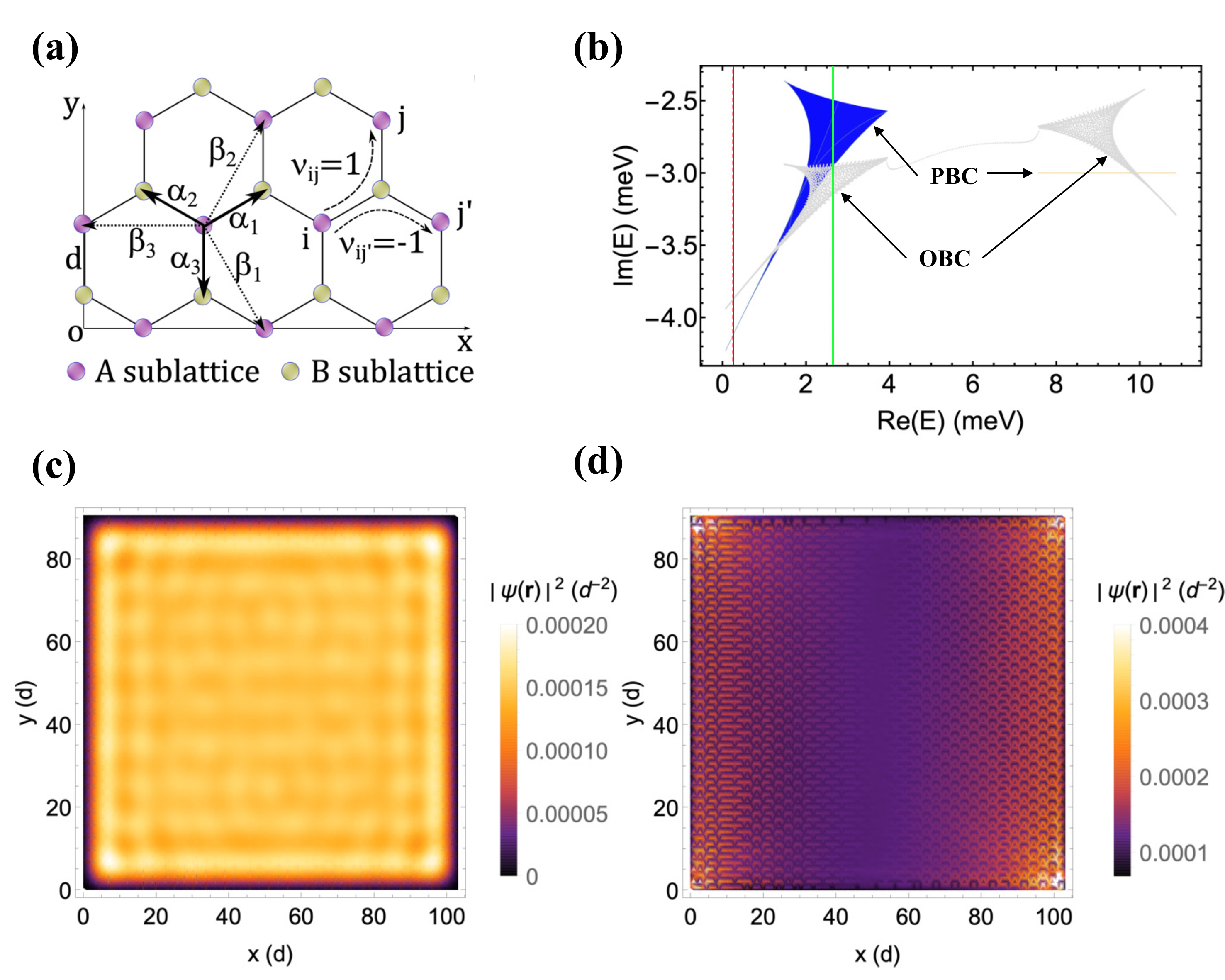}
\caption{Two-dimensional magnonic non-Hermitian skin effect in the ferromagnetic honeycomb lattice. (a) is the configuration. (b) shows the energy spectra under the PBC (blue and orange dots) and OBC (gray dots) for a real-space non-Hermitian Hamiltonian. The red and blue lines in (b) specify two energies.  Below these two energies, (c) and (d) show, respectively, the spatial distribution of the right eigenstates $|\psi(\boldsymbol{r})|^2$. The figures are adapted with permission from Ref.~\cite{deng2022non}.} 
		\label{fig:MSK2}
	\end{figure}

\textbf{van der Waals ferromagnetic monolayer}.---Recently, Deng \textit{et al.} propose a two-dimensional magnonic skin effect in a honeycomb lattice of van der Waals ferromagnetic monolayer \cite{deng2022non}, as shown in Fig.~\ref{fig:MSK2}(a). It is realized by introducing the Dzyaloshinskii-Moriya interaction and nonlocal magnetic dissipation or the dissipative hopping between sites, which the authors propose to realize by such as the spin-nonconserving magnon-phonon coupling \cite{liu2017magnon,wang2020magnon,wang2021magnon}. The unit cell of the honeycomb lattice is composed of $A$ and $B$ sublattices with the classic spin $\hat{\bf S}$ coupled ferromagnetically with nearest-neighboring exchange coefficient $J>0$ and next-to-nearest neighboring exchange coefficient $J_2>0$. The Dzyaloshinskii-Moriya interaction strength is parametrized by $D$ with $\nu_{ij}=-\nu_{ji}=\pm 1$ denoting its non-reciprocity as shown in Fig.~\ref{fig:MSK2}(a). Subjected to a magnetic field $B$ along the $\hat{\bf z}$-direction, the spin Hamiltonian reads
\begin{align}
\mathcal{H}=-J\sum\limits_{\langle i,j\rangle}\hat{\bf S}_i\cdot\hat{\bf S}_j-J_2\sum\limits_{\langle\langle i,j\rangle\rangle}\hat{\bf S}_i\cdot\hat{\bf S}_j-B\sum\limits_i \hat{S}_i^z+D\sum\limits_{\langle\langle i,j\rangle\rangle}\nu_{ij}
\hat{\boldsymbol{z}}\cdot\left(\hat{\bf S}_i\times\hat{\bf S}_j\right),
\label{eq:vdw}
\end{align}
where $\hat{S}_i^z$ denotes the $z$-component of spin  $\hat{\bf S}$. 
Equation \eqref{eq:vdw} is Hermitian and therefore the magnon spectrum is real. However, there exist many scattering channels for magnon dissipation such as magnon-magnon, magnon-electron, and magnon-phonon interactions, which cause a finite magnon lifetime or certain magnon band broadening \cite{woolsey1969theory,rezende1978spin}. In particular, at finite temperatures the phonon fluctuation affects the distance between coupled spins and thereby their interaction, leading to the broadening of the acoustic and optical modes of magnons \cite{liu2017magnon,wang2020magnon,wang2021magnon}.  
To reproduce such broadening, Deng and coauthors introduce extra non-Hermitian contributions phenomenologically to the magnon dissipation, including the on-site and nonlocal dissipations, and obtained a real-space non-Hermitian magnon Hamiltonian \cite{deng2022non}. As the authors addressed, these non-Hermitian terms can be fitted by $ab$ $initio$ results, providing a convenient and effective method of constructing an effective non-Hermitian Hamiltonian for band broadening. The general case with the microscopic origin of the non-local dissipation needs to be considered
in future studies.

The numerical results of magnon energy spectra by diagonalizing the non-Hermitian Hamiltonian under the OBC are represented by the gray dots in Fig.~\ref{fig:MSK2}(b). They are different from the calculated spectra under the PBC, as represented by the blue and orange curves, implying the breakdown of the bulk-edge correspondence. 
The average spatial distribution 
\begin{equation}
 |\psi(\boldsymbol{r})|^2=\frac{1}{M} \sum_{n=1}^M\left|\phi_n(\boldsymbol{r})\right|^2,
\end{equation}
where $|\psi(\boldsymbol{r})|^2$ describes the spatial distribution of the chosen $M$ eigenmodes $\phi_n(\boldsymbol{r})$ on the site ${\bf r}$, 
is different with different regions on the complex plane. 
For example, the energy spectra under the PBC are approximately an arc at the left of the red line in Fig.~\ref{fig:MSK2}(b), but cover an area in the region between the red and green lines. Averaging the states in the arc region, $|\psi(\boldsymbol{r})|^2$ is extended as shown in Fig.~\ref{fig:MSK2}(c). $|\psi(\boldsymbol{r})|^2$ becomes localized when the states belong to the area region as shown in Fig.~\ref{fig:MSK2}(d).
Thereby, the two-dimensional magnonic non-Hermitian skin effect reproduces the same corresponding relation: the non-Hermitian skin effect occurs when the spectra under PBC cover a finite area~\cite{zhang2022universal}. 

\begin{figure}[!htp]
\centering
\includegraphics[width=0.99\textwidth]{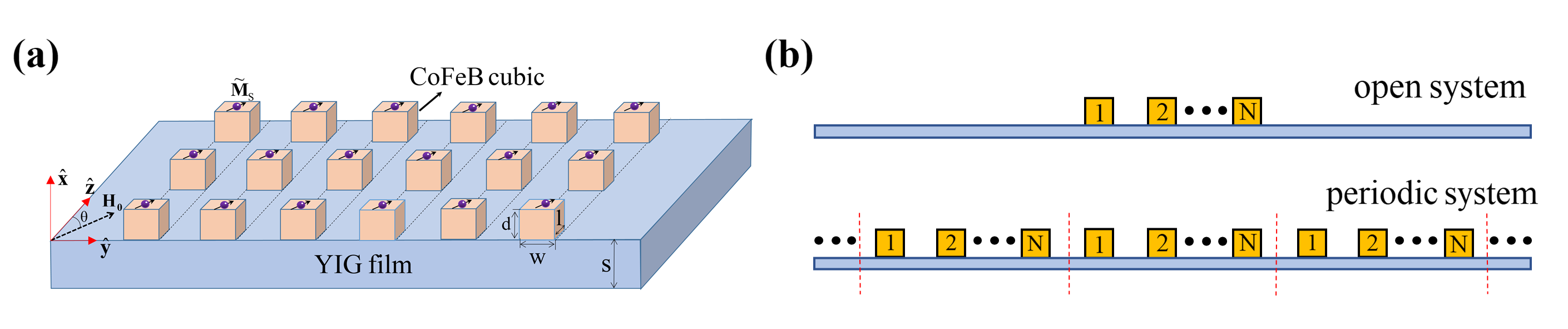}
\caption{ (a) illustrates the model of the two-dimensional array of nanomagnets that are coupled to the magnetic YIG thin film of thickness $s$ via the dipolar interaction. A magnetic field along the direction $\theta$ with respect to the {\color{red}$\hat{\bf z}$}-direction biases the magnetizations of the magnetic film and nanomagnets. (b) shows the 
periodic system built from the open system, i.e., the heterostructure composed of a finite array and a substrate, by repeating the array on the substrate an infinite number of times and imposing periodic conditions on the magnon operators. The figures are adapted with permission from Ref.~\cite{cai2023corner}.} 
\label{fig:2Dmagnonicmodel}
\end{figure}

\textbf{Magnetic array with new topological characterization}.---Of course, the phenomenological description in terms of the energy spectra distribution under the PBC is qualitative, without a precise characterization for different skin modes such as the corner and edge skin modes.
Recently, Cai \textit{et al.} proposed the  topological characterization in terms of a winding tuple
that successfully characterize the non-Hermitian skin effect of a two-dimensional array of magnetic blocks coupled to magnetic film via the dipolar interaction, as shown in Fig.~\ref{fig:2Dmagnonicmodel} for the configuration~\cite{cai2023corner}. They work out that the long-range chiral interaction between magnetic blocks mediated by the magnon in the film can lead to the two-dimensional non-Hermitian skin effect of magnons, and predict different non-Hermitian skin modes such as the edge modes or corner modes, \textit{i.e.}, the aggregations of all the magnon states at the boundary or corner, can be achieved by tuning the direction of the in-plane magnetic field with an angle $\theta$ with respect to the $\hat{\bf z}$ direction.

An important remaining issue in the heterostructure systems, e.g., a finite array of nanomagnets over an infinite substrate~\cite{yu2020chiral,yu2023chirality,yu2020magnon_accumulation,yu2022giant,zeng2023radiation}, is how to build the periodic boundary condition for the nanomagnet subsystem. Cai \textit{et al.} proposed that the periodic boundary condition can be built by repeating the finite nanomagnetic array an infinite number of times on the substrate and requesting the magnon operator to satisfy periodic conditions~\cite{cai2023corner}. For example, for an one-dimensional array with $N$ nanowires, the magnon operators of $l$-th nanowire should satisfy $\hat{\beta}_l=\hat{\beta}_{l+N}$ as addressed in Fig.~\ref{fig:2Dmagnonicmodel}(b), and for a 2D system with $N_y\times N_z$ nanomagnets, the magnon operator of the nanomagnet in the $a$-th column and $b$-th row should satisfy $\hat{\beta}(a,b)=\hat{\beta}(a+N_y,b)=\hat{\beta}(a,b+N_z)$. By this approach, when the analytical solution is not available beyond the one-dimensional situation~\cite{zeng2023radiation}, the authors can numerically obtain the eigenspectra of the magnetic subsystem under the periodic boundary conditions and find the winding tuple for the topological characterization~\cite{cai2023corner}.

Such a flexible control of the magnonic non-Hermitian skin effect can be intuitively explained by the angle-dependent chiral coupling $\Gamma({\bf r})$ between the magnetic cubics mediated by spin waves of the film, where ${\bf r}$ here is the distance vector between any two cubics, as shown in the first column of Table \ref{fig:2DMSE}. Such coupling drives magnonic states to pile up at the edge or corner. For example, when $\theta = -\pi/4$, the coupling constant implies that magnons tend to pile up in the $\hat{\bf y}$ and $\hat{\bf z}$ directions, leading to the aggregations of the magnon states at the upper right corner.  
When $\theta = 0$, the coupling constant only exhibits directional preference along the $\hat{\bf y}$ direction but not the $\hat{\bf z}$ direction, therefore the magnon wavefunction is skewed at the right edge of the array. Similar correspondence between the coupling constant and non-Hermitian skin effect is also found when $\theta = \pi/4$, where the directional preference of the coupling constant along the $\hat{\bf y}$ and $-\hat{\bf z}$ directions drives the magnons to the lower right corner.  These results are summarized in the second column of Table \ref{fig:2DMSE}.

 As addressed in the one-dimensional Hatano-Nelson model in Sec.~\ref{Nonhermitian_skin_effect}, the spectral winding number captures the emergence and the skewed direction of the non-Hermitian skin effect. This approach works in the one-dimension since the complex spectra $\omega(\kappa)$ form a curve under the PBC when the real wave vector $\kappa$ evolves by a period from $-\pi$ to $\pi$.  Here this approach can be generalized to the two dimensions when the complex spectra $\omega(\kappa_1,\kappa_2)$ are the function of two real wave vectors $\kappa_1$ and $\kappa_2$ under the PBC, thus having a complicated distribution on the complex plane.
Recently, Cai \textit{et al.} propose to characterize the two-dimensional nonreciprocal skin effect under the PBC by fixing one of $\kappa_1$ and $\kappa_2$ and varying the other, and observing the evolution of the energy spectra~\cite{cai2023corner}. This contributes to red a winding tuple. When both vanish, the two-dimensional non-Hermitian skin effect does not appear; when only one of them exists, the skin modes red are located on the edge, with four skin edges characterized by $\{W_1,W_2\}=\{1,0\}$, $\{-1,0\}$, $\{0,1\}$, and $\{0,-1\}$, respectively; when both exist, the skin modes pile up at the corner, with four skin corners characterized by $\{W_1,W_2\}=\{1,1\}$, $\{1,-1\}$, $\{-1,1\}$, and $\{-1,-1\}$, respectively. The third column of Table \ref{fig:2DMSE} shows the spectral winding when fixing one of $\kappa_y$ and $\kappa_z$. The evolution direction of the spectra, \textit{i.e.},  clockwise or anti-clockwise, is denoted by the red or green arrows that correspond to the spectral winding index of 1 or $-1$. When $\theta = -\pi/4$, the spectra with fixing either $\kappa_y$ or $\kappa_z$ evolve in a clockwise fashion, \textit{i.e.}, $\{W_1,W_2\}=\{1,1\}$, corresponding to the aggregations of all the magnon modes at the upper right corner. The situation changes when $\theta = 0$, where the spectra with fixing $\kappa_y$ evolve along an arc, corresponding to the topological winding tuple $\{W_1,W_2\}=\{1,0\}$, \textit{i.e.}, edge skin effect at the right boundary. The spectra with fixing $\kappa_y$ evolve along a closed loop again when $\theta = \pi/4$, but in an anti-clockwise fashion, corresponding to $\{W_1,W_2\}=\{1,-1\}$, \textit{i.e.}, a corner skin effect with aggregations at the lower right corner.

\begin{table}[htp]
\caption{Chiral coupling, two-dimensional non-Hermitian skin effect, and topological characterization with double winding indexes with different configurations $\theta=0,\pm \pi/4$ in the two-dimensional magnetic array that is coupled effectively via the dipolar interaction by the spin waves in the magnetic film. The figures are adapted with permission from Ref.~\cite{cai2023corner}. }\label{fig:2DMSE}
	\centering
		\begin{tabular}{cccc}
		\hline
		\toprule
	\makebox[0.07\textwidth][c]{~~Configuration} & \makebox[0.21\textwidth][c]{~~~~~~~~~Coupling constant} &
	 \makebox[0.21\textwidth][c]{~~~~~~~~~Skin effect}  & \makebox[0.24\textwidth][c]{Spectral winding} \\
		\midrule[0.5pt]
		\begin{minipage}[m]{0.06\textwidth}
      \centering
		$\theta= 0$
		\end{minipage} & \begin{minipage}[m]{0.12\textwidth}			\centering\vspace*{2pt}			\includegraphics[width=1.5\textwidth,trim=1.2cm 0cm 0cm 0cm]{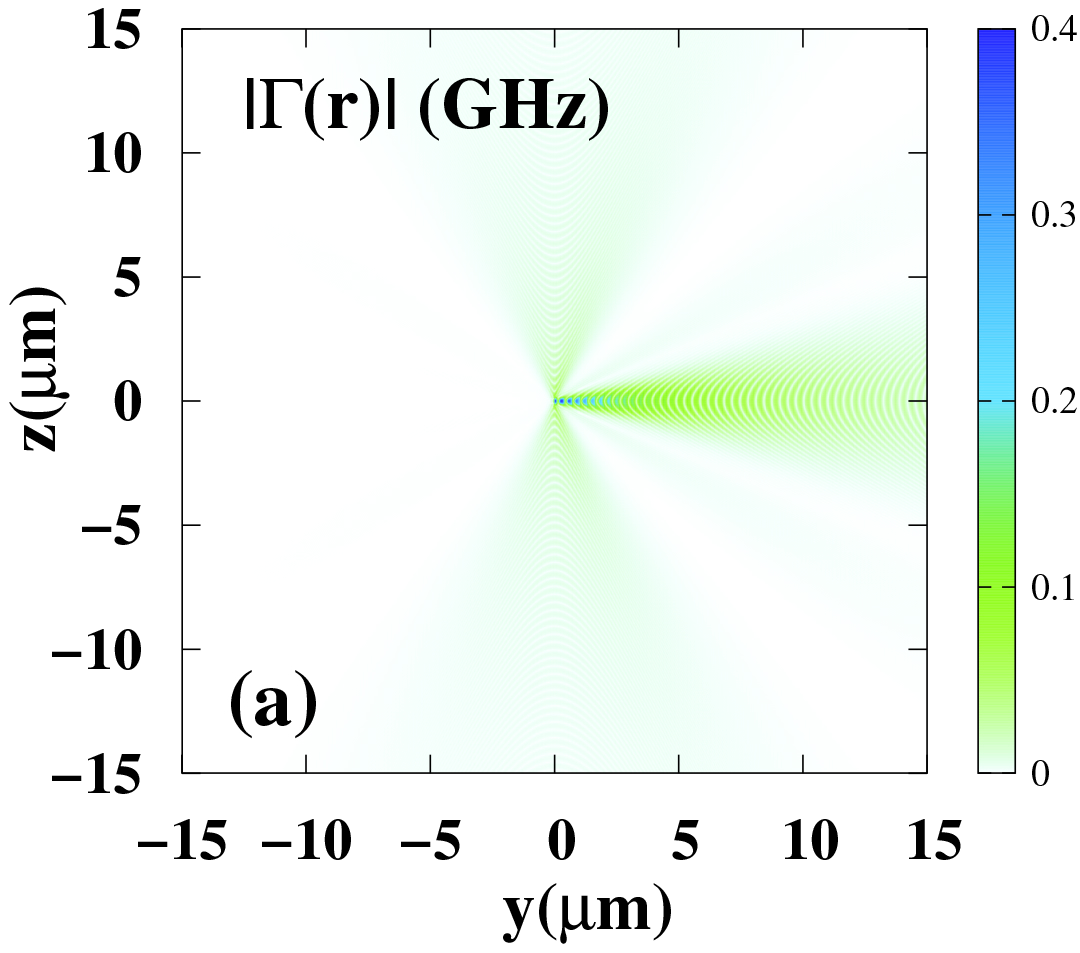}
		\end{minipage} &  \begin{minipage}[m]{0.12\textwidth}
			\centering\vspace*{2pt}
			\includegraphics[width=2.05\textwidth,trim=2cm 0cm 0cm 0.1cm]{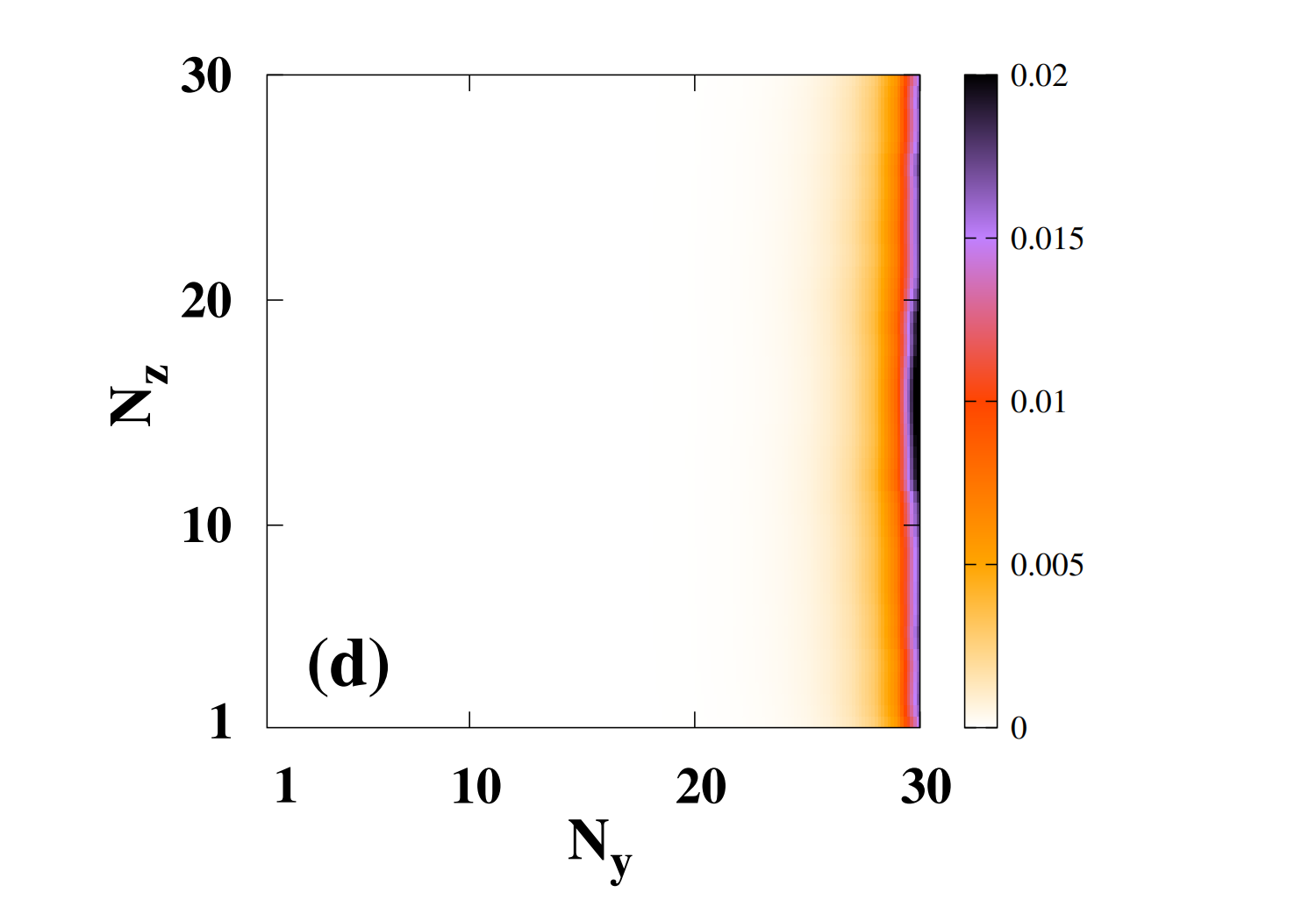}
		\end{minipage} &  \begin{minipage}[m]{0.4\textwidth}
			\centering\vspace*{2pt}
			\hspace*{3pt}
			\includegraphics[width=0.47\textwidth,trim=0cm 0cm 0cm 0.2cm]{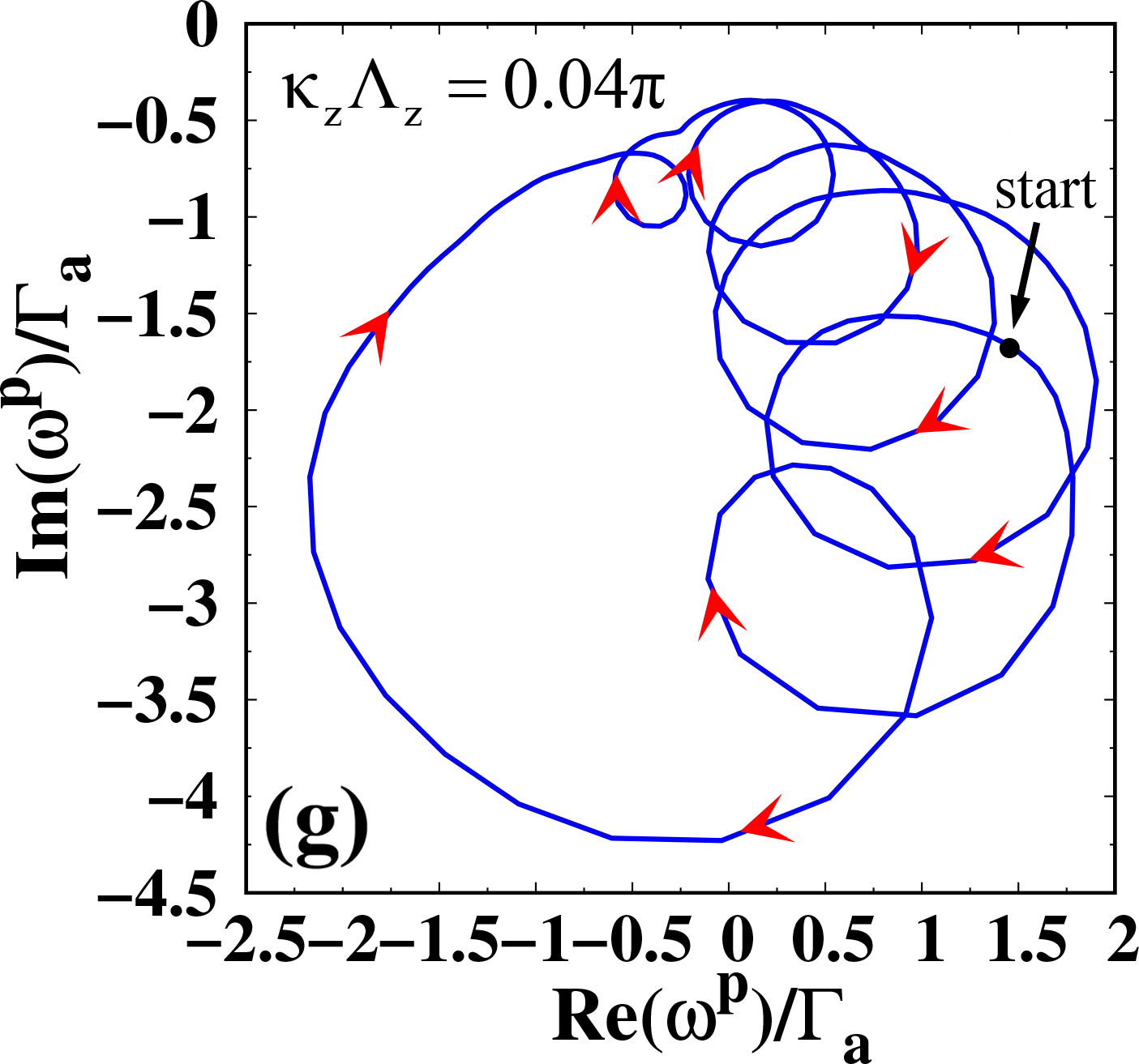}
   \includegraphics[width=0.473\textwidth,trim=0cm 0cm 0cm 0cm]{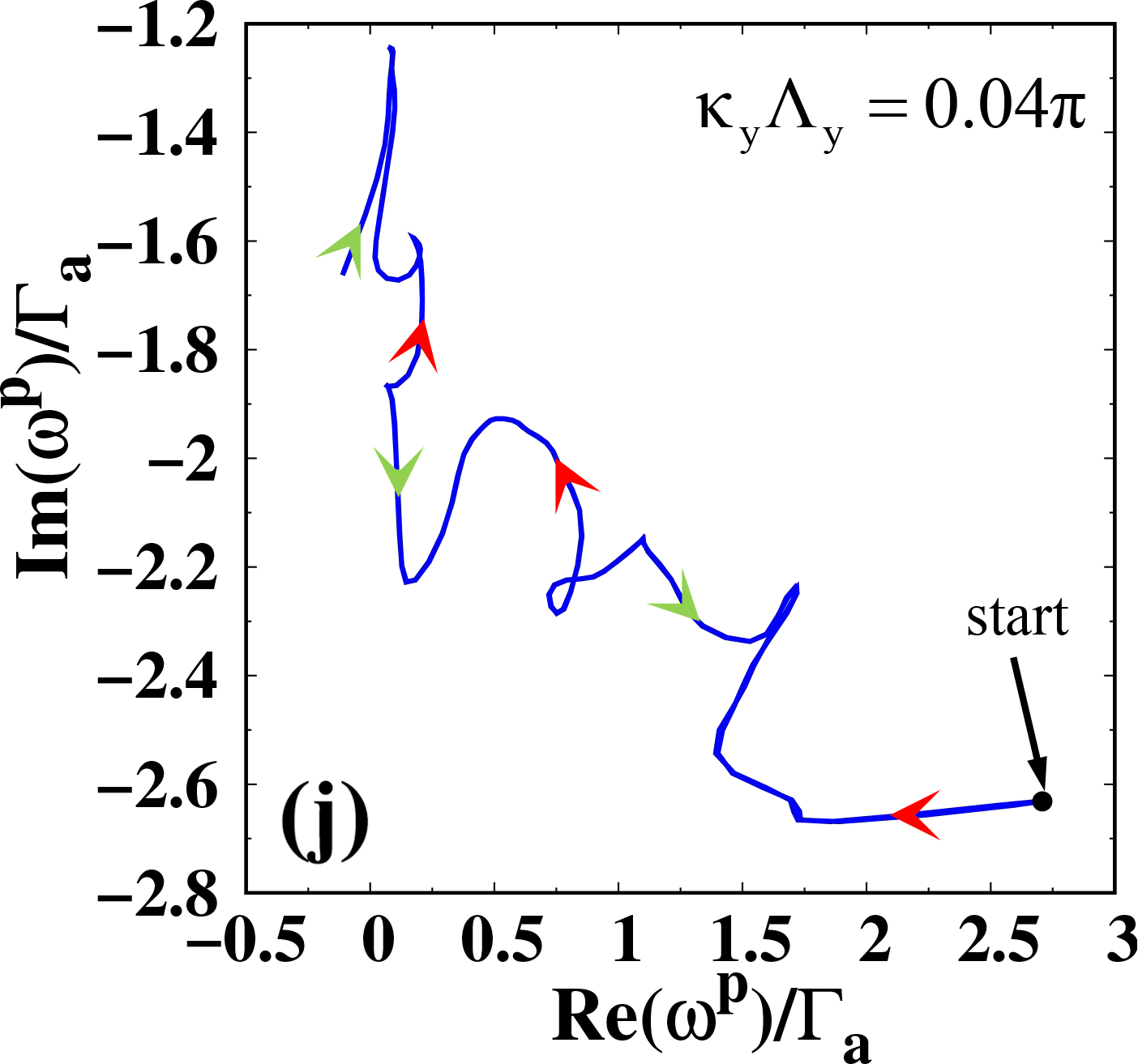}
		\end{minipage} \\
		\midrule[0.5pt]
		\begin{minipage}[m]{0.06\textwidth}
			\centering
		$\theta= \frac{\pi}{4}$
		\end{minipage} & \begin{minipage}[m]{0.12\textwidth}
			\centering\vspace*{2pt}
			\includegraphics[width=1.5\textwidth,trim=1.2cm 0cm 0cm 0cm]{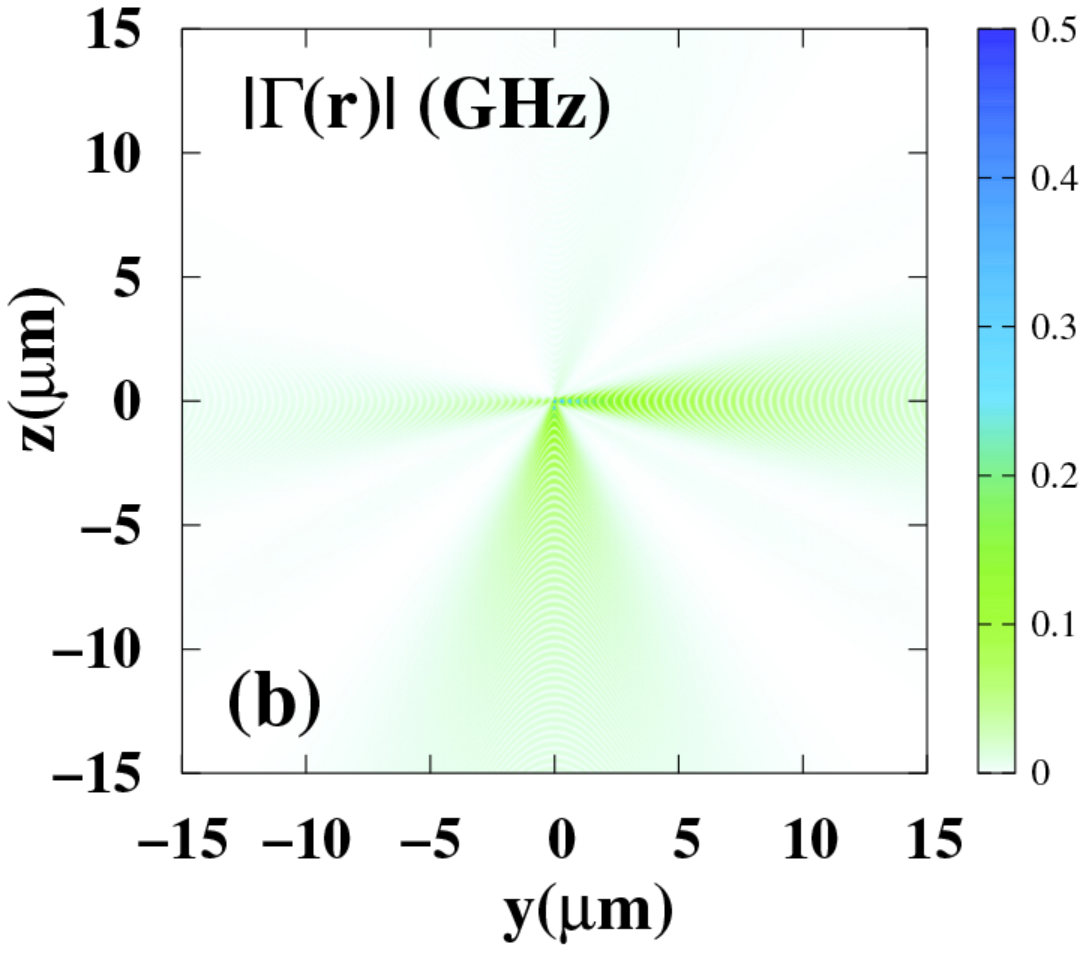}
		\end{minipage} &  \begin{minipage}[m]{0.12\textwidth}
			\centering\vspace*{2pt}
			\includegraphics[width=2.05\textwidth,trim=2cm 0cm 0cm 0.1cm]{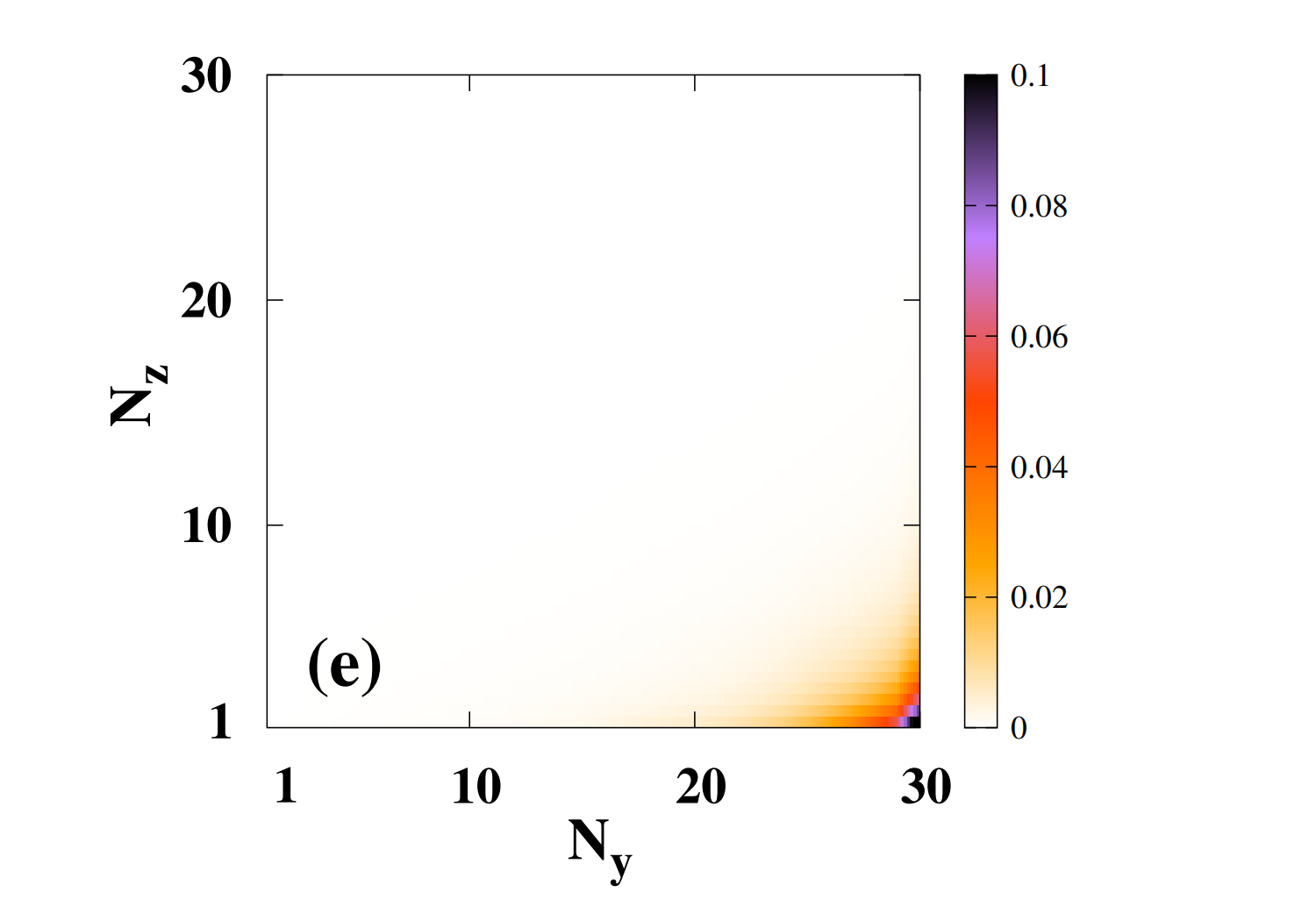}
		\end{minipage} &  \begin{minipage}[m]{0.4\textwidth}
			\centering\vspace*{2pt}
			\hspace*{3pt}
				\includegraphics[width=0.469\textwidth,trim=0cm 0cm 0cm 0.2cm]{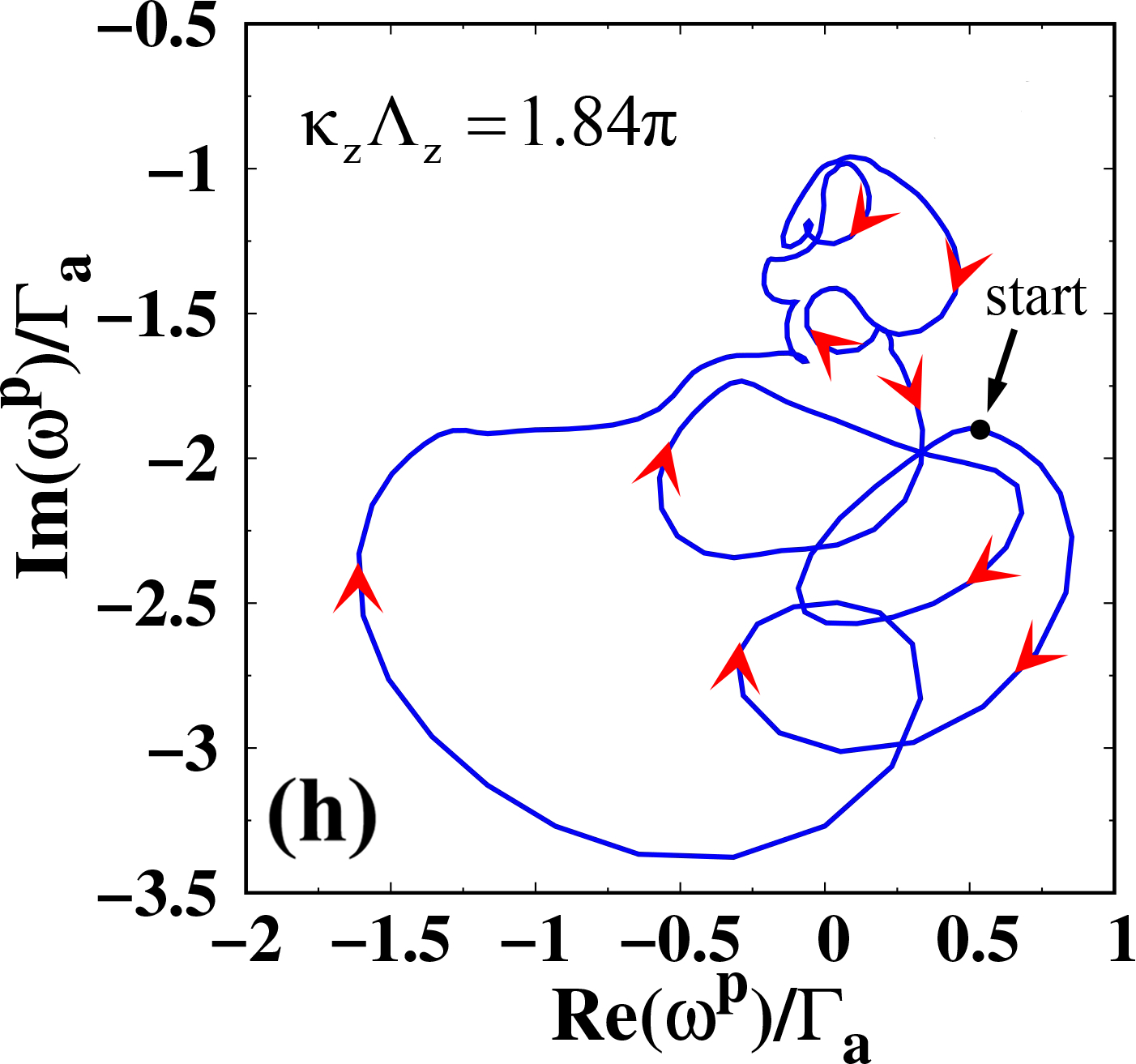}
   \includegraphics[width=0.475\textwidth,trim=0cm 0cm 0cm 0.2cm]{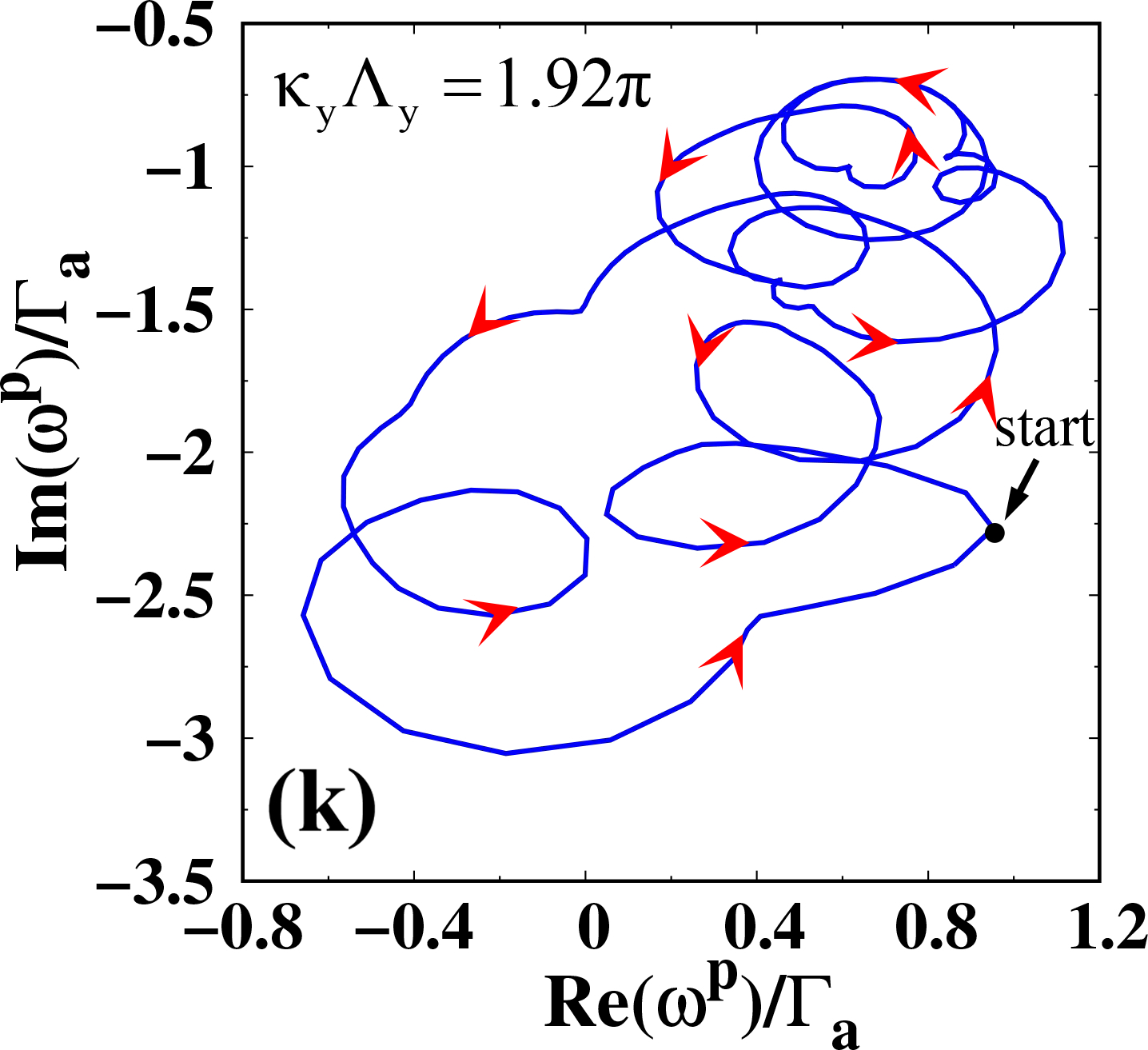}
		\end{minipage} \\
		\midrule[0.5pt]
	\begin{minipage}[m]{0.06\textwidth}
		\centering
		$\theta= -\frac{\pi}{4}$
	\end{minipage} & \begin{minipage}[m]{0.12\textwidth}
		\centering\vspace*{2pt}
		\includegraphics[width=1.5\textwidth,trim=1.2cm 0cm 0cm 0cm]{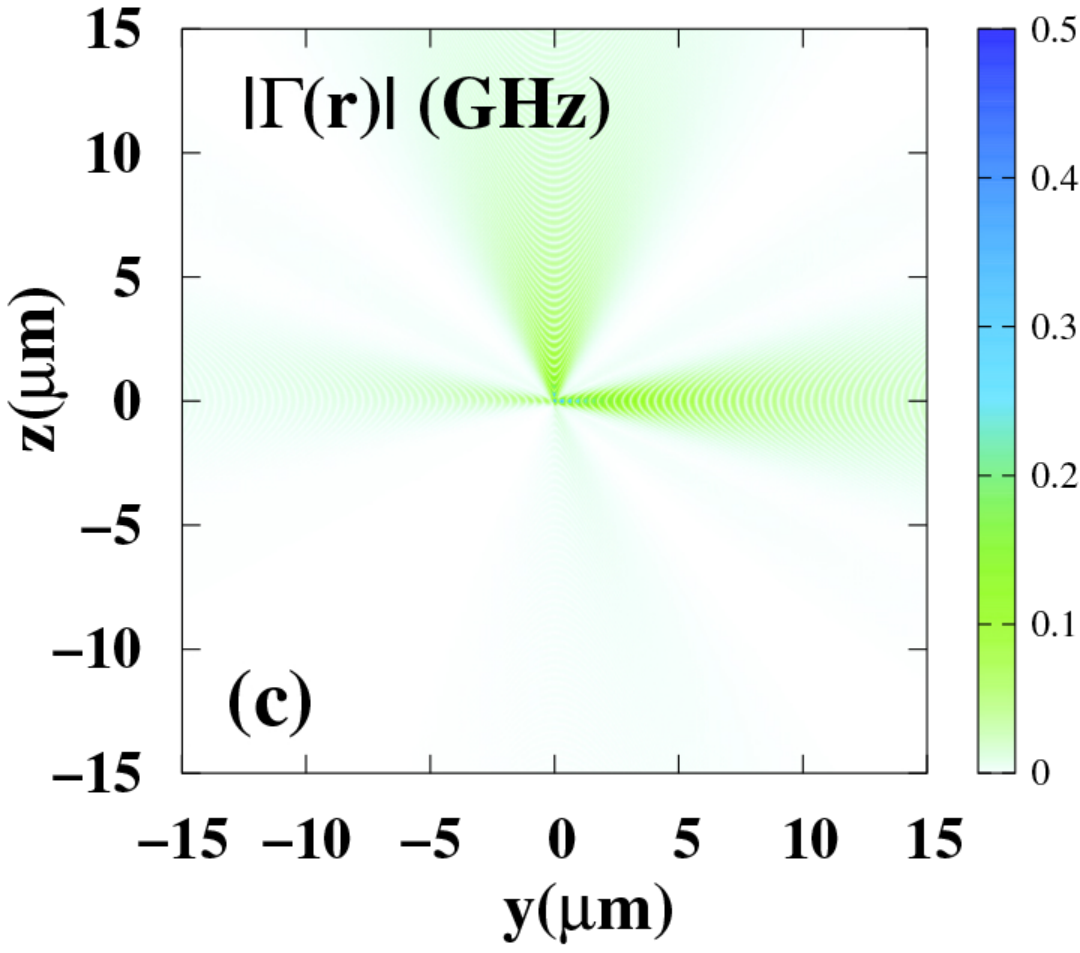}
	\end{minipage} &  \begin{minipage}[m]{0.12\textwidth}
		\centering\vspace*{2pt}
		\includegraphics[width=2.05\textwidth,trim=2cm 0cm 0cm 0.1cm]{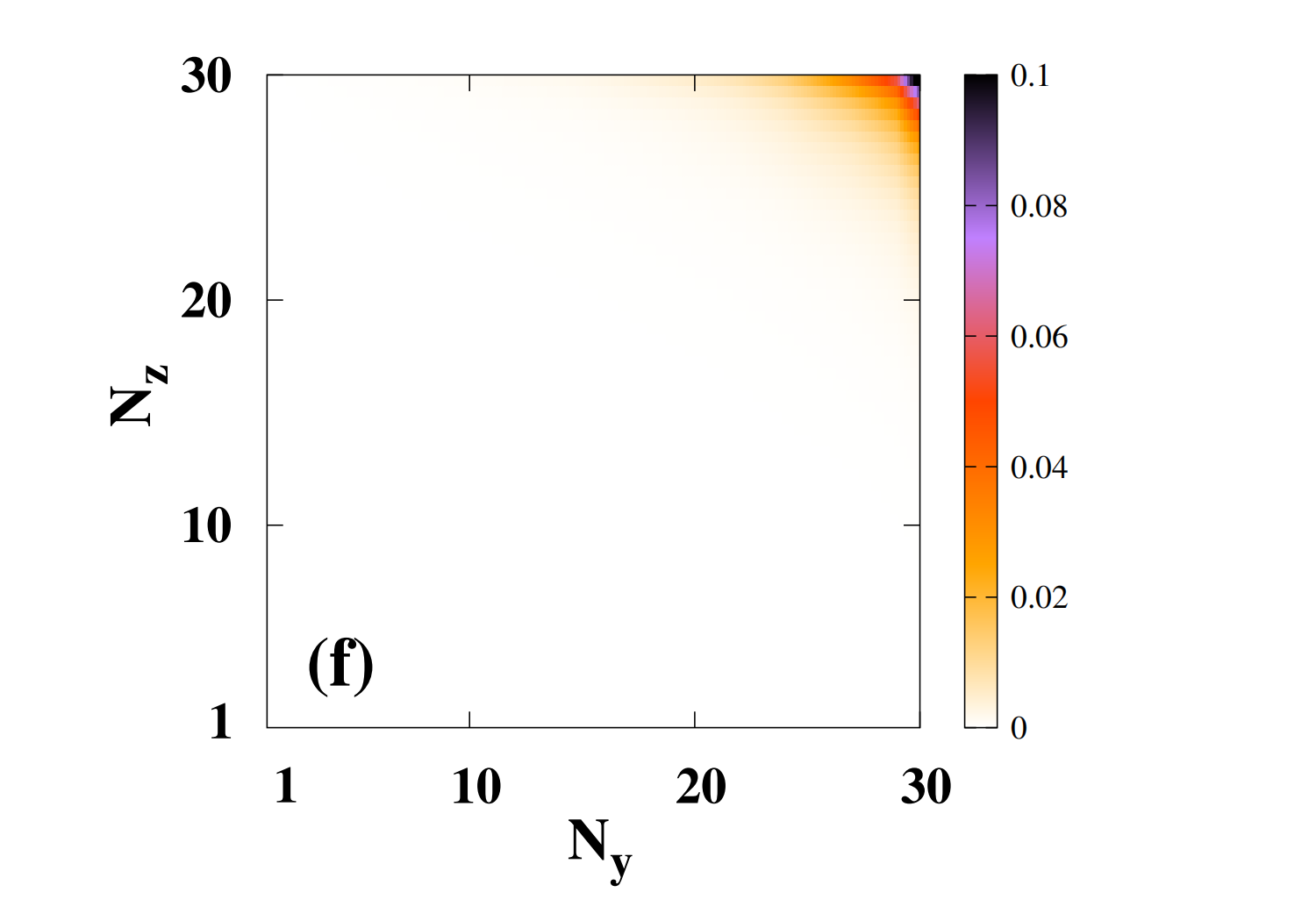}
	\end{minipage} &  \begin{minipage}[m]{0.4\textwidth}
		\centering\vspace*{2pt}
		\hspace*{3pt}
		\includegraphics[width=0.47\textwidth,trim=0cm 0cm 0cm 0.2cm]{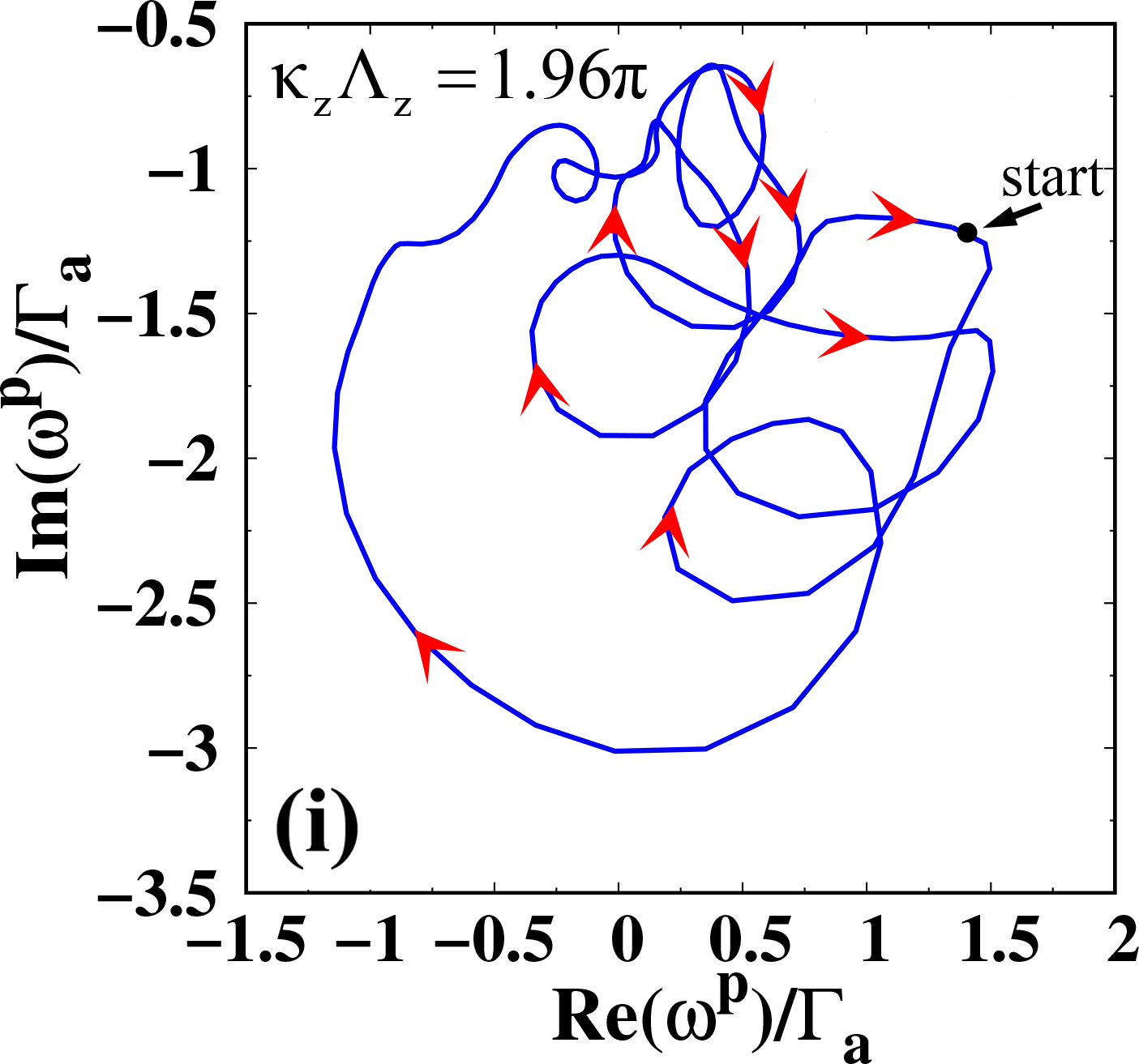}
   \includegraphics[width=0.475\textwidth,trim=0cm 0cm 0cm 0.3cm]{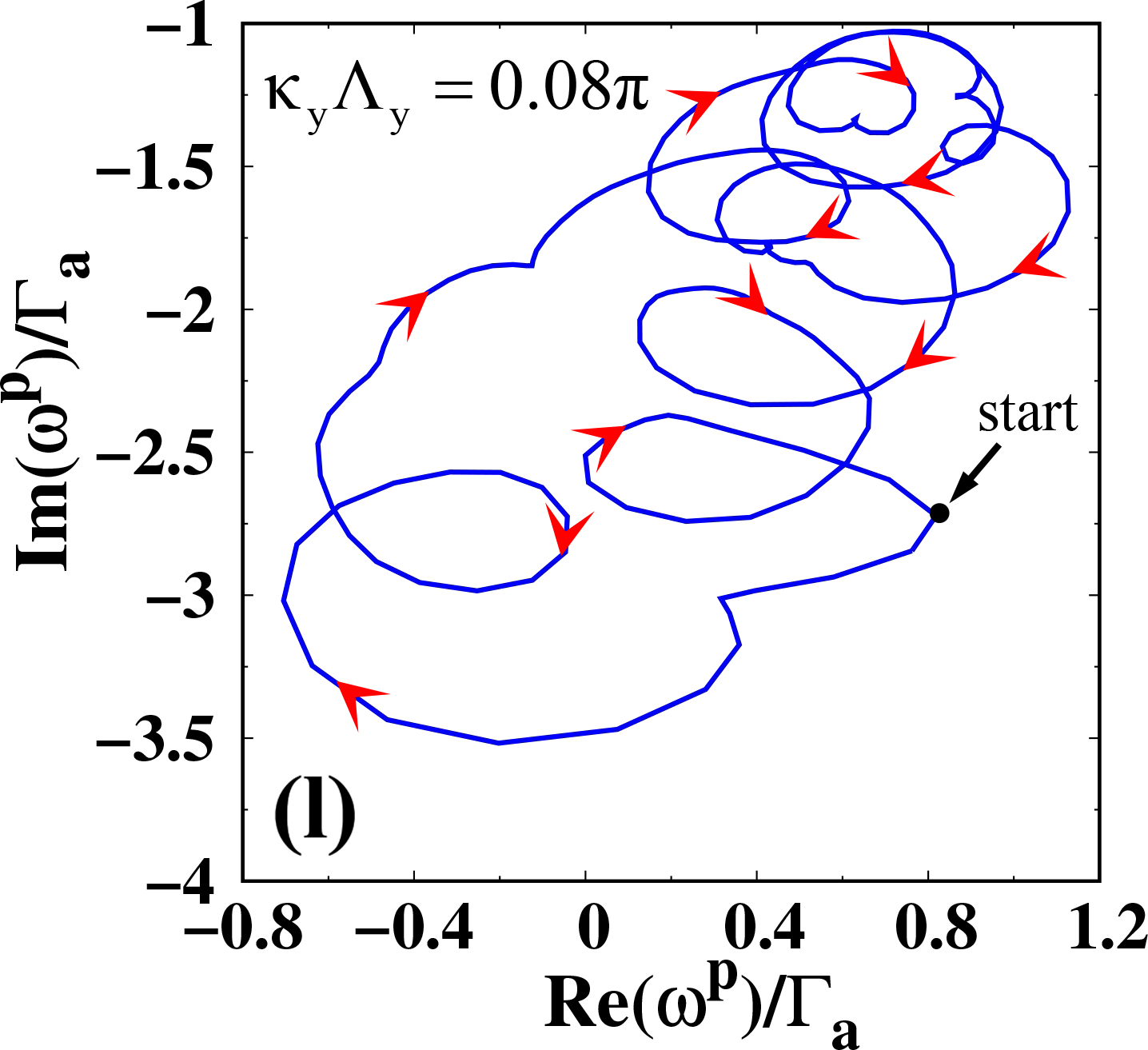}
	\end{minipage} \\
		\toprule
	\end{tabular}
\end{table}

\section{Summary and outlook}

\label{summary_outlook}

Non-Hermitian topological magnonics is an emerging field that seeks to realize functionalities beyond those achievable in the Hermitian scenario. In this sub-field, dissipations, which are usually considered detrimental in the application, can be harnessed as important resources for engineering the non-Hermitian topological states or properties when the magnonic device or system is hybridized with the other systems or driven to the nonlinear regime. 
We review the unified master equation~\cite{nielsen2002quantum,Heinz,Lidar2019,rivas2012open} and Green-function~\cite{bruus2004many,altland2010condensed,yu2019chiral,yu2020chiral,yu2020magnon_accumulation,yu2022giant,zeng2023radiation} approaches for the construction of the effective non-Hermitian Hamiltonian of a subsystem, which is valid in a short-time regime governed by the leakage (or decay) rates of the states when the quantum jump effect is insignificant, introduce the concept of spectral topology that is used in the topological characterization of the non-Hermitian systems \cite{shen2018topological,yang2020non,zhang2020correspondence,okuma2020,ding2022non,okuma2023non}, and review in detail the theoretical proposal and experimental realization of the non-Hermitian topology in magnonic systems, including the exceptional degeneracies~\cite{lee2015macroscopic,galda2016parity,PhysRevB.95.214411,zhang2017observation,galda2018parity,yang2018antiferromagnetism,wang2018magnon,xiao2019enhanced,PhysRevLett.123.237202,cao2019exceptional,huai2019enhanced,Liu2019Observation,yuan2020steady,yang2020Unconventional,zhao2020observation,tserkovnyak2020,yu2020higher,wang2020steering,wang2021coherent,lu2021exceptional,wang2022pt,nair2021enhanced,rao2021interferometric,PT_bilayer,deng2023exceptional,wang2023floquet}, non-Hermitian nodal phases~\cite{bergholtz2019non,mcclarty2019non,yang2021exceptional,li2022multitude}, non-Hermitian SSH model~\cite{flebus2020non,gunnink2022nonlinear}, and non-Hermitian skin effect~\cite{yu2020magnon_accumulation,yu2022giant,deng2022non,zeng2023radiation,cai2023corner,li2023reciprocal}. These bring useful functionalities such as the giant enhancement of magnonic frequency combs~\cite{rao2023unveiling,wang2023MFC},
magnon lasing or amplification~\cite{wang2018magnon,wang2020steering,wang2022pt}, (quantum) sensing of the magnetic field with unprecedented sensitivity~\cite{cao2019exceptional,wang2021enhanced},
edge and magnon accumulation \cite{yu2020magnon_accumulation,yu2022giant,deng2022non,zeng2023radiation}, and perfect absorption of microwaves~\cite{zhang2017observation,rao2021interferometric}. 
While the flourished building and development of the magnonic realization of the non-Hermitian topology in magnonic devices has been achieved in recent years, according to our knowledge, there are a lot of spaces for further theoretical and experimental explorations.

\textbf{Approaches}.---Magnons couple to many degrees of freedom including the other magnons~\cite{zhitomirsky2013colloquium,yang2018antiferromagnetism,mcclarty2019non,Liu2019Observation,yu2020higher,PT_bilayer,shiota2020tunable,yu2022giant}, microwave photons~\cite{kiselev2003microwave,zhang2014strongly,harder2018level,zhang2020broadband}, optical photons, phonons~\cite{liu2017magnon,zhang2020unidirectional,ruckriegel2020long,wang2021magnon,zeng2023radiation}, qubits~\cite{trifunovic2013long,flebus2019entangling,candido2020predicted,psaroudaki2021skyrmion,fukami2021opportunities,zou2022prb}, and electrons~\cite{auerbach2012interacting,bergholtz2019non,bertelli2021imaging,flebus2020non,deng2022non,deng2023exceptional}. These interactions cause dissipation. Non-Hermitian topological magnonics aims to engineer the dissipation, gain, and coupling to realize useful spin functionalities.

These couplings, nevertheless, can be strong or weak. 
In the subsystem of magnons, both the master equation and Green-function approaches integrate the other degrees of freedom. To this end, both approaches may rely on a prerequisite that the system is weakly coupled to other systems or environments, such that the weak couplings can be treated as perturbations. An effective description of the magnon subsystems when strongly coupled to other degrees of freedom or environment is a challenge that needs to be overcome in the future theory.

Nonlinearity brings about additional challenges since the approximated non-Hermitian description is often set up in the linear regime.  For example, the emergence of a strong nonlinear effect that suppresses the lasing mode in the $\mathcal{PT}$-symmetry broken regime \cite{lee2015macroscopic}. Recent experiments report the emergence of the EPs when a magnet is driven by strong microwaves~\cite{rao2023unveiling,wang2023MFC}, which calls for proper theoretical description.

Another opportunity, yet a challenge, is the exploration of the non-Markovian effects originating from the environment coupled with the magnon subsystem. Analogous to the non-local dissipation or dissipative coupling, which emerges from the nontrivial spatial correlations of the environment, non-Markovian effects result from nontrivial temporal correlations. Despite the development of several methodologies, such as the projection operator method (Nakajima-Zwanzig or time-convolutionless master equation)~\cite{Heinz,zou2023spatially} and reaction-coordinate approach~\cite{bulnes2016quantum,landi2022nonequilibrium}, these effects remain less well-understood and continue to be an active area of investigation~\cite{cerrillo2014non,zhang2022predicting,mutter2022fingerprints,groszkowski2023simple}. It would be compelling to harness the temporal correlations of the environment, in addition to the spatial correlations, to discover novel magnetic non-Hermitian phases and potential applications in spintronics.

\textbf{Materials and devices}.---Although the exceptional degeneracies with magnons have been experimentally realized in several systems~\cite{zhang2017observation,Liu2019Observation,PhysRevLett.123.237202,wang2020steering,wang2021coherent,rao2021interferometric,yang2020Unconventional},  the non-Hermitian nodal phase, non-Hermitian SSH model, and non-Hermitian skin effect with magnons still lie at the theoretical proposal stage. 
There are thereby high opportunities in future experiments and device applications after improvements in the material growth and micro-nano fabrication techniques.

To realize and exploit the non-Hermitian topological magnonic phases or properties in the experiments, magnetic materials with high magnetic quality and low dissipation appear to be essential for the construction of the magnonic devices. YIG~\cite{mallmann2013yttrium,sun2013damping,barker2016thermal} acts as the very first choice for current experiments in cavity magnonics and nano magnonics, which have high acoustic quality as well. The primary limitation of YIG lies in its insulating property, which, while preventing dissipation from conducting electrons, also renders it incompatible with direct integration into spintronic or electronic circuits. This constraint significantly narrows the range of application scenarios for YIG. Consequently, the common compromise in device design involves depositing normal metal layers to serve as spin sinks or sources. CoFeB~\cite{ikeda2010perpendicular,burrowes2013low,tacchi2017interfacial} and permalloy (FeNi)~\cite{bozorth1953permalloy,kaya1953uniaxial,shinjo2000magnetic} have the potential as the candidate materials due to their excellent magnetic properties and hybridization with the other materials. Nonetheless, in comparison to YIG, the damping rates of these soft ferromagnetic metals are much higher because of dissipation from conducting electrons. This poses a challenge in the design of magnonic systems, especially when coupling them with microwave photon modes. One reason is that a higher damping rate necessitates a larger material volume to achieve strong photon-magnon coupling. However, the skin depth of these metals imposes constraints on the sample dimensions, limiting the potential for increasing material volume.
The stacked FeNi and Co thin films separated by a normal metal layer have been used to confirm the band anticrossing near the EPs~\cite{Liu2019Observation} and the recent fabrication of ultrathin film and nanowire of YIG with very low damping may further improve the experimental performance~\cite{wang2020chiral,schmidt2020ultra}. While for such multilayer magnetic materials, global or local control of the interaction between layers or the dissipation for each layer remains a challenge. Moreover, the fabrication of the nanoscale magnetic heterostructures raises many challenges and one of which is how to control precisely the interface quality. Stacked van der Waals magnetic monolayer provides alternative choice for the non-Hermitian topological phase~\cite{li2022multitude,deng2023exceptional}. The magnetic nanowires array~\cite{yu2022giant,zeng2023radiation} and the STO array~\cite{flebus2020non,gunnink2022nonlinear} are used to study the non-Hermitian topological phase in the periodic crystal lattice, while the precise control of dissipation of each building block on the same level brings about the additional difficulty.

Magnetic systems have shown promising potential for a variety of applications, including energy storage~\cite{tserkovnyak2018energy,jones2020energy}, as well as efficient transmission, processing, and storage of classical information~\cite{parkin2008magnetic,ryu2013chiral,parkin2015memory,kim2017fast,yoshimura2016soliton,zou2020topological,zou2019topological,tserkovnyak2019quantum,tserkovnyak2020quantum,chumak2015magnon,cornelissen2015long}.
It should be noted that the most commonly addressed magnetic configurations for realizing non-Hermitian topological phases or properties are the ferromagnetic or antiferromagnetic ordered macro spins or magnetic multi-layers or magnetic arrays. In fact, there are a variety of magnetic interactions such as Dzyaloshinskii-Moriya interaction, dipolar interaction, and magnetic anisotropy interaction, which bring various kinds of magnetic configurations~\cite{bernevig2022progress,yu2023chirality}, such as skyrmion~\cite{psaroudaki2021skyrmion,psaroudaki2021skyrmion,xia2023universal,xia2023universal}, domain walls~\cite{yoshimura2016soliton,kim2017fast}, and chiral magnetic texture~\cite{tottori2012magnetic,tserkovnyak2018energy}. The extension of non-Hermitian descriptions and the exploration of non-Hermitian phenomena to these magnetic configurations is rarely mentioned, but it may bring unexpectedly rich physics and applications.
It has recently been proposed theoretically that magnetic systems are also suitable for quantum information processing, where both stationary and flying qubits could be realized based on magnetic textures, such as magnetic domain walls~\cite{zou2022domain,takei2018quantum} and skyrmions~\cite{psaroudaki2021skyrmion,xia2023universal,xia2022qubits}. By exploiting the quantum coherence in the magnetic system, magnetic-texture-based quantum computers can solve certain problems that are intractable for classical ones. However, this very feature also renders them highly sensitive to dissipation, which can lead to the loss of quantum information. Therefore, gaining a deeper understanding of the non-Hermitian aspect of magnetic systems, which is rooted in dissipation, is crucial for the development of magnet-based quantum processors.

\textbf{Urgent issues and opportunities}.---Compared to the photonic and phononic devices~\cite{regensburger2012parity,hodaei2014parity,peng2014parity,ruter2010observation,el2018non,ozdemir2019parity}, it is still a challenge to efficiently introduce the gain for magnons, which acts as an important design parameter for the non-Hermitian topological phases. The spin-polarized current could compensate a bit for the original dissipation for magnon via the spin transfer torque. However, overcoming the damping rate of magnon by the spin current is still difficult to realize in the experiment, i.e., a real ``gain'' remains wanting. This issue is related to many predictions in non-Hermitian topological magnonics~\cite{slonczewski1996current,berger1996emission,tserkovnyak2002enhanced,demokritov2006bose,lee2015macroscopic,zhang2016cavity,bracher2017parallel,yao2017cooperative,yang2020Unconventional,rao2021interferometric,cao2022negative,rameshti2022cavity}. 

The growth of high-quality magnetic materials, micro-nano fabrication techniques and effectively introducing gain for magnon would undoubtedly facilitate experimental performance, but these appear to be long-standing issues that are difficult to be solved in the short term. One feasible way is to develop techniques that can \textit{independently} and \textit{locally} manipulate the properties of ferromagnetic nanostructures, for example, by using bias voltage, field gradient, thermal gradient, laser, or microwave pump. These external drives can excite spin currents or magnon flows, which can transfer energy and torques among different magnetic layers or spatial positions. By controlling the flow of these spin currents and magnons, it may be possible to effectively engineer the dynamics in non-Hermitian magnonic systems.
The opportunity indeed arises in the nonlinear regime. A recent effort achieves a ``nonlinear magnon mode'' in a magnet when loaded in the waveguide and driven by a strong microwave pump. It displays a high level of tunability when driven to a nonlinear regime, which holds the potential to overcome current technological challenges \cite{rao2023unveiling,wang2023MFC}.  The opposite effect is reported in the literature as well, in which the anticrossing closes when a magnetic sphere is driven by the microwave in a cavity \cite{lee2023nonlinear}.

In addition, hybrid systems based on cavity magnonics may offer another feasible solution. Magnon modes in different magnetic materials can indirectly couple with each other in a long-range via the mediation of microwaves. In this scenario, the interfacial quality that is vital for magnetic heterostructures becomes less significant in a cavity magnonic system. Gains that may need delicate design of spintronic devices can easily be realized in a cavity magnonic device by embedding an amplifier or gain material in the microwave cavity~\cite{yao2017cooperative,yang2020Unconventional,rao2021interferometric}. However, an awkward reality in this approach is that nearly all current research on non-Hermitian cavity magnonics is implemented from the cavity side. The magnon mode is merely adopted as a tunable high-Q resonance, because the direct operation techniques on magnon mode, especially precisely controlling its dissipation and realization of its gain, are lacking. Given this, a significant opportunity is to develop local tuning or readout techniques that can access the photon-magnon coupling process from the magnon side.

\addcontentsline{toc}{section}{Declaration of competing interest}
\section*{Declaration of competing interest}
The authors declare no competing financial interests that could have appeared to influence the work reported in this paper.

\addcontentsline{toc}{section}{Acknowledgments}
\section*{Acknowledgments}
TY and collaborators have studied non-Hermitian physics in magnetism for seven years. He gratefully acknowledges many useful discussions with the students in his group Chengyuan Cai, Zubiao Zhang, Xiyin Ye, and Xi-Han Zhou. We thank the colleagues Michael Sentef, Dante Kennes, Bimu Yao, Vahram Grigoryan, Lihui Bai, Yu-Xiang Zhang, Jian-song Pan, Hanchen Wang, Haiming Yu, Qi Wang, Huajun Qin, Kun Yu, Weichao Yu, Di Wu, C.-M. Hu, Shu Zhang, Gerrit Bauer, Yaroslav Blanter, Yaroslav Tserkovnyak, and Daniel Loss. This work has been funded by the National Natural Science Foundation of China (Grants No.~12374109, No.~12204306, and No.~12088101), the startup grant of Huazhong University of Science and Technology (Grants No.~3004012185 and No.~3004012198), and the Georg H. Endress Foundation, STCSM No.~21JC1406200.

\addcontentsline{toc}{section}{References}

\bibliography{refs}
\bibliographystyle{apsrev4-2}
\end{document}